\documentclass[preprint,12pt,sort&compress]{elsarticle}  
\usepackage{amsmath}
\usepackage{amssymb}
\usepackage{comment}
\usepackage{indentfirst}
\usepackage[]{setspace}
\usepackage{enumerate}
\usepackage{chngcntr}
\usepackage{color,soul}
\usepackage{graphicx}
\usepackage{comment,subfigure,color}

\newcommand{\fg}[1]{\mbox{\pmb{$#1$}}}
\newcommand{\bey}{\begin{eqnarray}}
\newcommand{\eey}{\end{eqnarray}}
\newcommand{\vep}{\varepsilon}
\newcommand{\fvep}{\fg \varepsilon}
\newcommand{\sg}{\sigma}
\newcommand{\fsg}{\fg \sigma}

\newcommand{\bec}{\begin{center}}
\newcommand{\eec}{\end{center}}

\def\bc{{\\\bf !!!!![}}
\def\ec{{\bf ]!!!!!\\}}
\newcommand{\ssst}{\scriptscriptstyle}
\begin{document}
\setcounter{tocdepth}{2}
\baselineskip   15pt 
\belowdisplayskip14pt 
\belowdisplayshortskip10pt 
\renewcommand{\thefootnote}{\fnsymbol{footnote}}

\begin{frontmatter}

\title{  Review on phase transformations, fracture, chemical reactions,  and other structural changes in inelastic materials\footnote{Extended version of paper:
		Levitas V.I. Phase transformations, fracture, and other structural changes in inelastic materials. International Journal of Plasticity, 2021, Vol. 140, 102914, 51 pp., invited review.
		DOI 10.1016/j.ijplas.2020.102914}
}

\author{Valery I. Levitas  }

\address{Iowa State University, Departments of Aerospace Engineering and Mechanical Engineering,
 Ames, Iowa 50011, USA, vlevitas@iastate.edu}
\address{Ames Laboratory, Division of Materials Science and Engineering,  Ames, IA, USA}

\begin{abstract}
Review of selected fundamental topics on the interaction between phase transformations, fracture,  and other structural changes in inelastic materials is presented. It mostly focuses on the concepts developed in the author's group over last three decades and numerous papers that affected us. It includes a general thermodynamic and kinetic theories with sharp interfaces and within phase field approach. Numerous analytical (even at large strains) and numerical solutions illustrate the main features of the developed theories and their application to the real phenomena.
Coherent, semicoherent, and noncoherent interfaces, as well as interfaces with decohesion and with intermediate liquid (disordered) phase are discussed. Importance of the surface- and scale-induced phenomena on interaction between
phase transformation with fracture and dislocations as well as inheritance of dislocations and plastic strains is demonstrated. Some nontrivial phenomena, like solid-solid phase transformations via intermediate (virtual) melt,
virtual melting as a new mechanism of plastic deformation and stress relaxation under high strain rate loading, and
 phase transformations and chemical reactions induced by plastic shear under high pressure are discussed and modeled.
\end{abstract}
\vspace{1.5cm}


\end{frontmatter}

\newpage

\noindent
{\bf Notations}

\noindent
SCs  $\qquad$ Structural changes
\\
PTs $\qquad$ phase transitions
\\
TRIP $\qquad$ transformation-induced plasticity
\\
CRs $\qquad$ chemical reactions
\\
RIP $\qquad$ reaction-induced plasticity
\\
SMA $\qquad$ shape memory alloys
\\
PFA $\qquad$ phase field approach
\\
DAC $\qquad$   diamond anvil cell
\\
RDAC $\qquad$ rotational diamond anvil cell
\\
{\sl G} $\qquad$ graphite
\\
{\sl D} $\qquad$ diamond
\\
FEM  $\qquad$  finite element method
\\
IM  $\qquad$  intermediate melt(ing)
\\
VM  $\qquad$  virtual melt(ing)
\\
MD  $\qquad$ molecular dynamics
\\
HPP  $\qquad$ high-pressure phase
\\
$\sf A$, $\sf  M$,  and  ${\sf M}_i$  $\qquad$ austenite, martensite, and martensitic variant
\vspace{1.0cm}
\par
Direct tensor notations are used throughout this paper. Vectors and
tensors are denoted in boldface type;  $\, {\fg A} \, {\fg \cdot} \,
{\fg B} \, = \, \left(A_{ij} \, B_{jk}\right) \,$ and $\, {\fg A} \, {\fg :} \,
{\fg B} \, = \, A_{ij} \, B_{ji}\,$ are the contraction of tensors
over one and two nearest indices. Superscripts $\; - 1 \,$ and $T$ denote
inverse operation and transposition, respectively, $\; := \;$ means equals per definition, subscript
$\, s \,$ designates symmetrization of the tensors, the indices 1 and
2 indicate the values before and after the SC.
Subscripts $e$, $t$, and $p$ designate elastic, transformational, and plastic deformations or deformation gradients.
\\\\

\newpage

\tableofcontents

\begin{abstract}

\end{abstract}

\newpage
\section{Introduction}

Solid-solid phase transformations (PTs) are
broadly studied in physical, material, and mechanical experiments. They are utilized in modern
technologies (e.g., thermal and thermomechanical treatments of metals, shape memory and elastocaloric applications, and high-pressure technologies) and broadly spread in nature, e.g., in geophysical processes. In most cases, PTs are accompanied by plastic deformations; in some cases, they are caused by plastic deformations (plastic strain induced PTs).
Googling "phase transformation and plasticity" returns 13,600,000 results. It is clear that any attempt of review on this topic will miss many important aspects. We will focus on topics and papers, which we worked on for several decades, and on works  which have affected us. There are multiple  reviews related to the interaction between PT and plasticity (e.g., \cite{olson+cohen-1986,haezebrouck-87,lovey99} on material aspects, \cite{Chowdhury-Sehitoglu-17,Sittneretal-18} on shape memory alloys, on the continuum mechanical aspects \cite{levitas-kniga-1992,fischer+berveiller+etal-1994,Levitas-chapter-04}, transformation-induced plasticity (TRIP) \cite{fischer+sun+tanaka-1996,22}, PTs and structural changes (SCs) under high-pressure torsion
\cite{bridgman-1937,bridgman-1947,zharov-1984,zharov-1994,36,38,Blank-Estrin-2014,valiev+islamgaliev+alexandrov-2000,Zhilyaev-2008,valievetal-JOM-2016,valievetal-MRL-2016,Edalati-Horita-16,Mazilkin-etal-MT-19,Straumal-etal-2019,Levitas-MT-19},  and mechanochemistry \cite{koch-1993,Levitas-hidden-mech-chem-chaprter-10}), and we will avoid overlapping.

One of the broad topics of this review is the development of general thermodynamic and kinetic approaches to SCs in {\it inelastic} materials within a sharp interface approach. SCs under consideration include various PTs (martensitic, reconstructive, melting, sublimation, and others), twinning, chemical reactions (CRs), and fracture (crack and void nucleation and growth).
A well-known   formalism was developed
for  the description of the evolution and interaction of various defects or singularities
in {\it elastic} materials \cite{Eshelby-51,Eshelby-56,Eshelby-57,Eshelby-70,Rice-Chapter-68,Rice-JAM-68,Cherepanov-67,Cherepanov-79}. They include phase interfaces, grain and twin boundaries, crack tips and voids, and point and linear defects,
 participating in  such structural changes as PTs, grain evolution, damage, and plastic deformation. For each of these defects,  the rate of energy release or dissipation rate
can be written  as
$\,D \, = \, {\fg X}\, {\fg \cdot} \, \dot{\fg q} \geq 0\,$, where
$\, \dot{\fg q} \,$ is the defect velocity relative to the  material and $\, {\fg X} \,$ is the generalized material/driving
force acting on the defect \cite{Eshelby-51,Eshelby-56,Eshelby-70}. The formula for
$\, {\fg X} \, = \, \int_{\scriptscriptstyle \Sigma} {\fg Q} \,
{\fg \cdot} \, {\fg n} \; d \, \Sigma \,$ is derived, where $\, {\fg Q}  \,$ is  the Eshelby energy-momentum tensor,
$\, \Sigma  \,$ is an arbitrary surface with the unit normal $\, {\fg n}  \,$ surrounding
the defect and separating it from other defects  or  the  actual surface defect
(e.g., phase interface or grain boundary). This concept  was extended for or rediscovered
in various specific  fields. For example, in fracture mechanics, a driving force
$\, {\fg X} \,$
was introduced as a path-independent $\, J \,$-integral \cite{Rice-Chapter-68,Rice-JAM-68}
or $\, \Gamma \,$-integral \cite{Cherepanov-67,Cherepanov-79}; see also  \cite{Grinfeld-DAN,Grinfeld-1991,kaganova+roitburd-1988} for PTs
and \cite{Maugin-93,Maugin-95}
for  Eshelbian  mechanics. Since, for elastic materials, dissipation occurs due to
the motion of defects only, adding to $\, {\fg X} \,$
the integral over the volume without evolving defects does not
change the dissipation rate and $\, {\fg X} \,$--this explains
 the independence of $\, {\fg X}  \,$ of the surface
$\, \Sigma \,$ (or integration path).

\par
For {\it inelastic} materials, the
dissipation due to plastic deformation and change in internal variables contributes to the total dissipation
rate $\, {\fg X}\, {\fg \cdot} \, \dot{\fg x} \,$ in the volume
$\, v \,$ surrounded by surface $\, \Sigma \,$ and $\, {\fg X} \,$
depends on the choice of surface $\, \Sigma \,$. Even
for the volume $\, v \,$ tending to zero, i.e., when  it includes the defect
only, inelastic dissipation in the singular point or surface still takes place (e.g., at the crack tip or moving interface). It is not easy to split the dissipation
due to the defect evolution itself and plastic deformation. That is why it was accepted
in  works on fracture mechanics \cite{Rice-Chapter-68,Kolednik-etal-97,Atluri-97,Kolednik-Fischer-IJF-14} that
the strict thermodynamic criterion for ductile fracture
is not developed and other approaches like energy flow in an infinitesimal or finite-sized
process zone
\cite{Rice-Chapter-68,Cherepanov-79,Hutchinson-83,Atluri-97}, total dissipation rate  \cite{Turner-Kolednik-97,Kolednik-etal-97}, plastic work \cite{Kuang-Chen-97}, critical plastic strain
 \cite{McClintock-71,Rice-Chapter-68}, critical crack tip opening displacements
\cite{Harrison-80}, $\, {J}  \,$-integral  (or $\, \Gamma \,$- integral) without  separation of plastic dissipation \cite{Cherepanov-67,Cherepanov-79},
and others are used.
A similar situation took place in PT theory, see Section \ref{general-theory}.
\par
For {\it elastic}  materials,  the conditions for both the appearance of a defect ("nucleation") and its equilibrium
are described by  the principle of the
 minimum of Gibbs free
energy. That is why they coincide. Thus, from a thermodynamic (but not a
kinetic) viewpoint,  it is not necessary to treat a nucleation process separately;
it is sufficient to insert a nucleus and study its
equilibrium using the local condition, e.g., for an  interface or crack tip.
However, for {\it inelastic} materials,
defect nucleation and equilibrium  conditions are different. Thus, { a defect nucleation problem } within  inelastic materials has to be formulated and solved.
Defect nucleation and evolution in inelastic materials cannot be described  by the
principle of the minimum of Gibbs energy or with the help of an energy-momentum
tensor only. Thus, the new driving force and extremum principle for the determination of all unknown
parameters (e.g., position, shape, orientation, and internal structure of the nucleus of a product phase)
are required.
In contrast to elastic materials, since the constitutive behavior of inelastic materials is history dependent,
analysis  of the {\it entire transformation-deformation process in the transforming region}
is required.

Another broad topic of this review is the development of phase field approaches (PFAs) to different SCs in inelastic materials, see e.g.,  \cite{Khachaturyan-actamat-01,jin01let,Steinbach-09,wang01acta,Khachaturyan,wang10acta,Hakim-Karma-05,Karmaetal-01,Ruffini-Finel-15,Levitas-IJP-13,Levitas-Javanbakht-JMPS-14,Levitas-Javanbakht-15-1,Javanbakht-Levitas-JMPS-15-2,Levitasetal-fracture-IJP-18,Jafarzadeh-Levitas-etal-Nanosclae-19}.
PFA    is based on the concept of the order parameters that describe instabilities of the crystal lattice  during PT, twinning,  dislocation nucleation, and fracture, as well the evolution of phase and dislocational structures  in a continuous way by solving Ginzburg-Landau evolution equations for the order parameters. Typical solutions for these equations are propagating finite-width phase and twin interfaces, crack surfaces,  and dislocation core regions that describe the evolution of complex microstructures.
 Thermodynamic potential has as many minima in the space of the order parameters   as many phases and structural states system possesses. These minima are separated by energy barriers. Besides, the thermodynamic potential depends on the gradients of the order parameters, which are concentrated at the finite-width phase and twin interfaces, crack surfaces, and dislocation cores; this reproduces the interface, surface, and dislocation core energies.

 The sharp-interface approach gives specific expressions for the thermodynamic driving forces for the nucleation and evolution of defects. It is convenient for solving problems with relatively simple geometries of interfaces and defects, and allows analytical solutions for some problems. The PFA includes additional information about the stability and instability of phases and different states. It allows for studying the evolution of arbitrary complex geometries of interfaces and defects, without any computational cost on tracking interfaces.   Thus, both approaches have their advantages and disadvantages, and they supplement each other. Both will be reviewed in the current paper.

\par
 This review is organized as follows.
In Section \ref{background}, short background information on martensitic PTs in inelastic materials is presented, including some examples of interactions between PTs and plasticity.
 In Section \ref{universal-interface},  some universal kinematic and balance relationships for a coherent interface in an arbitrary medium  are presented.  Thermodynamic and kinetic descriptions of PTs in elastic materials are discussed in Section \ref{PT-elastic}.  Athermal resistance to interface propagation and its relationship to the yield strength are introduced in Section  \ref{athermal-resist}.
 General sharp-interface thermodynamic and kinetic theories
for SCs in inelastic materials are presented in Section  \ref{general-theory}.
They include the driving forces and thermodynamic criteria for nucleation and interface propagation, and the extremum
principle for determination of all unknown parameters that substitutes the minimum Gibbs energy principle for elastic materials. Three types of kinetics are described: athermal kinetics, thermally-activated kinetics, and "macroscale" thermally-activated kinetics. In addition, the global SC criterion based on stability analyses is introduced.
Peculiarities of the finite strain approaches are summarized in Section \ref{Finite strain}. All equations from last two Sections are summarized in ten boxes.
Sections \ref{sphere}, \ref{interface-analytic}, and \ref{ellips-incl} present
some analytical solutions based on the developed  theory, which illustrate the main points of the theory as well as
their application to
some real problems. They include SC in a sphere (with application to PT from graphite to diamond),  PTs and CRs in a shear band (with application to transformation- and reaction-induced plasticity (TRIP and RIP)),  consideration of a  propagating interphase, as well as PT in an ellipsoidal inclusion.
Various aspects of the nucleation and growth of a martensitic region  with coherent, semicoherent, and incoherent interfaces, as well as  interfaces with decohesion are analyzed with the finite element method (FEM) in Section \ref{interface-numerics}. Both the sharp-interface theory and PFA are utilized.
Sections \ref{virtual-melt} and \ref{VM-shock} are devoted to recently revealed nontraditional mechanisms of
PTs in solids via intermediate (virtual) melt and mechanism of plastic deformation and stress relaxation under high strain rate loading via virtual melting, both much below the melting temperature, when melting is not expected.
Strain-induced martensite nucleation at the shear-band intersection was studied with FEM in Section \ref{Shear-band-int}.  The importance of the application of the global  criterion for SCs based on stability analysis was demonstrated.
The appearance and growth of a martensitic plate in an elastoplastic material for temperature-induced PT was analyzed with FEM in Section \ref{sec-lath}. The effect of the inheritance of plastic strain on a large change in the plate's shape and  on the arrest of plate growth  (i.e., morphological transition from the plate to lath martensite) is demonstrated.
Nanoscale and microscale PFAs to the interaction between dislocation plasticity and PTs are described in Section \ref{Sec-PT-plast}.
 In Section \ref{RDAC},
a  multiscale theory of strain-induced SCs under high pressure is
presented for the interpretation of various phenomena
during compression and shear of materials in a rotational diamond anvil cell.
Scale transitions and phenomenological theories for the interaction between PT and plasticity are analyzed in Section \ref{phenomenology}.
Fracture and the interaction between fracture and PT in inelastic materials are described within both the sharp-interface theory and PFA in Section \ref{Sec-fracture}.
Concluding remarks are presented in Section    \ref{Concl-rem}.

\section{Martensitic transformations in inelastic materials: some background information}\label{background}

{ We will focus on
displacive PTs, which are dominated by the
deformation of a unit crystal cell of the parent phase into a unit cell  of the product phase that is described by
transformation deformation gradient $\fg F_t= \fg R_t  \cdot  \fg U_t  $, where $ \fg R_t$ is the orthogonal rotation tensor,  $ \fg U_t  =\fg I + \fvep_t $ is the transformational right stretch (Bain) tensor, and $\, \fvep_t \,$ is the
transformation strain tensor.  Note that  $ \fg U_t$ produces mapping of the stress-free crystal lattice of the parent phase into that for the product phase at a fixed temperature. Usually, the higher symmetry and higher temperature phase is called the austenite, and the lower symmetry and lower temperature phase is called the martensite.

Displacive PTs include
martensitic PTs during which atoms do not change
their neighbors and reconstructive PTs in the
opposite case.
In addition to $ \fg U_t$, displacive PT involves intra-cell
displacements or shuffles. The diffusion of species does not occur during martensitic PTs.

\begin{figure}[htp]
\centering
\includegraphics[width=0.5\textwidth]{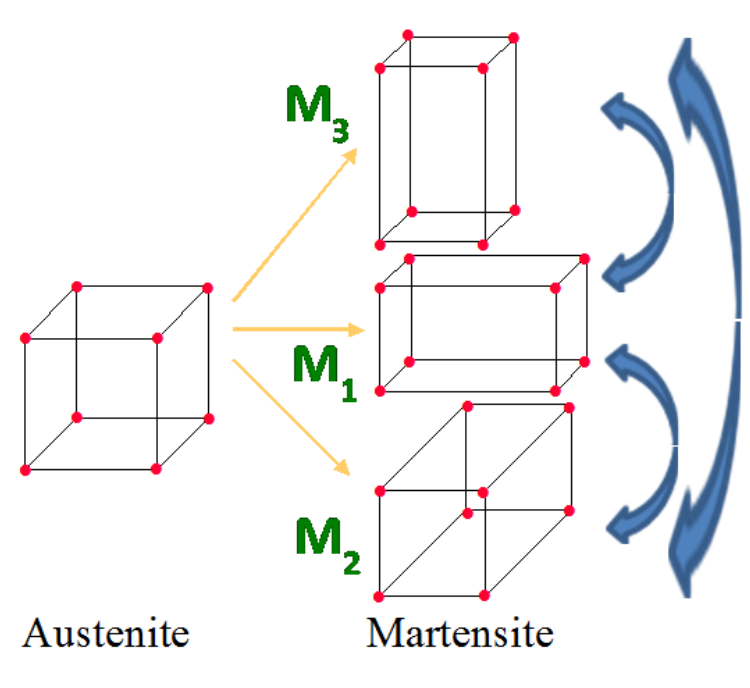}
\caption{ Schematics of the transformation between austenite and martensitic variants as well as between martensitic variants. \label{Fig-5_8}}
\end{figure}
Due to the symmetry of crystal lattice, there is a finite number (e.g., 3 for the cubic to tetragonal PT and
12 for the cubic to monoclinic PT) of crystallographically equivalent martensitic variants
$M_i$ (Fig. \ref{Fig-5_8}). Lists of the transformational right stretch tensors for various PTs and for all martensitic variants are presented, e.g., in
\cite{Bhattacharya-book,pitteri+zanzotto-03}.
The transformation strain can be quite large. For example, for cubic to tetragonal PT from Si I to Si II \cite{Zarkevichetal-18} and cubic to monoclinic PT from rhombohedral graphite to hexagonal diamond \cite{britun+kurdyumov-2002} tensors $ \fvep_t $ are
\bey
{\fvep}^{Si\, I-II}_{tj}= \left( \begin{array}{ccc}
0.243 & 0 & 0 \\
0 & 0.243& 0 \\
0      & 0 & -0.514
       \end{array} \right) ;
       \qquad
  {\fvep}^{G-D}_{tj}     \; = \;
     \left( \begin{array}{ccc}
 0.024 &  0& 0.105 \\
0 & -0.034& 0 \\
0.105       & 0 &- 0.35
       \end{array} \right).
\label{IJP-1}
\eey
The  components of $ \fvep_t $,  $\{a,a,c\}$,  for cubic to tetragonal PT from phase I to II in  Ge and GaSb, as well as  for $ \alpha \rightarrow \beta$ PT in Sn, are in the range $a= 0.254-1.281$ and $-c=0.502- 0.510$
\cite{malushitskaya}.
For the layer-puckering mechanism \cite{britun+kurdyumov-2000} for PT
from cubic diamond and boron nitride to rhombohedral graphite and BN,
$a= 0.98$ and $c=1.596 $.
Transformation strain  is generally much larger than the maximum elastic strain, which  varies from $10^{-3}$ for
steels to $10^{-2}$  for NiTi and CuZnAl shape memory alloys.

Most of the martensitic variants are in twin relationship to each other.
Twinning is a simple shear of one part of the crystal lattice with respect to another
up to a position in which it represents a mirror reflection of the initial lattice (Fig. \ref{Fig-5_10}). Thus, twinning is described by the transformation deformation gradient
${\fg F}_{t}={\fg I} + \gamma_t\,{\fg m}{\fg n}$, where $\gamma_t$ is the twinning shear strain that occurs in the direction ${\fg m}$ in the plane with the
normal ${\fg n}$, which is called the twinning plane.
Generally, twinning is a mechanism of plastic deformation in crystalline materials
whereby jump-like shear deformation of the crystal lattice occurs.
It both competes with and supplements dislocation plasticity.
 For the body-centered cubic (bcc) lattices (for example, in Mo, Na, and Cr) and the face-centered cubic (fcc)
metals (for example, in Al, Cu,  and Co), the magnitude of the transformation shear $\,\gamma\,=\,0.707\,$
is very large. For the hexagonal close-packed (hcp) lattice in Mg $\,\gamma_t\,=\,0.137\,$ and in Zr $\,\gamma_t\,=\,0.225\,$.
For the body-centered tetragonal lattice in NiAl $\,\gamma_t\,=\,0.150\,$. Various aspects of twinning can be found in the books
\cite{wayman-64,Bhattacharya-book,Klassen-Neklyudova,pitteri+zanzotto-03}. An example of the constitutive equations for twinning can be found in \cite{meyers+vohringer+lubarda-2001} and references herein.
As a mechanism of plastic deformation, in which twinning competes with dislocations, twinning usually occurs at a lower temperature, higher strain rates, and smaller grain size than the dislocation plasticity. Such   competition also takes place during PTs. Since a single martensitic variant is generally not compatible with austenite and generates large internal stresses, these stresses relax
either by twinning (for example, in shape memory alloys and some steels) or by plastic slip (for example, in steels), or both, producing an invariant plane strain variant with averaged transformation deformation gradient ${\fg F}_{t}={\fg I} + \gamma_{in}\,{\fg m}{\fg n}+ \vep_{in}\,{\fg n}{\fg n}$, where $\gamma_{in}$ is the invariant plane shear strain along the direction ${\fg m}$ in the invariant (or habit) plane with the
normal ${\fg n}$, and $\vep_{in}$ is the normal to habit plane strain.  Typical values of  $\vep_{in}$ are around zero for NiTi and CuZnAl  shape memory alloys, 0.02 to 0.05 for steels, and  0.2 for  $\;\;\delta\;\rightarrow\;\alpha $
PT in plutonium; typical values of $\gamma_{in}$ are 0.1-0.2 for shape memory alloys, 0.2 for steels, and 0.27 for
$\;\;\delta\;\rightarrow\;\alpha $
PT in plutonium.

\begin{figure}[htp]
\centering
\includegraphics[width=0.7\textwidth]{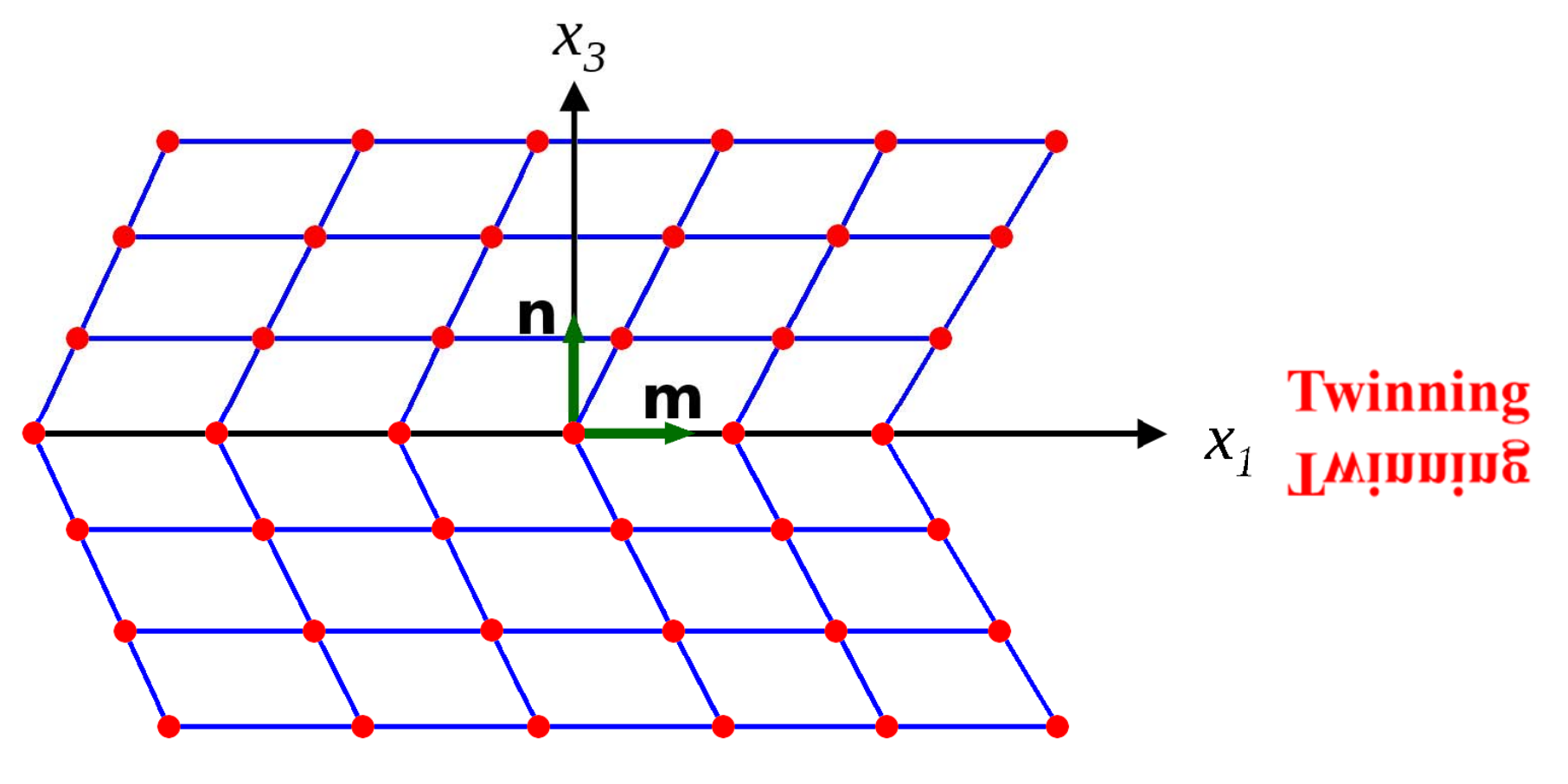}
\caption{ Schematics of the twinning. \label{Fig-5_10}}
\end{figure}


\par
\par
Usually,  three types of PTs are distinguished: temperature-induced, stress-induced,
and strain-induced.
Temperature-induced PTs occur without external stresses, and nucleation starts  at  pre-existing
defects (e.g., dislocations, point defects, grain and subgrain boundaries, stacking faults, and twins).
Stress-induced (or assisted) 
PTs take place under external stresses below the macroscopic yield strength $\, \sg_y \,$ by nucleation at the same
pre-existing defects
as temperature-induced PTs. 
Strain-induced PTs  occur during plastic deformation by nucleation at new defects generated by
plastic flow \cite{16,olson+cohen-72,olson+cohen-75,olson+cohen-1986}.
}

Formally, we define any type of the SCs enumerated above (PT, CRs, twinning, and fracture) as a thermomechanical deformation process of the
growth of transformation strain $\, \fvep_t \,$ from its initial value $\, \fvep_{t1} \,$ in
the parent phase to the final value $\, \fvep_{t2} \,$ in the product
phase, which is accompanied by a change in all the thermomechanical properties (elastic moduli of an arbitrary rank, thermal expansion tensor, specific heat, etc.).
\par
  Below we outline  some examples of interaction between SCs and inelasticity, which are interrelated.
\par {
1. PTs and other SCs, possessing a transformation strain, are processes of an inelastic deformation in materials. For some materials and applications, for which traditional (dislocation or twinning) plasticity is not desirable, like shape memory alloys (SMA), strong ceramics, and semiconductors (like Si and Ge),   PT is the main mechanism of inelasticity. For all other cases,
 transformation strain supplements
dislocation plasticity and twinning.
\par
2. Microplastic straining, which occurs in SMA during cyclic loading, accumulates and leads to defect generation,  damage, and  degradation of transformational properties (e.g., reduces recoverable strain and increases stress hysteresis and energy losses).
\par
3. Slip and twinning in martensite (a combination of two martensitic variants) are mechanisms for lattice-invariant shear.
Along with the transformation (Bain) strain and  crystal lattice rotation, they  produce
an invariant plane strain variant \cite{wayman-64} between   austenite and martensite. This process is driven by the reduction of the
energy of internal stresses; it promotes
nucleation and growth.
\par
4. Transformation strain in the transforming regions generates internal stresses, which in most cases exceed the yield strength and produce accommodational plastic
strains   within and outside of the transforming regions. Reduction of the internal stresses
increases the thermodynamic driving force for  nucleation. However, stress redistribution caused by plastic deformation
near the growing product phase   reduces the thermodynamic driving force for interface propagation.
In elastic materials, the martensitic region  is arrested by a strong obstacle
(e.g.,  grain, subgrain, or twin boundaries, or another martensitic region), producing plate martensite morphology.
Plasticity  stops the growth of martensitic units inside the grain before reaching an obstacle, leading to a morphological transition to the lath martensite. Numerous analytical and numerical results illustrating the above
statements are presented in the following Sections.
\par
5. Plastic deformation generates a group of defects (e.g., dislocation walls or pileups), and corresponding stress concentrators produce   nucleation sites for PT.
At the same time, chaotic defect structures like dislocation forests resist the interface propagation.
Preliminary plastic deformation generally suppresses martensitic PTs.
\par
6.  Plastic flow that occurs during PT   causes strain-induced
PT, which proceeds by nucleation at defects generated during plastic flow,
e.g., at a dislocation pileup or slip-band intersections. While for stress-induced PTs, the  PT stress
grows  linearly with the temperature increase, 
for strain-induced PTs, the PT stress reduces with the temperature
rise  due to the reducing  yield strength but  grows with increasing plastic strain due to strain hardening.
\par
7. Internal stresses caused by transformation strain  in superposition
with external stresses,  which may be well below  the yield strength,
cause plastic flow. This phenomenon is called TRIP for PTs \cite{fischer+sun+tanaka-1996,22,16,23,zackay+parker+fahr+busch-1967}
or RIP for CRs \cite{levitas+nesterenko+meyers-1998-1,levitas+nesterenko+meyers-1998-2}.
They serve as a relaxation mechanism for internal stresses and as  an additional mechanism of inelastic deformation. For cyclic direct-reverse PT under the
external stress, below the yield strength, TRIP  is accumulated in each cycle and may exceed hundreds of percent.
\par
8. The phase interface, similar to a twin interface, can be presented as an array of  partial dislocations \cite{olson+cohen-1986}. In such a representation, PT consists of the nucleation and motion of these dislocations, like dislocation plasticity.
\par
9. Some PTs, in addition to transformation strain, involve shuffles (intracell atomic motion) produced by the motion of partial dislocations, e.g., for bcc-fcc and fcc-hcp PTs \cite{Porter-1992}.
\par
10. The strong promoting effect of the plastic shear under high pressure on PTs and CRs will be discussed  in Section \ref{RDAC}.

Knowledge of the influence of plastic strain and applied and local
stress fields on SCs is very important for the understanding, simulation,
and improvement of the technical processes, as well as for the development
of new technologies and materials.
Examples include heat and thermomechanical treatment of materials; increasing toughness utilizing TRIP; severe plastic deformation technologies, including high pressure torsion and ball milling, friction, wear, surface treatment (polishing and cutting); as well as the interpretation of earthquakes.

}
\section{Some universal relationships for a coherent interface}\label{universal-interface}

Let the motion of the homogeneously deformed small vicinity
of a material point 
be described by the function $\, {\fg r} \, = \, {\fg r } \left(
{\fg r}_0 , \; t \right) $, where $\, {\fg r} $ and
$\; {\fg r}_0 \,$ are the positions vectors  in the actual
$\,  \Omega \,$  and reference $\, \Omega_0 \,$ configurations  and $t$ is time.
The deformation gradient is
$\, {\fg F} \, = \, \frac{\displaystyle \partial \,
{\fg r}}{\displaystyle \partial \, {\fg r}_0} \;$.
We will not focus on the multivariant structure of martensite here, and will instead consider a two-phase material.

 For a {\it coherent interface}, when a jump in displacement across an interface is absent, but the particle velocity vector ${\fg v}$ and  the deformation gradient ${\fg F}$ have a jump, the Hadamard compatibility
condition is valid in the reference configuration  $\Omega_0$ \cite{Truesdell-Toupin}
\bey
\left[{\fg F}\right] \, = \, - \left[{\fg v} \right] \,
{\fg n}_0 / v_n \, ,
\qquad {\rm whence} \qquad
\left[{\fg v} \right] \, = \, - \left[{\fg F} \right]
\, {\fg \cdot} \, {\fg n}_0 \,v_n
\qquad {\rm and} \qquad
\left[{\fg F} \right] \, = \left[{\fg F} \right]
{\fg \cdot} \, {\fg n}_0 \, {\fg n}_0 \; ,
\label{IJP-5}
\eey
where  $\, {\fg n}_0$ is the unit normal to the interface in  $\Omega_0$, $v_n$ is the interface velocity, and $[\fg A ]= \fg A_2-\fg A_2$ is the jump of parameters across the interface. The conservation of mass at the interface is expressed as
\bey
\left[\rho_0 c\right] \, = 0,
\label{IJP-6}
\eey
where $\rho_0$ is the mass density. Neglecting inertia, the traction continuity condition at the interface is
\bey
\left[\fg P \right]\cdot {\fg n}_0  \, = 0,
\label{IJP-7}
\eey
where $\fg P$ is the non-symmetric first Piola-Kirchhoff stress tensor.

A list of all universal equations for the moving interface, including energy balance, entropy inequality, and inertia, can be found in \cite{kondaurov+nikitin-1986,Abeyaratne-JMPS-90,raniecki-tanaka-94}.
Surface stresses
are included in \cite{Gurtin-Struthers-ARMA-97,Gurtin-Murdoch-75,Gurtin-2000,fischer-08}.

\section{Phase transformations in elastic materials}\label{PT-elastic}

As an initial step and for comparison, we will describe an approach to phase transformations in elastic
materials in the reference configuration.
Let us consider a volume $\, V_0 \,$ of a two-phase material with the prescribed traction $\fg p_0$
at the part of the boundary $S_{0p}$ and displacement $\fg u$   at  the rest of the boundary $\, S_{0u} \,$; $S_0 = S_{0p}  \cup S_{0u}$. The dissipation at the interface $\Sigma_0$ between phases is neglected. The problem is to find a two-phase configuration in thermodynamic equilibrium. For isothermal processes the total dissipation increment due to variation of the position of the interface $\Sigma_0$ is
 \bey
{\cal D} \delta t\, = \int\limits_{S_{0p}} \fg p_0 \cdot \delta \fg u dS_{0p} - \delta \int\limits_{V_0} \rho_0 \psi \left(\fg F, \, \theta  \right)
d \, V_0 - \delta \int\limits_{\Sigma_0}  \Gamma_0
d \, \Sigma_0=0,
 \label{IJP-2}
 \eey
where 
$\theta$ is the temperature, $\psi$ the Helmholtz free energy per unit mass, and $\, \Gamma_0 \,$ is the interface  energy
per unit reference area. Note that if $\Gamma_0=const$,  it does not  produce surface stresses and does not change
the traction continuity condition (\ref{IJP-7}) (in contrast to the case when the surface energy
per unit deformed area $\Gamma=const$   or when   $\Gamma_0$ depends on strain).
If $\fg p_0$ is fixed, then Eq. (\ref{IJP-2})  can be presented in the form
 \bey
{\cal D} \delta t\, = -\delta G =0; \qquad
G:= -\int\limits_{S_{0p}} \fg p_0 \cdot  \fg u dS_0 +  \int\limits_{V_0} \rho_0 \psi \left(\fg F, \, \theta  \right)
d \, V_0 + \int\limits_{\Sigma_0}  \Gamma_0
d \, \Sigma_0,
 \label{IJP-3}
 \eey
where $G$ is the Gibbs energy of the system body+loading. Thus, phase equilibrium for an elastic material, i.e., the geometry of the interface,  is determined by the stationary value of the Gibbs energy. If the Gibbs energy has a local minimum, phase equilibrium is stable; otherwise, it is unstable. If a stable interface does not exist under the prescribed boundary conditions, then only the single-phase solution is stable.

Let us consider each phase separately, without interfaces. The Gibbs energy can then be introduced for the parent phase (subscript 1) and product phase (subscript 2):
 \bey
G_i:= -\int\limits_{S_{0pi}} \fg p_0 \cdot  \fg u dS_{0i} +  \int\limits_{V_{0i}} \rho_0 \psi_{0i} \left(\fg F, \, \theta  \right)
d \, V_{0i} = \int\limits_{V_{0i}} \rho_{0i} g_i \left(\fg P, \, \theta  \right)
d \, V_{0i};
\qquad
  \label{IJP-4a}
 \eey
 \bey
 \rho_{0i} g_i \left(\fg P, \, \theta  \right)=  \rho_{0i} \psi \left(\fg F, \, \theta  \right) - \fg P^T \fg : \fg F.
 \label{IJP-4}
 \eey
Here,  $g_i$ is the local Gibbs energy per unit mass of each phase, expressed in terms of $\fg P$ with the help of the elasticity rule; the divergence theorem was used to transform surface integral into a volume integral, and the equilibrium  equation has been utilized.
Since analytical inversion of the elasticity rule is in most cases  impossible in practice,  one can keep the argument $\fg F$ in $g_i$, and $\fg F$ should numerically correspond to the given $\fg P$.

For a given traction $\fg p_0$ (or, for homogeneous stress, a given stress tensor $\fg P$) and temperature, the phase with smaller Gibbs energy is called the stable phase, and the other is called the metastable phase. The transition from a metastable to stable phase is accompanied by the reduction in Gibbs energy and a positive dissipation increment (see Eq. (\ref{IJP-3})), i.e., it is thermodynamically possible. However, it does not mean that this transition will occur because there is usually an energy barrier between phases.
When an  energy barrier disappears due to change in traction or temperature, the metastable phase becomes unstable and barrierless PT occurs. PT from the phase with the lower Gibbs energy to the higher energy is thermodynamically impossible.
 Phases with equal Gibbs energy are considered to be in thermodynamic equilibrium. This definition has a physical sense for hydrostatic media, liquids, and gases under prescribed pressure, because (with neglected surface tension (stresses)) the pressure is the same in both phases. For solids, phase equilibrium should be considered across an interface and not the entire stress tensor; only traction remains the same across an interface (Eq. (\ref{IJP-3})). Of course, phase equilibrium conditions are different for different interface orientations. Still, in high-pressure research, phase equilibrium is defined by equality of the Gibbs energies of phases under prescribed pressure.

   It is also clear that the definition of the Gibbs energy and related definitions depend on the chosen stress measure and boundary conditions, i.e., which components of stresses are fixed at the boundary. Integrating Eq. (\ref{IJP-3})
   over the time $t_s$ for the appearance of a nucleus of   phase 2 in a finite volume $V_n$,
   we obtain
 \bey
X_v= \int\limits^{t_s}_{0} {\cal D} dt  = - (G_2-G_1)= - \Delta G=0,
 \label{IJP-3a}
 \eey
i.e., the total dissipation increment is equal to the negative  difference between the Gibbs energy of the final and initial states, and represents the global thermodynamic driving force for the PT.  Eq. (\ref{IJP-3a})  will be used as the main hint and limiting case for checking when we will develop SC theory for inelastic materials.

   Let us transform Eq. (\ref{IJP-2}) for the neglected interface energy using the divergence theorem and the rule of differentiation for the volume integral, both  for a volume with moving surfaces with discontinuous velocity or a deformation gradient:
 \bey
 \int\limits_{S_{0p}} \fg p_0 \cdot \delta \fg u dS_{0p}
 =
 \int\limits_{V_0} \fg P^T \fg : \delta \fg F   d \, V_0 - \int\limits_{\Sigma_0}  \left[{\fg v} \right] {\fg n}_0 \fg :  \fg P^T \delta d \, \Sigma_0
 =
 \int\limits_{V_0} \fg P^T \fg : \delta \fg F   d \, V_0 + \int\limits_{\Sigma_0}  \left[{\fg F} \right]  \fg :  \fg P^T v_n d \, \Sigma_0\delta t;
 \label{IJP-8}
 \eey
 \bey
 \delta \int\limits_{V_0} \rho_0 \psi 
d \, V_0 = \int\limits_{V_0} \rho_0 \delta \psi 
d \, V_0 +\int\limits_{\Sigma_0} \rho_0 \left[ \psi \right] v_n
d \, \Sigma_0 \delta t,
 \label{IJP-9}
 \eey
where Eqs. (\ref{IJP-5}) and (\ref{IJP-7}) were used. Substituting  Eqs. (\ref{IJP-8}) and (\ref{IJP-9})  into Eq. (\ref{IJP-2}), we obtain
 \bey
{\cal D} \delta t\, =
\int\limits_{V_0} \left(\fg P^T \fg : \delta \fg F - \rho_0 \delta \psi \right)  d \, V_0 + \int\limits_{\Sigma_0}  \left( \fg P^T \fg : \left[{\fg F} \right]  -  \rho_0 \left[ \psi \right] \right) v_n d \, \Sigma_0\delta t
=0.
 \label{IJP-10}
 \eey
Due to the independence of both integrals and arbitrariness of $V_0$ and $\Sigma_0$, both integrands in Eq. (\ref{IJP-10}) are equal to zero. The first integrand results in the nonlinear elasticity rule for points of the volume. The second results in
the phase equilibrium condition at the interface
 \bey
X_\Sigma:= \fg P^T \fg : \left[{\fg F} \right]  -  \rho_0 \left[ \psi \right] ={\fg n}_{0} \, {\fg \cdot} \, {\fg P}^T  \, {\fg \cdot}
\left[ {\fg F} \right] {\fg \cdot} \, {\fg n}_{0} -
\rho_0 \left[\psi \right] = \,
 {\fg n}_{0} \, {\fg \cdot} \left[{\fg H}_0 \right]
{\fg \cdot} \, {\fg n}_{0} = 0;
\qquad
{\fg H}_0 \, := \, {\fg P}^T \, {\fg \cdot} \, {\fg F} -
\rho_0 \, \psi \, {\fg I},
 \label{IJP-11}
 \eey
where $X_\Sigma$ is the thermodynamic driving force for the interface motion. Eq. (\ref{IJP-5}) was utilized in the derivations. The expression for $X_\Sigma$ in Eq. (\ref{IJP-11}) is also called the Eshelby driving force for the interface propagation per the celebrated  work  \cite{ehselby-1970}, where it was derived.
The chemical potential tensor in the reference
configuration $\, {\fg H}_0 \,$
was introduced in \cite{Grinfeld-DAN,Grinfeld-1991}, where various aspects of phase equilibrium and different chemical potentials are discussed (see also  \cite{kondaurov+nikitin-1986,kaganova+roitburd-1988}
and references).
For geometrically linear approximation, ${\fg F} =\fg I + \fvep + \fg \omega$, where $\fvep$ and $\fg \omega$ are the small symmetric strain and antisymmetric spin tensors, respectively, $ \fg P=\fsg $ is the symmetric stress tensor  and   Eq. (\ref{IJP-11}) simplifies
 \bey
X_\Sigma:= \fsg  \fg : \left[{\fvep} \right]  -  \rho \left[ \psi \right] ={\fg n} \, {\fg \cdot} \, {\fsg }  \, {\fg \cdot}
\left[ {\fvep} \right] {\fg \cdot} \, {\fg n} -
\rho \left[\psi \right] = \,
 {\fg n}  \, {\fg \cdot} \left[{\fg H}  \right]
{\fg \cdot} \, {\fg n}  = 0;
\qquad
{\fg H}  \, := \, {\fsg } \, {\fg \cdot} \, {\fvep} -
\rho \, \psi \, {\fg I}.
 \label{IJP-11a}
 \eey
When isotropic surface energy is taken into account, Eq. (\ref{IJP-11}) generalizes to
 \bey
 {\fg n}_{0} \, {\fg \cdot} \left[{\fg H}_0 \right]
{\fg \cdot} \, {\fg n}_{0} = 2\Gamma_0 \kappa_{av},
 \label{IJP-12}
 \eey
where $\kappa_{av}$ is the mean interface curvature.

The following static problem formulations are usual for PTs in elastic materials:

1. For the prescribed boundary conditions, find a two-phase solution within a  body.
Solutions (possibly multiple) correspond to the stationary value of the Gibbs energy (Eq. (\ref{IJP-3}))  or local thermodynamic equilibrium condition Eq. (\ref{IJP-12}) for each point of a phase interface.

2. For the neglected interface energy for the problem in item 1, one can find external tractions and displacements at which the two-phase equilibrium
becomes  possible for the first time. This is the initiation of the PT. Examples include the minimum pressure for PT from a low-pressure to high-pressure phase or  the maximum pressure for PT from a high-pressure to low-pressure phase.
The critical conditions for PT initiation can be found from the solution for a small inclusion of the product phase  in item 1
when the characteristic size of the inclusion tends to be zero while keeping a shape that minimizes the Gibbs energy.
This zero size explains why surface energy should be neglected.

\begin{figure}[htp]
\centering
\includegraphics[width=0.5\textwidth]{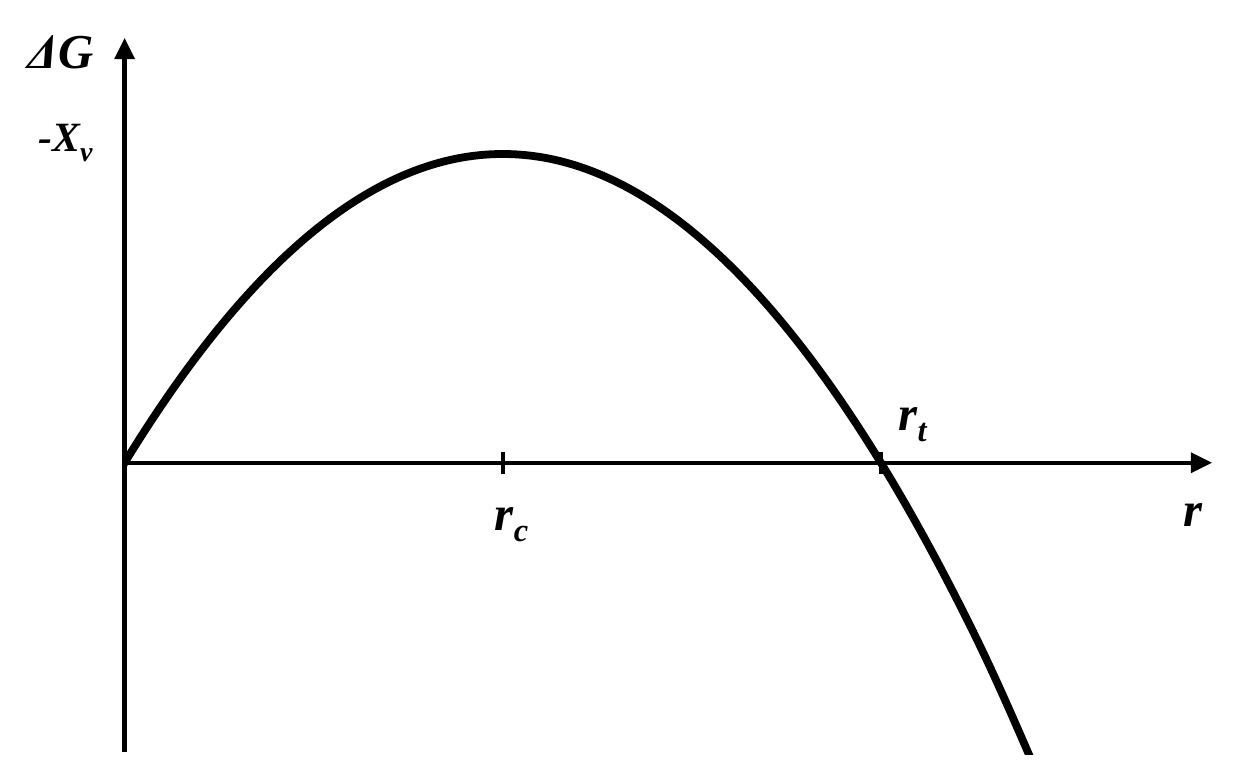}
\caption{ Gibbs free energy of the nucleus vs.  nucleus' size $r$ and definition of the size $ r_c$ of the critical
nucleus, which corresponds to the maximum (generally, minimax) of the energy (and negative thermodynamic driving force $-X_v$) and the size $r_t$ of the thermodynamically admissible nucleus, corresponding to the  zero driving force $X_v$. \label{Fig-energy-crit-nucl}}
\end{figure}

3. When surface energy is included, an unstable stationary solution called the critical nucleus plays an important role. It corresponds to the minimax of the Gibbs energy, the minimum with respect to the shape (and all internal parameters, like a twinned structure, if included) and the maximum with respect to the size or volume of the nucleus $V_{n}$ (Fig. \ref{Fig-energy-crit-nucl}). The energy increment for  the critical nucleus, $\Delta G_n=G_n-G_{par}$ is the difference between the energy of the critical nucleus and the energy of the parent phase,
which can be any phase the material is initially in.  The energy of the critical nucleus
is higher than the energy of the parent phase and {\it its appearance causes a negative dissipation increment.}
Since the   appearance of the critical nucleus is not thermodynamically favorable and
requires fluctuations, it cannot be described by traditional continuum thermodynamics based on two thermodynamics laws.
While it was well-known in material literature that the energy grows during the appearance of a critical nucleus (i.e., dissipation increment is negative), this statement caused  a psychological problem for the continuum mechanics
community, which required some tutorial explanations in \cite{Levitas-sublim-IJP-12}.
Earlier,   in the treatment of nucleation in  elastic materials in \cite{lusk-94}, it was specified that
nucleation produces, rather than dissipates, energy and that the subcritical nuclei cannot grow due to the
restriction produced by dissipation inequality. In \cite{levitas-ijp2000-1,levitas-ijp2000-2}, after stating that the appearance of the
critical nucleus is accompanied by a negative dissipation increment, thereby requiring thermal
fluctuations, the concept of the thermodynamically admissible nucleus was suggested. In particular, the nucleus
of the radius $r_t$ in Fig. \ref{Fig-energy-crit-nucl} corresponds to $X_v=-\Delta G=0$, i.e., it is thermodynamically admissible. However,
such a treatment is not consistent with the existing textbook approaches on nucleation, and it does not explicitly
determine the activation energy. This concept appears naturally within a phenomenological theory of the appearance of  a finite region of a product phase without going into detail about complex nucleation-growth processes. The
approach in \cite{levitas-ijp2000-1,levitas-ijp2000-2} illustrates that, in general, there is nothing unusual in the nucleation and
that the second law is satisfied for the event, averaged over the size $r> r_t$ or the corresponding time interval. The problem arises because we are interested in the event that occurs during a shorter period of time, and
it is not surprising that the second law of thermodynamics is not applicable for such a scale.

Indeed, continuum thermodynamics
operates with parameters that are averaged over some  time and volume, and any fluctuations
are filtered out. As a result, the nucleation process is generally described by statistical theories (as seen in \cite{Kashchiev}), which we will not discuss here.

General ideas of the  thermally activated kinetics can be found in  classical nucleation theory \cite{fisher,turnbull+fisher-49,Porter-1992}.
The Arrhenius-type equation for the nucleation time
 in a sample with volume $V$ is
\bey
t_s \, = \, t_0 \, \exp \left(\; - \;
\frac{Q}{k \, \theta} \right)
 .
\label{meyers1-126a1-0}
\eey
   Based on the probability consideration, the pre-exponential factor $t_0$ is usually considered
 to be inversely proportional to the volume of the entire sample $V$. The activation energy $Q$ for thermally activated nucleation is equal to the energy of the critical nucleus $\Delta G_n$.  We write
\bey
Q=\max_{{V_{0n}}} \min_{{shape}} \min_{{structure}}\min_{{position}}\Delta G .
\label{IJP-13}
\eey
The first maximum in Eq. (\ref{IJP-13}) belongs to the definition of the critical nucleus. All minima are motivated as follows: the smaller the activation energy, the smaller the nucleation time. Consequently, a nucleus with the smallest activation energy should appear first. That is why in Eq. (\ref{IJP-13})  activation energy is minimized with respect to all relevant parameters.

The traditional criterion for thermally activated nucleation is
\bey
Q= \beta k\theta, \qquad \beta= 40-80,
\label{IJP-14}
\eey
which is determined from the
condition that, for larger $Q$, the nucleation time exceeds any realistic time
 of observation for any choice of pre-exponential parameters in the Arrhenius-type kinetic equation  \cite{Porter-1992,lin+olson+cohen-1993,Levitas-PRB-04,Levitas-chapter-04}. Here  $\, k\, =  1.380  \cdot 10^{-23}  J/K \,$
is the Boltzmann constant. While criterion (\ref{IJP-14}) is from a material science textbook on PTs \cite{Porter-1992}
and broadly used in the treatment of PTs by material scientists   \cite{Porter-1992,lin+olson+cohen-1993}, its application was limited in the continuum mechanics community \cite{Levitas-PRB-04,Levitas-chapter-04,Levitas-Alt-VM-PRB-09,Levitas-Alt-void-ActaMat-11,Levitas-sublim-IJP-12,Levitas-Alt-sublim-IJP-12,Levitas-Samani-14}.
Note that definition (\ref{IJP-13})  includes an  extremum principle for the determination of all unknown parameters of the nucleus.

Since the critical nucleus cannot appear thermodynamically, it should be
introduced "by hand" into the problem for continuum treatment. This allows the nucleus slightly larger than the critical one (supercritical nucleus) to grow in a thermodynamically consistent way. The nucleus slightly smaller  than the critical one (subcritical nucleus) shrinks and disappears. In the phase-field approach, nucleation  is modeled by including a stochastic term  in the
Ginzburg-Landau or Cahn-Hilliard equations, which satisfies the dissipation-fluctuation theorem (see, e.g., \cite{khachaturyan-01,Lookman-prl-08,Khachaturyan-actamat-97,Khachaturyan-actamat-01}).

\section{Athermal resistance to interface propagation}\label{athermal-resist}

Athermal resistance for interface propagation is analogous to dry friction
in classical mechanics: interface propagation
occurs only if the driving force $X_\Sigma$
exceeds a rate-independent threshold  $K_\Sigma$. The athermal resistance is responsible
for deviation of the actual SC stress or/and temperature from their thermodynamical equilibrium values, and consequently, for stress or/and temperature hysteresis during forward-reverse SC, and for energy
 dissipation, when interface velocity is small and viscous friction is negligible.
It is observed for SCs both in elastic (e.g., in shape memory alloys) and elastoplastic materials. Inclusion of the athermal dissipation in the description of
PT in elastic materials significantly and conceptually complicates the description of the  transformation process, similar to the mechanics of the system with dry friction. Thus, the entire behavior becomes loading-path dependent and should be treated incrementally. The  thermodynamic equilibrium does not correspond to the minimum of the Gibbs energy, and the principle of stationary Gibbs energy cannot be applied for searching a two-phase equilibrium microstructure.  Such   features are closer to those for PTs in elastoplastic materials.
Extremum principle substituting the principle of the stationary Gibbs energy for SCs in elastic materials with athermal friction were obtained in  \cite{levitas-ijss1998} as a particular case of the corresponding principle for PTs in elastoplastic materials. Mathematic treatment of the problem for SCs in elastic materials with athermal interfacial resistance was initiated in \cite{mielke+theil+levitas-2002}.

There are several sources of athermal interfacial friction $\, K_\Sigma \,$
\cite{ghosh+olson-1994-1}:

\begin{enumerate}
\item Peierls barrier, which appears due to the discrete periodic structure of the crystal lattice, similar to that for dislocations.
\item Interaction of a moving interface with a long-range stress field of various
defects, e.g., point defects (solute and impurity atoms, vacancies),
dislocation forest, stacking faults, grain, subgrain, and twin boundaries, and precipitates.
\item Emission of acoustic waves.

\end{enumerate}
When SC is considered in a finite volume, an additional contribution to the athermal dissipation and SC hysteresis is related to a nucleation barrier.
An athermal threshold $K_v$ is determined as a minimum value of the driving force at which SC occurs, i.e.,
\bey
{X}_v \; = \;  K_v=  \; \int_{V_n} \rho \, {K} \; d \, V_n  = K^0 m_n.
\label{ij1-19a}
\eey
where $\, K \,$ is a locally determined  athermal threshold per unit mass $m_n$ and $K^0$ is an athermal threshold averaged over a nucleus per unit mass. If the volume is transformed by continuous interface propagation, then $X_\Sigma= \rho \, {K}$.

By definition, the athermal threshold cannot be overcome by thermal fluctuations.
While the magnitude of  $\, K \,$ can be different 
for direct and reverse SC and based on a variety of the mechanisms, $K$ is expected to be dependent on  the entire deformation-SC process and the evolving material
microstructure. However, a simple relationship
\bey
K \, = \, \, L \, \sg_y \left(\theta, \, \fvep_p, \, {\fg g} \right) \, \vep_o
\label{003-a}
\eey
was suggested in \cite{Levitas-JMPS-I-97,levitas-ijss1998} based
on the comparison of some high-pressure experiments summarized in
\cite{Blank-Estrin-2014} with the solution of the corresponding boundary-value problems.
Here  $\, \sg_y \,$
is the yield strength, which is the function of the temperature $\theta$, plastic strain $\fvep_p$, and set of internal variables
${\fg g}$, $\, L \,$ is the proportionality factor, and
$\vep_o$ is the volumetric transformation strain, which can be included in $L$. The values $L$ for some PTs are determined in
 \cite{Levitas-JMPS-I-97,levitas-ijss1998}.
Physically,  the proportionality between $\, \sg_y \,$
and  $\, K \,$ is caused by the fact that $\, \sg_y \,$
and  $\, K \,$  characterize resistance to the dislocation and interface motion, respectively,
through the same material microstructure consisting of various point, linear, planar, and bulk defects.
For  large plastic deformations,
based on the regularity revealed in \cite{levitas-book96}, $\, \sg_y \,$
and consequently $\, K \,$ reach their maximum and are independent of plastic strain and strain-history.
Since experiments in \cite{Blank-Estrin-2014} were performed in 1980-1983, utilization of modern diagnostic, like x-ray diffraction with synchrotron radiation, may change our understanding of the effect of plastic strain on the athermal threshold.

For SMAs,  the stress hysteresis and dissipation are proportional to $\, K \,$.
Since both the stress hysteresis  and $\, \sg_y \,$  linear depend on plastic strain,
as seen in  experiments  in \cite{hornbogen-1997,spielfeld+hornbogen-1998,spielfeld+hornbogen+franz-1998,treppmann+hornbogen-1998},
this also supports Eq. (\ref{003-a}).
A slightly different but close relationship between $K$ and the yield strength of an austenite $\sg_y^A$ for   steel is suggested in \cite{ghosh+olson-1994-1,ghosh+olson-1994-2} and presented in Eq. (\ref{ol-3}).

\section{Theory of structural changes in inelastic materials with an unstable intermediate state }\label{general-theory}

\par
For the description of SC in an elastic solid, the explicit equation for the thermodynamic driving force
for the interface propagation and corresponding local thermodynamic equilibrium condition (Eqs. (\ref{IJP-11})  and (\ref{IJP-12})) for each point of a phase interface are known. They correspond to
the stationary value of the Gibbs energy (Eq. (\ref{IJP-3})). The activation energy for a critical nucleus Eq. (\ref{IJP-13})
is also determined in terms of the Gibbs energy.  Furthermore, each solution of the elastic problem (including cases where there are multiple solutions for given boundary conditions) is independent of the process and depends on the final boundary conditions.

 For inelastic materials,  such results are lacking. All processes in plasticity are loading-history dependent
 and accompanied by energy dissipation, even for infinitesimally slow processes. For this reason, a change in the Gibbs energy  alone does not define the driving force, and the stationarity of the Gibbs energy does not determine the phase equilibrium or critical nucleus.
{\it A conceptually different thermodynamic approach was required to determine the thermodynamic driving force for nucleation and interface propagation, the definition of the critical nucleus, the expression for the activation energy for the critical nucleus, and the extremum principle for the determination of all unknown parameters of the nucleus.}
 Eq. (\ref{IJP-3a})  will be utilized  as the main hint for the formulation of the driving force for SC in inelastic materials:
 it should have a sense of the dissipation increment due to SC only for the appearance of a complete nucleus.

 We will present  our theory developed in   \cite{levitas-ijss1998,levitas-ijp2000-1,levitas-ijp2000-2} and
\cite{Levitas-sublim-IJP-12,Levitas-Alt-sublim-IJP-12}.
Initially   \cite{53,levitas-mrc1995,levitas-1995-phys-III-41,Levitas-JMPS-I-97,Levitas-JMPS-II-97},
the  nucleation condition was postulated in the form of the dissipation
increment due to PT only (excluding plastic and other types of
dissipation) reaching its experimentally determined value related to the athermal threshold. Later in  \cite{levitas-mrc1996,54,levitas-ijss1998,levitas-ijp2000-1,levitas-ijp2000-2},
 a local description of PTs was developed  for better justification, which we will use here.

\subsection{Definition of the structural changes
without a stable  intermediate state and local equations}

For simplicity and transparency of the main ideas, we will start with the geometrically linear formulation.
The main local equations describing SC are
collected in Box 1.
\par
Traditional additive kinematics  (\ref{ijss1-1}) is utilized. It is convenient to introduce a scalar
parameter $\xi$ describing SC: SC starts at $\, \xi = 0 \,$ and completes
at $\, \xi = 1 \,$. The parameter $\, \xi \,$ is an internal variable similar to a volume
fraction of the product phase in the mixture theory or order parameter in the PFA.
The parameter $\xi$ can be determined by Eq. (\ref{ijss1-1+}) or a similar equation based on any other material property of the phases, such as elastic moduli or entropy.
For the Helmholtz free energy, we assume Eq. (\ref{ijss1-12-00}),
where $\, {\fg g} \,$ is a set of internal variables describing plasticity, e.g., back stress or
dislocation  density. Traditional thermodynamic treatment results in the constitutive equations  (\ref{ijss1-12})
for stress and entropy, the yield condition  (\ref{ijss1-12a}), and the evolution equations for plastic strain and internal variables (\ref{ijss1-12b}). It is assumed that $\xi$-dependence of all material properties is known, e.g., based on linear interpolation between properties of two phases, similar to the simplest mixture theory.

Note that in contrast to the PFA or mixture theory, we do not prescribe the local evolution equation for $\xi$ describing the SC.
 We assume  that the SC process at each point of the transforming
volume {\it cannot be stopped in an intermediate stage}. Consequently, the
material in each material point  can only be in either  phase 1  or phase 2.
Such   SC is coined in  \cite{levitas-ijp2000-1,levitas-ijp2000-2} as the {\it  SC
without a stable local intermediate state}, and the following definition is used.
\par
{\it The SC will be considered as a process of variation of the
transformation deformation gradient and some or all thermomechanical
properties in an infinitesimal or finite transforming volume from the
initial to the final value. This process cannot be stopped at an intermediate
state in any transforming point. Thermodynamic equilibrium for an
intermediate value of the transformation deformation gradient or material
properties is impossible.}
\par
Such a definition excludes from consideration the Landau-Ginzburg model
(see Section \ref{PFA-martensite}), in which a smooth
transition from phase 1 to phase 2 occurs due to a nonlocal term, and any intermediate state
can be stable inside a diffuse interface
of finite thickness. We will
focus on the case with a sharp interface and local constitutive
equations describing the deformation in each material point.

\noindent
\underline{\hspace*{17cm}}

\begin{center}
{\large Box 1.} {\bf Local equations describing phase transitions and plasticity
                     \cite{levitas-mrc1996,54,levitas-ijss1998,levitas-ijp2000-1,levitas-ijp2000-2}}
\end{center}
\noindent
\underline{\hspace*{17cm}}

\noindent
{\bf 1. Additive  decomposition of strain into elastic $\fvep_e$, plastic $\fvep_p$, and transformational $\fvep_p$ parts}
\begin{eqnarray}
\fvep \; :=   (\fg \nabla \fg u)_s =\; \fvep_e \, + \, \fvep_p  \, + \, \fvep_t \; .
\label{ijss1-1}
\end{eqnarray}
{\bf 2. Internal time $\xi$}
\begin{eqnarray}
\xi \; := \;
\frac{\mid \fvep_t - \fvep_{t1} \mid}{\mid \fvep_{t2}-\fvep_{t1} \mid}
 \qquad  0 \, \leq \, \xi \, \leq \, 1 .
\label{ijss1-1+}
\end{eqnarray}

\noindent
{\bf 3. Local constitutive equations}

{\bf 3.1. Helmholtz free energy per unit mass $\psi$, elastic $\psi^e$ and thermal $\psi^\theta$ energies}
\bey
\psi = \psi \left(\fvep_e, \, \theta,  \, {\fg g}, \,
  \xi \right) =\psi^e \left(\fvep_e, \, \theta, \, \fvep_p, \, {\fg g},  \, \xi \right)+\psi^\theta \left(\theta, \, \fvep_p, \, {\fg g}, \,
 \xi \right) .
\label{ijss1-12-00}
\eey

{\bf 3.2. Elasticity rule  and  entropy-temperature relationship}
\bey
{\fsg } \; = \rho\; \frac{\partial \, \psi
\left(\fvep_e, \, \theta, \, \fvep_p, \, {\fg g}, \,
  \xi  \right)}{\partial \, \fvep_e} \; ;
\qquad
s \; = \; - \, \frac{\partial \, \psi
\left(\fvep_e, \, \theta, \, \fvep_p, \, {\fg g}, \,
  \xi \right)}{\partial \, \theta} \; .
\label{ijss1-12}
\eey

{\bf 3.3. Dissipation rate per unit mass due to plastic flow $D_p $ and variation of the internal variable $D_g$}
\bey
D_p:= {\fg X}_p \; :  \dot{\fvep}_p ; \qquad
D_g:= {\fg X}_g \; \cdot  \dot{\fg g}^T .
\label{ijss1-12a-0}
\eey

{\bf 3.4. Dissipative forces for plastic flow ${\fg X}_p $ and variation of the internal variable ${\fg X}_g$}
\bey
{\fg X}_p \; := \; \frac{1}{\rho}{\fsg }\, - \; \frac{\partial\,\psi}{\partial\,\fvep_p}; \qquad
{\fg X}_g \; := \; - \, \frac{\partial \, \psi}{\partial \, {\fg g}^T} \; .
\label{ijss1-12a-01}
\eey

{\bf 3.5. Yield condition}
\bey
f \left({\fg X}_p \, , \, \theta, \, \fvep_p, \, {\fg g}, \,
  \xi \right) = \, 0 \; .
\label{ijss1-12a}
\eey

{\bf 3.6. Evolution equations for plastic strain and internal variables}
\bey
&&\dot{\fvep}_p \, = \, {\fg f}_p \left({\fg X} \, , \, \theta, \, \fvep_p, \, {\fg g}, \,
  \xi  \right)\, \quad {\rm if} \quad    f= \, 0; \qquad   {\rm and } \quad    \dot{f} >  0;
\qquad   \dot{\fvep}_p \, = 0 \quad {\rm otherwise}.
\nonumber\\
&& \dot{\fg g} \; = \; {\fg f}_g \left({\fg X}_g \,  , \, \theta, \, \fvep_p, \, {\fg g}, \,
 \xi \right) \, .
\label{ijss1-12b}
\eey
\noindent
\underline{\hspace*{17cm}}

\subsection{Thermodynamic driving forces for nucleation and interface propagation}
\label{Drive-force}

\begin{figure}[htp]
\centering
\includegraphics[width=0.3\textwidth]{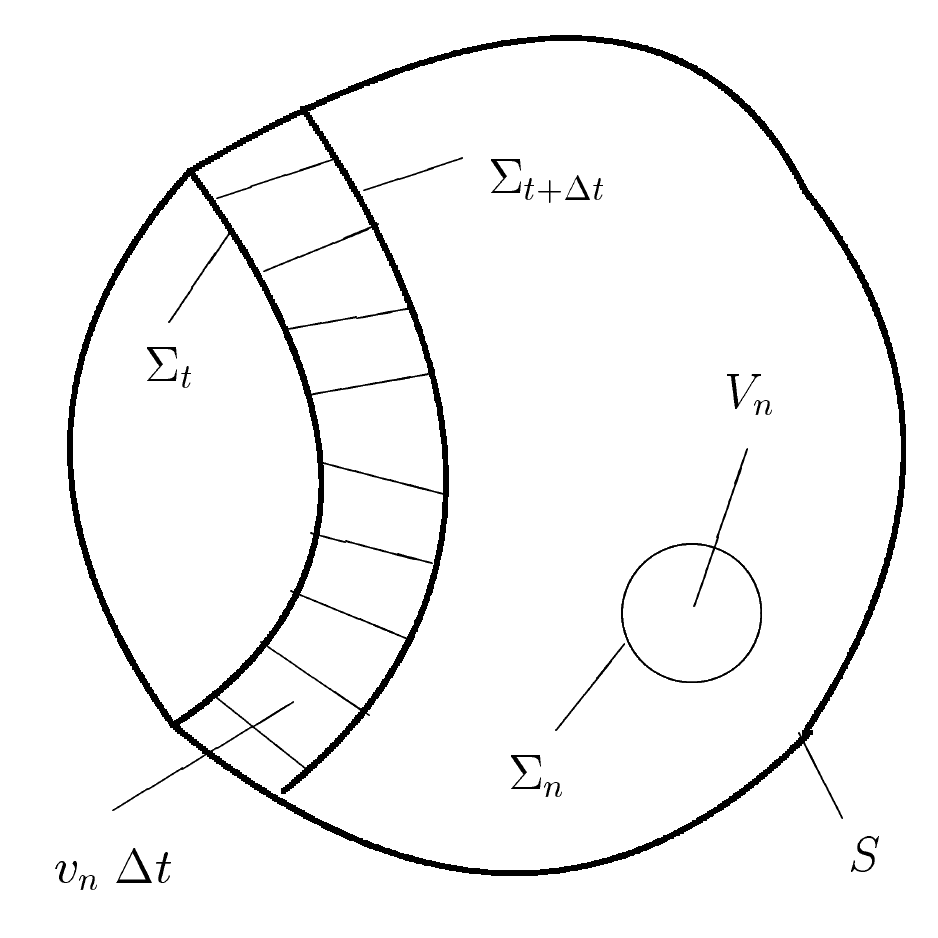}
\caption{ Schematics of the region $\, V \,$ of a material undergoing SCs with  a boundary $\, S \,$. The
region $\, V_n \,$ with the boundary $\, \Sigma_n \,$ represents a new nucleus.  SC within a volume covered by a phase interface $\, \Sigma_t \,$ propagating with velocity $v_n$  during the
time $\, \Delta t \,$ is shown as well.
\label{Fig-schematics}}
\end{figure}

Our goal here is to generalize the thermodynamic driving forces for nucleation and interface propagation derived for elastic materials (Eqs.(\ref{IJP-3a}) and  (\ref{IJP-11})) for those in inelastic materials.

{\it Problem Formulation.}
Let us consider some  region $\, V \,$ of a material undergoing SCs under the prescribed
boundary conditions in tractions and displacements at a boundary $\, S \,$ (Fig. \ref{Fig-schematics}). Nucleation of phase 2 will be considered as  the SC that starts and completes  in some
region $\, V_n \,$ with the boundary $\, \Sigma_n \,$ during the nucleation
time $\,  t_s \,$.  In practice, we change the field of the parameter $\, \xi  \left({\fg r} \right) \,$ from 0 to 1   in the nucleus, homogeneously or heterogeneously, which   introduces  into the nucleus the transformation strain field
$\, \fvep_t \, \left({\fg r} \, , \; \xi \right) \,$ and variation of all material
properties from properties of phase 1 to properties of phase 2---
all without changing  the boundary  conditions. This process represents the prescribed deformation process, for which the inelastic boundary  value problem and heat conduction equation should be solved incrementally, producing  fields
 $\, {\fsg } \,$, $\; \fvep_p \,$, $\fg g$, $\; \theta \,$, and all other participating fields.
 Note that depending on the goal, we will use a different definition of a nucleus. In addition to the traditional critical nucleus, we will also consider the appearance of  a macroscopic region that
is formed  during nucleation and growth processes.
 Having this  information, one must answer the following four questions:
\par
1. What is the thermodynamic driving force for the nucleation and phase interface propagation?

2. Is SC possible for the chosen  boundary and initial conditions, i.e., what is the SC criterion?
\par
3. How do we  determine  all the unknown parameters of a nucleus and the transformation process in it, i.e. the position, volume, shape, and orientation of a   nucleus, its internal structure (i.e., actual field $\, \fvep_t \; ({\fg r} \, , \; \xi ) \,$?
All unknown parameters will be collectively  designated as
   $\,{\fg b}\,$.
\par
4. If time-dependent kinetics is considered, what is  the nucleation time $t_s$?

\par
The traditional thermodynamic procedure based on two thermodynamic laws has been applied to find the thermodynamic driving forces presented in Box 2.   Similar to the local dissipation rates due to plastic flow $D_p$ and the change in the internal variable
$D_g$ (Eqs.(\ref{ijss1-12a-0}) and
(\ref{ijss1-12a-01})), a similar local dissipation rate $\, {\cal D}_{\xi} \,$ and dissipative force  $ X_\xi $ for structural changes Eq. (\ref{ijss1-11})  have been derived.  One of the main assumptions is that irreversible processes of plastic flow,
a variation of internal variables and SC are thermodynamically independent; they only interact through stress fields.
However, since we do not want to describe the kinetics of SC in terms of $\dot{\xi} $, rather complete SC in the point
(such as for the point of an elastic nucleus), we introduce the local thermodynamics driving force for a complete structural change $X$ by integrating the dissipation rate over the entire SC in the point, see Eq. (\ref{ijss1-9}).
This expression can be transformed into Eq. (\ref{ijss1-180}), which is physically clear: the local dissipation increment $X$ due to SC alone is equal to the total dissipation increment (the first three terms in Eq. (\ref{ijss1-180})) minus the
dissipation increment due to all other dissipative processes except SC, namely plastic flow and evolution of the internal variable, which are
the two last terms in Eq. (\ref{ijss1-180}).   At $\, 0 < \xi < 1 \,$ a
nonequilibrium process takes place, which requires  energy and stress
fluctuations.  It is necessary to average the thermodynamic parameters,
related to SC, over the SC duration $\, t_s \,$ in order to filter off these
fluctuations, which results in consideration of the dissipation increment.

Since we want to find the thermodynamic driving force for nucleation in a volume $V_n$, we integrate $X$ over the volume of a nucleus $V_n$ and add a change in the surface energy, see Eq. (\ref{ijss1-19}). If the initial surface energy is zero, then the change in the surface energy is just the  surface energy of a nucleus. Finally, the global dissipation rate for nucleation  Eq. (\ref{mrc95kor-44e}) is the dissipation increment due to SC only, divided by nucleation time $t_s$; thus, the generalized rate is $\dot{\chi} \, := \; 1 \, / \, t_s \,$.

It is easy to show that an {\it integral in Eq.(\ref{ijss1-19}) can be evaluated
over an arbitrary volume $\, v  \,$ containing a single transforming region
$\, V_n  \,$ with SC}, see Eq.(\ref{ijss1-19-r}).
Indeed, integration over the region $\, v - V_n  \,$ without SC gives zero
contribution to the driving force $\, X_v \,$, which
is the dissipation rate due to SC only.
The change in surface energy should be evaluated over the nucleus
surface $\, \Sigma_n \,$.
Thus, {\it instead  of a surface-independent
Eshelby integral in the theory of defects and path-independent
J-integral in fracture mechanics, we introduced
a region-independent integral for arbitrary inelastic materials}.
In contrast to the $\, \Gamma \,$-integral in \cite{Cherepanov-67,Cherepanov-79} and a region-dependent $\, T  \,$-integral in \cite{Atluri-97}, the region-independent
integral (\ref{ijss1-19-r}) separates the dissipation increment due to
SC only from other dissipation contributions and is the thermodynamic
driving force for SC. It also includes the temperature variation in the process of SC.

\par
It is shown in \cite{levitas-ijss1998} that Eq. (\ref{ijss1-19}) for the thermodynamic driving force for nucleation in a volume $V_n$  reduces for SC in elastic materials to the negative increment of the  Gibbs free energy Eq. (\ref{IJP-3a}),
i.e., our approach is noncontradictory.

Note that in \cite{levitas-ijp2000-1,levitas-ijp2000-2} a concept of SC without a stable intermediate state for $\, 0 < \xi < 1 \,$ was introduced and used for justification that we should not prescribe  an evolution equation for
$\, \dot{\xi} \, = \, f_\xi \left(X_\xi , \; \xi \right) \,  \,$, like for other internal variables.
Since for any  $\, 0 < \xi < 1 \,$ one can chose $X_\xi $ to enforce $\, \dot{\xi} \, = \,0$ and arrest any $\, 0 < \xi < 1 \,$
(which contradicts the definition that only states  $ \xi = 0 \,$ and $ \xi = 1 \,$ should be in thermodynamic equilibrium), this led to the conclusion that  an evolution equation for
$\, \dot{\xi} \,$ should not be prescribed. However, thermodynamic equilibrium for $\, 0 < \xi < 1 \,$  may be unstable (e.g., like in Landau theory,  see Section \ref{PFA-martensite}), and contradiction is seeming.
The real reason for avoiding an evolution equation for
$\, \dot{\xi} \,$  is the desire to describe the complete SC in some region $V_n$ and to generalize the approach, as developed for elastic materials in Eq. (\ref{IJP-3a}), for an inelastic material.

\begin{center}
{\large Box 2.} {\bf Dissipation rate due to SC and thermodynamic driving forces
                     \cite{levitas+idesman+stein-ijss1998,levitas-ijp2000-1,levitas-ijp2000-2}}
\end{center}
\noindent
\underline{\hspace*{17cm}}

\noindent
{\bf 1. Local dissipation rate $\, {\cal D}_{\xi} \,$ per unit mass and dissipative force  $ X_\xi $ for structural changes }
\bey
{\cal D}_{\xi} \, := X_\xi \dot{\xi} ; \qquad
X_\xi \, = \, \frac{1}{\rho} \,
           \fsg \, {\fg :} \, \frac{\partial \fvep_t }{\partial \, \xi} \,
          - \, \frac{\partial \, \psi}{\partial \, \xi} \,.
\label{ijss1-11}
\eey
{\bf 2. Local thermodynamics driving force per unit mass for a complete structural change $X$ }
\bey
X \, :=\int\limits^{t_s}_{0}
             {\cal D}_{\xi} dt \; =  \int\limits^{t_s}_{0}
             X_\xi \, \dot{\xi} dt \; = \int\limits^{1}_{0}
             X_\xi d\xi \,
\label{ijss1-9}
\eey
or
\bey
 \rho X \, := \int\limits^{\fvep_2}_{\fvep_1}
{\fsg } \, {\fg :} \, d \, \fvep \, - \, \rho \left( \psi_2 - \psi_1 \right) -
\int\limits^{t_s}_{0}  \rho \left( s \, \dot{\theta} \, + \,
{\fg X}_p \, {\fg :} \, \dot{\fvep}_p +
{\fg X}_g \, {\fg :} \, \dot{\fg g} \right) \, d \, t \; .
\label{ijss1-180}
\eey
{\bf 3. Global thermodynamic driving force $X_v$ for nucleation in a volume $V_n$, i.e., the total dissipation increment due to SC
only during the complete SC in the transforming region}
\bey
X_v :=  \bar{X} m_n \, =  \int\limits_{V_n} \rho X
d \, V_n -  \Delta\, \int\limits_{\Sigma_n} \Gamma \; d \, \Sigma_n \, .
\label{ijss1-19}
\eey
{\bf 4. Global thermodynamic driving force for nucleation $X_v$ in terms of a region-independent integral }
\bey
X_v :=  \bar{X} m_n \, =  \int\limits_{v} \rho X
d \, v -  \Delta\, \int\limits_{\Sigma_n} \Gamma \; d \, \Sigma_n \, .
\label{ijss1-19-r}
\eey
{\bf 5. Global dissipation rate $\, {\cal D}_v \,$ for nucleation in a volume $V_n$}
\bey
{\cal D}_{v} \, = \; X_{v} \, \dot{\chi} \; \geq 0;
\qquad\qquad
\dot{\chi} \, := \; 1 \, / \, t_s \; .
\label{mrc95kor-44e}
\eey
{\bf 6. The thermodynamic driving force for a phase interface propagation }
\bey
X_{\Sigma }: = \, \fsg \, {\fg :} \, \left[\fvep \right] \,-
\rho \left[ \psi \right] \, -
\int\limits^{\theta_2}_{\theta_1} \, \rho \, s \; d \, \theta \, - \,
\int\limits^{t + \Delta \, t}_t \rho \, \left({\fg X}_p \, {\fg :} \,
\dot{\fvep}_p \, + \, {\fg X}_g \, {\fg :} \, \dot{\fg g} \right)
\, d \, t \,.
\label{ijss1-123}
\eey
{\bf 7. Local dissipation rate $\, {\cal D}_\Sigma \,$ per unit area for a phase interface propagation}
\bey
{\cal D}_{\Sigma } \,:  = \; X_{\Sigma } \, v_n \; \geq 0.
\label{dis_sigma}
\eey
\par
If $\, \psi \,$ depends only on $\, \fvep_e \,$ , $\; \theta\;$, and
$\; \xi \,$, and if the elastic properties of phases are the same (i.e., $\, \psi^e \,= \psi^e(\fvep_e \,,\, \theta) $),
$\psi^\theta =\psi^\theta \left(\theta\right)$,
surface energy is negligible for isothermal approximation and
homogeneous $\, \theta \,$ 
in the nucleus, then Eqs. (\ref{ijss1-180}) and (\ref{ijss1-19})  result in
the following very simplified expression for the thermodynamic driving force for nucleation  
\bey
\bar{X}\, = \, \varphi \, - \,  \; \Delta \, \psi^\theta \left(
\theta\right) 
\qquad\qquad
\varphi \, := \; \frac{1}{m_n} \,
\int_{V_n} \int^{\fvep_2^t}_{\fvep^t_1}
{\fg \sg} \, {\fg :} \, d \, \fvep^t \; d \, V_n \; .
\label{eq:6}
\eey
It looks strange that Eq. (\ref{eq:6}) does not explicitly contain any plastic strain and looks similar to that for an elastic material. However,
stress variation within the nucleus during the SC depends on the evolution of the plastic strain field, which essentially affects the transformation work $\varphi $  and $\bar{X}$.

Let us consider SC within a volume covered by a phase interface $\, \Sigma_t \,$ propagating with velocity $v_n$ during the
time $\, \Delta t \,$ (Fig. \ref{Fig-schematics}). After some transformations that include  universal conditions (\ref{IJP-5})-(\ref{IJP-7}) for a coherent interface, one arrives at Eqs.(\ref{ijss1-123}) and (\ref{dis_sigma}) that
determine the thermodynamic driving force for a phase interface propagation and the corresponding dissipation rate.
Note that for evaluation of the integrals in Eq.(\ref{ijss1-123})
one has to perform the same procedure as for SC in a nucleus $V_n$: produce $V_n$ by infinitesimally advancing an interface, change $\xi$ incrementally from 0 to 1 in an infinitesimal layer $V_n$, and solve the thermomechanical boundary-value problem for $V_n$. For such a volume $V_n$, we obtain $\rho \bar{X}=X_{\Sigma } $.
Examples for such analytical solutions can be found in  \cite{Levitas-JMPS-I-97,levitas-ijss1998,levitas-ijp2000-1,levitas-ijp2000-2} for small strain,  in \cite{levitas+nesterenko+meyers-1998-1,levitas+nesterenko+meyers-1998-2,levitas-zamm1998} for large strains (Section \ref{interface-analytic}),  and for FEM solutions
in \cite{ides-lev-CMS-1997,levitas+idesman+stein-ijss1998,idesman-levitas+stein-ijp-2000,31} for FEM solutions (Sections \ref{semicoherent-int} and \ref{growth-lath}). The important point is that for the isothermal process and neglected plastic dissipation, and dissipation due to internal variable,  Eq. (\ref{ijss1-123}) transforms into the Eshelby driving force (\ref{IJP-11a}) for interface propagation in an elastic material.

\subsection{Three types  of  kinetic descriptions}

\par
The following kinetic descriptions will be considered.
\begin{itemize}
\item Athermal or rate-independent kinetics, for which
real time and rate are irrelevant. SC occurs in the chosen region instantaneously when
SC criterion is satisfied.  To some extent, this idealization is similar to the rate-independent plasticity.
\item Thermally-activated kinetics of the appearance of the critical nucleus, similar to that in classical homogeneous nucleation theory. Like a classical theory, for which the typical size of a critical nucleus is a few to tens of nanometers, this theory is applicable at the nanoscale.

\item "Macroscale" thermally-activated kinetics, for which time for the appearance of a macroscale region is postulated.
The main assumptions here are quite different than in classical nucleation theory.
\end{itemize}
All kinetic approaches should include some extremum principle to determine all unknown parameters of a nucleus,
like its position, shape, internal structure, etc. They should substitute the principle of the stationary Gibbs energy (\ref{IJP-3})  and the extremum principle incorporated in the definition of the activated energy (\ref{IJP-13}).

\subsection{Athermal kinetics}

We consider the appearance of an arbitrary macroscopic region of phase 2 by some
nucleation and growth process.
It is accepted in  SC criterion (\ref{ijss1-19a}) in Box 3 that SC in the chosen volume occurs when
the thermodynamic driving force per unit mass $\bar{X}$ is equal to the athermal threshold
$\;K^0 \,$.
Note that the same criterion is applied for SC in elastic materials.

SC condition (\ref{ijss1-19a})  does not require fulfillment of  any condition for each point of the nucleus, i.e., it has nonlocal nature caused by surface energy. However, for large volumes, the effect of surface energy is negligible, and an
integral SC condition   without surface energy was applied in many papers, including \cite{roitburd+temkin-1986,kaganova+roitburd-1989,fischer+berveiller+etal-1994,53,levitas-mrc1995,levitas-1995-phys-III-41,Levitas-JMPS-I-97,Levitas-JMPS-II-97}.
 This,  in particular, means that for $\, K=0 \,$ the
dissipation increment due to SC may be negative in some points, both in elastic and inelastic materials.

Similar criteria (\ref{ijss1-19b}) and  (\ref{ijss1-19-b1}) have to be satisfied for points of the interface, both at time $t$ and $t+\Delta t$, where the subscript $\Delta$ denotes that a parameter is
determined at time  $\, t + \Delta \, t \,$. Two equations for the propagating interface are a consequence of the lack of a kinetic equation for interface propagation and the fact that SC occurs in the infinitesimal volume covered by a moving interface during the time interval $[t; t+\Delta \, t ]$. Similarly, for time-independent plasticity, the yield condition should be satisfied
at  time $t$ and $t+\Delta t$, which results in the consistency condition in
addition to the yield condition.

However, the SC criterion (\ref{ijss1-19a}) is just one scalar equation that is not sufficient for the determination of all
unknown parameters $\; {\fg b} \,$ (e.g., position and shape of the nucleus, internal structure, and transformation path in the  nucleus or at the interface, etc.) among all possible parameters
${\fg b}^*$.
To resolve this problem, the extremum principles  (\ref{ijss1-179}) and (\ref{ijss1-179-a}) were derived
for the nucleus and propagating interface. This was done
using the postulate of realizability, see below. For $K=0$ and elastic materials,  principle  (\ref{ijss1-179})  reduces to the principle of the minimum of Gibbs energy.
The extremum principle  (\ref{ijss1-179-a}) is considered for time  $\, t + \Delta \, t \,$ only because for time $t$ it was met at the previous time step.

\subsection{The postulate of realizability and  extremum principle for determination of all unknown parameters for a nucleus and interface}\label{post-realiz}

In order to derive the extremum principle (\ref{ijss1-179}),
the plausible assumption that we called the
postulate of realizability \cite{Levitas-postulate-I-1995,Levitas-postulate-II-1995,levitas-ijss1998,levitas-ijp2000-1,levitas-ijp2000-2}
was formulated. For time-independent kinetics, it consists of two steps:
\begin{itemize}
\vspace*{-2mm}
\item It is evident, that if strict  inequality (\ref{ijss1-179})$_1$ is
      valid for all variable parameters $\bf b^*$, then SC cannot occur,
      because the SC criterion (\ref{ijss1-19a}) is not met.
\item Let us vary boundary conditions and check the inequality (\ref{ijss1-179}) for all variable parameters $\bf b^*$ for each of them.
          The main assumption is  that if  the SC criterion is satisfied for the {\it first time}
      for some parameter $\, {\fg b} \,$ and  SC can occur, this SC will occur.
\vspace*{-2mm}
\end{itemize}
The postulate of realizability is quite a  natural assumption
expressing  the stability concept. If some
dissipative process (PT, fracture, plastic flow, contact sliding, etc.) can occur from an energetic point
of view, it must occur, since
various perturbations  provoke the initiation of a process.
Recollecting   Murphy's law that "anything that can go wrong will go wrong," the postulate of realizability can be considered its optimistic and thermodynamic version.

Numerous
applications of the postulate of realizability to derive constitutive equations for plastic flow and  plastic spin for anisotropic plasticity, friction, nonlinear nonequilibrium
thermodynamics, PTs, twinning, CRs,
fracture, as well as for stability analysis
\cite{levitas+idesman+stein-ijss1998,Idesman-CMAME-99,Levitas-postulate-I-1995,Levitas-postulate-II-1995,levitas-ijss1998,57,levitas-ijp2000-1,levitas-ijp2000-2,63,levitas+nesterenko+meyers-1998-1,levitas+nesterenko+meyers-1998-2,
levitas+idesman+olson-1999,65,31} lead to  the impression that this
postulate catches  a general essential property of dissipative systems.
Mathematical treatment of the extremum principle for PT in elastic materials with athermal friction
that follows
from the postulate of realizability was performed in \cite{mielke+theil+levitas-2002}. This paper has initiated significant
mathematical literature on the study of the rate-independent and hysteretic systems, including PT in ferroelastic, ferromagnetic, ferroelectric, and multiferroic materials (e.g., SMA), elastoplasticity, damage, crack propagations, friction, delamination,  etc., which can be found from citations of the paper
\cite{mielke+theil+levitas-2002}.

{The formulation based on SC condition (\ref{ijss1-19a}) and extremum
principle (\ref{ijss1-179}) is consistent in the limiting case of elastic materials and $K=0$ with the
classical description based on the principle
of the stationary or    minimum of the  Gibbs free energy (\ref{IJP-3}). }

{ When simplified expression (\ref{eq:6})  is used for the driving force for nucleation, extremum principle (\ref{ijss1-179}) results in the maximum
of transformation work $\, \varphi \,$.
Even in this simplest case, the SC criterion  (\ref{ijss1-19a}) includes
the history of stress variation $\, {\fsg } \,$ in the nucleus during SC, i.e.,
we cannot define the SC condition using only the initial stresses before SC.
We have to solve the elastoplastic problem and determine the variation of
stresses in the nucleus during SC in order to calculate the transformation work
$\, \varphi \,$ in Eq. (\ref{eq:6}).

There is a major problem in the application of the time-independent kinetics
for homogeneous materials under prescribed stresses. The minimum of Gibbs
free energy for elastic materials with $K=0$ and the maximum driving force for elastoplastic materials
may be reached when the entire volume transforms homogenously
after meeting the SC criterion $\bar{X}=K$. Indeed, as a corroborating argument,  if the energy of the external surface does not change during PT,
then the negative surface term disappears from $\bar{X}$. Also, the negative contribution to $\bar{X}$ due to the positive energy of internal stresses during nucleation  disappears as well.
Our statement  can be easily proved for  the nucleation of a spherical product
phase within a parent sphere.

However, homogenous PT in a large volume is unphysical (with some exceptions \cite{Levitas-Ravelo-PNAS-2012,Levitasetal-PRL-17,Babaei-Levitas-IJP-18,Babaei-Levitas-AM-19}, some which are discussed in Sections \ref{VM-shock}, \ref{Atomistic studies}) because of local barriers, which we filtered out while integrating $X$ over $\xi$ from 0 to 1.
That is why the nucleus of some not-strictly-justified size is considered in analytical or numerical solutions, in many cases equal to a single finite element \cite{levitas+idesman+stein-ijss1998}, region \cite{levitas+idesman+olson-1999,31}, band \cite{ides-lev-CMS-1997,levitas+idesman+stein-ijss1998,wen+denis+gautier-1999}. In some cases, plasticity arrests PTs \cite{31}, see also Section \ref{growth-lath}.
For heterogeneous boundary conditions or fields in   bulk, or prescribed displacements, which control the amount or transformed phase via total transformation strain, the size of the nucleus is determined via SC conditions and extremum principle.

 Generally, only the kinetic treatment,
due to its accounting for the activation  barrier and energy,
results in the finite size of the nucleus. It will be described in Sections \ref{therm-act-kin}  and \ref{macro-act-kin}.

\subsection{Global  criterion for structural changes based on stability analyses }\label{global}

Time-independent problem formulation simplifies a solution, but nothing comes for free.
Indeed, it leads in some cases to the necessity of some additional principles.
In particular, for some problems  under the given increment of boundary  conditions
the  SC criterion and extremum principle (\ref{ijss1-179}) allow
several different solutions, e.g., nucleation in different places or propagation
of different interfaces. In most cases, at least two following solutions are  possible:
(a) the solution without SC (since all equations of continuum
mechanics can be satisfied without SC as well) and (b) the solution
with SC. That is, it is possible that SC will not occur even when  the
 SC criterion and extremum principle (\ref{ijss1-179}) are met.
Such a problem  was first revealed in  \cite{Levitas-postulate-I-1995,Levitas-postulate-II-1995}.
It was proposed to use the stability consideration to choose the unique solution among all possible solutions.
Remarkably that the {\it postulate
of realizability} was utilized again to formulate the stability criterion and corresponding extremum principle.
Since the general extremum principle \cite{Levitas-postulate-I-1995,Levitas-postulate-II-1995} is too bulky,
we will
utilize its simplified version (\ref{eq:yyyyy}) either for the
prescribed displacements $\, {\fg u} \,$ or   traction vector $\, {\fg  p} \,$ at the external  boundary $\, S \,$.

It follows from the principles  (\ref{eq:yyyyy}) or (\ref{eq:yyyyy-u})
that the stable solution minimizes the work of external stresses for prescribed
displacements and maximizes the work of external stresses at given
tractions.
 Thus, the
fulfillment of the SC criterion and extremum principle (\ref{ijss1-179}) is not sufficient for the occurrence
of SC, and only the extremum principle (Eq. (\ref{eq:yyyyy}) or (\ref{eq:yyyyy-u}))--- called the
global SC criterion---offers the final solution.   Application of stability
analysis to strain-induced nucleation at a shear-band intersection can be
found in \cite{levitas+idesman+olson-1999} and in Section \ref{Shear-band-int} and in \cite{idesman-levitas+stein-ijp-2000} and in Section  \ref{PT-fracture}  for competition between PT and fracture.

\noindent
\underline{\hspace*{17cm}}

\begin{center}
{\large Box 3.} {\bf Athermal kinetics of SCs
                     \cite{levitas+idesman+stein-ijss1998,levitas-ijp2000-1,levitas-ijp2000-2}}
\end{center}

\underline{\hspace*{17cm}}

\noindent
{\bf 1. Thermodynamic SC criterion for a nucleus }
\bey
\bar{X} \; = \;  K^0 \; .
\label{ijss1-19a}
\eey
{\bf 2. Thermodynamic SC criterion for interface propagation }

{\bf 2a. At time $t$ }

\bey
{X}_{\Sigma} \; = \; K_\Sigma \; .
\label{ijss1-19b}
\eey

{\bf 2b. At time $t+\Delta t$ }

\bey
{X}_{\Sigma \Delta} \; = \; K_{\Sigma\Delta} \; .
\label{ijss1-19-b1}
\eey
{\bf 3. Extremum principle for determination of all
unknown parameters $\; {\fg b} \,$ among all possible ${\fg b}^*$}

{\bf 3a. For a nucleus }
\bey
\bar X \,\left({\fg b}^*\right) - \, K^0 \left({\fg b}^*\right)
\, < \; 0 \; = \;
\bar X \,\left({\fg b} \right) - \, K^0 \left({\fg b}\right) \, .
\label{ijss1-179}
\eey

{\bf 3a. For an interface }

\bey
X_{\Sigma \Delta} \,\left({\fg b}^*\right) - \,   K_{\Sigma\Delta} \left({\fg b}^*\right)
\, < \; 0 \; = \;
 X_{\Sigma \Delta}  \,\left({\fg b} \right) - \, K_{\Sigma\Delta} \left({\fg b}\right) \, .
\label{ijss1-179-a}
\eey
{\bf 4. Athermal threshold $\; K^0 \,$}
\bey
K^0 \, = \,  \, L \, \sg_y \left(\theta, \, \fvep_p, \, {\fg g} \right) \, \vep_o \; .
\label{003}
\eey
{\bf 5. Extremum principle for choosing the stable solution, i.e.,
      the global SC criterion}
\bey
\int\limits_S
\int\limits^{{\fg u}_2}_{{\fg u}_1} {\fg p} \, {\fg \cdot} \,
d \, {\fg u} \; d \, S \, \Rightarrow \; {\sl min}
\quad \mbox{(for  prescribed } \; {\fg u} \;
\mbox{ at the external surface } \; S \, ) \, .
\label{eq:yyyyy}
\eey
\bey
\int\limits_S
\int\limits^{{\fg u}_2}_{{\fg u}_1} {\fg p} \, {\fg \cdot} \,
d \, {\fg u} \; d \, S \, \Rightarrow \; {\sl max}
\quad \mbox{(for  prescribed } \; {\fg p} \;
\mbox{ at the external surface } \; S \, ) \, .
\label{eq:yyyyy-u}
\eey
\underline{\hspace*{17cm}}
\vspace{1cm}

Let us summarize the main steps for the solution of the boundary-value problems for the athermal kinetics based on equations in Boxes 1-3.
The initial stages of the solution are the same as described in the problem formulation in Section \ref{Drive-force}.
We consider a body under prescribed  boundary conditions that do not change during the nucleation event (Fig. \ref{Fig-schematics}).
We choose a potential nucleation region  $V_n$, changing the field of the parameter $\, \xi  \left({\fg r} \right) \,$ from 0 to 1   in it homogeneously or heterogeneously, which   introduces  into a nucleus the transformation strain field
$\, \fvep_t \, \left({\fg r} \, , \; \xi \right) \,$ and variation of all material
properties from properties of the parent phase 1 to properties of the product phase 2. For such a prescribed deformation process in $V_n$, the  inelastic boundary  value problem and heat conduction equation should be solved incrementally, producing  fields
 $\, {\fsg } \,$, $\; \fvep_p \,$, $\fg g$, $\; \theta \,$, and all other participating fields.
Then we
 calculate the driving force $\, \bar{X} \,$ (Eq.(\ref{ijss1-19})) and athermal resistance
$\, K^0 \,$ (Eq.(\ref{003})).
Next,
we vary the possible SC region and way of variation of the transformation strain
and properties
from initial to final values in it and find such a PT region and way
of varying   transformation strain and properties, which maximizes the net driving force $\, \bar{X} \, - \, K^0 \,$.
It is important for the athermal kinetics to begin solving for the boundary conditions for which $\, \bar{X} \, - \, K^0 <0 \,$, i.e., the SC is impossible. Then we change boundary conditions incrementally to increase the net driving force until we find boundary conditions for which $\, \bar{X} \, - \, K^0 = 0 \,$ for the first time for the "optimized" nucleus.
Then we compare the obtained solution with the solution without the nucleus for the same load increment (or with other solutions with SCs, if they exist) and find out using extremum principle (\ref{eq:yyyyy}), which solution is more stable. If with the nucleus,
 we keep this nucleus in the transformed state with all corresponding fields and use as the initial conditions for the next loading increment, for which we repeat the same procedure.
If without a nucleus, then we use the solution without the nucleus as the initial conditions for the next loading increment, for which we repeat the same procedure. We do the same for all following load increments. Interface propagation does not require  special treatment. If under a given load increment, the next transformed volume is in touch with the previous one,
then the interface propagation occurs; otherwise, this is the appearance of a new nucleus. For interface propagation, $K_\Sigma$ should be used instead of $K^0$, but in most cases, this difference is neglected.

\subsection{Thermally-activated kinetics }\label{therm-act-kin}

In the equations presented in Box 4, we follow Eqs.  (\ref{meyers1-126a1-0})-(\ref{IJP-14}) as much as possible
for SC in elastic materials.  The main problem is that since the change in the Gibbs energy is not the thermodynamic driving force for SC, it cannot be used for the definition of the activation energy.  Since according to Eqs.  (\ref{ijss1-19}) and  (\ref{ijss1-179})  $(\bar{X}- K^0)m_n$ is the net thermodynamic driving force for SC, it is utilized in the definition (\ref{IJP-13-1}).  Also, Eq. (\ref{IJP-13-1}) includes additional minimization with respect to the transformation path along which a critical nucleus appears.  For example, one can consider a homogeneous SC process in the entire critical nucleus of a fixed radius, or the formation of a critical nucleus by the motion of a sharp interface from zero to the critical radius, or a homogenous SC process in the region of the smallest size, which can be treated as a nucleus within a  continuum approach and then grows to the critical size by interface propagation, or any general  scenario involving heterogeneous fields \cite{Levitas-sublim-IJP-12,Levitas-Alt-sublim-IJP-12}.  For SC in elastic materials,
the transformation path is irrelevant for the definition of  $\, Q \,$ because $\Delta G_n$ is
path-independent.  This is not the case for nucleation in an inelastic material, for which
 stress variation depends on the entire transformation process, and so does $\bar{X}$.

This fact has one more consequence. In elasticity,  if a supercritical nucleus appears,
it will grow because the Gibbs potential reduces and the dissipation rate is positive during  growth.
However,  due to the path-dependence of plastic solutions, one
cannot say whether the supercritical nucleus will grow, disappear, or remain unchanged after nucleation.
Thus, an additional condition concerning growth after appearance should be checked.

Due to the path-dependence, the driving force
for interface propagation for forward $X_\Sigma^f$ and reverse $X_\Sigma^r$
PT can be different. The driving force for   reverse SC is defined in the same way as for forward SC, but when $\xi$
varies from 1 to 0.  The kinetic equation for the interface propagation may be based on experiments, atomistic simulations, or
some dislocation models  \cite{grujicic+olson+owen-1985,grujicic+olson-1998,olson+cohen-1986,haezebrouck-87,ghosh+olson-1994-1,ghosh+olson-1994-2}.
The following scenarios  are presented in item 5 of  Box 4.

The cases 5.1 and 5.2
are similar to
PT in elastic materials, i.e., the supercritical  nucleus will grow, and the subcritical nucleus will shrink. Other cases do not have
the counterparts for PTs in elastic materials. For case 5.3, the evolution of the nucleus is determined
by the sign of the resultant interface velocity, where $f^f(X_\Sigma^f)$ and $f^r(X_\Sigma^r)$ are the kinetic equations for forward and reverse PTs, respectively.
 Case 5.4 describes the nucleus that cannot evolve, i.e., it is stable rather
than the critical nucleus. Finally, in 5.5, a gaseous  nucleus may grow independent of the interfacial driving force due to loss of mechanical stability if the pressure in the gas exceeds the resistance of the material $-\sg_n$ to plastic expansion and surface stress. For gaseous or any other hydrostatic media, athermal resistance is zero.

 The main steps for the solution of the boundary-value problems for the thermally activated kinetics are similar to those for
 athermal kinetics with the following difference. The actual nucleus is chosen by the minimization of activation  energy instead of the net driving force, and the nucleation time is defined. In addition, growth condition are checked to choose the next transforming region. The growth conditions are based on parameters at time $t+\Delta t$, i.e., again, the next transformed volume is checked like for the athermal kinetics.

The  examples of the application of equation in Box 4 for SC in an inelastic material are given in   \cite{Levitas-sublim-IJP-12,Levitas-Alt-sublim-IJP-12}
for sublimation (and melting and chemical reaction), i.e. when a nucleus is a hydrostatic medium, and in \cite{Levitas-PRB-04,Levitas-chapter-04} for nucleation of a high-pressure solid phase in a low-pressure solid phase.

Note that the typical size of a critical nucleus determined by the nucleation criterion (\ref{meyers1-126c1}) is in the range of a few to few tens of nanometers. The effect of surface energy is the leading one for such sizes.

\begin{center}
{\large Box 4.} {\bf Thermally-activated kinetics
                     \cite{Levitas-sublim-IJP-12}}
\end{center}
\noindent
\underline{\hspace*{17cm}}

{\bf 1. The Arrhenius-type kinetic equation for nucleation time}
\bey
t_s \, = \, t_0 \, \exp \left(\; - \;
\frac{Q}{k \, \theta} \right)
 .
\label{meyers1-126a1}
\eey

{\bf 2. Activation energy for the appearance of a critical nucleus}
\bey
Q=\max_{{m_{0n}}} \min_{{shape}}\min_{{position}}\min_{{path}} \min_{{structure}}(- (\bar{X}- K^0)m_n).
\label{IJP-13-1}
\eey

{\bf 3. Criterion for thermally activated nucleation}
\bey
Q= \beta k\theta, \qquad \beta= 40-80 .
\label{meyers1-126c1}
\eey

{\bf 4. Interface propagation kinetics}
\begin{eqnarray}
v_n^f = f^f(X_\Sigma^f- K_{\Sigma}^f)   \qquad {\rm for}   \qquad  X_\Sigma^f> K_{\Sigma}^f;
\nonumber\\
v_n^r = f^f(X_\Sigma^r- K_{\Sigma}^r)   \qquad {\rm for}   \qquad  X_\Sigma^r> K_{\Sigma}^r.
\label{int-prop}
\end{eqnarray}

{\bf 5. Growth conditions for a critical nucleus}
\begin{eqnarray}
{\bf 5.1.} \; X_{\Sigma\Delta}^f > K_{\Sigma\Delta}^f \qquad {\rm and} \qquad  X_{\Sigma\Delta}^r <  K_{\Sigma\Delta}^r \quad \Rightarrow \quad {\rm growth \; (supercritical \;  nucleus) } \label{6bo1}
\end{eqnarray}
\begin{eqnarray}
{\bf 5.2.} \; X_{\Sigma\Delta}^f<  K_{\Sigma\Delta}^f \qquad {\rm and} \qquad  X_{\Sigma\Delta}^r >  K_{\Sigma\Delta}^r \quad \Rightarrow \quad {\rm shrinking \; (subcritical \; nucleus) }
\label{6bo1-1}
\end{eqnarray}
\begin{eqnarray}
{\bf 5.3.} && X_{\Sigma\Delta}^f \geq  K_{\Sigma\Delta}^f \;\; {\rm and} \;\;  X_{\Sigma\Delta}^r \geq  K_{\Sigma\Delta}^r \;\; \Rightarrow \;\;
\label{6abo1}\\
&& v_{res}=   f^f(X_{\Sigma^f \Delta}- K_{\Sigma\Delta}^f)-f^r(X_{\Sigma^r \Delta}- K_{\Sigma\Delta}^r)   \quad \Rightarrow \quad
{\rm competing  \; kinetics } \
\nonumber
\end{eqnarray}
\begin{eqnarray}
{\bf 5.4.} \; X_{\Sigma\Delta}^f \leq  K_{\Sigma\Delta}^f \qquad {\rm and} \qquad  X_{\Sigma\Delta}^r \leq  K_{\Sigma\Delta}^r \quad \Rightarrow \quad {\rm stable \; (rather \; than \; critical) \; nucleus}
\label{6bbo1}
\end{eqnarray}
\begin{eqnarray}
{\bf 5.5.} \; p_g> -\sg_n+ 2\Gamma_2/R \qquad  \Rightarrow \quad {\rm growth \; of \; a \; gaseous \; nucleus \; by \; mechanical \; instability \; }
\label{sa6abo1}
\end{eqnarray}

\noindent
\underline{\hspace*{17cm}}

\subsection{ "Macroscale" thermally-activated kinetics for structural changes}\label{macro-act-kin}

As it is shown in \cite{levitas-ijp2000-1,levitas-ijp2000-2} and Section \ref{post-realiz}, the time-independent model can lead
to some contradictions. That is why development of phenomenological time-dependent kinetics
 is necessary. It is not related to the appearance of a critical nucleus, for which size and energy
(i.e., activation energy) can be calculated. We consider plausible phenomenological kinetics for the time of appearance of
an arbitrary region of a product phase of volume $V_n$ or mass $m_n$, because
 in irreversible thermodynamics, the kinetic equation between
rate and force $\;\dot{\chi}\,=1/t_s =\, f\left(X_v\, , \,{\fg .}{\fg .}{\fg .}
\right) \,$ has to be formulated. We coin this region as a macroscale nucleus.
Motivated by the approach in Section \ref{therm-act-kin}, we consider in Box 5 a {\it size-dependent}
Arrhenius-type kinetics, which includes both {\it thermal} activation and an
{\it athermal} threshold $\, K^0 \,$, see Eq.(\ref{meyers1-126a}). Here,
$\, E_a \,$ is the activation energy per unit mass at
$\, \bar{X} - K^0 \, = \, 0 \,$, which is an experimentally-determined parameter (in contrast to Eq.(\ref{IJP-13-1}) for traditional thermal activation), $\, N \, = \, 6.02 \cdot 10^{23} \,$ is
Avogadro's number (number of atoms in 1 {\sl mol}),
$\; R \, = \, 8.314 \,$ ${\sl J}/\left({\sl K \; mol}\right)$
is the gas constant, and
$\, n \,$ is the number of atoms in the volume 
which undergoes thermal fluctuations during the entire macroscopic nucleation process (an experimentally fitting parameter).
The actual activation
energy per unit mass $\; \bar{E}_a \, := \, E_a \, - \, \bar{X} \, + \, K^0 \geq\,0 $, otherwise, there is no need for thermal fluctuations, and
barrierless nucleation occurs. The characteristic time $\, t_0 \,$ has a meaning of a nucleation time for $ \bar{E}_a \, =0$.
 Since $k=R/N$, without $n$  using the same arguments as for formulating Eq.(\ref{IJP-14}), we obtain that  $\; \bar{E}_a m_n \leq (40-80)  k\theta$, i.e., the nucleus size would be similar to the size of the critical nucleus, i.e., few to tens of nanometers.
The effective temperature $\, \theta_{ef} \,$  takes into account
that the temperature may vary significantly during the SC, e.g., during CR \cite{levitas+nesterenko+meyers-1998-1,levitas+nesterenko+meyers-1998-2}.
As the simplest assumption, we define the effective temperature either as a temperature
averaged over the transformation process and transforming volume or use a similar definition in terms of inverse temperature.    As a nucleation criterion, we accept that the nucleation time is shorter or equal to the  accepted observation time $t_{ob}$.
With the help of  the postulate of realizability \cite{Levitas-postulate-I-1995,Levitas-postulate-II-1995,levitas-ijp2000-1,levitas-ijp2000-2},
{\it the principle of the \it minimum
of transformation time}, Eq.(\ref{meyers1-126c}), (or the  maximum of transformation rate) is obtained.

We used  a specific model for the interface velocity kinetics developed in \cite{ghosh+olson-1994-1,ghosh+olson-1994-2,olson+cohen-1986} and combined it with our continuum thermodynamic treatment  in \cite{31}. It is taken into account that the athermal threshold $K^0$ in Eq. (\ref{ol-3})  consists of two  parts
due to solute atoms $\, K_{\mu} \,$ and dislocation forest
hardening $K_d=B [\sg_y^A (\bar{q})-\sg_y^A (0)]$,   where $B$ is the proportionality factor, $\bar{q}$ is the accumulated plastic strain averaged over the small volume covered by a moving interface during time $\Delta t$, $\sg_y^A (\bar{q})$ is the plastic strain dependence of the yield strength of the austenite, and $A$ is a parameter in this dependence.  Dependence of $K^0$ on $\bar{q}$ is consistent with the relationship (\ref{003}) and dependence of the yield strength on $q$ for a specific steel in \cite{ghosh+olson-1994-1,ghosh+olson-1994-2}.  In the kinetic equation for the interface propagation (\ref{ol-4})
$\, v_{n\,0} \,$ is the characteristic velocity on the order of the shear-wave velocity, $\, Q_0 \,$ is the activation energy,
 $\, W_0 \,$ is the height  of the  barrier above which thermal fluctuations are not required,  and $\, p \,$ and $\, b \,$ are
constants. All material parameters estimated for the alloy {\sl Fe} -
22.31 {\sl Ni} - 2.888 {\sl Mn} are presented in \cite{ghosh+olson-1994-1,ghosh+olson-1994-2,31}, and the
application of the kinetics for the lath martensite growth is presented in \cite{31} and Section \ref{growth-lath}.

\begin{center}
{\large Box 5.} {\bf "Macroscale" thermally-activated kinetics for SCs
                     \cite{levitas-ijp2000-1,levitas-ijp2000-2,31}}
\end{center}
\noindent
\underline{\hspace*{17cm}}

{\bf 1. Thermodynamic SC criterion for  a macroscale nucleus}
\bey
\bar{X} \, \geq \, K^0 \, .
\label{meyers1-1126a}
\eey

{\bf  2. Arrhenius-type kinetic equation and nucleation criterion}
\bey
&&t_s \, = \, t_0 \, \exp \left(\; - \;
\frac{\left( { \bar{X} \, - \, K^0} \, - \, E_a \right)
m_{\, n}}{R \, \theta_{ef}} \; \frac{N}{n} \; \right) \leq t_{ob}
\qquad \mbox{at} \qquad
0 \, \leq \, \bar{X} \, - \, K^0 \, \leq \, E_a \,
\label{meyers1-126a}\\
&& t_s \, = \, \infty
\qquad \mbox{at} \qquad
\, \bar{X} \, \leq \, K^0 \, .
\nonumber
\eey

{\bf 3. Principle of the  minimum
of transformation time}
\bey
t_s \; = \; t_0 \, \exp \; - \;
\frac{ \left( \bar{X} \left({\fg b}^* \right) \, - \,
K^0 \left({\fg b}^* \right) - \, E_a \left( {\fg b}^{*} \right)
\right) m_n\left( {\fg b}^{*} \right)}{R \, \theta_{ef}\left( {\fg b}^{*} \right)} \; \frac{N}{n} \quad
\longrightarrow \;\; {\sl min} \;\, .
\label{meyers1-126c}
\eey

{\bf 4. Interface propagation criterion }
\bey
\bar{X} \geq K^0 = K_{\mu} + K_d= K_{\mu} + B [\sg_y^A (\bar{q})-\sg_y^A (0)] = K_{\mu} +  A\bar{q}^{0.5}.
\label{ol-3}
\eey

{\bf 5. Kinetic equation for an interface propagation}
\bey
v_n \, = \, v_{n\,0} \, \mbox{\sl exp} \left[ - \, \frac{Q_0}{k \, \theta}
\left(1 - \left(\frac{\bar{X}
- K^0}{W_0} \right)^p\; \right)^b \right]  \quad {\rm for } \quad 0\leq \bar{X}
- K^0\leq W_0.
\label{ol-4}
\eey

\noindent
\underline{\hspace*{17cm}}
}

A straightforward way to reduce the transformation time is to reduce the mass (i.e., size) of the nucleus as much as possible.
However, the increasing contribution of the surface energy to $\bar{X}$ will then lead to a violation of the SC criterion
(\ref{meyers1-1126a}).  In this case, the minimum nucleus size will be determined by thermodynamic "static"
constraint  $\;\, \bar{X} \left({\fg b}^* \right)
- \, K^0 \left({\fg b}^* \right) = \, 0 \;$, which should be explicitly imposed. That is why we called this nucleus a thermodynamically admissible nucleus. Corresponding equations are presented in Box 6.  The expression for the transformation time (\ref{meyers1-126a-c})   and for the principle of the minimum of the transformation time
(\ref{meyers1-126c-c}) are becoming significantly simpler.
For  mutually independent $\, E_a \,$, $\, m_n \,$ and
$\, \theta_{ef} \,$, the principle of the minimum of the transformation time  results in three principles,
namely in {\it the principle of the minimum of transforming mass}, {\it the
minimum of activation energy per unit mass}, and {\it the maximum of effective temperature}.
Since $\, E_a \,$ is a fitting constant independent of ${\fg b}^*$, and if the temperature variation is neglected,
then the main principle is {\it the principle of the minimum of transforming mass}.

\begin{center}
{\large Box 6.} {\bf "Macroscale" thermally-activated kinetics for SCs when the minimum in   principle (\ref{meyers1-126c}) violates the thermodynamic criterion of SC (\ref{meyers1-1126a}), and it is included as a constraint
                     \cite{levitas-ijp2000-1,levitas-ijp2000-2}}
\end{center}

\underline{\hspace*{17cm}}
\noindent
{\bf 1. Thermodynamic "static" SC criterion for  a macroscale nucleus}
\bey
\bar{X} \, = \, K^0 \, .
\label{meyers1-1126a-c}
\eey
{\bf  2. Arrhenius-type kinetic equation  and nucleation criterion for a thermodynamically admissible nucleus satisfying criterion  (\ref{meyers1-1126a-c})}
\bey
t_s \, = \, t_0 \, \exp \left(
\frac{  \, E_a m_{ n}}{R \, \theta_{ef}} \; \frac{N}{n} \; \right) \leq t_{ob}.
\label{meyers1-126a-c}
\eey
{\bf 3. Principle of the  minimum
of transformation time}
\bey
t_s   =   t_0 \, \exp
\frac{  E_a \left( {\fg b}^{*}
\right) m_n\left( {\fg b}^{*} \right)}{R \, \theta_{ef}\left( {\fg b}^{*} \right)}   \frac{N}{n} \quad
\longrightarrow \;\; {\sl min}  \quad   \Rightarrow \quad  \frac{E_a \left({\fg b}^{*} \right) \, m_n\left( {\fg b}^{*} \right)}{\theta_{ef}\left( {\fg b}^{*} \right)}
\; \rightarrow \; {\sl min}.
\label{meyers1-126c-c}
\eey
{\bf 4. For mutually independent $\, E_a \,$, $\, m_n \,$ and
$\, \theta_{ef} \,$}
\bey
E_a \left({\fg b}^{*} \right) \; \rightarrow \; {\sl min} \; ;
\qquad\qquad
m_n \left( {\fg b}^{*} \right) \; \rightarrow \; {\sl min} \; ;
\qquad\qquad
\theta_{ef} \left( {\fg b}^{*} \right) \; \rightarrow \; {\sl max} \; .
\label{ijss1-178++}
\eey
\noindent
\underline{\hspace*{17cm}}

For homogenous fields within a macroscale nucleus, the formulated equations are further elaborated in Box 7.
Here, we consider a nucleus of  arbitrary shape with a surface area $\Sigma_n$
and  characteristic size $\, \displaystyle{ \bar{r} \, = \; \frac{V_n}{\Sigma_n}} \;$.
The thermodynamic SC criterion for  a macroscale nucleus (\ref{meyers1-1126a-h}) (which is equality, like for time-independent kinetics) allows us to introduce the new concept of  a {\it thermodynamically admissible nucleus } (see also Fig.
\ref{Fig-energy-crit-nucl}), for which characteristic size is determined by Eq.(\ref{1126a-h}).
The size of a thermodynamically admissible nucleus is larger than the size of a critical nucleus, which has an equal probability
of growing and disappearing.
It may be qualitatively related to the concept of the operational nucleus in \cite{olson+cohen-1986}, which reached the size required for fast growth. However, as it is described in Box 4, due to the path-dependence plastic flow theory and  the driving
forces for nucleation and growth, there is no guarantee that our thermodynamically admissible nucleus will grow; this should be checked for each specific case.

Eq. (\ref{meyers1-1126a-h}) allows us to elaborate Eq.(\ref{meyers1-126a-h}) for the time of SC. While applying the principle of the minimum of transformation time (\ref{meyers1-126c-h}), we,
for simplicity, assume constant temperature, $K^0$, and $E_a$.
This principle then results in the principle of minimum volume (or mass) of the nucleus.

\begin{center}
{\large Box 7.} {\bf "Macroscale" thermally-activated kinetics for SCs in Box 6 for a thermodynamically admissible nucleus for homogeneous fields
                     \cite{levitas-ijp2000-1,levitas-ijp2000-2}}
\end{center}

\underline{\hspace*{17cm}}

\noindent
{\bf 1. Thermodynamic "static" SC criterion for a nucleus}
\bey
\left(\, X \, - \, K^0 \,\right) \, \rho \,
\Sigma_n \, \bar{r} \, = \,\Delta  \Gamma \, \Sigma_n \quad \Rightarrow  \left(\, X \, - \, K^0 \,\right) \, \rho \,
  \bar{r} \, = \,\Delta \Gamma \,   .
\label{meyers1-1126a-h}
\eey
{\bf  2.  The characteristic size of a thermodynamically admissible nucleus}
\bey
 \bar{r}= \frac{\Delta\Gamma }{\rho \left(\, X \, - \, K^0 \,\right) } \,   .
\label{1126a-h}
\eey
{\bf  3. Arrhenius-type kinetic equation  and nucleation criterion for a thermodynamically admissible nucleus satisfying criterion  (\ref{meyers1-1126a-h})}
\bey
t_s  =  t_o   {\sl exp} \; \left(
\frac{E_a  \rho
\Sigma_n  \bar{r} }{R  \theta_{ef} }
\frac{N}{n}  \right) =t_o   {\sl exp} \; \left(
\frac{E_a \Delta\Gamma \Sigma_n  }{ \left(\, X \, - \, K^0 \,\right)  R  \theta_{ef} }
\frac{N}{n}  \right)  \leq t_{ob}.
\label{meyers1-126a-h}
\eey
{\bf 4. Principle of the  minimum
of transformation time}
\bey
&& t_s   = t_o   {\sl exp}   \left(
\frac{E_a  \rho
\Sigma_n\left( {\fg b}^{*} \right)  \bar{r}\left( {\fg b}^{*} \right) }{R  \theta_{ef} }
\frac{N}{n}  \right)  =  t_o   {\sl exp}   \left(
\frac{ E_a  \Delta \Gamma \Sigma_n\left( {\fg b}^{*} \right) }{ \left(\, X\left( {\fg b}^{*} \right) \, - \, K^0 \,\right)  R  \theta_{ef} }
\frac{N}{n}  \right)
 \longrightarrow  {\sl min}   \quad \Rightarrow
 \nonumber\\
 &&\Sigma_n\left( {\fg b}^{*} \right)  \bar{r}\left( {\fg b}^{*} \right)
\longrightarrow {\sl min}.
\label{meyers1-126c-h}
\eey

\noindent
\underline{\hspace*{17cm}}

When a change in surface energy $\Delta \Gamma $ is very small or even zero,
according to Eq.(\ref{1126a-h}), it is possible  that the transforming mass is becoming smaller  than the mass of
a single  atom or molecule
or  the mass of $\, n \,$ atoms, which undergo thermal fluctuations. This should be avoided by the constraint
(\ref{n m_a}), which changes the equations in Box 5 to those in Box 8. It is taken into account in Eqs. (\ref{meyers1-126anm}) and (\ref{meyers1-126cmn})  where $\, k\, = \, \frac{R}{N}$
and the mass of the nucleus is fixed.

\begin{center}
{\large Box 8.} {\bf "Macroscale" thermally-activated kinetics for SCs in Box 5 when a mass of a nucleus is smaller than mass of $n$ atoms  $n m_a$
                     \cite{levitas-ijp2000-1,levitas-ijp2000-2}}
\end{center}

{\bf 1. The constraint for   a mass of a nucleus }

\bey
m_n=n m_a .
\label{n m_a}
\eey

{\bf 2. Thermodynamic SC criterion for  a macroscale nucleus}
\bey
\bar{X} \, \geq \, K^0 \, .
\label{meyers1-1126ana}
\eey

{\bf  3. Arrhenius-type kinetic equation}
\bey
&&t_s \, = \, t_0 \, \exp \left(\; - \;
\frac{\left( { \bar{X} \, - \, K^0} \, - \, E_a \right)
m_{\, a}}{k \, \theta_{ef}} \;  \; \right)
\qquad \mbox{at} \qquad
0 \, \leq \, \bar{X} \, - \, K^0 \, \leq \, E_a \,
\label{meyers1-126anm}\\
&& t_s \, = \, \infty
\qquad \mbox{at} \qquad
\, \bar{X} \, \leq \, K^0 \, .
\nonumber
\eey

{\bf 4. Principle of the  minimum
of transformation time}
\bey
t_s \; = \; t_0 \, \exp \; - \;
\frac{ \left( \bar{X} \left({\fg b}^* \right) \, - \,
K^0 \left({\fg b}^* \right) - \, E_a \left( {\fg b}^{*} \right)
\right) m_a}{k \, \theta_{ef}\left( {\fg b}^{*} \right)} \;   \quad
\longrightarrow \;\; {\sl min} \;\, .
\label{meyers1-126cmn}
\eey
\noindent
\underline{\hspace*{17cm}}

\par
A general scheme for the application of SC criterion and the principle of the minimum of transformation time
for the macroscale nucleation kinetics is similar to that for thermally activated kinetics. For the interface propagation,
a small shift of the interface is produced at each step and the volume covered as an interface is considered as a new transformed region. After calculating $X$ and $K^0$ and the maximization of $X-K^0$ with respect to internal structure (if any), interface velocity is calculated using Eq.(\ref{ol-4}).

\subsection{Comparison with alternative approaches }

Since we are not aware of  any previous work on CRs in plastic materials within a framework of the materials without an intermediate stable state but papers by  \cite{levitas-ijp2000-1,levitas-ijp2000-2,levitas+nesterenko+meyers-1998-1,levitas+nesterenko+meyers-1998-2},
we will focus on the PTs only. However, as it is demonstrated in  \cite{levitas-ijp2000-1,levitas-ijp2000-2,levitas+nesterenko+meyers-1998-1,levitas+nesterenko+meyers-1998-2},
formal continuum theory for PTs and CRs is practically the same.
\par
\subsubsection{"Macroscale nucleation"}
As it is usual in physical literature, instead of a general theory, analytical solutions for some tractable model
problems were found first.
Melting of a small spherical particle in an elastoplastic space was presented in the first publication \cite{lifshitz+gulida-1952} on the topic. The
appearance of the spherical nucleus in a finite-size sphere under  external
pressure was treated in \cite{roitburd+temkin-1986}.
The ellipsoidal nucleus
in an infinite space with the stress-free boundary was analyzed in  \cite{kaganova+roitburd-1989}. The shape of an inclusion
corresponding to minimum energy losses during growth was found, much like the counterpart of a "critical macroscopic nucleus," while surface energy was not included in this and most of the other works on PTs in elastoplastic materials. A Landau-type theory was applied in \cite{baryachtar+enilevskiy+etal-1986} to study the emergence  of spherical and plate-like regions of the product phase in
an infinite elastoplastic space without external stresses and neglecting surface energy.
In these works, the deformation theory of plasticity was utilized, which is
thermodynamically equivalent to a nonlinear elasticity instead of elastoplasticity. In these papers, the PT criterion and extremum principle for the determination of some
unknown parameters are the same as for PT in elastic materials, i.e.,
the total Gibbs  energy    is minimized. As it is discussed above,
this principle is not applicable  for elastoplastic materials.
Still, due to specific simple problems, some results are either correct or give reasonable hints
on some new effects. Thus, it was found in
\cite{roitburd+temkin-1986} that (in contrast to elastic materials)
nucleation and interface propagation conditions in elastoplastic
materials are not equivalent.
\par
Only in  \cite{roitburd+temkin-1986}, for  the appearance of a spherical nucleus, was it hypothesized that  the mechanical work  should be smaller than
the change in thermal energy, which for some simplifications coincides with the driving force in Eq. (\ref{eq:6}). Unfortunately,  in the next paper \cite{kaganova+roitburd-1989}  the principle of the minimum of
free energy was implemented again. Numerous investigations of PT in elastoplastic materials
in \cite{fischer+berveiller+etal-1994,marketz+fischer-1994,marketz+fischer-cms1994,marketz+fischer-1995}
were also based on  the comparison of Gibbs free energy before and after PT.
A very different approach was developed in \cite{olson+cohen-72,olson+cohen-76,olson+cohen-1986}
based on  the dislocation representation of the martensitic interfaces and considering heterogeneous nucleation at
the dislocation wall and corresponding overall nucleation kinetics, as well as martensite growth in an elastic and elastoplastic material.
Various physical, material, and
mechanical features of the nucleation theory were critically reviewed in
\cite{olson+roytburd-1995}.

In \cite{53,levitas-mrc1995,levitas-1995-phys-III-41,Levitas-JMPS-I-97,Levitas-JMPS-II-97}
we formulated a nucleation condition in the form that the dissipation
increment due to PT only (excluding plastic and other types of
dissipation) reaches its experimentally determined value (see
Eq.(\ref{ijss1-19a})). To justify this
condition,  a local description of PTs, similar to the ones presented in Boxes 1-3,  was developed in \cite{levitas-mrc1996,54,levitas-ijss1998}.
The  postulate of realizability
\cite{Levitas-postulate-I-1995,Levitas-postulate-II-1995,Levitas-JMPS-I-97,Levitas-JMPS-I-97,levitas-ijss1998} was formulated and applied to the  derivation of  the
extremum principle (\ref{ijss1-19}) for finding all unknown
parameters of the  nucleus, see \cite{53,levitas-mrc1995,levitas-1995-phys-III-41,levitas-mrc1996,54,Levitas-JMPS-I-97,levitas-ijss1998} for athermal kinetics. The "macroscale"  thermally activated kinetics equations presented in Boxes 5-8, were suggested in \cite{levitas-ijp2000-1,levitas-ijp2000-2} for small strains with application to some analytical solutions and in \cite{idesman-levitas+stein-ijp-2000,31} for finite strains. An approach in line with the classical nucleation theory but for elastoplastic material presented in Box 4 was developed in
\cite{Levitas-PRB-04,Levitas-chapter-04} for nucleation at a dislocation pileup, in \cite{Levitas-Alt-PRL-08,Levitas-Alt-VM-PRB-09,Levitas-sublim-IJP-12,Levitas-Alt-sublim-IJP-12} for sublimation, melting, CR, and sublimation via the virtual melting within elastoplastic material, and in \cite{Levitas-Alt-void-ActaMat-11} for void nucleation due to fracture, sublimation, sublimation via the virtual melting, and melting and evaporation within an elastoplastic space.

\par
\subsubsection{Interface propagation}
In material science literature \cite{grujicic+olson+owen-1985,grujicic+olson-1998,olson+cohen-1986,haezebrouck-87,ghosh+olson-1994-1,ghosh+olson-1994-2}, various contributions to the driving
force for interface propagation were discussed.  The growth of a thin ellipsoidal
martensitic region in a viscoplastic material was approximately modeled in \cite{olson+cohen-1986,haezebrouck-87}.
Despite  very approximate stress-strain fields,
some very important features were revealed. In particular, it was found that plastic deformation near the tip of an  inclusion stops its lengthening. The first FEM treatment  of the appearance and thickening of a martensitic
plate with fixed ends was performed  in
\cite{marketz+fischer-1994,marketz+fischer-cms1994}.
In \cite{roitburd+temkin-1986},
 growth of a spherical
region  in a spherical elastoplastic matrix was solved.  The defect heredity for the same problem was treated for the first time in \cite{kaganova+roitburd-1987}. The driving force for PT was not localized to an interface, but it represented  the variation of the Gibbs free
energy combined with  the plastically  dissipated heat  in the whole
body. In the next paper by  \cite{kaganova+roitburd-1989},
an alternative approach for the thermodynamic equilibrium for an elastoplastic
ellipsoidal region  in the elastic matrix was utilized.
This means that the correct approach to the interface propagation and final equation were not known in physical literature.  All the balance equations for the points
of a propagating phase interface in a viscoplastic material, including PT criterion,
are obtained in \cite{kondaurov+nikitin-1986}. However, the  PT is assumed to be much
faster than the  plastic relaxation, and the plastic strain
increment does not have a jump at the phase interface. In this case, the PT
conditions are the same as for elastic materials, which does not explain
the strong influence  of plastic strain  on PT.
In this case, the plastic dissipation at the
interface is absent, and our  driving force (\ref{ijss1-123}) for isothermal processes coincides with the Eshelby
driving force \cite{ehselby-1970}, i.e., like for elastic materials.
\par
The Eshelby thermodynamic
driving force was suggested to be used for the interface propagation  condition in elastoplastic materials in  \cite{levitas-kniga-1992} and independently in
\cite{fischer+berveiller+etal-1994,marketz+fischer-1994,marketz+fischer-cms1994}. This was similar to the condition for elastic materials with an athermal threshold, but different from   Eq. (\ref{ijss1-123}) even for the isothermal process.
Note that the Eshelby driving force represents the total dissipation increment at the moving interface, including plastic dissipation and dissipation due to changes in internal variables not related to SCs.

An approach in which the thermodynamic driving force for
nucleation and interface propagation represent the dissipation increment
due to PT only, i.e., excluding all other types of dissipation---in particular,
plastic dissipation  has been developed in  \cite{53,Levitas-postulate-I-1995,Levitas-postulate-II-1995,levitas-mrc1995,levitas-1995-phys-III-41,levitas-ijss1998}.  The result coincides with the isothermal version of Eq. (\ref{ijss1-123}).
This expression was justified in \cite{levitas-mrc1996,54,levitas-ijss1998},
utilizing a local description of PTs.
\par
Thus, two different expressions for the driving force for interface propagation in plastic materials were used in literature:
based on the Eshelby driving force in
\cite{cherkaoui+berveiller+sabar-1998,cherkaoui+berveiller-aam2000,fischer+reisner-1998,fischer+reisner+werner+etal-2000} and based on the
dissipation increment due to PT only in
\cite{Idesman-CMAME-99,levitas+idesman+stein-ijss1998,levitas-ijss1998,levitas-ijp2000-1,levitas-ijp2000-2,31}.
It was shown in \cite{Levitas-IJP-02} that
using the Eshelby driving force, along with its maximization, leads to a conceptual contradiction for plastic materials. In contrast,  Eq. (\ref{ijss1-123}) does not exhibit this type of contradiction. Recently, the correctness of Eq. (\ref{ijss1-123})
and phase equilibrium condition  (\ref{ijss1-19b})  was confirmed utilizing the phase field approach to coupled  PTs and discrete dislocations at the nanoscale in \cite{Javanbakht-Levitas-PRB-16,Javanbakht-Levitas-JMS-18} and coupled PTs and slip bands at the microscale in \cite{Levitas-etal-PRL-18,Esfahani-Levitas-AM-20}.
\par
A combination  of the strict
equation for the driving force  with physically-based expressions for
athermal and thermal parts of the interfacial friction from
\cite{ghosh+olson-1994-1,ghosh+olson-1994-2,olson+cohen-1986} was suggested in
\cite{31}, and it was applied to the martensite growth problem, see  Section \ref{growth-lath}.
\par
\subsubsection{Extremum principle for PT in plastic materials}
Maximizing the mechanical work in order to find the habit-plane
(or invariant plane strain) variant was suggested in  \cite{patel+cohen-1953}. When the invariant plane strain includes dislocation plastic shear in addition to the transformation
(Bain) strain and   lattice rotation, this represents the maximization
of the Eshelby thermodynamic driving force for the  interface propagation and total dissipation increment for nucleation.
However, when the lattice invariant shear includes twinning only, this represents the maximization of  the transformation work,
since twinning in martensite represents the appearance of two twin-related martensitic variants, which is a part of the PT process.
This case is equivalent to the maximization
of  the  total dissipation increment for nucleation and interface propagation.
The Eshelby driving force was also maximized with
respect to the  habit-plane variant in
\cite{fischer+berveiller+etal-1994,marketz+fischer-1994,marketz+fischer-cms1994,fischer+reisner-1998,cherkaoui+berveiller+sabar-1998,
cherkaoui+berveiller-aam2000}.
\par
The transformation work uses local stress in place of nucleation
before the PT was maximized in \cite{ganghoffer+etal-2001,marketz+fischer-1995}.
This generalizes the
extremum principle in \cite{patel+cohen-1953} for the  presence of internal
stresses. However, this contradicts for the limiting case of elastic
materials with the principle of the minimum of Gibbs energy, because the variation of stresses
in the course of  the PT is not taken into account. It leads to significant errors because
stresses change drastically during the PT (i.e., growth of transformation strain)
and even change a sign (see, e.g. \cite{Levitas-JMPS-I-97,levitas-ijp2000-1,levitas-ijp2000-2,roitburd+temkin-1986}
and Section \ref{ellips-incl}). In \cite{fischer+berveiller+etal-1994,marketz+fischer-1994,marketz+fischer-cms1994,marketz+fischer-1995},
the difference in the Gibbs  energies after and before the  PT governed the PT.
In  \cite{wen+denis+gautier-1999}, an alternative potential was suggested
to be maximized. The common point of  the above-mentioned works  is that
the PT conditions are not directly related to the second law of
thermodynamics and dissipation due to the PT. That is why they were not justified, and it is
problematic  to comprehend the physical sense of the proposed criteria and
extremum principles and strictly compare them.
\par
 A plausible
assumption called the postulate of the realizability  was formulated in  \cite{Levitas-postulate-I-1995,Levitas-postulate-II-1995,Levitas-JMPS-I-97,levitas-ijss1998}, which results in the
maximization of the net driving force $\, \bar{X} - K \,$, both for macroscale nucleation and
interface motion (Eq. (\ref{ijss1-179}) for athermal kinetics. It involves the justified expressions   (\ref{ijss1-180}) and  (\ref{ijss1-19}) for $\bar{X}$ and  (\ref{ijss1-123})  for $X_\Sigma$, i.e., based on the dissipation increment due to PT only.
At constant $\, K$, this
principle simplifies to  the maximization of $\, \bar{X} \,$ or $X_\Sigma$, and, in the simplest case, to the principle
of maximum   transformation work, which takes into account  the stress variation in
the nucleus during the transformation process. Such transformation work  significantly differs
from $\, {\fsg } \, {\fg :} \, \fvep_t \,$ and for some problems can possess the opposite sign
\cite{levitas+idesman+stein-ijss1998}. For thermally activated kinetics, the principle of the minimum transformation time was derived  in  \cite{levitas-ijp2000-1,levitas-ijp2000-2} using the postulate of realizability
(see Eq. (\ref{meyers1-126c})). One benefit is that such a principle was intuitively formulated and routinely used
in material science \cite{fisher,turnbull+fisher-49,Porter-1992,olson+cohen-1986} for any nucleation events as a part of a concept of a critical nucleus (see Eqs. (\ref{meyers1-126a1-0}) and (\ref{IJP-13})). Indeed, the critical nucleus corresponds to the minimax
of the energy, maximum with respect to the size and minimum  with respect to all other parameters (shape, position, etc.)
From a practical point of view, our main conceptual result for thermally activated kinetics is the expression (\ref{IJP-13-1}) for the activation energy for inelastic materials, which subtracts all types of dissipations, but due to PT, as well as athermal resistance.
At the same time, arriving at an intuitively  known principle of the minimum of SC time using the postulate of realizability
increases its plausibility for other and more general systems. Furthermore, the postulate of realizability
can be used to derive some known and new extremum principles in various fields,
like the plastic flow rule and the problem of plastic spin in plasticity, nonlinear nonequilibrium thermodynamics, twinning, ductile fracture, CRs, and stability analysis
\cite{levitas+idesman+stein-ijss1998,Idesman-CMAME-99,Levitas-postulate-I-1995,Levitas-postulate-II-1995,levitas-ijss1998,57,levitas-ijp2000-1,levitas-ijp2000-2,63,levitas+nesterenko+meyers-1998-1,levitas+nesterenko+meyers-1998-2,
levitas+idesman+olson-1999,65,31}.
\par
The idea that the solution without PT (even if PT criterion could be met) is one of the possible solutions to be compared with the solution with PT was formulated in  \cite{Levitas-postulate-I-1995,Levitas-postulate-II-1995,levitas-ijss1998}.
This led to the formulation of the global SC criterion based on stability analysis and its specific expression (\ref{eq:yyyyy}),
derived again with the help of the postulate of realizability
in \cite{Levitas-postulate-I-1995,Levitas-postulate-II-1995,levitas-ijss1998}. The nontrivial choice of the stable solution among several possible ones  was
demonstrated in  \cite{Idesman-CMAME-99,Levitas-postulate-I-1995,Levitas-postulate-II-1995,levitas+idesman+olson-1999} for PT and in
\cite{idesman-levitas+stein-ijp-2000} for fracture.

\section{Finite strain formalism}\label{Finite strain}

As discussed in the Introduction, for various PTs the transformation strains are quite large.
High pressure can cause large elastic strains.
For perfect crystal deviatoric strains are often finite. For example, based on the first principle simulations, under uniaxial compression elastic strain reaches 0.16 for Si I \cite {Zarkevichetal-18} and 0.38 for graphite \cite{Gaoetal-17}  before lattice instability and initiation transformation
to Si II and diamond, respectively.
For nanotechnology applications, large elastic strains can be reached in defect-free volumes (e.g., nanofilms, nanotubes, and nanowires) and near strong stress concentrators (like dislocation pileups) and lattice misfits, when plastic relaxation is suppressed.
TRIP strain can be very large, theoretically infinite, in a thin layer (see Eq. (\ref{meyers1-207}).
One of the great examples of large plastic deformations during thermally-induced lath martensite nucleation in steel, for which
 invariant plane shear strain is 0.2 and volumetric transformation strain is just 0.02, is presented in \cite{31} (see Section \ref{growth-lath}).
Also, for plastic strain-induced PTs, applied plastic shear during high-pressure torsion can be $1 - $10 and larger \cite{Blank-Estrin-2014,levitasetal-SiC-12,Edalati-Horita-16,bridgman-1937,bridgman-1947,08,zharov-1984,zharov-1994}.
Additionally, lattice and material rotations can be finite even for small transformation strains.
The crystallographic theory of the martensite \cite{wayman-64,Ball-James-92,Bhattacharya-book} includes finite transformation strains
and rotations, as well as lattice-invariant shears due to twinning (or slip) but neglects elastic strain. Continuum thermodynamic theory of PTs in elastic materials that also includes interfacial instabilities but does not discuss crystallography of the martensite is presented by \cite{Grinfeld-1991}.

Thermodynamic theory for large elastic, transformational, and plastic deformations and rotations was drafted in \cite{54} and developed in \cite{levitas-ijss1998}, and is used for various analytical \cite{levitas-ijss1998,levitas+nesterenko+meyers-1998-1,levitas+nesterenko+meyers-1998-2,levitas-zamm1998,Levitas-Alt-PRL-08,Levitas-sublim-IJP-12,Levitas-Alt-sublim-IJP-12,Levitas-IJP-02} and numerical \cite{31,levitas+idesman+olson-1999,Idesman-CMAME-99} solutions.
A summary of this theory is contained in Boxes 9 and 10. A few equations in Box 9 are similar to those in Box 11 in Section \ref{PT-disl} for the PFA to PTs and discrete dislocations.

Multiplicative decomposition of the deformation gradient ${\fg F}$ (\ref{n1ag-2-0}) is justified in \cite{Levitas-Javanbakht-15-1}  as a noncontradictory and most economical way to describe discrete dislocation plasticity
in {\sf A} and {\sf M}$_i$ and inheritance of dislocation during cyclic direct-reverse PTs. Specific sequence of the terms in
Eq. (\ref{n1ag-2-0}) is important. The simplest way to justify the sequence in \cite{levitas-ijss1998,Levitas-Javanbakht-15-1} was that ${\fg F}_e {\fg \cdot}{\fg U}_t$
represents deformation of a crystal lattice and in some theories \cite{Lookman-prl-08,01,falk-83,reid+olson+moran-98,falk+konopko,Barsch-84,Jacob-84,abeyaratne-05} it is not decomposed into elastic and transformational parts. Thermoelastic deformation gradient ${\fg F}_e$ can be multiplicatively decomposed into elastic and thermal parts  \cite{levitas-ijss1998,levitas-book96}.
 Theory in \cite{levitas-ijss1998} includes additional terms in the multiplicative decomposition, namely, plastic deformation gradients in {\sf A},   {\sf M}$_i$, and during the transformation process. However, they never were used for the solution of a specific problem. As analyzed in \cite{Levitas-Javanbakht-15-1}, splitting plastic deformation gradients into several components makes this theory unnecessararily  sophisticated and introduces undesired features.
Some general approaches for justification of kinematic decomposition based on physical principles are presented in
\cite{levitas-ijss1998,levitas-book96}.

\begin{figure}[htp]
\centering
\includegraphics[width=0.45\textwidth]{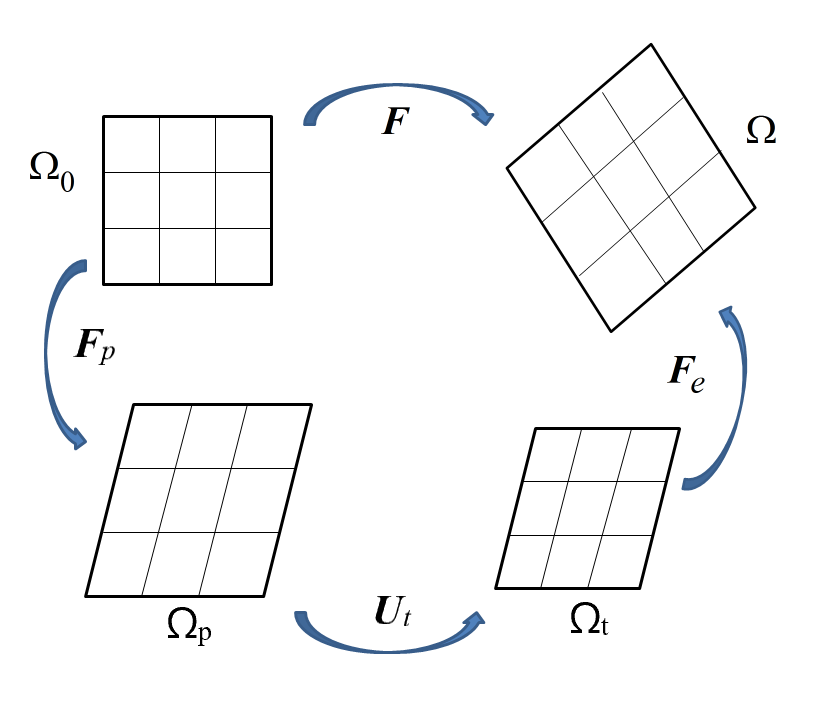}
\caption{ Multiplicative decomposition of
the deformation gradient into elastic ${\fg F}_e $,  transformational ${\fg U}_t $, and plastic ${\fg F}_t $ parts. Besides the reference stress-free configuration $\Omega_0$ corresponding with $\fg F=\fg I$ and the  deformed configuration $\Omega$, additional stress-free configuration  $\Omega_t$, obtained after elastic unloading without reverse PT from the current configuration $\Omega$, and the stress-free configuration  $\Omega_p$, obtained after elastic unloading with reverse PT from $\Omega$ are  introduced. Parameters defined in these configurations will be marked with corresponding subscripts.
\label{kinematics}}
\end{figure}

 In addition to the reference stress-free configuration $\Omega_0$ corresponding with $\fg F=\fg I$ and the current deformed configuration $\Omega$, additional stress-free configuration  $\Omega_t$, obtained after elastic unloading without reverse PT from the current configuration $\Omega$, and the stress-free configuration  $\Omega_p$, obtained after elastic unloading with reverse PT from $\Omega$ are  introduced  (Fig. \ref{kinematics}). Parameters defined in these configurations will be marked with corresponding subscripts.
Elastic and plastic Lagrangian strain tensors  used in the constitutive equations are introduced in Eq. (\ref{n1ag-2-01-}). Eq. (\ref{jac-0}) introduces Jacobian determinants describing ratios of elemental volumes (and mass densities) in different configurations.  Internal time $\xi$ is introduced in Eq. (\ref{ijss1-1+0}) the same way as for small strains, but in terms of the finite-strain measure of the transformation strain. Multiplicative decomposition (\ref{n1ag-2-0}) of the deformation gradient leads to the additive decomposition of the deformations rate Eq. (\ref{hF98g9a}).
The Helmholtz free energy (\ref{helm-2-0}) consists of elastic and thermal parts.
Note the elasticity rule is determined experimentally or with atomistic simulations in configuration $\Omega_t$, which is the reference state for ${\fg F}_e$. To transform it to the energy per unit mass, the multiplier $J_t /\rho_0$ is used:
$J_t $ transforms it in the energy per unit reference volume and $1\rho_0$ in the energy per unit mass.
Such a finite-strain correction was introduced in \cite{Levitas-Attariani-SciRep-13,Levitas-Attariani-JMPS-14}
 for large anisotropic compositional expansion during lithiation-delithiation of silicon and then in \cite{Levitas-14b} for PFA to PTs.
The simplest expression for elastic energy (\ref{hF98g9a}) is assumed, while higher-order terms can be added \cite{levitas-ijss1998,Levitas-Javanbakht-JMPS-14}.
Application of thermodynamic laws results, in particular, to the expressions (\ref{Elrul1-2-0}) and (\ref{Elrul2-2-0}) for the first Piola-Kirchhoff ${\fg P}$ and Cauchy $\fsg$ stress tensor, which are presented in the general form and for elastic energy (\ref{hF98g9a}).  Expressions (\ref{ijss1-12a-0})-(\ref{ijss1-12b-0}) for dissipation rate due to plastic flow $D_p $ and variation of the internal variable $D_g$, corresponding dissipative forces, the yield conditions and evolution equations for plastic strain and internal variables appear similar to those at small strain but using corresponding finite-strain measures. Specification for single crystals will be presented in Box 11.

\bec
{\large \bf Box 9. }{\bf Local equations describing phase transitions and plasticity
                     \cite{levitas-mrc1996,54,levitas-ijss1998,Idesman-CMAME-99,idesman-levitas+stein-ijp-2000,Levitas-Javanbakht-JMPS-14}}
\eec

{\bf 1. Kinematics}

1.1. Multiplicative decomposition of the deformation gradient ${\fg F}$
into thermoelastic ${\fg F}_e$, transformation ${\fg U}_t$, and plastic ${\fg F}_p$ parts
\bey
{\fg F}  ={\fg F}_e {\fg \cdot}{\fg U}_t{\fg \cdot}{\fg F}_p. 
\label{n1ag-2-0}
\eey
1.2.  Elastic and plastic Lagrangian strain tensors
\bey
\fg E_e= 0.5({\fg F}_e^T \cdot {\fg F}_e - {\fg I}) ; \qquad \fg E_p= 0.5({\fg F}_p^T \cdot {\fg F}_p - {\fg I}).
\label{n1ag-2-01-}
\eey

1.2. Jacobian determinants
\bey
&& J:=\frac{dV}{dV_0}=\frac{\rho_0}{\rho}=det {\fg F};  \quad J_e:=\frac{d V}{dV_t} =\frac{\rho_t}{\rho_0}=det {\fg F}_e;
\nonumber\\
&& J_{tp}:=\frac{d V_t}{dV_p} =\frac{\rho_p}{\rho_t}=det {\fg U}_t \, det{\fg F}_p = det {\fg U}_t = J_t; \quad J_p:=  det {\fg F}_p=1;
\qquad J=J_e  J_t,
\label{jac-0}
\eey
where  $dV_0$ ($\rho_0$), $dV_t$ ($\rho_t$),  $dV_p$ ($\rho_p$), and $dV$ ($\rho$) are the elemental volumes (mass densities) in the reference $\Omega_0$, transformed $\Omega_t$, plastically deformed $\Omega_p$,   and the actual ($\Omega$) configurations, respectively.

{1.3. Internal time $\xi$}
\begin{eqnarray}
\xi \; := \;
\frac{\mid \fvep_t - \fvep_{t1} \mid}{\mid \fvep_{t2}-\fvep_{t1} \mid}
 \qquad  0 \, \leq \, \xi \, \leq \, 1 \qquad   \Rightarrow \qquad {\fg U}_t =  \fg I + \xi (\fvep_{t2}-\fvep_{t1}).
\label{ijss1-1+0}
\end{eqnarray}

1.4. Decomposition of the deformation rate ${\fg d}$
\bey
&&\fg d := \left(\fg l \right)_s =\left(\dot{\fg F} \cdot {\fg F}^{-1}\right)_s = \left(\dot{\fg F}_{e}\cdot {\fg F}_{e}^{-1}\right)_s + \fg d_t  + \fg d_p ;
\nonumber\\
&& \fg d_t = \left({\fg F}_{e}\cdot \dot{\fg U}_t
\cdot {\fg U}_t^{-1}\cdot {\fg F}_{e}^{-1}\right)_s= \left({\fg F}_{e}\cdot (\fvep_{t2}-\fvep_{t1})
\cdot {\fg U}_t^{-1}\cdot {\fg F}_{e}^{-1}\right)_s \dot{\xi};
\nonumber\\
&&\fg d_p= \left({\fg F}_e{\fg \cdot}{\fg U}_t  {\fg \cdot} \dot{\fg F}_p  \cdot  {\fg F}_{p}^{-1} \cdot {\fg U}_t^{-1} \cdot {\fg F}_{e}^{-1}\right)_s.
 \label{hF98g9a}
\eey

\noindent
{\bf 2. Helmholtz free energy per unit mass}
\bey
\psi = \psi \left(\fg E_e, \, \theta,  \,\fg E_p,\,  {\fg g}, \,
  \xi \right) =\frac{J_t}{\rho_0}\psi^e \left(\fg E_e, \, \theta, \, \fg E_p, \, {\fg g},  \, \xi \right)+\psi^\theta \left(\theta, \, \fg E_p, \, {\fg g}, \,
 \xi \right) .
\label{helm-2-0}
\eey

{\bf 3. Elastic energy per unit volume in $\Omega_t$}
\bey
&&
 \psi^e=\frac{1}{2} {\fg E_e} {\fg :}{\fg C}(\xi) {\fg :} {\fg E_e},
\label{Eeneg-2-0}
\eey
where   $\fg C$ is the fourth-rank tensor of elastic moduli.

\noindent
{\bf 4. First Piola-Kirchhoff ${\fg P}$ and Cauchy $\fsg$ stress tensor}

\bey
&&{\fg P}= J_t {\fg F_e}\fg\cdot \frac{\partial \psi^e}{\partial {\fg E_e}} \fg\cdot {\fg U^{-1}_t} \fg\cdot {\fg F^{T-1}_p} =  J_t
{\fg F_e}\fg\cdot {\fg C}{\fg :} {\fg E_e}  \fg\cdot {\fg U^{-1}_t} \fg\cdot {\fg F^{T-1}_p};
\label{Elrul1-2-0}
\eey
\bey
&& \fg \sigma = \frac{1}{J_e} {\fg F_e}\fg\cdot \frac{\partial \psi^e}{\partial {\fg E_e}} \fg\cdot {\fg F^T_e} = \frac{1}{J_e} {\fg F_e}\fg\cdot {\fg C}{\fg :} {\fg E_e} \fg\cdot {\fg F^T_e}
\label{Elrul2-2-0}
\eey
{\bf 5. Dissipation rate per unit mass due to plastic flow $D_p $ and variation of the internal variable $D_g$}
\bey
D_p:= {\fg X}_p \; : \fg d_p ; \qquad
D_g:= {\fg X}_g \; \cdot  \dot{\fg g}^T .
\label{ijss1-12a-0-0}
\eey

{\bf 6. Dissipative forces for plastic flow ${\fg X}_p $ and variation of the internal variable ${\fg X}_g$}
\bey
{\fg X}_p \; := \; \frac{1}{\rho}{\fsg }\, - \; \frac{\partial\,\psi}{\partial\, \fg E_p}; \qquad
{\fg X}_g \; := \; - \, \frac{\partial \, \psi}{\partial \, {\fg g}^T} \; .
\label{ijss1-12a-01-0}
\eey

{\bf 7. Yield condition}
\bey
f \left({\fg X}_p \, , \, \theta, \, \fg E_p, \, {\fg g}, \,
  \xi \right) = \, 0 \; .
\label{ijss1-12a-0-y}
\eey

{\bf 8. Evolution equations for plastic strain and internal variables}
\bey
&&\fg d_p \, = \, {\fg f}_p \left({\fg X} \, , \, \theta, \, \fg E_p, \, {\fg g}, \,
  \xi  \right)\, \quad {\rm if} \quad    f= \, 0; \qquad   {\rm and } \quad    \dot{f} >  0;
\qquad  \fg d_p  \, = 0 \quad {\rm otherwise}.
\nonumber\\
&& \dot{\fg g} \; = \; {\fg f}_g \left({\fg X}_g \,  , \, \theta, \, \fg E_p, \, {\fg g}, \,
 \xi \right) \, .
\label{ijss1-12b-0}
\eey

Application of the same thermodynamic procedure as for small strains but within finite-strain formalism allows us to
generalize equations in Box 2 for the dissipation rate due to SC and thermodynamic driving forces for finite strains, see Box 10. Note that in Eq. (\ref{ijss1-123-f}) the transformation work (first term) can be substituted with the corresponding term in Eq. (\ref{IJP-11}).

\begin{center}
{\large Box 10.} {\bf Dissipation rate due to SC and thermodynamic driving forces at finite strains
                   \cite{levitas-mrc1996,54,levitas-ijss1998,Idesman-CMAME-99,idesman-levitas+stein-ijp-2000,Levitas-Javanbakht-JMPS-14}}
\end{center}
\noindent
\underline{\hspace*{17cm}}

\noindent
{\bf 1. Local dissipation rate $\, {\cal D}_{\xi} \,$ per unit mass and dissipative force  $ X_\xi $ for structural changes }
\bey
{\cal D}_{\xi} \, := X_\xi \dot{\xi} ; \qquad
&& X_\xi \, = \,  \frac{1}{\rho_0} {\fg P}^{\ssst T} {\fg \cdot}  {\fg F}_e {\fg :} \frac{\partial {\fg U}_t}{\partial \xi}
 {\fg \cdot}  {\fg F}_p - \frac{J_{t} }{\rho_0} {\fg U}_{t}^{-1} \fg : \frac{\partial {\fg U}_t }{\partial \xi} \psi^e  -\frac{J_{t} }{\rho_0} \frac{\partial \psi^e}{\partial \xi} \,
          - \, \frac{\partial \, \psi^\theta}{\partial \, \xi} \,
          =
          \nonumber\\
&&            \frac{1}{\rho} {\fg T} {\fg :}  \left({\fg F}_{e}\cdot  \frac{\partial {\fg U}_t}{\partial \xi}
\cdot {\fg U}_t^{-1}\cdot {\fg F}_{e}^{-1}\right)_s
- \frac{J_{t} }{\rho_0} {\fg U}_{t}^{-1} \fg : \frac{\partial {\fg U}_t }{\partial \xi} \psi^e  -\frac{J_{t} }{\rho_0} \frac{\partial \psi^e}{\partial \xi} \,
          - \, \frac{\partial \, \psi^\theta}{\partial \, \xi} \, .
\label{ijss1-11-f}
\eey
{\bf 2. Local thermodynamics driving force per unit mass for complete structural change $X$ }
\bey
X \, :=\int\limits^{t_s}_{0}
             {\cal D}_{\xi} dt \; =  \int\limits^{t_s}_{0}
             X_\xi \, \dot{\xi} dt \; = \int\limits^{1}_{0}
             X_\xi d\xi \,
\label{ijss1-9-f}
\eey
or
\bey
   X \, := \int\limits^{t_s}_{0}  \frac{1}{\rho}
{\fsg } \, {\fg :} \, \fg d  d t \, - \,   \left( \psi_2 - \psi_1 \right) -
\int\limits^{t_s}_{0}   \left( s \, \dot{\theta} \, + \,
{\fg X}_p \, {\fg :} \, \fg d_p +
{\fg X}_g \, {\fg :} \, \dot{\fg g} \right) \, d \, t \; .
\label{ijss1-180-f}
\eey
{\bf 3. Global $X_v$ thermodynamic driving force for nucleation in a volume $V_n$, i.e., the total dissipation increment due to SC
only during the complete SC in the transforming region}
\bey
X_v :=  \bar{X} m_n \, =  \int\limits_{V_n} \rho X
d \, V_n -  \Delta\, \int\limits_{\Sigma_n} \Gamma \; d \, \Sigma_n \, .
\label{ijss1-19-f}
\eey
{\bf 4. Global dissipation rate $\, {\cal D}_v \,$ for nucleation in a volume $V_n$}
\bey
{\cal D}_{v} \, = \; X_{v} \, \dot{\chi} \; \geq 0;
\qquad\qquad
\dot{\chi} \, := \; 1 \, / \, t_s \; .
\label{mrc95kor-44e-f}
\eey
{\bf 5. The thermodynamic driving force for a phase interface propagation per unit reference area}
\bey
X_{\Sigma }: = \, {\fg P}^{\ssst T} \, {\fg :} \, \left[{\fg F} \right] \,-
\rho_0 \left[ \psi \right] \, -
\int\limits^{\theta_2}_{\theta_1} \, \rho_0 \, s \; d \, \theta \, - \,
\int\limits^{t + \Delta \, t}_t \rho_0 \, \left({\fg X}_p \, {\fg :} \,
\fg d_p \, + \, {\fg X}_g \, {\fg :} \, \dot{\fg g} \right)
\, d \, t \,.
\label{ijss1-123-f}
\eey
{\bf 6. Local dissipation rate $\, {\cal D}_\Sigma \,$ per unit area for a phase interface propagation}
\bey
{\cal D}_{\Sigma } \,:  = \; X_{\Sigma } \, v_n \; \geq 0.
\label{dis_sigma-f}
\eey
\par
For a particular case when  $\, \psi \,$ depends on $\, \fg E_e \,$ , $\; \theta\;$, and
$\; \xi \,$ only and for equal elastic properties of phases  (i.e., $\, \psi^e \,= \psi^e(\fg E_e \,,\, \theta) $),
surface energy is negligible, for isothermal approximation and
homogeneous $\, \theta \,$ 
in the nucleus, Eqs. (\ref{ijss1-180-f}) and (\ref{ijss1-19-f}) result in
the following very simplified expression for the thermodynamic driving force for nucleation 
\bey
\bar{X}\, = \, \varphi \, - \frac{ \left [J_t \right]}{\rho_0}\psi^e  - \,  \; \left [ \, \psi^\theta \left(
\theta\right) \right] ; 
\qquad\qquad
\varphi \, := \; \frac{1}{m_n} \,
\int_{V_n} \int^{t_s}_{0}
{\fg \sg} \, {\fg :} \, \fg d_t  dt \; d \, V_n \; .
\label{eq:6-0}
\eey
Note that in all applications, the term related to $ \left [J_t \right]$ was neglected.
Algorithms for finite element solutions at finite strains are presented
in \cite{Idesman-CMAME-99,levitas+idesman+stein-ijss1998}. Various numerical solutions can be found in our papers
\cite{Idesman-CMAME-99,levitas+idesman+stein-ijss1998,idesman-levitas+stein-ijp-2000,levitas+idesman+olson-1999,31}.

\section{Spherical elastic nucleus within an  elastic - perfectly plastic sphere}\label{sphere}

We consider a spherical nucleus of the radius
$\, r \,$ of a product phase  within an infinite elastoplastic sphere without strain hardening under
action of the  external pressure $\, p \,$. Such a problem was considered
from a thermodynamic point of view using different criteria in several
papers, see \cite{kaganova+roitburd-1987,fischer+berveiller+etal-1994,Levitas-JMPS-II-97,roitburd+temkin-1986,Fischer-Oberaigner-JAM-00,Fischer-Oberaigner-AAM-01}. To illustrate our thermodynamic approach presented in Box 3 in the simplest way, we will follow \cite{Levitas-JMPS-II-97}, where this
solution was, in particular, applied to PT from graphite to diamond and to PT in steel.
To present the simplest example of our kinetic approach, summarized in Boxes 5-7 , we will sketch the solution from
\cite{levitas-ijp2000-1,levitas-ijp2000-2}.

\subsection{Pressure variation and athermal kinetics}
\par
 We assume   (a)
a spherical transformation strain $\; \fvep_t \, = \, 1/3 \, \vep_o \, {\fg I} \, \xi \,$ with a volumetric transformation strain $\; \vep_o \,$ (negative if compressive) and (b) that isotropic elastic properties do not change during PT for simplicity.
Then the pressure $\, \tilde{p} =-1/3 (\sg_1+\sg_2+\sg_3)\,$ in a nucleus is
determined by equations from  \cite{roitburd+temkin-1986}:
\bey
{\rm in the } & & {\rm elastic} \quad {\rm regime} \qquad\quad
\tilde{p}_e \; = \; p + \, \frac{\vep_o \, \xi}{3 \, C} \; ,
\quad\quad
 \xi  \, \leq \,  \xi{'} \; ,
\quad\quad
\xi{'} \, := \; \frac{2 \, \sg_y \, C}{|\vep_o|} \; ;
\label{eq:5-y} \\
\nonumber \\
{\rm in the } & & {\rm elastoplastic} \quad {\rm regime} \qquad
\tilde{p}_p \; = \; p +\, \frac{2}{3}
\sg_y \left( \ln \; \frac{|\vep_o| \, \xi}{2 \, \sg_y \, C} \; + \, 1
\right) sing (\vep_o)\; ,
\quad\quad
  \xi  \, > \,   \xi{'}  \; .
\label{eq:6-x}
\eey
Here, $\sg_i$ are the principle stresses, the elastic constant $\; \displaystyle{ C \; = \; \frac{3 \, \left(1 - \nu \right)}{2 \,
E}} \,$ is expressed in terms of the Young's modulus $\; E \,$ and  the Poisson's ratio $\, \nu \,$,
$\, \sg_y \,$ is the yield strength of the parent phase, and $\, \vep_o \, \xi{'} \,$
is the transformation strain corresponding to the onset of plastic deformation in the parent phase.
While the equations are valid for any sign of $\, \vep_o $, to be specific, we will discuss SC with compressive $\, \vep_o<0 $ under action of compressive pressure $p>0$.
Then, based on  Eqs. (\ref{eq:5-y})  and (\ref{eq:6-x}),
the pressure in the
nucleus  reduces with increasing $ \xi$ during the SC. In the elastic region, pressure reduces
linearly and can even change a sign. Plasticity retards the pressure reduction. Let us evaluate the  transformation work:
\bey
\rho \varphi \; = \, \int\limits^{\fvep_{t \, 2}}_{\fvep_{t \, 1}}
{\fsg } \, {\fg :} \,  d \, \fvep_t \; = \,-
\int\limits^1_0 \tilde{p} \, \vep_o \; d \, \xi \; =  \;- p \, \vep_o \; - \;
\frac{\vep_o^2}{6 \, C}
\label{eq:35a}
\eey
in the elastic region and
\bey
\rho \varphi \; = \,
-
\int\limits^1_0 \tilde{p} \, \vep_o \; d \, \xi \; = \, -
\int\limits^{\xi{'}}_0 \tilde{p}_e \, \vep_o \; d \, \xi \, - \,
\int\limits^{1}_{\xi{'}} \tilde{p}_p \, \vep_o \; d \xi \; = \,
-p \, \vep_o - A_m \;
\label{eq:35}
\eey
\bey
{\rm with} \qquad\qquad
A_m \; := \; \frac{2}{3} \, \sg_y^2 \, C
 +
 \frac{2}{3} \, \sg_y \,| \vep_o|
 \; \ln \;
\frac{|\vep_o|}{2 \, \sg_y \, C}
\label{xxxxx}
\eey
in the elastoplastic region.
 Apparently, Eq.(\ref{eq:35})
reduces to Eq.(\ref{eq:35a}) at $\; \displaystyle{\sg_y \; = \; \frac{|\vep_o|}{2 \, C}} \;$ (i.e. at
$\, \xi^{'} \, = \, 1 \,$). For simplicity, we will neglect surface energy in this Subsection, assuming a large-size "macroscale nucleus"
(allowing surface energy in this problem to be trivial). Then,
substituting $\, \varphi \,$ in Eqs.
(\ref{eq:6}) and (\ref{ijss1-19a}) and introducing the thermodynamically equilibrium pressure
$\, p_e \, = -\,\rho \Delta \, \psi / \vep_0\,$, we resolve for  the PT pressure in the elastic and elastoplastic
regimes, respectively:
\bey
p \; = \; p_e - \frac{\vep_o}{6 \, C} \, - \; \frac{\rho K^0 }{\vep_o} \,
 ,
\label{eq:37}
\eey
\bey
p \; = \; p_e - \frac{2}{3} \, \frac{\sg_y^2 \, C}{\vep_o}
 -
 \frac{2}{3} \, \sg_y \,  
 \ln
\frac{|\vep_o|}{2 \, \sg_y \, C} sign (\vep_o) \, - \;
\frac{\rho K^0 }{\vep_o}  \, .
\label{203}
\eey
Thus, the SC pressure may significantly exceed $\, p_e$ due to   the work of internal stresses and athermal threshold $\, K^0 \,$. Plastic relaxation
decreases the work of elastic internal stresses and, consequently, the SC pressure. Note that Eqs.
(\ref{eq:37}) and (\ref{203}) at $K^0=0$ are identical with the solution in \cite{roitburd+temkin-1986} but
are different from solution in \cite{fischer+berveiller+etal-1994} because of different expressions for the thermodynamic driving force.  Adiabatic heating due to the SC latent heat and athermal dissipation was considered in  \cite{levitas-ijp2000-1,levitas-ijp2000-2}. Allowing for adiabatic heating for graphite-diamond ({\sl G}-{\sl D}) PT was presented in \cite{Levitas-chapter-04}.
Large strain formulation, which also include surface stresses, was applied for sublimation, melting, void nucleation
due to fracture, sublimation, and melting and evaporation  in \cite{Levitas-Alt-PRL-08,Levitas-Alt-VM-PRB-09,Levitas-sublim-IJP-12,Levitas-Alt-sublim-IJP-12,Levitas-Alt-void-ActaMat-11},
see Sections \ref{void-sph} and \ref{void-altern}.

\subsection{ Phase transformation from graphite  to diamond}
The obtained solution for a spherical nucleus was applied
for interpretation of nontrivial experimental results on PT from graphite ({\sl G}) to diamond ({\sl D}) .
The equilibrium pressure-temperature
line is obtained with the help  of chemical thermodynamics
\cite{bundy-1989,kurdyumov-1980,novikov+fedoseev+shul'zhenko+bogatireva-1987} is approximated by the following relationship:
\bey
p_{e q} (GPa) \; = \; 1.2575 \, + \, 0.0025 \, \theta .
\label{204}
\eey
In experiments,
  martensitic PT {\sl G}-{\sl D}
      occurs at significantly larger pressure
      \cite{bundy-1989}, e.g., at 70 GPa at room temperature. However,   in the presence of some liquid metals (e.g.,
      {\sl NiMn, Fe,  Co, Ni} \cite{novikov+fedoseev+shul'zhenko+bogatireva-1987}),
      PT {\sl G}-{\sl D} can be observed very close to the equilibrium pressure.

There are numerous  theories attempting to qualitatively explain
these results by  hypothesizing some  CRs,
catalytic or solvent properties of liquid metals, etc.
 \cite{kurdyumov-1980,novikov+fedoseev+shul'zhenko+bogatireva-1987}.
 It was shown in \cite{Levitas-JMPS-II-97} that  our
  theory   with athermal kinetics  explains these experiments without involving
   additional physical mechanisms or CRs.
\par
Based on our theory and numerical estimates, there are three  reasons causing the  PT pressure
in the experiment to significantly exceed the phase equilibrium pressure:
(a) the pressure reduction during PT, which reduces the transformation work;
(b) the athermal resistance  PT $K^0$; and
(c) the adiabatic temperature increase due to transformation heat released during the short time of martensitic PT.

In the presence of a liquid metal, {\sl G} dissolves in the   metal and recrystallizes from an  oversaturated solution as a {\sl D}
when pressure slightly exceeds the {\sl G}-{\sl D}  phase equilibrium pressure. {\sl D} grows in an atom-by-atom
mechanism from the oversaturated melt.
The effect of a liquid is three-fold.
\begin{itemize}
\item It changes
      martensitic PT into diffusive PT \cite{bundy-1989,kurdyumov-1980,
      novikov+fedoseev+shul'zhenko+bogatireva-1987}. Furthermore, liquid, as a hydrostatic medium, does not contain defects and does not interact with the stress field  of defects in the growing diamond. All these reduce the athermal threshold $K^0$ down to zero.
\item Second, the  pressure variation in the
      transforming particle within the liquid is negligible because the volume of a liquid is much larger than the volume
      of  a {\sl D}.
\item Third, the adiabatic process in a nucleus is replaced by an isothermal one due to slow diffusional  growth.
\end{itemize}
Thus,  there is no cause for  the actual PT
pressure to
exceed the phase equilibrium pressure. Additional aspects of {\sl G}-{\sl D} PT can be found in \cite{Levitas-JMPS-II-97,Levitas-chapter-04}. Modeling of the industrial process of the {\sl D} synthesis
in a high-pressure apparatus is performed in \cite{levitas+idesman+leshchuk+polotnyak-1989,novikov+levitas+leshchuk+idesman-1991,leshchuk+novikov+levitas-2001}.

One must mention that the above consideration of the {\sl G}-{\sl D} PT within the melt was based on the
athermal kinetics (as described
in Box 3), neglected nucleation barrier, and actual thermally activated nucleation.  Due to the large surface energy of a diamond with any surroundings, activation energy for nucleation is
very high and the nucleation criterion (\ref{meyers1-126c1}) is not met for experimental temperature.
Some possible mechanisms are discussed in \cite{novikov+fedoseev+shul'zhenko+bogatireva-1987}, but
they do not resolve the nucleation problem, either. Thus, the kinetics of diamond nucleation is currently still a mystery.

\par
\subsection{Critical nucleus}
Let us consider the kinetics of the appearance of a thermally activated spherical nucleus using Box 4.
We have
\bey
(\bar{X}-K^0) m_n= \left(\, X \, - \, K^0 \,\right) \,
\frac{4}{3}\rho \; \pi \, r^3 \, - \,
\Gamma \, 4 \, \pi \, r^2 \;;
\label{mrc98-1-3-o}
\eey
\bey
{X}\, = \, \varphi \, - \,  \; \Delta \, \psi^\theta \left(
\theta\right);  \qquad X \, - \, K^0 >0, 
\label{eq:6-a}
\eey
where $\varphi $ is defined by Eq.(\ref{eq:35}) and is independent of radius $r$.
According to Eq.(\ref{IJP-13-1}),    maximization of $-(\bar{X}-K^0)m_n$  with respect to $r$ leads to the critical radius
\bey
r_c \; = \; \frac{2 \, \Gamma}{\rho(X-K^0)} \; .
\label{mrc98-1-3b}
\eey
Substituting $r_c$ into the expression for $-(\bar{X}-K^0)m_n$ in Eq. (\ref{mrc98-1-3-o}),  we obtain the activation energy
\bey
Q \; = \; \frac{16}{3} \; \frac{\pi \, \Gamma^3}{
\left(\rho(X-K^0) \right)^2}  .
\label{mrc98-Q}
\eey
Then, $Q$ can be substituted in Eq.  (\ref{meyers1-126a1}) for nucleation time and in  Eq. (\ref{meyers1-126c1}) for
kinetic nucleation criterion. If, for a given $X$ the criterion (\ref{meyers1-126c1}) is met, then nucleation time in Eq. (\ref{meyers1-126a1}) is realistic. Otherwise, it is much larger than any reasonable observation time. An increase in $X$ reduces the critical  radius and mass (volume) of a nucleus and, consequently, activation energy and nucleation time.
Classical nucleation theory for elastic materials can be obtained by changing $\rho(X-K^0) $ in Eqs. (\ref{mrc98-1-3b}) and (\ref{mrc98-Q}) with the difference in the bulk energy of the initial and final states.

\subsection{ Macroscale nucleation  kinetics}
We will start with an approach in Box 5. With Eqs. (\ref{mrc98-1-3-o}) and (\ref{eq:6-a}) for $(\bar{X}-K^0)m_n$,
Eq. (\ref{meyers1-126a})  specifies to
\bey
t_s \; = \; t_0 \, \exp \; \left( \; - \,
\frac{ \left(\, X \, - \, K^0 - \, E_a \,\right) \,
\frac{4}{3} \;\rho  \pi \, r^3 \,  \, - \,
4 \, \Gamma \, \pi r^2}{R \, \theta_{ef}} \; \frac{N}{n} \; \right)
\label{mrc98-1-2} \\
{\rm at} \qquad\qquad 0 \; \leq \left(\, X \, - \, K^0 \,\right) \,
\frac{4}{3} \, \rho \pi \, r^3 \,  \, - \,
4 \, \Gamma \, \pi \, r^2 \, \leq   \, E_a \,
\frac{4}{3} \;\rho  \pi \, r^3  .
\label{mrc98-1-2a}
\eey
We must also impose
\bey
 0 \; \leq \, X \, - \, K^0  \, \leq   \, E_a \, ,
\label{mrc98-1-2a-lo}
\eey
otherwise, either SC is thermodynamically impossible or activation energy for a large enough nucleus is negative.
Application of the principle of minimum time $\, t_s \,$  (\ref{meyers1-126c}) with respect to the radius $\, r \,$
under constraint  (\ref{mrc98-1-2a-lo}) results in  $\, r \; \rightarrow
\; {\sl min} \,$, since nucleation time is a monotonously decreasing function  of $\, r \,$.
Then, we need to switch to Box 6 or 7 and find the thermodynamically admissible
radius  $\, r_t \,$
 from the SC criterion $\, \bar{X} \, = \, K^0 \,$, i.e.
\bey
\left(\, X \, - \, K^0 \,\right) \,
\frac{4}{3}  \rho \pi \, r^3_t \, - \,
\Gamma \, 4 \, \pi \, r^2_t \; = \; 0
\qquad\qquad \mbox{or} \qquad\qquad
r_t \; = \; \frac{3 \, \Gamma}{ \rho \left(X \, - \, K^0 \right)} \;\, .
\label{mrc98-1-3}
\eey
Substitution of (\ref{mrc98-1-3}) in  Eq.(\ref{mrc98-1-2}) results in
\bey
t_s \; = \; t_0 \; {\sl exp} \left( \,
\frac{\, E_a}{R \, \theta_{ef}} \; \frac{N}{n} \; \frac{4}{3} \rho
\pi \; r^3_t \, \right) \; =  \; t_0 \; {\sl exp} \left( \,\frac{36\pi }{3\rho^2}
\frac{\, E_a}{R \, \theta_{ef}} \; \frac{N}{n} \;
\; \left(\frac{ \Gamma}{ \left(X \, - \, K^0 \right)} \, \right)^3 \right)  .
\label{mrc98-1-4}
\eey
The radius of the thermodynamically admissible nucleus is 1.5 times larger than the critical nucleus (see also Fig. \ref{Fig-energy-crit-nucl}). While  the activation energy of a critical nucleus is fully determined in terms of
$\rho \left(X \, - \, K^0 \right)$ and $\Gamma $, the activation energy of the thermodynamically admissible nucleus has a fitting parameter $E_a/n$, which should be calibrated from a macroscale experiment.
Also, the activation energy of the thermodynamically admissible nucleus reduces stronger  with the increase in $\left(X \, - \, K^0 \right)$ than that for the critical nucleus.


\section{Stress- and strain-induced chemical reactions and phase transformations in a thin layer: propagating interface, shear band, and TRIP and RIP phenomena}\label{interface-analytic}

Problems on an SC in a thin layer in a half-space or space within rigid-plastic formulation allows simple analytical
solutions at small strains \cite{Levitas-JMPS-I-97,levitas-ijss1998,levitas-ijp2000-1,levitas-ijp2000-2} and large strains \cite{levitas+nesterenko+meyers-1998-1,levitas+nesterenko+meyers-1998-2,levitas-zamm1998}, in isothermal
and adiabatic consideration, for athermal and thermally activated kinetics. These solutions can be applied to
analyses of a phase interface propagation, SCs in a shear band and in surface layer caused by friction,
and for deriving equations for TRIP and RIP.

\subsection{Analytical solution}
\par
\begin{figure}[htp]
\centering
\includegraphics[width=0.5\textwidth]{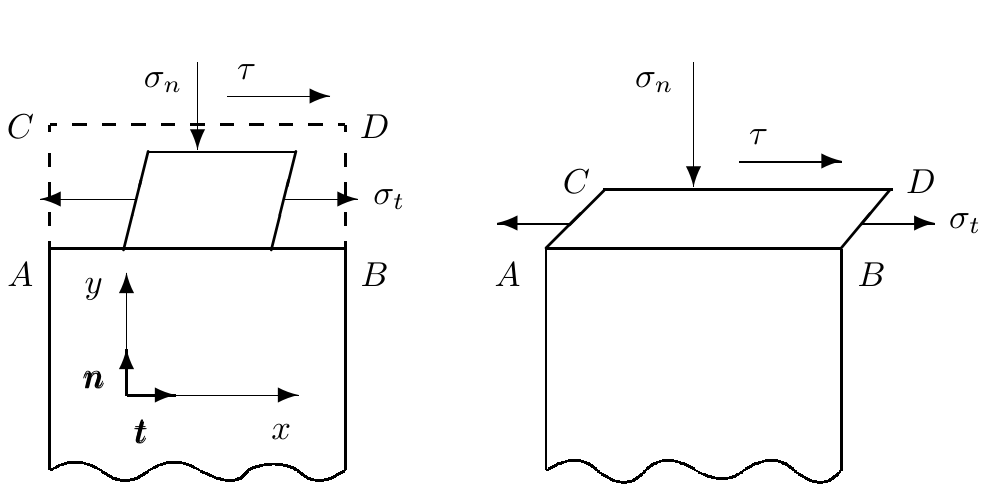}    
\caption{ Schematic illustration of SCs in a shear band $ABCD$ within underformed half space (below the line $AB$): (a) initial state (dashed line) and state after applying transformation strain (solid line); (b) the final state after adding TRIP or RIP. Reproduced with permission from
\cite{levitas-zamm1998}.
\label{Fig-shear-band-TRIP}}
\end{figure}

Consider an infinite rigid-plastic half-space or full-space with prescribed normal
$\, \sg_n \,$ and shear $\, \tau \,$ stresses on the entire horizontal surface
(Fig. \ref{Fig-shear-band-TRIP}). Plane strain formulation is assumed both for plastic and transformational strain.
We consider an infinite thin layer in which localized SC and plastic flow occur, while the material outside the layer is rigid.
Displacements are continuous across the interface, i.e., phase interfaces are coherent.
All strain and stress fields are homogeneous within a layer, which also means the total strain represents
an invariant plane strain, i.e., shear strain along the layer and normal strain along the normal to the layer.
This layer can be obtained by a plane phase interface moving along the normal and producing an SC within a layer,
or by localized plastic shear deformation (i.e., coinciding with a shear band) or can be a part of a shear band.
Additionally, stresses are assumed to be time-independent (or plastic strain-independent), restricting solution to a perfectly plastic model. The Tresca yield condition is utilized.

For simplicity, we consider a 2D spherical transformation strain with the volumetric strain $\, \vep_o \,$, while allowing for an arbitrary transformation strain without plane strain assumption is performed in \cite{levitas-ijss1998}.
Geometrically, in Fig. \ref{Fig-shear-band-TRIP}a, a transformed
Particle, is presented after the transformation strain (which also includes shear).
To satisfy the invariant plane strain conditions, an additional
plastic strain (TRIP or RIP) further deforms a particle to the configuration shown Fig. \ref{Fig-shear-band-TRIP},b.
Adiabatic heating is included in the solution.

\par
Expressions for $\,X\,$, for the temperature $\theta$ during and $\theta_2$ after
the end of the SC, and for the effective temperature
look as \cite{levitas-ijp2000-2,levitas-zamm1998,levitas+nesterenko+meyers-1998-2}
\bey
\rho X \, = \,
 \sg_n \,\vep_0 - \; 0.5 \,|\vep_0|\sqrt{\sg^2_y \, - \, 4 \, \tau^2}
\; + \; 0.5 \;\rho   \Delta \, s_o \,
\left(\theta_1 \, + \, \theta_2 \right) \, - \; \rho \Delta \, U_o \; ,
\label{meyers1-97*}
\eey
\bey
\theta \; = \; \theta_1 \, + \; \frac{A}{\nu} \; \xi \; ;
\qquad\qquad
\theta_2 \; := \; \theta \, \left(1\right) \; = \;
\theta_1 \, + \; \frac{A}{\nu} \; ;
\qquad\qquad
\theta_{ef} \; = \; 0.5 \, \left(\theta_1 \, + \, \theta_2 \right) \; ;
\label{meyers1-96}
\eey
\bey
A \; := \;\frac{1}{\rho } \left( \sg_{n}{\vep_0}  \; + \;
 |\vep_0| \frac{2 \, \tau^2}{\sqrt{\sg^2_y \, - \, 4 \, \tau^2}}\right) \, -
\; \Delta \, U_o \; .
\label{meyers1-92}
\eey
Here $\, \Delta \, U_o \,$ and $\, \Delta \,s \,$ are the change in internal energy
and entropy per unit mass, respectively, 
$\, A \,$ is the heat source due to the SC heat
$\, \Delta \, U_o \,$, part of transformation work $\sg_{n}{\vep_0}$, and TRIP or RIP (see below).
\par
We consider the macroscale kinetics from Boxes 5 and 6. Minimizing the SC time with respect to the thickness of the layer $\, h \,$, we obtain
\bey
\frac{n}{N} \; R \; \theta_{ef} \; \ln \; \frac{t_0}{t_s} \; = \;
\rho \, \left( X \, - \, K^0 \, - \, E_a \right) \, h \,\Sigma\, - \,
\Gamma \; 2 \, \Sigma \;\; \rightarrow \;\; {\sl \max_h} \; ,
\label{esh1-2++}
\eey
where $\, \Sigma \,$ is the interface
area.  Maximum in Eq. (\ref{esh1-2++}) results in $\, h \; \rightarrow \; {\sl min} \,$. Then thermodynamically admissible $h$ is determined
from the thermodynamic SC criterion (\ref{meyers1-1126a-c})
\bey
h \; = \; \frac{2 \, \Gamma}{\rho \, \left(X \, - \, K^0 \right)} \;\, .
\label{1-3}
\eey
With this expression, the kinetic Eq. (\ref{meyers1-126a-c})
is
\bey
t_s & = & t_0 \, {\sl exp} \left(\, \frac{\, E_a}{R \, \theta_{ef}} \;
\frac{N}{n} \; \frac{2 \, \Gamma \Sigma}{ \left(X \, - \, K^0 \right)}  \, \right) \; \leq t_d,
\label{1-4}
\eey
where  $\, t_d \,$ is the time of
deformation in the shear band.

\par
By integrating flow rule, we found the plastic shear $\, \gamma \,$, which is TRIP for PTs and RIP for CRs, is determined by the equation
\bey
\gamma \; = \;  \mid \, \vep_o \, \mid \;\;
\frac{\tau}{\sqrt{\sg^2_{y} \, - \, 4 \, \tau^2}} \; ,
\label{meyers1-207}
\eey
Due to variation
in transformation strain (which generates internal stresses and forces plastic strain to restore displacement continuity conditions across an interface), plastic flow takes place at arbitrary (even infinitesimal)
shear stress, below
the yield shear strength in shear $0.5 \sg_{y}$.  For relatively small $\tau$, the relationship between $\gamma$ and $\tau$ is approximately linear, as shown in known experimental or theoretical expressions for TRIP \cite{fischer+sun+tanaka-1996,22,16,23,zackay+parker+fahr+busch-1967,fischer+berveiller+etal-1994}. For larger $\tau$, plastic shear grows faster than linear.
The most important property of this solution (which was not present in any previous solutions) is that for $\tau \rightarrow 0.5 \sg_{y}$ (e.g.,  in a shear band), plastic shear tends to infinity. In reality, it may reach extremely large values, especially for large $\vep_o$ \cite{Levitasetal-JCP-BN-06,Kulnitskiyetal-16}.
It follows from
Eqs.(\ref{meyers1-97*})-(\ref{meyers1-92}) that for $\; \tau \, \rightarrow
\, 0.5 \, \sg_y \;$ the heat source due to TRIP/RIP tends to infinity as well, and the
temperature tends to infinity, or to melting temperature, above which the solution does not have a sense. However, if the SC is relatively slow, then the adiabatic condition is not met, and an increase in temperature is much lower. The increase in temperature suppresses martensitic PTs and accelerates some PTs and CRs, which are promoted
thermodynamically or kinetically by temperature.
\par
 Note that an interesting consequence of a finite-strain formulation is a definite {transformation path}, i.e., sequences
of dilatational and shear transformation strain variation during the PT \cite{levitas-zamm1998}.

\subsection{ Strain-induced chemical reactions  in a shear band }

\begin{figure}[htp]
\centering
\includegraphics[width=0.5\textwidth]{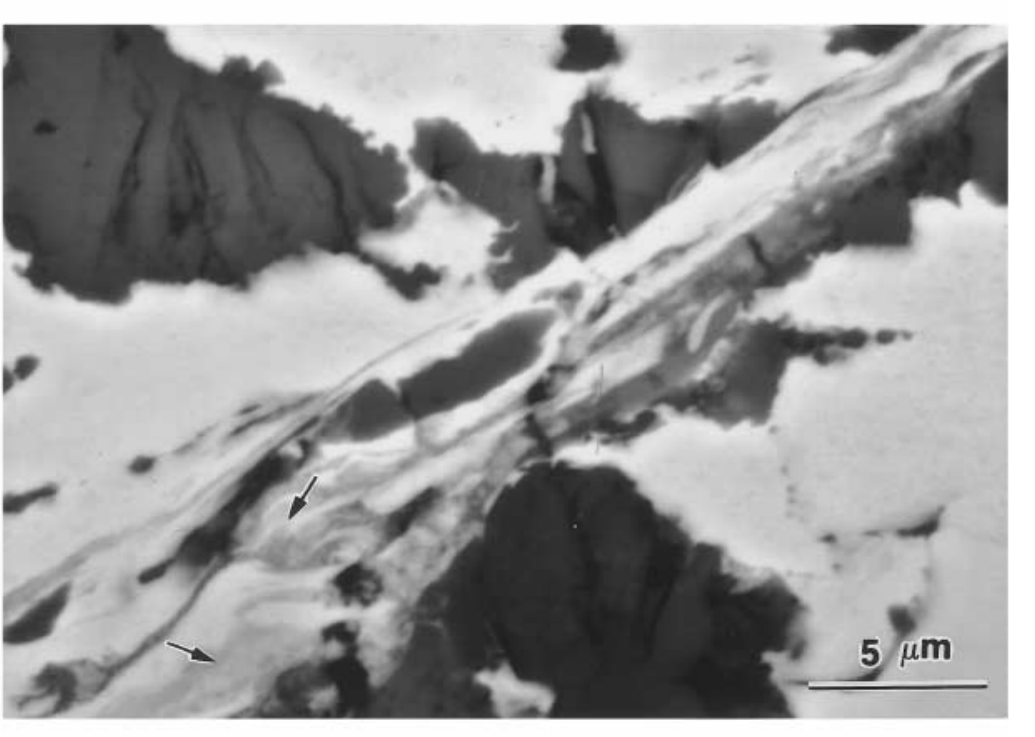}    
\caption{ Initiation of CR within a dynamically formed shear band in {\sl Ti-Si} and {\sl Nb-Si} powder mixture.     Reproduced with permission from
\cite{levitas+nesterenko+meyers-1998-1}.
\label{CR-shear-band-picture}}
\end{figure}

Such exothermic CRs in {\sl Ti-Si} and {\sl Nb-Si} powder mixtures were studied in
\cite{nesterenko+meyers+chen+lasalvia-94,nesterenko+meyers+chen+lasalvia-95}, see Fig. \ref{CR-shear-band-picture}.
The first theoretical modeling of continuum thermomechanical aspects of the problem
and interpretation of the possible mechanisms for the
promotion of the CRs due to large plastic shear is published in \cite{levitas+nesterenko+meyers-1998-1,levitas+nesterenko+meyers-1998-2,levitas-ijp2000-1,levitas-ijp2000-2}.
In these papers,
 the RIP phenomenon, which is similar to
TRIP for PTs, was predicted and used as one of the main accelerators of the CRs. The macroscopic similarity between RIP and TRIP allows one use of the existing knowledge in TRIP \cite{fischer+sun+tanaka-1996,22,16,23,zackay+parker+fahr+busch-1967,fischer+berveiller+etal-1994}
for studying RIP.  The concept of the {\it effective temperature} was introduced as well. Paper
 \cite{levitas+nesterenko+meyers-1998-1,levitas+nesterenko+meyers-1998-2} has utilized a simplified kinetic
equation. In
papers \cite{levitas-ijp2000-1,levitas-ijp2000-2},  a kinetic approach for a macroscale nucleus presented on Boxes 5 and 6
 was applied, which led to significantly different results.

\par
One problem of interest was  that in \cite{thadhani+graham+royal+durbar+anderson+holman-97},
an averaged pressure to shock-initiate the CR in $\, Ti - Si\,$ mixture was surprisingly low, just several {\sl GPa}.
This was explained by the "ease of plastic deformation" that leads to improved mixing.
However, both materials have quite a high yield strength. The authors assumed the volume change during PTs in $\, Ti \,$ and
$\, Si \,$ (that occur at a pressure above  $\, 10$ {\sl GPa}) promotes the plastic deformation.
While the averaged pressure to initiate the CR is just several {\sl GPa}, however, the
authors assume that local pressure in contact between particles may be sufficient for these PTs.
As an alternative, it was stated in \cite{levitas+nesterenko+meyers-1998-1,levitas+nesterenko+meyers-1998-2}
that since the
volume reduction for CR $\, 5 {\sl Ti} \, + \, 3 {\sl Si} \; \to \;
{\sl Ti}_5 {\sl Si}_3 \;$ is large,  $\, -0.278 \,$
\cite{thadhani+graham+royal+durbar+anderson+holman-97},
revealed RIP, as a new mechanism of plastic flow in solids, and  Eq. (\ref{meyers1-207}) may offer more plausible way for  interpreting  large
plastic shear below the yield strength, and how it may reduce stress to initiate CR.
\par
Eq. (\ref{1-4}) predicts the exponential influence
of $X$ and increase in $\theta_{ef}$ due to
RIP on the reduction of the transformation time.  The RIP
significantly increases the reaction finish temperature $\, \theta_2 \,$, by more than 1500 K, i.e., increases $\theta_{ef}$ by 750 K.
According to Eqs.(\ref{meyers1-97*})-(\ref{meyers1-92}), $X$ increases
due to shear stress and corresponding RIP by two macroscopic mechanisms: increase in $\theta_{ef}$ by 750 K and effective "reduction" of the yield strength
$\, \sqrt{\sg^2_y - 4 \, \tau^2} \,$ related to satisfaction of the yield
condition.

\subsection{Mechanochemical feedbacks and the effect of TRIP/RIP on the strain-induced structural changes}
\par
TRIP and RIP may cause positive thermomechanochemical feedback.
Let a SC occurring fluctuationally in a small part of a layer causing TRIP/RIP. Corresponding heating (if $X$ grows with the increasing
 temperature) accelerates PT and CR, which leads to the
intensification of TRIP and RIP and, consequently, of PT and CR, and so on.
With such a process, SC can be a cause of shear banding (especially, for the lower yield strength of the product phase)
instead of vice versa. This mechanism may be important for solid-gas CR
in explosives, e.g., in HMX or nanocomposite formulations \cite{Levitas-chapter-04}.

When $\tau=0.5 \sg_y$ (i.e., for plastic strain-induced SCs), it is difficult to separate traditional plastic strain due to external loading and TRIP/RIP.
This was approximately done, in particular, in high-pressure experiments \cite{Levitasetal-TRIP-HP-BN-APL-05,Levitasetal-JCP-BN-06} on PT from the hexagonal hBN to superhard wurtzitic wBN under compression and torsion in a rotational diamond anvil cell.
It was found by in-situ x-ray diffraction measurements that the evolution equation for the plastic strain-induced concentration of the turbostratic stacking
faults (which is considered as a physical measure of the plastic strain) has two terms. One is proportional to the applied twisting angle (i.e., traditional plastic strain), and the second is proportional to the volume fraction of wBN. The second component was interpreted as the contribution due to TRIP, which was the first revealing of TRIP in high-pressure experiments. It appears that due to the large $\vep_0=-0.39$, TRIP is 20 times larger than traditional plastic strain.
TRIP also resolves some puzzles in these experiments, see \cite{Levitasetal-TRIP-HP-BN-APL-05,Levitasetal-JCP-BN-06}.

Based on recent works on plastic strain-induced SCs under high pressure \cite{Levitas-PRB-04,Levitas-chapter-04,Levitas-MT-19}, there are additional reasons for the intensification of SCs due to plastic straining. For the CR, large plastic deformation produces fragmentation and mixing of reactants similar to that in
liquid phase reaction \cite{zharov-1984}. Then plastic straining promotes both PTs and CRs by producing defects with strong stress concentrations (like dislocation pileup or shear-band intersection), like for any strain-induced SCs. This can reduce the PT pressure by one to two orders of magnitude both in the experiment \cite{Blank-Estrin-2014,Ji-Levitasetal-12,Gaoetal-17} and in the theory or simulations \cite{Levitas-PRB-04,Levitas-chapter-04,levitas-javanbakht-Nanoscale-14,Javanbakht-Levitas-PRB-16,Javanbakht-Levitas-JMS-18,Levitas-MT-19}.
As a result, a microscale kinetic equation for the volume fraction of a high-pressure phase $c$ of the type of
\bey
\frac{dc}{dq}=f (p,q,c)
\label{str-ind-kineq-ss-a}
\eey
 is derived,
where $\, q \,$ is the
accumulated plastic strain (Odqvist parameter) defined as
$\; \dot{q} \, = \, \left(2/3 \, {\fg d}_p \, {\fg :} \,
{\fg d}_p \right)^{0.5} \;$, and  ${\fg d}_p$ is the plastic part of the deformation rate. An example of a specific kinetic equation is given in Eq. (\ref{l-g-4}). Such a kinetic equation can be used for the above problem on SC in a shear band to determine evolution of the volume fraction $c$ with the increase of plastic shear, instead of determination of transformation time for complete PT. However, it is directly applicable when the plastic strain occurs prior to and during the SC, i.e., for $\tau=0.5 \sg_y$.

The fundamental question arises: {\it should TRIP/RIP be included in the accumulated plastic strain $q$ that governs the kinetic equation for} $dc/dq$?
Based on experiments on PT in BN discussed above \cite{Levitasetal-TRIP-HP-BN-APL-05,Levitasetal-JCP-BN-06},
 TRIP (and, consequently RIP) is not distinguishable from a traditional plasticity
 from the point of view of dislocation and twinning mechanisms, and the generation
 of strong stress concentrators at the tip of defects.
It is caused by internal stresses produced by the transformation strain combined with the external stresses rather
than by solely external stresses. This explains why TRIP/RIP, similar to the traditional plasticity, generates new nucleating defects (along with new turbostratic stacking faults in hBN) that promote the SC. Thus, TRIP/RIP should be included in $q$ participating in the kinetic equation for $dc/dq$. This, however, was not done in the literature.
Note that for temperature-induced PTs, an autocatalytic effect (i.e., formation of a martensitic unit promotes the nucleation of other units via stress- and strain induced mechanism) is an important part of the kinetic equation, see \cite{olson+cohen-1986}.

The entire process represents another positive mechanochemical feedback, which is called in \cite{Levitasetal-TRIP-HP-BN-APL-05,Levitasetal-JCP-BN-06} the cascade
mechanism of structural changes during the twisting of an
anvil. Thus, prescribed plastic deformation produces both turbostratic stacking faults
that suppress the martensitic PT, and nucleating defects (e.g., dislocation pileups) that promote the PT.
PT under shear stress generates strong TRIP; TRIP, in a similar way as traditional plastic flow, produces
the additional turbostratic stacking fault and nucleating defects; the new nucleating defects again promote the PT that
induces TRIP, etc.

In \cite{Kulnitskiyetal-16}, the shear transformation-deformation bands have been revealed
in the fcc phase of the $C_{60}$ after compression and shear in an RDAC.  The bands consisted of the shear-induced nanocrystals of linearly-polymerized fullerene and polytypes, the triclinic, hcp, and monoclinic, $C_{60}$, and amorphous structures. Thus, plastic straining arrests five high-pressure phases under normal pressure, which may be potentially important for their practical applications. Localized shear deformation appears counterintuitive because high-pressure phases of $C_{60}$ possess greater strength than the parent low-pressure phase. However, this was explained by TRIP during localized PTs, which occurs because of a combination of applied stresses (below the yield strength) and internal stresses due to large volume reduction during PTs. Eq. (\ref{meyers1-207}) was used for qualitative analyses. Localized PTs and plastic shear deformation promote each other, producing positive mechanochemical feedback and cascading structural changes.
Thus, our solution for a shear band is instrumental for the interpretation of SCs in various systems.

{\it PT of a thin inclined plastic layer.} A problem on PT of a thin inclined plastic layer within a rigid-plastic half-space
under the action of uniform normal and shear stresses
was solved in \cite{Levitas-JMPS-I-97}. The inclination angle was determined explicitly by maximization of the net thermodynamic driving force with allowing for the anisotropy of the athermal threshold $K$.  The final expression for the PT criterion was derived. The yield condition for the parent
phase was considered as a constraint. The effect of the ratio of yield strengths of the parent and product phases
on the PT in the layer was analyzed, and nontrivial behavior was revealed.

\section{Phase transition in ellipsoidal inclusion}\label{ellips-incl}

\begin{figure}[htp]
\centering
\includegraphics[width=0.5\textwidth]{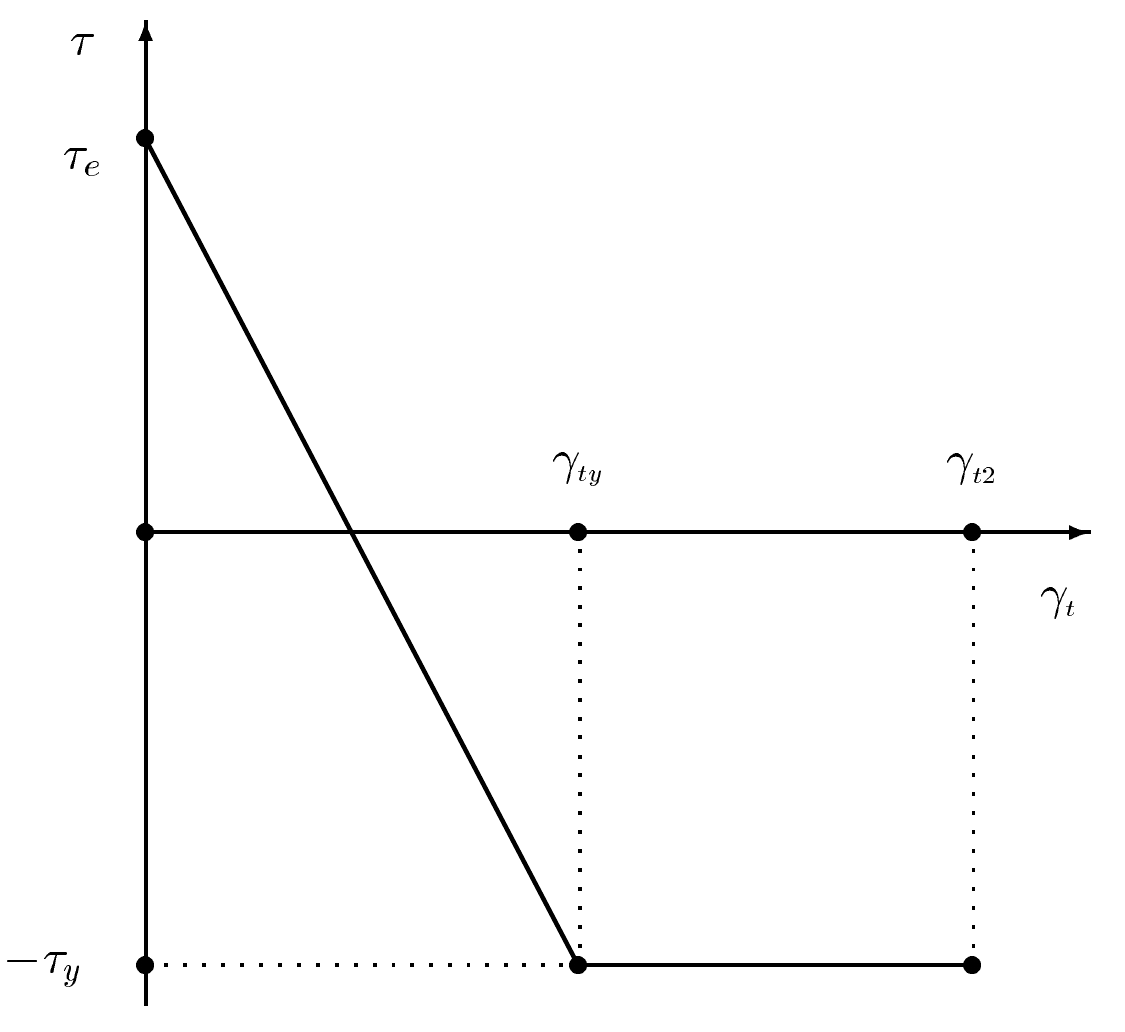}    
\caption{ Shear stress variation in the ellipsoidal nucleus vs. transformation strain.  Reproduced with permission from
\cite{levitas-ijp2000-2}.
\label{Fig-shear-stress-inclusion}}
\end{figure}
Consider an infinite elastic space under the horizontal shear stress
$\, \tau_e \,$. It is instructive to solve the simplest problem on the PT within a penny-shape ellipsoidal region $\, V_n \,$ with
the radius $\, r \,$ and semi thickness $\, b \,$, $\;\, r \, \gg \, b \,$, within the space, allowing for elastoplasticity within the transforming region, see \cite{levitas-ijp2000-2}. Assuming homogeneous fields in
$\, V_n \,$ and horizontal shear component of the transformation strain
$\, 0.5 \, \gamma_t \,$ only, one can use the Eshelby solution \cite{mura-1987}
 for the shear stress in $\, V_n \,$  (Fig.  \ref{Fig-shear-stress-inclusion}):
\bey
\tau \; = \; \tau_e - m \, \gamma_t \; \frac{b}{r}
\qquad {\rm at} \quad
\gamma_t \, < \, \gamma_{ty } \; ;
\qquad\qquad
m \; := \; \frac{\mu \, \pi \left(2 - \nu \right)}{4
\left(1 - \nu \right)} \; ;
\nonumber
\eey
\bey
\tau \; = \; - \tau_y
\qquad {\rm at} \quad
\gamma_{ty} \, \leq \, \gamma_t \leq \gamma_{t2} \; ;
\qquad\qquad
\gamma_{ty} \; := \; \frac{\tau_e + \tau_y}{m} \; \frac{r}{b} \; .
\label{esh1-29+}
\eey
During the increase in $\, \gamma_t \,$, the shear stress reduces linearly, changing the
sense, and at $\, \gamma_t \,= \gamma_{ty} \,$ reaches the yield strength in shear $\, \tau_y \,$ in the {\it direction opposite
to the applied shear stress. }
Then, the transformation work is
\bey
\int^{\gamma_{t2}}_0 \tau \; d \, \gamma_t
\, = \; - \, \tau_y \; \gamma_{t2} \, + \, 0.5 \, \left(\tau_e \, + \,
\tau_y \right)^2  \; \frac{r}{m b} \; .
\label{tr-work}
\eey
Both stresses and transformation work are very different than in \cite{patel+cohen-1953}  and \cite{ganghoffer+etal-2001,marketz+fischer-1995},
where the transformation work was evaluated based on the global or  local stress in the nucleus
before the PT, respectively, i.e., at $\tau_e $ for the current problem.
We consider the macroscale kinetics from Boxes 5 and 6. Minimizing the SC time with respect to $\, r \,$ and $\, b \,$, we obtain
\bey
\left( M \, + \, B \; \frac{r}{b} \, \right) \,
\frac{4}{3} \; \pi \, r^2 \, b \, - \, \Gamma \; 2 \, \pi \, r^2
\; \rightarrow \; {\sl \max_{r,b}} \; ,
\qquad\qquad M \; := \; A - E_a \; ,
\label{esh1-23a++}
\eey
\bey
A \; := \; - \, \tau_y \; \gamma_{t2} \, - \,
\Delta \, \psi \, - \,  \, K^0 \; ;
\qquad\qquad
B \; := \; 0.5 \left(\tau_e \, + \, \tau_y \right)^2 \, / m \; ,
\label{esh1-32+}
\eey
for determination of the actual $\, r \,$ and $\, b \,$. For
$\; \displaystyle{ M + B \; \frac{r}{b} \; < \, 0} \;$ and
$\; A \, > \, 0 \;$, the principle (\ref{esh1-23a++}) results in  $\; r \; \rightarrow \; {\sl min} \,$,
$\; b \; \rightarrow \; {\sl min} \,$.  Applying these two principles under
constraint of the thermodynamic PT criterion $\; \left(X - K^0 \right)
\, V_n \, - \, \Gamma \, S_n \, = \, 0 \;$ or
\bey
\left(A \, + \, B \; \frac{r}{b} \, \right) \, \frac{4}{3} \;
\pi \, r^2 \, b \, - \, \Gamma \; 2 \, \pi \, r^2 \; = \; 0 \; \;   \rightarrow \;\; 2 \, A \, b \; =\; 3 \, \Gamma \, - \, 2 \, B \, r \, ,
\label{esh1-31+}
\eey
we obtain
\bey
r \; = \; \frac{\Gamma}{B} \; , \qquad\qquad
b \; = \; \frac{\Gamma}{2 \, A} \; , \qquad\qquad
V_n \; = \; \frac{2 \, \pi}{3} \;\, \frac{\Gamma^3}{A \, B^2} \; .
\label{esh1-35+}
\eey
For $\, A < 0 \,$,   the semithickness $\, b \,$ in Eq.(\ref{esh1-35+}) is getting negative. Then the minimum $\, b \,$ is equal  to the lattice parameter $\, a \,$
in the $\, b \,$ direction, i.e., $\, b \, = \, a \,$. The radius of the macroscale nucleus $\, r  \,$ is determined from the
thermodynamic PT condition
\bey
r \; = \; \frac{3 \, \Gamma - 2 \, A \, a}{2 \, B} \; ,
\qquad\qquad
\frac{r}{b} \; = \; \frac{3 \, \Gamma - 2 \, A \, a}{2 \, B \, a} .
\label{esh1-36a+}
\eey
Some other cases have been considered in \cite{levitas-ijp2000-2}. With known expression for the transformation work, it is easy to find a solution for the critical nucleus using equations from Box 4.

\section{Nucleation and growth of martensite with coherent, semicoherent, and incoherent interfaces, and interface with decohesion}\label{interface-numerics}



\subsection{Semicoherent interface and interface with a decohesion}\label{semi-decoh}

\par
Several types of interfaces will be considered \cite{Porter-1992}. For a {\it coherent interface}
 displacements, $\fg u$ are continuous across the
interface (i.e., $\fg u^2 ={\fg u}^1$, where subscripts designate phase 1 and 2), which usually generates large internal stresses.  Atomic positions in contacting lattices are continuous across an interface as well. For a {\it semicoherent interface},
 displacements are discontinuous across the
interface, producing relaxation of internal stresses by sliding along the interface (i.e., dislocation generation, in particular, misfit dislocations) and cracks or decohesion (i.e., jump in normal to the interface displacements).
It is necessary to note the conceptual difference in definition of interface type  for PTs in inelastic materials  in continuum and atomistic approaches. At the atomic level, plasticity means there is a presence of dislocations and implies
semicoherence. In continuum approaches, plasticity is described in a continuous way and
both coherent (with continuous displacements) and semicoherent (with discontinuous displacements) interfaces are considered, see
\cite{Levitas-postulate-II-1995,Levitas-JMPS-II-97,levitas-ijss1998,levitas+idesman+stein-ijss1998,ides-lev-CMS-1997}.

 For a {\it incoherent interface}, there is no lattice correspondence across an interface from an atomistic point of view and shear
stresses are assumed  to be zero in continuum theories. Typical incoherent interfaces appear when one of the phases is molten or amorphous,
or for precipitates with a very different crystal structure than the matrix. Due to zero shear stresses at the incoherent interface, incoherent inclusions
are under hydrostatic stress states \cite{Nabarro-40,Porter-1992,Lee-Johnson-78}.
This is clear  for solid-liquid
interfaces and is reasonable for solid-solid interfaces under stress-free conditions.
Indeed, if there is no correspondence of atomic planes across an interface, the product phase can minimize the energy of internal deviatoric stresses by altering its atomic position within a transforming region.
For solid-solid incoherent interfaces under external loading,
shear stresses still can be supported by an interface and the assumption of zero shear stress is contradictory.
Indeed, incoherent high-angle grain boundaries and interfaces between amorphous and crystalline phases
can support shear stresses.
At the continuum level, one must use a theory similar to that for the semicoherent interface discussed below with finite maximum shear stress at the incoherent interface.
However, stress relaxation for incoherent solid-solid interfaces is more pronounced. It can be described by additional stress relaxation within transforming region rather than at interface.

In material science books \cite{christian-1965,Porter-1992}, spacing between misfit dislocations at the semicoherent interface is introduced from the geometric conditions to completely eliminate the misfit between lattices in the averaged sense. Such spacing is also confirmed within PFA  \cite{Levitas-Javanbakht-PRB-12,levitas-javanbakht-APL-13}  with the stationary solution of the evolving interfacial dislocations.
In analytical approaches (see \cite{Porter-1992,boguslavskiy-1985,christian-1965,roitburd-1972}), the
initiation of semicoherence in elastic materials was determined by
equaling the energy of  coherent
and  semicoherent nuclei with allowing for  the energy of
dislocations. A semicoherent interface was energetically favorable above some critical nucleus size.
The detailed dislocation  model of a semicoherent nucleus
was suggested  in \cite{olson+cohen-76,olson+cohen-1986}.
Continuum derivations of the conditions at an incoherent interface (i.e., assuming zero shear stresses at the interface)  were performed in
\cite{Grinfeld-1991,leo+sekerka-1989,Larche-Cahn-AM-78,Jasiuketal-IJSS-87,mura-1987,Leo-Hu-CMT-95,Hu-Leo-JMPS-97} using the energy minimum principle.
A  sophisticated kinematic approach to semicoherent interfaces was suggested in  \cite{cermelli+gurtin-1994}.

It is clear for a semicoherent interface  that the glide along the interface is a
dissipative process.
The dissipation-based approach in the theory of semicoherent PTs within
inelastic materials assuming small sliding  and in the reference
configuration was developed in
\cite{Levitas-postulate-II-1995,Levitas-JMPS-II-97,levitas+idesman+stein-ijss1998},
which was generalized for  arbitrary sliding and in the
actual configuration   in \cite{levitas-ijss1998}.
Since discontinuities at the interface were treated in these works as the {\it contact problem} at the moving interface,
solutions in the actual configuration
were much simpler.
It is assumed that SC, interfacial sliding, and decohesion
 are thermodynamically independent processes which  interact through the stress fields only. While in the general theory \cite{Levitas-postulate-II-1995,Levitas-JMPS-II-97,levitas-ijss1998} the change in the free energy due to interfacial sliding was included, it was neglected in all applications.
 The conditions at the interface are then described by the following equations:
\bey
\mbox{\rm coherent interface:}    \;\;   \;\mid \sg_n \mid \, < \, \sg_c \;\, \mbox{and} \;\,
f_s (\fg \tau ) \, < \, 0  & \Rightarrow &
\dot{\fg u}^2 =\dot{\fg u}^1 ;
\label{eq:y1} \\
\mbox{\rm semicoherent or incoherent interface:} \;\; \,  f_s (\fg \tau ) \, = \, 0 & \Rightarrow &
\dot{\fg u}^2_s - \dot{\fg u}^1_s = \fg q ( \fg \tau ) \, ;
\label{eq:yy1}
\eey
\bey
\mbox{interface with decohesion}: \;\;\mid  \sg_n  \mid \, = \, \sg_c & \Rightarrow &
\dot{\fg u}^2\neq \, \dot{\fg u}^1 \; ,
\quad \sigma_n \, = \,\fg  \tau \,= \, 0
\, .
\label{eq:yyyy1}
\eey
Here $\, \fg \tau \,$ and $\, \sg_n \,$ are the shear stress vector and normal stress
at the interface; $\, \sg_c \,$ is the
critical normal stress for decohesion; and $\, {\fg u}_s \,$ is the tangential to an interface displacements.
Function $f=0$ describes the limit curve in the plane of the shear stress $\fg { \tau}$, which characterizes athermal resistance to slip, within which relative sliding is prohibited, and $\boldsymbol{q}$ is the function that determines the kinetics of sliding. These functions, in general, incorporate crystallographic anisotropy, the magnitude of sliding rate, and $\, \sg_n \,$, as well as other features from the theory of crystal slip or discrete or continuously distributed dislocations  \cite{TheoDisl,Boikoetal-94,Asaro}.
The relative sliding $\,  {\fg u}^2_s -  {\fg u}^1_s \,$ can characterize continuously distributed dislocations \cite{TheoDisl,Boikoetal-94} within an interface and number dislocations is equal to $\, |{\fg u}^2_s -  {\fg u}^1_s|\,$ divided by the magnitude of the Burgers vector  $|{\fg b}|$.  The description of the interfacial sliding is formally similar to the
flow theory in plasticity, where $f$ is for the yield surface and $\fg q$ is for the flow rule.  In particular, in
\cite{Levitas-postulate-II-1995,levitas-ijss1998} an associated sliding rule was derived using the extremum principle similar to that in plasticity, that was derived using the postulate of realizability.

{\it Remark.} Note that such a continuum description of the incoherence does not completely reflect real dislocation processes, because the  Burgers vector of dislocations and their sliding are limited to the interface only, which is typical for misfit dislocations. In general,  the  Burgers vector and sliding
can be inclined to the interface and dislocations may slide together with the propagating interface (glissile interface), thus producing much less resistance to the interface motion than the misfit dislocations.

For isotropic functions $f$ and $q$ Eqs. (\ref{eq:y1})-(\ref{eq:yy1})  simplify to
\bey
\mbox{\rm coherent interface:}    \;\;   \;\mid \sg_n \mid \, < \, \sg_c \;\, \mbox{and} \;\,
\mid \fg \tau \mid  \, < \, \tau_s & \Rightarrow &
\dot{\fg u}^2 =\dot{\fg u}^1 ;
\label{eq:y} \\
\nonumber \\
\mbox{\rm semicoherent or incoherent interface:} \;\; \,  \mid \fg \tau \mid \, = \, \tau_s & \Rightarrow &
\dot{\fg u}^2_s - \dot{\fg u}^1_s =k \fg \tau  \, ,
\label{eq:yy}
\eey
where $\, \tau_s \,$ is  critical  shear stress (or athermal threshold) for sliding and $k>0$ is a scalar, which is determined from the condition $|{\tau}| = \tau_c$.
Note that the last condition corresponds to the main equilibrium equation for continuously distributed  dislocations \cite{TheoDisl,Boikoetal-94}.

If during the increase in  $\, \fvep_t \,$ and
changing  thermomechanical  properties in a nucleus, a chosen decohesion condition
is satisfied in some point of the interface, the crack appears or grows. If during the
same process the sliding criterion is met, we admit glide in
this point up to a value, at which the criterion is violated. After finishing
the SC, we use  the SC criterion to determine if SC is a thermodynamically admissible process.

\subsection{Propagation of a semicoherent interface }\label{semicoherent-int}

\begin{figure}[htp]
\centering
\includegraphics[width=0.5\textwidth]{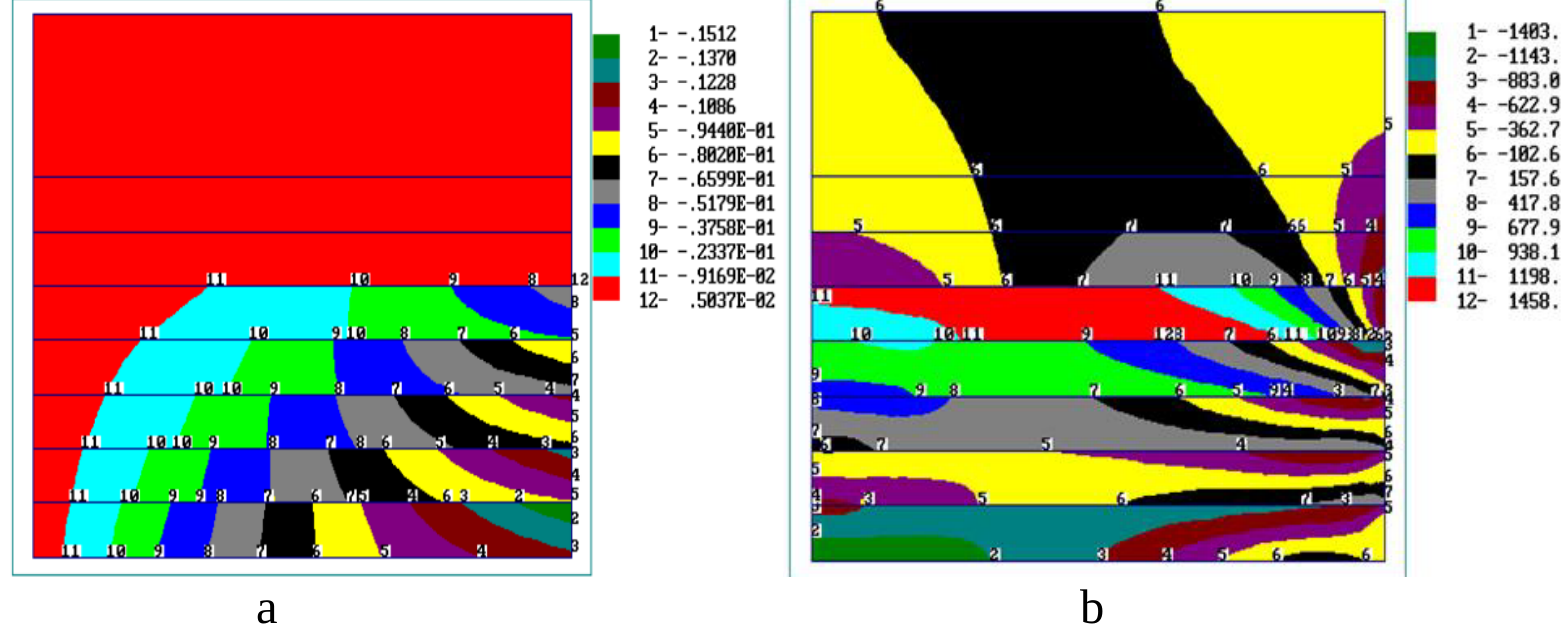}    
\caption{ Isobands of distribution of radial displacements (a) and mean stress (b)  after the semicoherent interface, propagating layer-by-layer from the bottom until reaching the middle of a sample (a).  Reproduced with permission from
\cite{levitas+idesman+stein-ijss1998}.
\label{Fig-incoherent}}
\end{figure}

Propagation of  coherent and semicoherent interfaces in elastic and elastoplastic cylinders
 under an axial stress of 100 MPa is investigated in
\cite{ides-lev-CMS-1997,levitas+idesman+stein-ijss1998,idesman-levitas+stein-ijp-2000},
see Fig.  \ref{Fig-incoherent} for a semicoherent interface. Interface
propagation is simulated by layer-by-layer transformation.
For a coherent interface in an elastic material at
$\, K \, = \, {\sl const} \,$, the transformation
work  $\, \varphi \,$ grows for each next layer, i.e., after PT  in the
first layer propagation should occur with increasing velocity.
Phase
equilibrium could be achieved if  $\, K \,$
grows sufficiently  with the growing volume fraction $\, c \,$ of the product
phase or if it is distributed heterogeneously. For a semicoherent interface,  it is assumed that the sliding displacements  (dislocations) at the layer's interface do not vary after finishing PT in
the layer, so they represent a memory about semicoherent PT. Discontinuity in displacements causes discontinuity in pressure (Fig.  \ref{Fig-incoherent}) and all stresses except normal and shear stresses at the interface.
We did not consider further sliding within the product phase, assuming that the yield strength in shear is much larger for inherited dislocations, because in most cases dislocations do not belong to the main slip systems of the product phase. Sliding of dislocations inherited by propagating interface is taken into account within PFA to the interaction between the PT and dislocations in \cite{levitas-javanbakht-APL-13,Levitas-Javanbakht-15-1,Javanbakht-Levitas-JMPS-15-2}.

Stress relaxation due to semicoherence increases $\, \varphi \,$ and consequently the driving force for  PT  in
the first layer; i.e., PT can start at a higher temperature than for the coherent interface. The transformation work $\, \varphi \,$
slightly grows for PT in the second layer, i.e., the interface will propagate at a fixed applied stress and temperature.
However, the driving force for the PT in the third layer is smaller than for the first layer, and it continues decreasing for the fourth and fifth layers. Thus, there is a tendency for interface arrest because of stress relaxation due to semicoherence.

\par
Next, after PT in the first layer, the second and third layers were compared  as the next transforming
region \cite{levitas+idesman+stein-ijss1998}, based on the extremum principle
(\ref{ijss1-179}).  For a coherent interface, continuous interface propagation through the sample
was obtained. For a semicoherent interface, the transformation work was larger for PT in the third layer (for relatively small $\tau_s$),  which leads to formation of discrete martensitic microstructure.
Thus, a semicoherent interface
has lower mobility than a coherent one and  can be arrested more easily, in accordance with experiments.
 \par
Plastic deformation in the parent phase in this problem was considered in
\cite{ides-lev-CMS-1997,levitas+idesman+stein-ijss1998, idesman-levitas+stein-ijp-2000}.
For a semicoherent interface, plastic deformation  is quite small and weakly affects PT because
$\tau_s$ is much smaller than the yield strength in shear.
For a coherent interface, plasticity, as the stress relaxation mechanism,  produces effects similar to those for  sliding along the interface.
Thus, plasticity increases the transformation work for PT in the first layer (in comparison with elastic parent phase) but then leads to PT in the third and fifth layers, leading to a discrete martensitic microstructure and possible arrest of the interfaces.
\par
The effect of strain hardening, leading, according to Eq. (\ref{003}), to the increase in the athermal  threshold $ \, K \,$,
was analyzed in \cite{ides-lev-CMS-1997,idesman-levitas+stein-ijp-2000}.
 Since plastic deformation and increased  $ \, K \,$  are localized more near an interface, it
promotes the formation of a discrete microstructure.
Note that different scenarios of interface propagation strongly depend on the chosen material parameters
($K \,$, $\tau_s$, $\sg_y$,  $\vep_0$, etc.)
and applied stresses.

\subsection{Stress-induced PT in a spherical particle within a matrix and its interaction with
plasticity, semicoherence and adhesion}\label{sphere-FEM-int}

\begin{figure}[htp]
\centering
\includegraphics[width=0.5\textwidth]{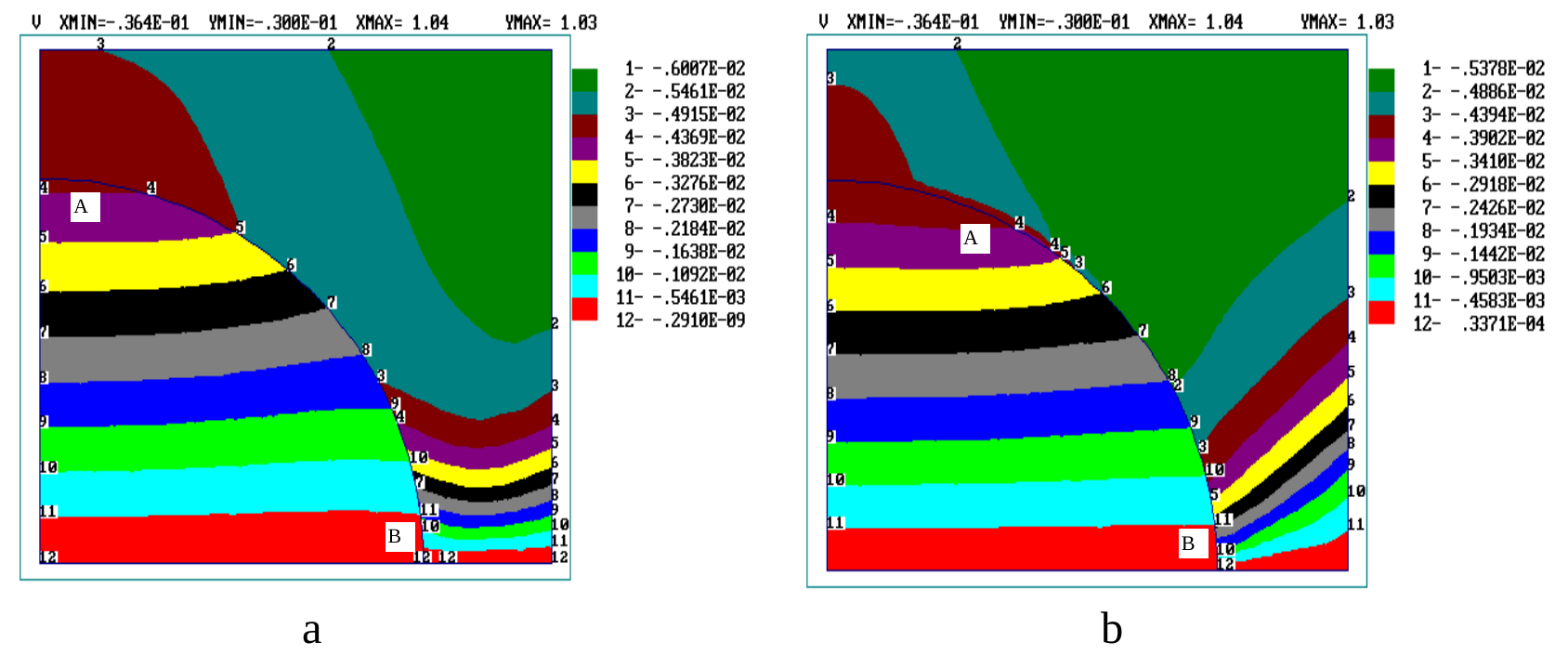}   
\caption{ Isobands of  vertical displacement
       distributions (mm) at volumetric transformation strain
$\, \vep_{o2} \, = \, -0.005 \,$
and compressive axial stress $P \, = \, 150 \, MPa$  for a semicoherent interface (critical shear
       stress $ \tau_s \, = \, 40 \, MPa$) (a)  and with the decohesion at the interface for the critical normal stress
       $ \sg_c \, = \, 50 \, MPa$ (b). AB is the sliding zone in (a) and decohesion zone in (b).
 Reproduced with permission from
\cite{levitas+idesman+stein-ijss1998}.
\label{Fig-cyliner-isobands}}
\end{figure}

\begin{figure}[htp]
\centering
\includegraphics[width=0.8\textwidth]{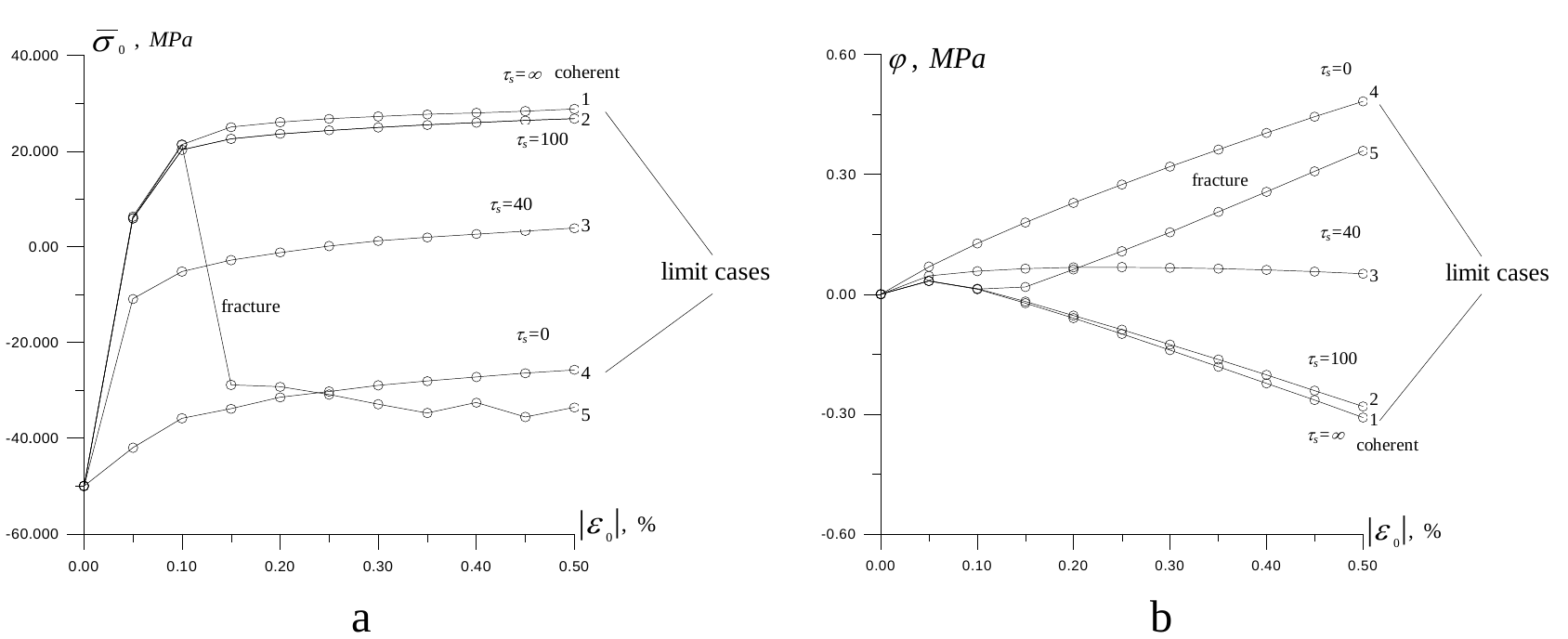}   
\caption{ Relationships between the mean stress  $\, \bar{\sg}_o \,$,
averaged over the nucleus (a) and the transformation work $\varphi$  (b) vs. the magnitude of the  transformation volumetric strain $ |\vep_o|$  at    axial stress    $P \, = \, 150 \, MPa$.  1-coherent interface, 2, 3, and 4 - semicoherent
       interfaces with the  critical shear stresses $ \tau_s \, = \, 100,
       \, 40, \; {\rm and }\; 0 \, MPa$, respectively; 5 - interface with decohesion for the
 critical normal stress $ \sg_c \, = \, 50 \, MPa$.
 Reproduced with permission from
\cite{levitas+idesman+stein-ijss1998}.
\label{Fig-cyliner-curves}}
\end{figure}

Following \cite{levitas+idesman+stein-ijss1998}, we will present results for PT with  volumetric transformation
strain  $\, \vep_0 =-0.005\,$ within  a spherical particle embedded in a nontransforming  cylindrical matrix (Figs. \ref{Fig-cyliner-isobands} and \ref{Fig-cyliner-curves})
at a fixed temperature and compressive axial stress $P=150\, MPa$.  Solutions are found for coherent
and semicoherent interfaces and an interface with decohesion, with different magnitudes of $\tau_c$ and $\sg_c $.
Examples of discontinuity of displacements for a semicoherent interface and an interface with decohesion are shown in Fig. \ref{Fig-cyliner-isobands}.
The averaged mean stress  $\, \bar{\sg}_o \,$ and transformation work
$\, \varphi \,$ as functions of  transformation strain $\, |\vep_o |\,$ for different interface conditions are presented  in Fig. \ref{Fig-cyliner-curves}.   The mean initial compressive stress in a sphere of $50 MPa$ reduces in magnitude and changes the sign during PT (increase in $\, |\vep_o |\,$)  due to internal stresses, which reduces the transformation work and makes it negative for a coherent interface and semicoherent interface with $\tau_c=100\, MPa$.  Interfacial sliding and decohesion relax internal stresses and increase the transformation work.  Remarkably, initiation of  decohesion leads to a large drop in tensile mean stress in a particle and a corresponding sharp change in slope of the plot for $\, \varphi \,$.
  Thus, the minimal $\, \varphi \,$ and consequently the driving force for SC is for the coherent
interface, the maximum $\, \varphi \,$ is for $\tau_c=0$. However, even for  $\tau_c=0$ the stress state in a spherical particle is nonuniform and nonhydrostatic, so it can hardly be called an incoherent inclusion.
The interface with decohesion shows the second largest transformation work.
\par
 Let us assume that the  external stress is small enough (in particular, zero) to cause plasticity,
semicoherence, and decohesion without PT.  We can choose a temperature at which PT criterion  (\ref{eq:6})  is not met without plasticity, semicoherence, and decohesion, because the transformation work  is too small
without stress relaxation.  Thus, none of the inelastic  processes can occur separately under the chosen conditions.
At the same time,  when at least two   inelastic stress relaxation mechanisms proceed simultaneously, they assist  each other via  the field
of internal stresses, relax internal stresses,  and can  all occur thermodynamically and based on sliding, decohesion, and/or yield criteria.

\subsection{Semicoherent interface within a phase field approach}\label{PFA-semicoherent}

The first work on introducing incoherence in the diffuse interface within PFA was presented by
\cite{Appolaireetat-PhilMag-10}. Because of the finite-width interface, localized sliding was substituted with additional eigen strain fields within an interface that relaxes interfacial shear stresses. An additional order parameter $\eta_{in}$ for the description of incoherence is introduced, which is equal to zero and one for the coherent and incoherent interfaces, respectively. The evolution equation for the order parameter is derived thermodynamically, and the driving force
for the evolution of incoherence is the shear stress along the interface minus the change in the interfacial energy $\gamma$
due to change in the order parameter $\gamma(\eta_{in})$, which was approximated by some function. From a thermodynamic point of view,
the theory in \cite{Appolaireetat-PhilMag-10} is the finite-width counterpart of the theory for sharp interfaces in
 \cite{Levitas-postulate-II-1995,levitas-ijss1998}. At the same time, change in the interfacial energy with incoherence was taken into account in the solution of all problems, which eliminated complete relaxation of the shear stresses. However, function $\gamma(\eta_{in})$ was not included in the expression for the interfacial energy that governs the evolution of the main order parameter $\eta$ describing evolution of the two-phase system.
 This did not affect correctness of the results of the solution of various problems for different geometries of the interfaces  in   \cite{Appolaireetat-PhilMag-10} because they were presented for fixed interfaces only.
 Also, results were presented for the stress-free external surfaces,
 i.e., the problem of  the finite strength for the incoherent interfaces was not discussed.

\subsection{New approach to incoherent interface }\label{incoherent}

A method to relax stresses within the moving finite-width solid-melt interface within PFA was suggested in \cite{Levitas-Samani-PRB-11}.  While displacement continuity was assumed, due to the zero shear moduli of
the melt, deviatoric stresses in the melt are zero. The problem was that stresses within the finite-width interface, within which shear modulus varies from that for a solid to zero, due to the traditional pure spherical transformation strain for solid-melt PT, ${\vep}_{0t}$, were unreasonably high.  This did not allow for reproduction of the experimental size dependence of the melting temperature of the Al nanoparticle and  temperature-dependent thickness of the surface molten layer. To reduce interfacial stresses, the deviatoric part of the transformation strain, ${e}_t$ was introduced, which evolved according to thermodynamically consistent evolution equation during the PT
\bey
\fg {\dot{e}}_t = \Lambda | {\vep}_{0t} \dot{\phi}(\eta) |{\textit{\textbf{S}}},
 \label{hF9uy9}
\eey
where $\Lambda\geq 0$ is the kinetic parameter and ${\phi}(\eta)$ is an interpolation function of the order parameter $\eta$ for the volumetric transformation, ${\phi}(0)=0$, ${\phi}(1)=1$.
This led to significant reduction of the interfacial stresses, controlled by a kinetic parameter $\Lambda $, and the quantitative description of the above experimental data for Al.

A similar approach was developed in a thermodynamic theory in \cite{Levitas-Attariani-SciRep-13,Levitas-Attariani-JMPS-14} to relax internal stresses during a diffusion-driven compositional expansion/contraction in an amorphous material, in particular, during the lithiation-delitiation of silicon.
Instead of the isotropic compositional expansion typical for an isotropic amorphous material,  anisotropic deviatoric-stress-dependent compositional expansion was introduced and described by the following equation
\bey
 {\fg d}_{c }^S =
 \Lambda (  x){\fg S} \frac{d J_c}{d x}
 |\dot{x}|.
\label{i+1hgjh}
\eey
Here, ${\fg d}_{c }^S$ is the deviatoric part of compositional deformations rate, $J_c(x)$ is the third invariant of the compositional part of the deformation gradient (which describes volumetric compositional expansion) as a function of concentration of saturating atoms, $x$.  This led to  obtaining very good correspondence with experimental and atomistic results on the biaxial stress relaxation in $Li_x Si$ on a
rigid substrate in the course of lithiation-delithiation, utilizing just a single fitting kinetic parameter $ \Lambda$.

The results obtained allow us to {\it conceptually reconsider the definition and way of description of the incoherent interfaces in solids}, in order to include finite shear strength at least after transformation with strong stress relaxation during PT or chemical reaction. Since there is no atomic correspondence  across an incoherent interface, it is unphysical to describe this process through dislocation generation or sliding along the interface. We postulate that the lack of atomic correspondence
across an incoherent interface is due to the reconstructive atomic motion in the transforming volume covered by a moving interface.
 Then we keep
displacement  continuity across an interface but add a mechanism of stress relaxation within the transformed phase by introducing a deviatoric-stress-dependent deviatoric part of the transformation strain rate, described by equations of the type of Eqs.  (\ref{hF9uy9}) or  (\ref{i+1hgjh}), for the PFA,  or description in terms of volume fraction of phases or concentrations of the diffusing species, or the sharp interface approach.
Then the internal deviatoric stresses significantly (or almost completely) relax during PT or reaction, but after stopping transformation,
the two-phase material and incoherent interface can carry deviatoric stresses, which are determined by plasticity/strength of bulk phases and interfaces.

\subsection{Semicoherent interface within a phase field approach with discrete dislocations}\label{PFA-interface-dislocations}

\begin{figure}[htp]
\centering
\includegraphics[width=0.8\textwidth]{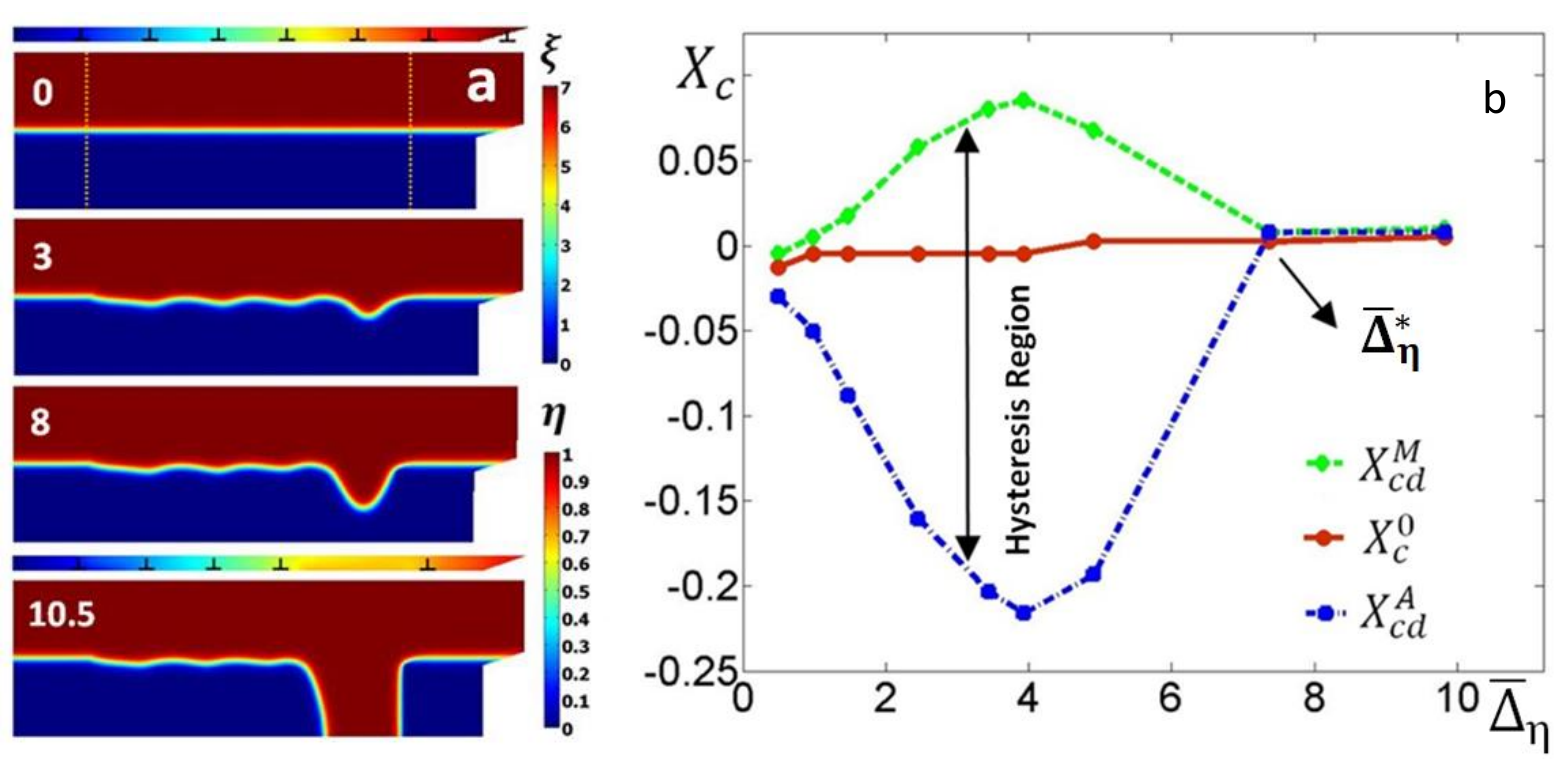}
\caption{ (a) PFA solution for the coupled   austenite (blue) - martensite (red) interface and misfit interfacial dislocation evolution. The thin band above the specimen demonstrates evolving dislocations  along the initial austenite-martensite interface, in order to exclude overlapping with the phase interface. (b)
Dependence of the critical dimensional thermal driving force $X_c$ for starting the interface propagation leading to complete PT in a sample  vs. relative interface width $\bar{\Delta}_\eta$.
The upper and lower lines for semicoherent interface correspond to initiation of the PT to martensite and austenite, correspondingly. The interface is arrested  for the driving forces between these lines, thus, producing  scale-dependent athermal hysteresis region. Hysteresis and athermal interface friction is absent for the coherent interface (middle red line). Reproduced with permission from  \cite{Levitas-Javanbakht-PRB-12,levitas-javanbakht-APL-13}.
\label{Fig-semicoh-PFA}}
\end{figure}

While PFA to coupled PTs and discrete dislocations  will be described in more detail in Section \ref{PT-disl}, here we will focus on semicoherent interfaces  \cite{Levitas-Javanbakht-PRB-12,levitas-javanbakht-APL-13}.
Initially, a stationary solution for the  coherent finite-width austenite-martensite interface was obtained  (Fig. \ref{Fig-semicoh-PFA}a)  with a misfit strain along the interface $\vep_m=0.1$.
The coherent interface could be kept stationary at some temperature close to the phase equilibrium temperature (and corresponding  normalized thermal driving force $X$) for a stress-free material, see the red line in Fig. \ref{Fig-semicoh-PFA}b designated as $X_c^0$. This small deviation is caused by nonsymmetric sample geometry due to dislocations and  energy of internal stresses, which, due to small sample size, slightly depends on the interface position.
At any other temperature, the interface propagates until completion of direct or reverse PT, depending on the temperature, which was not a surprise. It is known (see \cite{levitasetal-prl-07,levitasetal-IJP-10}) that PFA for PTs does not include an athermal threshold to the interface propagation, and one needs to develop some special ways (e.g., oscillating distribution of internal stresses or some heterogeneities) to introduce it into PFA.

Internal stresses due to a misfit strain led to   nucleation of the misfit dislocations at the intersection of the interface with a sample free surface, which  propagated along the interface  producing the stationary distribution of dislocations. Equilibrium spacing between misfit   dislocations, $s$, was found to be in perfect agreement with an analytical expression $s=|{\fg b}|/\vep_m $.

  For a semicoherent interface,  temperature and normalized thermal driving force $X$ can be varied in some range, limited by the critical values $X_{cd}^M$  for  martensitic PT and  $X_{cd}^A$ for martensite$\rightarrow$austenite PT (Fig. \ref{Fig-semicoh-PFA}b), without interface motion. These critical values of $X$ represent an athermal threshold (or interfacial friction) for the interface propagation due to misfit dislocations. After exceeding these thresholds, the   entire sample transforms into a single phase--austenite or martensite--for different signs of $X$.
   What is very surprising is that the {\it athermal thresholds  strongly depend on the ratio of two nanosize parameters } $\bar{\Delta}_\eta=\Delta_\eta/H$, where $\Delta_\eta$
is the interface width and $H=2|\boldsymbol b|$ is the dislocation height, see Fig. \ref{Fig-semicoh-PFA}b.
 Indeed,
    the  athermal interfacial friction is  a macroscale characteristic which can be measured in a macroscale test on a single or multiple interface propagation, and for macroscale treatment all nanoscale parameters are usually neglected. However,  the dimensionless ratio of two nanosize parameters
is a finite number and, according to Fig. \ref{Fig-semicoh-PFA}b, should be taken into account even in the macroscale treatment. Also, (almost) zero athermal friction for very small $\bar{\Delta}_\eta$ and for   $\bar{\Delta}_\eta >7$ is a very nontrivial result. The obtained result that the athermal hysteresis for  a broad enough interface is zero is intuitively
acceptable, because the resultant driving force $X$ from the interaction of all infinitesimal layers, producing the broad interface, with "thin" misfit dislocations disappears due to mutual compensation.
Since the dislocation height is approximately equal to  the crystal lattice parameter, this result also implies  that for broad  enough interface  the Peierls barrier due to discreteness of the crystal lattice tends to be zero  as well.

However, the case for an (almost) sharp interface looks contradictory and requires further study.
Indeed, for 1D plane interface propagation through  the oscillating stress field, which was used in  \cite{levitasetal-prl-07,levitasetal-IJP-10} in order to introduce an athermal barrier in PFA, the hysteresis was finite and was  determined by extreme (positive and negative) values of the oscillating stresses. Since dislocations also produce the oscillating stress field, the same is expected for the interactions of very thin interfaces with dislocations.
 One of the possible reasons of this discrepancy  is that
 in \cite{Levitas-Javanbakht-PRB-12,levitas-javanbakht-APL-13} the interface does not move as a plane. Instead, the thin interface penetrates between dislocations, pushes them away, increasing spacing between dislocations, and finally loses its stability and  propagates laterally through the entire sample (Fig. \ref{Fig-semicoh-PFA}a).
The finite size of the sample may contribute to this phenomenon, so larger scale simulations are required.

Note that many other examples within PFA were found for which the ratio of two nanoscale parameters (e.g., the width of two different interfaces or width of the interface and the external or internal surfaces) drastically affects
PT nanoscale  and macroscale behavior, see review by \cite{Levitas-ScriptaMat-18}.
 This ratio produces new phenomena, changes PT parameters and mechanisms, and should be considered as a new dimension in a "phase diagram."
Additional examples include PT between two solids via an intermediate phase (melt) and surface-induced melting of nanoparticles and martensitic PTs.

While the results above were limited to misfit dislocations that nucleate and evolve along the interface only, a more general situation was treated in \cite{levitas-javanbakht-Nanoscale-14,Javanbakht-Levitas-PRB-16,Javanbakht-Levitas-JMS-18}, where dislocations move along the natural slip system of evolving phases. In those examples, one of the initially coherent phase interfaces adjusted itself to the stress field of a dislocation pileup, and a  high-pressure phase was arrested at the dislocation side with an extra atomic plane because it could not propagate to the side with missing atomic planes and tensile pressures (Fig. \ref{Fig-PFA-pressure-shear}).  The results for this interface are quite similar to those for the interface with misfit dislocations.

An alternative approach to the coherency loss of the stationary spherical precipitate with volumetric transformation strain using PFA to discrete dislocations was suggested by \cite{Geslinetat-AM-14}. Their PFA is developed in terms of shear strain along the specific glide planes  $\{100\}$ and directions   $<100>$ as  order parameters. Periodic dependence of the  energy on this shear strain is accepted, which allows to reproduce punching of prismatic dislocations with edge and screw components from the precipitate. While small strains are assumed, it is written that generalization for finite strains is not a problem.

\section{Solid-solid phase transformations via intermediate (virtual) melt}\label{virtual-melt}

\begin{figure}[htp]
\centering
\includegraphics[width=0.6\textwidth]{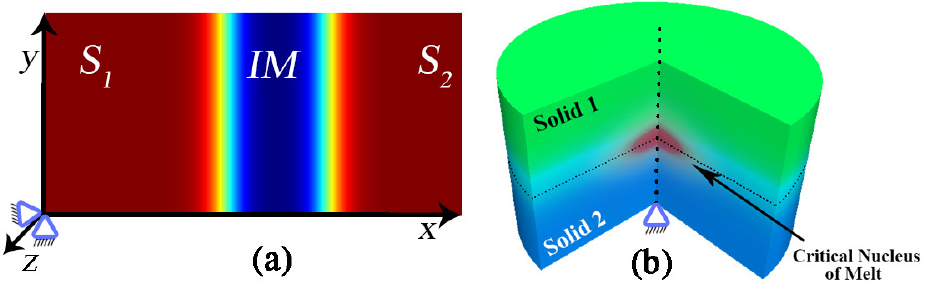}
\caption{Phase field solutions for a propagating solid 1-intermediate melt-solid 2 ($S_1-IM-S_2$) interface (a)
and for the critical nucleus of the $IM$ within $ S_1S_2 $ interface (b). Reproduced with permission from \cite{Momeni-Levitas-PCCP-16}.}
\label{figVM}
\end{figure}

When traditional mechanisms of plastic relaxation of elastic stresses through  dislocation motion and twinning are inhibited, nature finds  alternative ways to relax internal stresses.
Thus, solid-solid PTs through a nanometer-size liquid  layer, hundreds of Kelvins below the bulk melting temperature, was  predicted by continuum  thermodynamic estimates  and confirmed directly or indirectly in experiments for various material systems,
see \cite{Levitas-Henson-PRL-2004,Levitas-PRL-2005,Levitas-Henson-JPCB-2006,Levitas-Ren-PhysRevB-2012,Bowlan-etal-JCP-19}.
Instead of traditional propagating solid 1- solid 2 ($S_1-S_2$) interface,    solid 1-intermediate melt-solid 2 ($S_1-IM-S_2$) interface propagates through material with a thin layer of melt (Fig. \ref{figVM}a). This means that the solid $S_1$  partially or completely melts and recrystallizes into  $S_2$. Complete (or partial) melt within PFA for melting means that the order parameter $\eta_m$ describing melting and varying between 0 for bulk melt and 1 for solid, is equal to 0 (or is between 0 and 1). Complete melt fully relaxes deviatoric stresses, which appear at $S_1-S_2$ interface
due to transformation strain. This reduction in elastic energy due to melting increases the thermodynamic driving force for melting and leads to melting below the bulk melting temperature.
$IM$ transforms a coherent stressed interface in a stress-free noncoherent interface. Since melt does not interact with the stress field of defects and does not possess a Peierls barrier, the  athermal friction $K=0$.  Elimination of elastic energy and athermal friction may lead to PT with zero transformation hysteresis and energy dissipation,
which are ideal property for  shape memory alloys for actuation or medical applications  \cite{cuietal-2006,Chlubaetal-15,Song-etal-2013} or caloric materials \cite{Song-etal-2013,Takeuchi-2015,Houetal-Science-19}.

Another reason for   melting  significantly below the bulk melting temperature is the reduction in the total interface energy,
i.e. when the energy of two solid-melt interfaces is smaller than the energy of
a coherent solid-solid interface. There is the following difference between the $IM$ and virtual melt ($VM$).
The $IM$ is thermodynamically stabilized by a reduction in surface and elastic energies and can exist for a resting interface.
The $VM$ is an unstable transitional phase along the transformation path between two solid phases. It disappears in a material point quickly after it appears, and does not exist in the stationary interface but can
exist within a propagating interface. Short review of the virtual melting phenomena is given in
\cite{Levitas-hidden-mech-chem-chaprter-10}; see also \cite{Levitas-ScriptaMat-18}.

Below are some examples of solid-solid PTs via virtual or intermediate melt.

(a) The concept of the virtual melt was first introduced for the description of numerous counterintuitive experimental results for  reconstructive $\beta \leftrightarrow \delta$ PTs  in the organic energetic crystals HMX \cite{Levitas-Henson-PRL-2004,Levitas-Henson-JPCB-2006}, which were considered major puzzles
 for decades.
In total, sixteen   theoretical
predictions based on $VM$ are in
qualitative and quantitative agreement with experiments
\cite{Levitas-Henson-PRL-2004,Levitas-Henson-JPCB-2006,levitas-etal-06-APL-nucleation}.
In particular:

-  melting could indeed occur 120 K below the melting temperature. The energy of internal stresses due to volumetric transformation
strain $\vep_0=0.08$ is sufficient to reduce the melting
temperature from 551 K to 430 K
 for the $\delta$ phase during the  $\beta \rightarrow \delta$ PT studied at  430 K
 and from 520 K to 400 K for the $\beta$
phase during the  reverse $\delta \rightarrow \beta$ PT. Change in surface energy was neglected.

- Zero energy of elastic internal stresses and athermal friction  for both $\beta\leftrightarrow\delta$ PTs explain
 the experimentally observed lack of the temperature hysteresis, which usually exists for all known solid-solid PT.

 - Activation energies for direct
and reverse PTs are equal to the corresponding
melting energy. Temperature dependence of the rate
constant is determined by
the heat of fusion, both like in the experiment.

\begin{figure}[htp]
\centering
\includegraphics[width=0.6\textwidth]{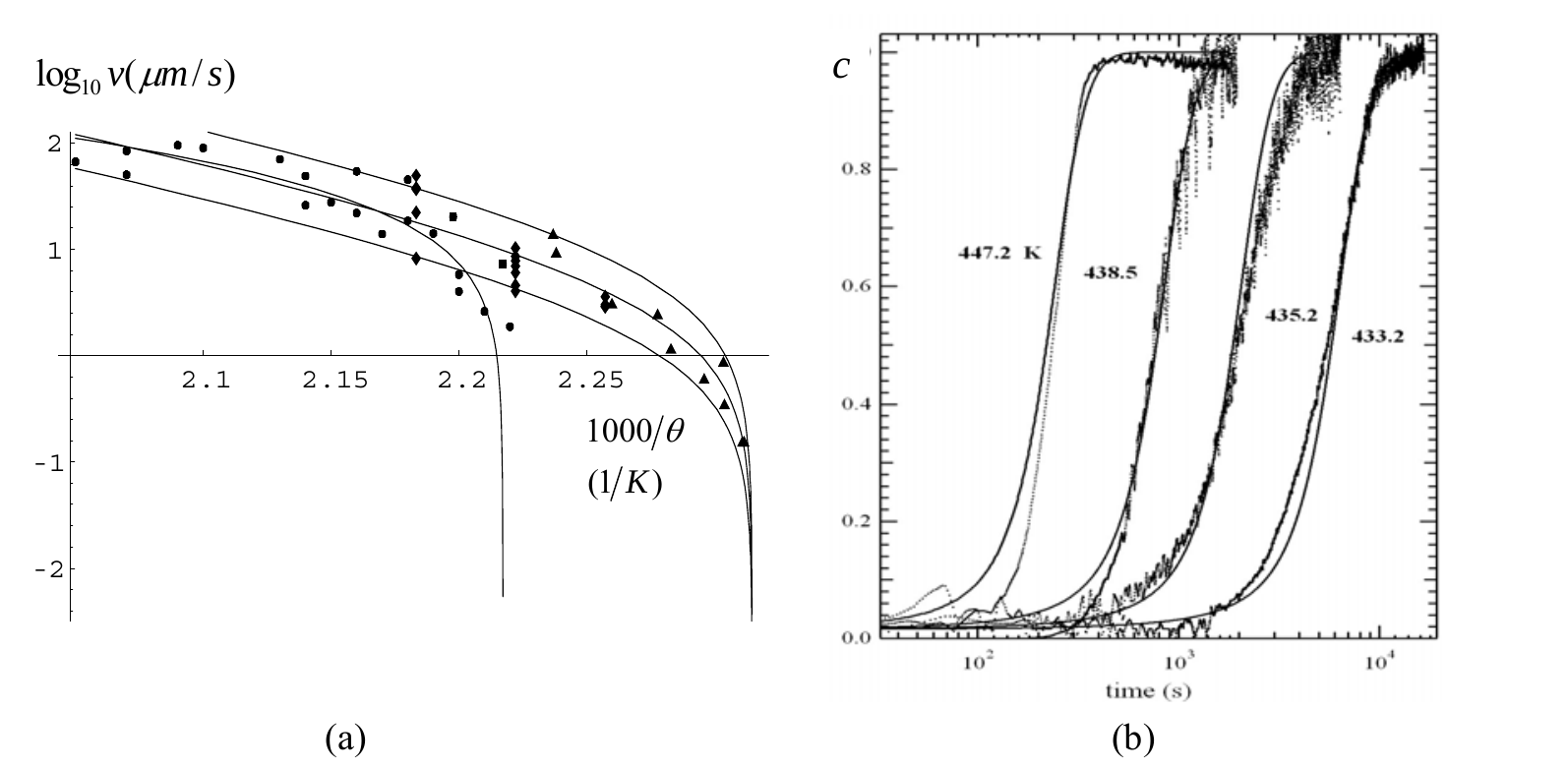}
\caption{A comparison between theoretical prediction and experimental data for
kinetics of the isothermal $\beta \leftrightarrow \delta$ transformation in the HMX energetic crystal \cite{Levitas-Henson-PRL-2004,Levitas-Henson-JPCB-2006}. (a) The $\beta - \delta$ phase interface velocity vs. temperature. Symbols represent experimental data obtained at LANL and LLNL using three
different methods.  The three lines are theoretical predictions with three slightly different pre-exponential factors based on the
$VM$ mechanism. (b) Theoretical prediction (solid
curves) in comparison with  experimental data for volume fraction of the $ \delta$ phase. Reproduced with permission from \cite{Levitas-Henson-JPCB-2006}.}
\label{fig-HMX}
\end{figure}

- Kinetics of phase interface propagation and a physically-based
 kinetic model in terms of volume fraction of the $\delta$ phase are in
good correspondence with  experiments (Fig. \ref{fig-HMX}).

- Nanovoids in the  transformed
material that accompany the PT do not affect  the PT
 thermodynamics and kinetics for the cyclic $\beta \leftrightarrow \delta$ PTs, like in experiments.

Kinetics of  $\gamma$-$\delta$ PT in HMX was also explained by $VM$, but a reduction in interfacial energy during PT was involved as well \cite{Bowlan-etal-JCP-19}.

(b) Virtual melting was suggested as the mechanism for crystal-crystal and crystal-amorphous PTs in materials with melting temperature decreasing with pressure,  e.g., in Si, Ge,  ice, geological material (e.g., quartz and coesite), and superhard materials (hexagonal BN and graphite) \cite{Levitas-PRL-2005}. When pressure-induced PT of phase 1 to phase 2 is suppressed due to large transformation strain and athermal friction, with increasing pressure, the melting line extrapolated to lower temperatures can be crossed before the line for PT $1\rightarrow 2$. After phase 1 melts,  the melt finds itself  in the region of the phase-temperature diagram, much below the melting temperature of phase 2. Therefore,  material solidifies in a stable phase 2. Below the glass formation temperature, solidification occurs into the amorphous phase, and above the glass formation temperature, crystalline phase 2 appears. An alternative scenario consists of the nucleation of phase 2, which causes large internal stresses, which relax via $VM$ followed by solidification into amorphous phase 2 (amorphization via internal stress induced $VM$). This model was applied to the explanation of melting, crystal-crystal, and crystal-amorphous PT  mechanisms in ice Ih.  Note that the $VM$ occurs in  Si and Ge at a temperature of more than 1000 K below the thermodynamic melting temperature!

(c) A new mechanism for crystal-crystal PTs via surface-induced $VM$ is justified
thermodynamically and confirmed experimentally for the PT in $PbTiO_3$ nanofibers \cite{Levitas-Ren-PhysRevB-2012}. For nanofibers, surface melting starts at a temperature much below the bulk melting temperature. When the thickness of the
surface melt  exceeds the size
of the critical nucleus of the product-phase, nucleation, and growth of the product phase takes place. For nanofibers, surface melting
starts near the smallest size, and hydrodynamic flow caused by a reduction of the external surface and, consequently,  its energy leads to a large shape
change   towards a cube and additional  promotion of crystal-crystal PT. In the course of  the product crystal growth, $VM$ is experimentally observed within the crystal-crystal interface using transmission electron microscopy.

Two  different PFAs to the $IM$ were developed  using two order parameters: one describing solid-solid PT and another one for melting, see Fig. \ref{etai_etaj_plane}(c) for approach in \cite{Momeni-Levitas-PRB-14,Levitas-Momeni-ActaMat-14,Momeni-Levitas-15,Momeni-Levitas-PCCP-16} and Fig. \ref{etai_etaj_plane}(f) for approach in \cite{levitas-roy-ActaMat-16}, as well as Sections \ref{PFA-sphere} and \ref{PFA-pen} for details. Also, papers
\cite{Levitas-Momeni-ActaMat-14,Momeni-Levitas-Warren-15,Momeni-Levitas-15} include coupling with elasticity
and  \cite{Momeni-Levitas-PCCP-16}  include interfacial stresses.
Internal elastic stresses   promote the existence and persistence of the $IM$. In particular, in \cite{Momeni-Levitas-Warren-15},   the internal stresses  decreased the activation energy of $IM$ critical nucleus (Fig. \ref{figVM}b) by a factor of 16 for the HMX,  making thermally activated nucleation of the $IM$ possible.

PFA solutions and obtained nanostructures, in addition to expected parameters, like the ratio of energies of solid-solid to solid-melt interfaces   and initial state, were found to strongly and nontrivially depend on the  ratio of widths of solid-solid to solid-melt interfaces, $k_{\delta}$.
Depending on $k_{\delta}$ (and other parameters), several types of $IM$ behavior are found:

(a) for small $k_{\delta}$, jump-like (first-order PT) nucleation of the interfacial disordering and then continuous (second-order PT) and reversible increase in disordering with temperature;

(b) for larger $k_{\delta}$, coexistence of $S_1-S_2$ and $S_1-IM-S_2$ nanostructures and jump-like  PTs between them for increasing and decreasing temperature;

(c) retaining of $IM$ as a metastable interfacial phase significantly below the bulk melting temperature, even  when the energy of solid-solid interface is smaller than the energy of two solid-melt interfaces,
and

(d) unstable $IM$, which is a critical nucleus between $S_1S_2$ and $S_1MS_2$ nanostructures  (Fig. \ref{figVM}b).
Increase in $k_{\delta} $ suppresses barrierless $IM$ nucleation but promotes retaining of $IM$ at much lower temperatures.

There are various follow-up works on virtual melting. Amorphization via virtual melting in Avandia (an important
antidiabetic pharmaceutical)  was proved experimentally using microcalorimetry \cite{randzio+kutner-08}.
Chemical reaction transforming Si into SiC via intermediate state, consisting of dilatational dipoles in Si, was independently studied in \cite{Kukushkin-Osipov-14}, including developing PFA.

Convincing direct experimental proof that
some reconstructive crystal-crystal  PTs (namely, the transition between square $\Box$ and
triangular $\triangle$ lattices of colloidal films of microspherical particles) can occur through nucleation via an intermediate liquid nucleus and
grow via an intermediate liquid layer that was presented in \cite{peng-NatMat-14}.
While it is stated in \cite{peng-NatMat-14} that crystal-crystal PT occurs below the bulk melting temperature $T_m$, the bulk thermodynamic  force for melting is considered to be positive. 
This is possible only if the temperature  is above the $T_m$  of the $\Box$ phase and below that for the $\triangle$ phase. In this case, it is not surprising that when both melting and crystal-crystal PT are thermodynamically possible, the nucleus with smaller activation energy (and, consequently, interface energy) appears first and then transforms to a more stable phase.

Also, this work was not be properly placed within the existing literature. Opposite  to the statement in \cite{peng-NatMat-14}, PTs between crystalline phases via intermediate and virtual melting were discussed for a decade and under much more surprising conditions, namely, significantly below the $T_m$, see the references in this Section.
Processes observed in \cite{peng-NatMat-14} and thermodynamic treatment are identical to particular cases of those
discussed in \cite{Levitas-PRL-2005,Levitas-Ren-PhysRevB-2012}. PFAs to the $IM$ discussed above are, of course,
much more powerful and informative than sharp interface approaches.
Additionally, it is stated in \cite{peng-NatMat-14} that the effect of anisotropic stresses on intermediate melt is worthy of study.
This topic was addressed in   \cite{Levitas-Ravelo-PNAS-2012}, where $VM$ in Al and Cu $4000 K$ below the $T_m$ under very high-strain-rate uniaxial compression was predicted thermodynamically and confirmed by MD simulations, see Section \ref{VM-shock}.

Despite these drawbacks, the
results in \cite{peng-NatMat-14} make valuable contribution by direct confirmation and visualization of the crystal-crystal PT via $IM$ for $\Box-\triangle$ reconstructive PT in colloidal films. They in-situ confirm, specify, and quantify  main statements in
\cite{Levitas-Henson-PRL-2004,Levitas-PRL-2005,Levitas-Henson-JPCB-2006,Levitas-Ren-PhysRevB-2012,Momeni-Levitas-PRB-14,Levitas-Momeni-ActaMat-14}, including fast growth kinetics for $S_1MS_2$ interface (consistent with the absence of the athermal friction) while a coherent crystal-crystal interface is arrested due to the athermal threshold.

Void nucleation due to sublimation within elastoplastic material via $VM$ is considered thermodynamically and kinetically  in \cite{Levitas-Alt-VM-PRB-09} and is compared with other scenarios (due to fracture, sublimation, and melting, and evaporation)  in \cite{Levitas-Alt-void-ActaMat-11}.

Some other processes that may occur via $VM$ and virtual amorphization, namely crystal reorientation during the nanofriction via the $VM$ \cite{hammerberg-03},  plastic deformation at a high-strain-rate tension of metallic
nanowires via the virtual amorphization \cite{ikeda-99}, fracture  \cite{okamoto-98,Lynden-Bell-95}, and grain boundary sliding and migration \cite{mott-48,shoenfelfer-97,wolf-01} are discussed in \cite{Levitas-hidden-mech-chem-chaprter-10}.

\section{Virtual melting as a new mechanism of plastic deformation and stress relaxation under high strain rate loading}\label{VM-shock}

Generation and motion of dislocations, twinning, and crystal-crystal
PTs are the main mechanisms of plastic deformation
and relaxation of non-hydrostatic stresses that are reflected in the
deformation-mechanism maps \cite{frost-89,meyers-01}.

Several large-scale non-equilibrium MD simulations  for
 metallic fcc single crystals found unexpected result
that for   propagation of a shock wave along the $<$110$>$ and $<$111$>$ directions,
melting occurs at temperatures below the equilibrium melt temperature
$T_m(p)$ at the  shock pressure $p$; for example, for Cu by
20\% in \cite{ravelo:2006} and by 7-8\% ($\pm 4\%$) in
\cite{luo:2008}. Usually, traditional superheating is observed. This suppression in $T_m(p)$ was interpreted  in terms of solid-state disordering due to high defect-densities, but significant dissatisfaction was remained.
The decrease in the  melting temperature   caused by deviatoric stresses was estimated by traditional approaches \cite{Grinfeld-1991,Sekerka-Cahn-04}
to be just 1 K. That is why it was
not considered for interpretation of results in \cite{ravelo:2006,luo:2008}.

\begin{figure}[htp]
\centering
\includegraphics[width=0.75\textwidth]{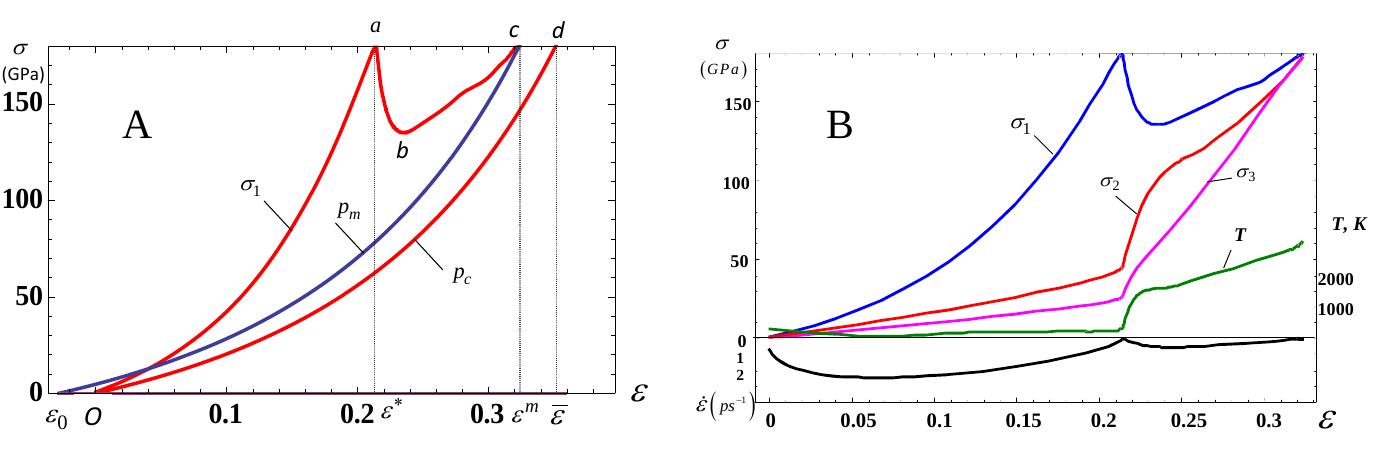}
\caption{(A) Stress-uniaxial strain curve for copper single crystal until melting
  ($\sg_1$), and equations of state of liquid  ($p_m $) and crystal
  ($ p_c $)  under hydrostatic pressure. Melting initiates at strain
$\vep=\vep^*$  and  completes at strain
$\vep=\vep^m$. Area between curves $\{Oabcd\}$ and $\{Od\}$ is the
  additional driving force for melting due to non-hydrostatic loading.
  (B) Evolution of normal Cauchy stresses $\sg_i$, temperature, and
  prescribed strain rate  vs. uniaxial strain
  from MD simulations for $<$110$>$ shock loading of Cu.  Reproduced with permission from \cite{Levitas-Ravelo-PNAS-2012}. }
\label{Fig-VM-shock}
\end{figure}

We proposed in \cite{Levitas-Ravelo-Plast-08,Levitas-Ravelo-PNAS-2012} the $VM$
as a new plastic deformation and deviatoric stress relaxation  mechanism  at
temperatures thousands Kelvin below $T_m(p)$.
Thermodynamics of melting under homogenous uniaxial deformation $\vep= U-1$ (equal to volumetric strain; in a planar shock, lateral strains are absent) was developed as a  generalization of the approach in \cite{levitas-ijss1998,levitas-ijp2000-1} for melting, see Boxes 1 and 2. Stress-strain curves required for evaluation of the thermodynamic driving force $\bar{X}$ for
complete melting were obtained  from quasi-isentropic MD
simulations (Fig. \ref{Fig-VM-shock}).  Surface energy and dissipation were excluded. Temperature increase after initiation of melting (Fig. \ref{Fig-VM-shock}B) was
neglected,  underestimating $\bar{X}$.
The condition $\, \bar{X}=0 \,$
after
some transformations leads to the following
expression for the equilibrium melt temperature under uniaxial straining $T_m^{nh}$:
\bey T_m^{nh} = T_m (\sg_1) - \left( \int_0^{\vep^m}\sg_1 d\vep -
\int_{0}^{\bar{\vep}}p_c d\vep + \sg_1
(\bar{\vep}-\vep^m)\right)/\Delta s,
\label{nat-6ab}
\eey
where all parameters and geometric interpretation are given in Fig. \ref{Fig-VM-shock}A.
  The magnitude of the negative mechanical
part of the thermodynamic driving force for melting under hydrostatic pressure  is
equal to the area $\{O\vep_0cd\}$ between the equation of state for melt  $p_m (\vep)$
and crystal $p_c (\vep)$.  This area characterizes the increase in the melting
temperature for Cu under hydrostatic loading from $T_m(0)=1357\, K$ to $T_m(179.1)=5087\, K$. The difference
between the areas under the stress-strain curve $\sg_1(\vep)$ $\{Oabcd\}$
and the equation of state of the crystal $p_c(\vep)$ $\{Od\}$
provides an additional driving force for melting due to non-hydrostatic
loading (the term in parentheses in Eq.(\ref{nat-6ab})). This
area is about three times of the area $\{O\vep_0cd\}$ and
produces about  three-fold reduction in $T_m(p)$ in  comparison with
the raise  due to the hydrostatic pressure.

\begin{figure}[htp]
\centering
\includegraphics[width=0.75\textwidth]{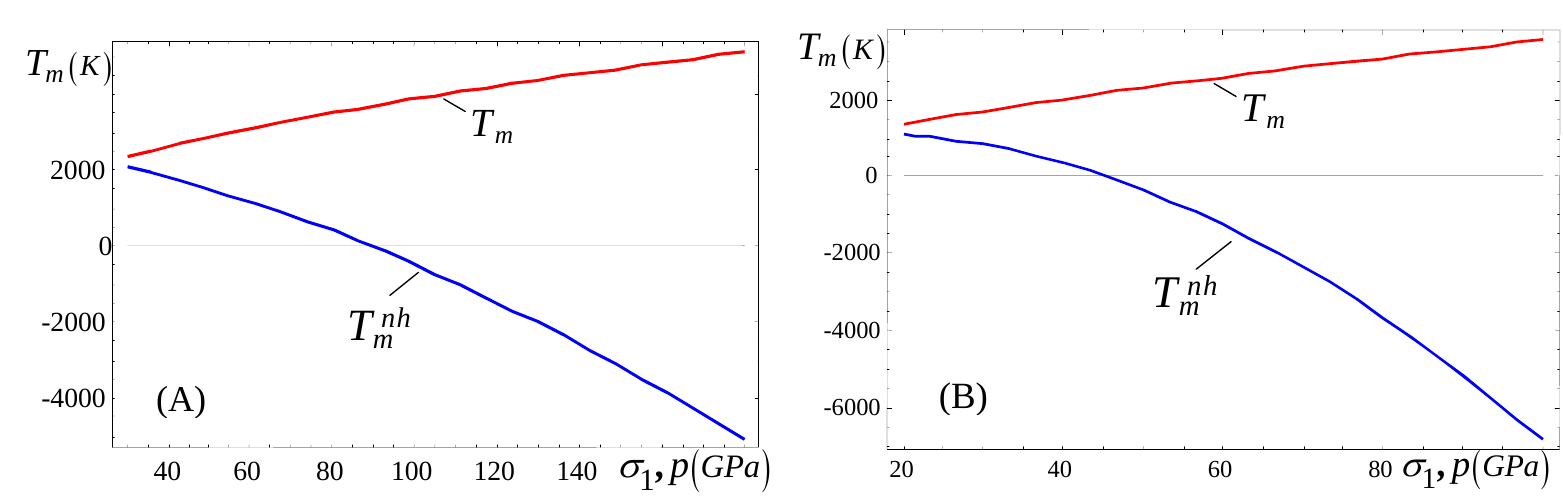}
\caption{Calculated equilibrium melting temperature under hydrostatic $T_m (p)$ and
  non-hydrostatic $ T_m^{nh} (\sg_1)$ conditions for $<$110$>$ shock loading of Cu (A) and Al (B). Reproduced with permission from \cite{Levitas-Ravelo-PNAS-2012}. }
\label{Fig-melt-nonhydr}
\end{figure}
The reduction in the thermodynamic melt temperature based on  Eq.(\ref{nat-6ab})
is presented in Fig. \ref{Fig-melt-nonhydr}, and it
 is enormous at high stresses, like $10^4\,K$.

 The above derivations are applicable for perfect crystal when dislocations and twins do not have time to nucleate,
 or for crystal with some defects, which produce stress relaxation slower than melting,
 i.e., at high strain rate of 10$^{9}$-10$^{12}$ s$^{-1}$ and higher. At lower strain rates, traditional dislocation or/and twinning  plasticity takes place. Since
   melting occurs on the  ps time scale, significant  overheating is
required.  Melting starts when crystal lattice loses its stability, with assistance of thermal fluctuations. Numerous MD simulations in \cite{Levitas-Ravelo-Plast-08,Levitas-Ravelo-PNAS-2012} have confirmed the $VM$  at least at $0.2 T_m(p)$, i.e., 4000 K below  $  T_m(p)$ for Cu
for $<$110$>$ loading and even at $T\simeq 0.055 T_m(p)$ (i.e., at 300 K)
for isothermal $<$111$>$ loading of defective Cu.
After melting, deviatoric stresses relax and hydrostatically loaded melt is deeply in the region of stability of a solid phase. Melt recrystallizes  at ps time scale. Since $VM$  competes with traditional
defect-based mechanisms,  it should be incorporated in
the deformation-mechanism maps \cite{frost-89,meyers-01} for strain rates of $\dot{\vep} \sim 10^{9}-10^{12}$ s$^{-1}$ in
metals, and high shear stresses. For materials
with inhibited plasticity (covalent  crystal Si, Ge, SiC, high-strength
materials and alloys, or complex organic molecules like HMX), VM may be observed at lower strain rates and stresses.

The VM phenomenon in a shock wave was confirmed and further elaborated in MD simulations for single crystal Cu, Al, Ta, Pb in
\cite{Budzevich-etal-VM-shock-12,He-etal-VM-shock-14,Wang-etal-VM-shock-16,Wang-etal-VM-shock-19} and
for polycrystalline Be in \cite{Dremov-etal-VM-shock-15}.

In \cite{Levitas-Ravelo-Plast-08} virtual amorphization was suggest as the first stage of $VM$.
Difference between amorphous and liquid phases from mechanical point of view is that amorphous phase has finite shear modulus and yield strength, which are both zero for liquid. After crystal lattice instability and initiation of disordering at strain $\vep^*$ (Fig. \ref{Fig-VM-shock}), material still keeps the yield strength (difference between $\sg_1$ and two other stresses) and crystal anisotropy  (difference between $\sg_2$ and $\sg_3$). For strain exceeding strain, for which $\sg_2=\sg_3$, material is completely disordered and isotropic but keeps the yield strength, i.e., it is amorphous solid.  It is unstable and transforms with further loading continuously (i.e., via second-order transition)
to VM, when all three stresses are getting equal.

For different materials (e.g.,  organic $\alpha$-HMX crystal, Si, SiC, and $SmCo_5$)
plastic deformation in shock or dynamic loading occurs by generation and motion of dislocations or twin boundaries up to some pressure or strain rate and via formation of amorphous shear nanobands
at higher pressure or strain rate   \cite{sewell-PRB-07,Meyers-etal-Si-SciRep-16,Meyers-etal-Si-EurPhysJ-16,Meyers-etal-Si-ActaMat-16,Meyers-etal-ExteMechL-15,Meyers-etal-SiC-ActaMat-18,Szlufarska-etal-VM-amorph-band-20}.
This amorphization may occur via VM, like pressure-induced amorphization in \cite{Levitas-PRL-2005}.
Since free energy of melt is usually considered equal to the free energy of amorphous phase, thermodynamic theory developed in \cite{Levitas-Ravelo-PNAS-2012} can be applied for uniaxial loading or modified for other loadings.

\section{Strain-induced nucleation at shear-band intersection.  Application of the global  criterion for structural changes}\label{Shear-band-int}

The main experimental results on strain-induced PTs in TRIP steels \cite{olson+cohen-72,16,olson+cohen-75} can be summarized as follows:
\begin{enumerate}
\item Intersections of  the shear bands serve as    the main  nucleation sites;
\item PT takes place during the shear-band intersection process;
\item Subsequent growth of the martensite beyond the shear-band intersections  is quite restricted;
\item Not  every shear-band intersection causes martensite nucleation.
\end{enumerate}
\begin{figure}[htp]
\centering
\includegraphics[width=0.35\textwidth]{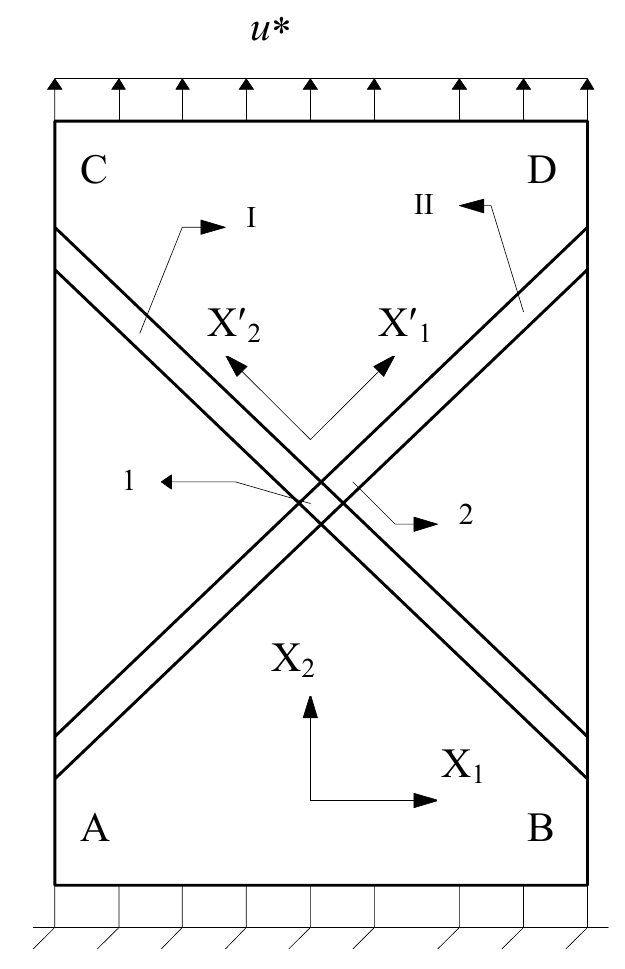}
\caption{ Schematics of a sample with martensitic nucleus
1 at two shear-band  (I and II) intersection. \label{Fig-intersection-scheme}}
\end{figure}

Problem of PT at shear-band intersections with athermal kinetics presented in Boxes 2 and 3 was formulated and solved at finite strains  in \cite{levitas+idesman+olson-1999}.
The schematics of the sample with prescribed vertical displacement $\, u^{*}\,$ as the boundary condition
is shown in Fig. \ref{Fig-intersection-scheme}.  Two orthogonal shear
bands are introduced as the regions where material deforms plastically, while deformation is elastic in the rest of the sample. The yield strengths of the austenite and martensite, as well as transformation shear and normal strains in  an invariant-plane strain are:
\bey
\sg_y^A=250 \,MPa; \quad  \sg_y^M=800 \,MPa; \quad     \gamma_t=0.2; \quad  \vep_n=0.026
\label{75tr}
\eey
Normal and shear directions of an invariant-plane strain were directed along the shear bands, because this corresponds
to the maximum of the transformation work.
 Transformation strain is introduced in the nucleus  proportionally to the increasing   displacement $\, u^{*}\,$,
for several  maximum displacements $\, u^*_{\sl max}\,$. By dividing $\, u^{*}\,$ by the initial length of the sample $l_0$, we can characterize prescribed displacement in terms of averaged vertical strain $\vep$.

\begin{figure}[htp]
\centering
\includegraphics[width=0.7\textwidth]{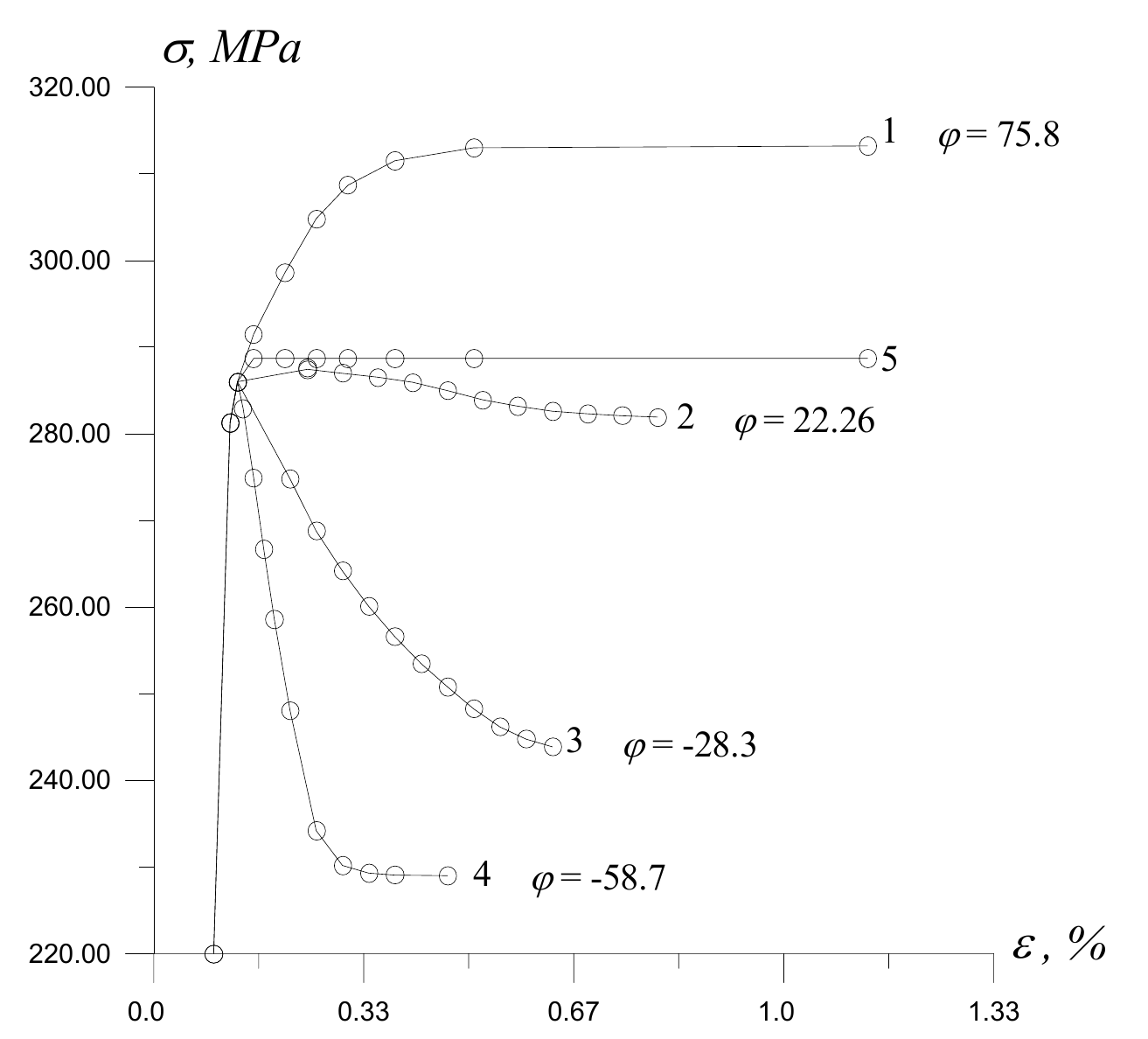}
\caption{ Averaged axial stress $\, \sg \,$ versus strain
$\, \vep \;$ (for  various values of the final strain  at finishing PT) for the process of
appearance of martensitic nucleus at shear-band intersection. Values of
transformation work $\, \varphi $ in {\sl MPa} are shown near curves. Curve 5 corresponds to the straining
 without PT. Reproduced with permission from \cite{levitas+idesman+olson-1999}.
 \label{Fig-shear-band-intersection-s-s}}
\end{figure}

Dependence  between averaged axial stresses $\, P \,$ and vertical strain $\vep$, for different $\vep_{max}$ during which complete PT in a nucleus occurs, is presented in  Fig. \ref{Fig-shear-band-intersection-s-s}.  The values of the calculated transformation
work $\, \varphi \,$ are shown near each curve. Stress-strain curve 5 corresponds to the deformation
without PT, and we will call the corresponding stress a macroscopic yield strength of an austenitic sample. For known  temperature and $K^0$,  one determines the  required transformation
work $\, \varphi \,$ from the PT criterion
(\ref{eq:6})-(\ref{ijss1-19a}) and finds with the help of  Fig. \ref{Fig-shear-band-intersection-s-s} at which $\vep$
PT may occur. However, as we discussed in Section \ref{global},  the solution without PT is also acceptable,
since    all equations of continuum thermomechanics are met.
Based on the global stability criterion (\ref{eq:yyyyy}), we conclude that if  $\, P -\vep \,$
curve for the solution with PT is higher than the  $\, P -\vep \,$
the macroscopic yield strength  without transformation 5 (like the curve 1 in Fig. \ref{Fig-shear-band-intersection-s-s}), the transformation is impossible, despite the potential
fulfilment of the thermodynamic  PT criterion. Thus, deformation will occur along the curve 5. In the opposite case
(for curves 2 - 4 in Fig. \ref{Fig-shear-band-intersection-s-s}),  the deformation-transformation  process with PT will be realized. This problem represents one of the nontrivial applications of the global SC criterion based on the stability analysis.
\par
It was found that the transformation work is much lower (a) when
nucleation occurs at any other place but  shear-band
intersection; (b)
nucleus  grows beyond the
shear-band intersection, and (c) if nucleation occurs after shear-band intersection without PT at fixed $P$ or $\vep$.
These findings explain the experimental results in items 1 to 3 above.

Next, the same loading and processes were considered for the sample, consisting of four samples shown in Fig. \ref{Fig-intersection-scheme} connected consecutively in the vertical direction. While all four  shear-band intersections
are fully equivalent, we consider four cases with one to four nuclei that appear during the same prescribed $\vep$. The
transformation work reduces with  the increasing number of nuclei at  shear-band
intersections. However, for a single nucleus, which can appear for the prescribed  $\vep$ at the highest temperature, stress-strain curve exceeds that for the deformation without PT. That means that such a PT is impossible due to
the global SC criterion (\ref{eq:yyyyy}). However, for multiple elements, such a PT is also impossible because plastic flow in the untransformed
shear-band intersections does not allow stresses above the macroscopic yield strength of an austenitic sample. Thus, the treatment of the sample  with the multiple samples shown in Fig. \ref{Fig-intersection-scheme}  connected in series gives us  the corroborating
arguments for the validity of the stability analysis and the global SC criterion based on the extremum principle (\ref{eq:yyyyy}).

Since the
transformation work reduces with  the increasing number of nuclei at  shear-band
intersections that appear during prescribed $\vep$,
this results in appearance of one nucleus during part $\vep^\diamond$ of the  prescribed $\vep$, for which the thermodynamic PT criterion is met, and stress does not exceeds the macroscopic yield strength for the austenite. Within next strain increment $\vep^\diamond$, the stress increases and reaches the  macroscopic yield strength for the austenite, and then nucleus appears at other shear-band intersection,  and so on. The maximum temperature for the strain-induced PT, $M_d$, corresponds to the process, for which nucleation of a single nucleus occurs during $\vep^\diamond$ that gives the stress-strain curve slightly lower than the macroscopic yield strength of the austenite.
That means that $M_d$  is determined not by thermodynamics (since the
larger transformation work can be obtained), but by
the impossibility of PT due the global PT criterion based on stability analysis.
\par
Thus,  experimental phenomena 1 to 4 enumerated at the beginning of this Section are at least
qualitatively described by our theory presented in Boxes 2 and 3  without
involving additional physical mechanisms. More quantitative description should include
discrete dislocations or twin bands  or $\vep$-martensite bands, leading to high stress concentration
 due to difficulty to transmit shear band through the intersecting band.
\par
It was also found in \cite{levitas+idesman+olson-1999}  that
the transformation work is a function   of the ratio $\; \displaystyle{
\frac{\Delta \,c}{\Delta \,\vep_p} \;\simeq\;\frac{d\, c}{d\,\vep_p}}\;$,
where $\, \Delta \, c \,$  is an increment of the volume fraction
of strain-induced martensite. For macroscopically
uniaxial strain, $\; \vep_p \, \,$  is the
accumulated plastic strain  $\, q \,$.
This finding led to the thermodynamic derivation in \cite{Levitas-PRB-04,Levitas-chapter-04}
of the
 microscale kinetic equation for the volume fraction of a plastic strain-induced high-pressure phase $c$ of the type of Eq. (\ref{str-ind-kineq-ss-a}), see example in Eq. (\ref{l-g-4}).
 \par
Simulations show a significant difference between transformation conditions
for displacement- and stress-controlled loadings. For prescribed stresses, the more martensite is, the larger the overall transformation
strain  and work are. The principle of the maximum transformation work results in the complete transformation in the entire
band, without plasticity. Thus, stress-induced PT produces a plate-like nucleus, in contrast to strain-induced PT, which corresponds to known experiments. The adiabatic heating and its effect on PT at the shear-band intersection was analyzed in \cite{idesman-levitas+stein-ijp-2000}.
\par
Stability analysis should be applied for any problem with boundary conditions in displacement,
when PTs competes with plasticity. For example, similar treatment was performed for  PT in
a spherical particle imbedded in  a cylindrical sample \cite{idesman-levitas+stein-ijp-2000}. The extremum principle for determination of the stable deformation
process is  applicable for the analysis of the competition between other
 inelastic mechanisms, e.g., twinning, damage,  and others.
For the softening behaviour during PT, the stability analysis
plays significant role for the PT in elastic materials as well, see \cite{Levitas-postulate-I-1995,Levitas-postulate-II-1995}.

\section{Appearance and growth of a martensitic plate in elastoplastic material}\label{sec-lath}

\subsection{Macroscale nucleation of a martensitic plate}\label{nucl-lath}

Appearance of a small rectangular (in the reference state) temperature-induced martensitic plate
of the length $l$ and height $h$ within a much larger rectangular austenitic sample was studied
in \cite{Idesman-CMAME-99,idesman-levitas+stein-ijp-2000} for finite elastoplastic and transformation strains.
Plane-strain formulation and the invariant plane transformation strain with shear along the length $l$ were considered.  The transformation strain  and the yield strengths are given in Eq. (\ref{75tr}).
Incrementally increasing the transformation strain components and solving the corresponding elastoplastic problem, 
 the  transformation work $\,\varphi\,$ was evaluated and approximated as
\bey
\varphi  =  \bar{A}  +  B  x  +  C  x^2  ,
\quad       \bar{A}   =   - 72.11 \, {\sl MPa}  ,
\quad       B   =  6.40  \, {\sl MPa}   ,
\quad        C  =   -0.29  \, {\sl MPa} ;  \quad  x  = \l/h .
\eey
\par
The  martensitic plate deforms
 elastically  except of narrow regions near short
sides. In the austenite the plastic strains are localized around
the transformed plate (Fig. \ref{Fig-martensite-plate}(a)).
The "macroscale" thermally activated kinetics presented in Boxes 5 and 6 was applied.
The  principle of the minimum of transformation
time
\bey
\left( A_1 \, + \, B \, x \, + \, C \, x^2 \right) \, l \, h \, - \,
2 \, \Gamma \, \left( l \, + \, h \right)
\;\; \rightarrow \;\; \max_{x \, , \; l} \; ,
\eey
where $\; A_1 \, := \, \bar A - K^0 - \, \Delta \psi - E_a \,$,
lead to two  equations
\bey
A_1 \, + \, 2 \, B \, x \, + \, 3 \, C \, x^2 \; = \;
\frac{2 \, \Gamma}{h}  ; \qquad\qquad
A_1 \, x \, - \, C \, x^3 \; = \; \frac{2 \, \Gamma}{h},
\eey
which combination results into a cubic equation for $\, x \,$.
  The thermodynamically admissible
solution to this equation exists for large $A_1$ only, i.e., for large net thermodynamic   force and small activation energy.
Otherwise, the principle of minimum of transforming volume (mass)
along  with the
thermodynamic SC criterion
\bey
\left( A \, + \, B \, x \, + \, C \, x^2 \right) \, l \; = \;
2 \, \Gamma \, \left( 1 \, + \, x \right) ; \qquad\qquad\qquad
A \, := \, \bar{A} - K^0 -  \, \Delta \psi,
\label{plate-crit}
\eey
results in
\bey
V_n \; = \; l \, h \; = \; \frac{l^2}{x} \; = \;
4 \, \Gamma^2 \; \frac{\left(1 \, + \, x \right)^2 }{x
\left(A \, + \, B \, x \, + \, C \, x^2 \right)^2}
\;\; \rightarrow \;\; \min_{x  } \; .
\label{V-n}
\eey
Principle (\ref{V-n}) leads to the explicit equation  for $x$ and then Eq. (\ref{plate-crit}) gives the length $l$. For  $\, x \; \gg \; 1 \;  \,$ obtained equations simplify to
\bey
x \; = \; - \;\frac{B\, + \,\sqrt{B^2\, + \,12\, A\, C}}{6\, C} \; ,
\qquad\qquad\qquad
l \; = \; \frac{2 \, \Gamma \left(1 \, + \, x \right)}{
\left(A \, + \, B \, x \, + \, C \, x^2 \right)} \; .
\eey
The higher the  net thermodynamic driving force $ - \, \Delta \psi  - K^0 $  (i.e., $A$) is, the smaller the nucleus size, the aspect ratio $x$, and consequently, the
transformation time are. For example, for $\; A \, = \, -30 \,$ we evaluate  $\; x \; = \; 10.61 \,$, $\; l \; = \;
4.40 \, \Gamma \,$ but for  $\; A \, = \, 0 \,$ we obtain
$\; x \; = \; 7.36 \,$, $\; l \; = \; 0.533 \, \Gamma \;$.


\subsection{ Growth of martensitic lath within the austenite: effect of inheritance of plastic strain}\label{growth-lath}

\par
Plate martensite increases its length $r$ until it is stopped by some  inhomogeneity,
like  a grain or twin boundaries, stacking fault, or other martensite plate. This process is related to the increase in the transformation work
(\ref{tr-work}) with increasing aspect ratio $r/b$ of the plate. In contrast, the growth of the lath martensite stops
 inside the grain and is not related to such obstacles.
 It was suggested in \cite{haezebrouck-87,olson+cohen-1986} that plastic
accommodation of the transformation strain (see also \cite{datta+ghosh+raghavan-1986,ghosh+raghavan-1986}) causes arrest of the lengthening and plate to lath
morphological transition in steels, which is important for optimization of steels mechanical properties and steel design.
This was done, however, utilizing relatively simple model.
\par
The problem on the growth of a temperature-induced rectangular
martensitic unit in an austenitic sample based on formulation   in Box 5 (in particular, interface propagation
Eqs. (\ref{ol-3}) and (\ref{ol-4})) was presented in \cite{idesman-levitas+stein-ijp-2000,31}. This was a natural continuation of the "macroscale" nucleation of the plate presented in Section \ref{nucl-lath}, based on the same finite-strain constitutive theory and FEM simulations. Both growth in the bulk and close to the free surface of a sample  were studied.

\begin{figure}[htp]
\centering
\includegraphics[width=\textwidth]{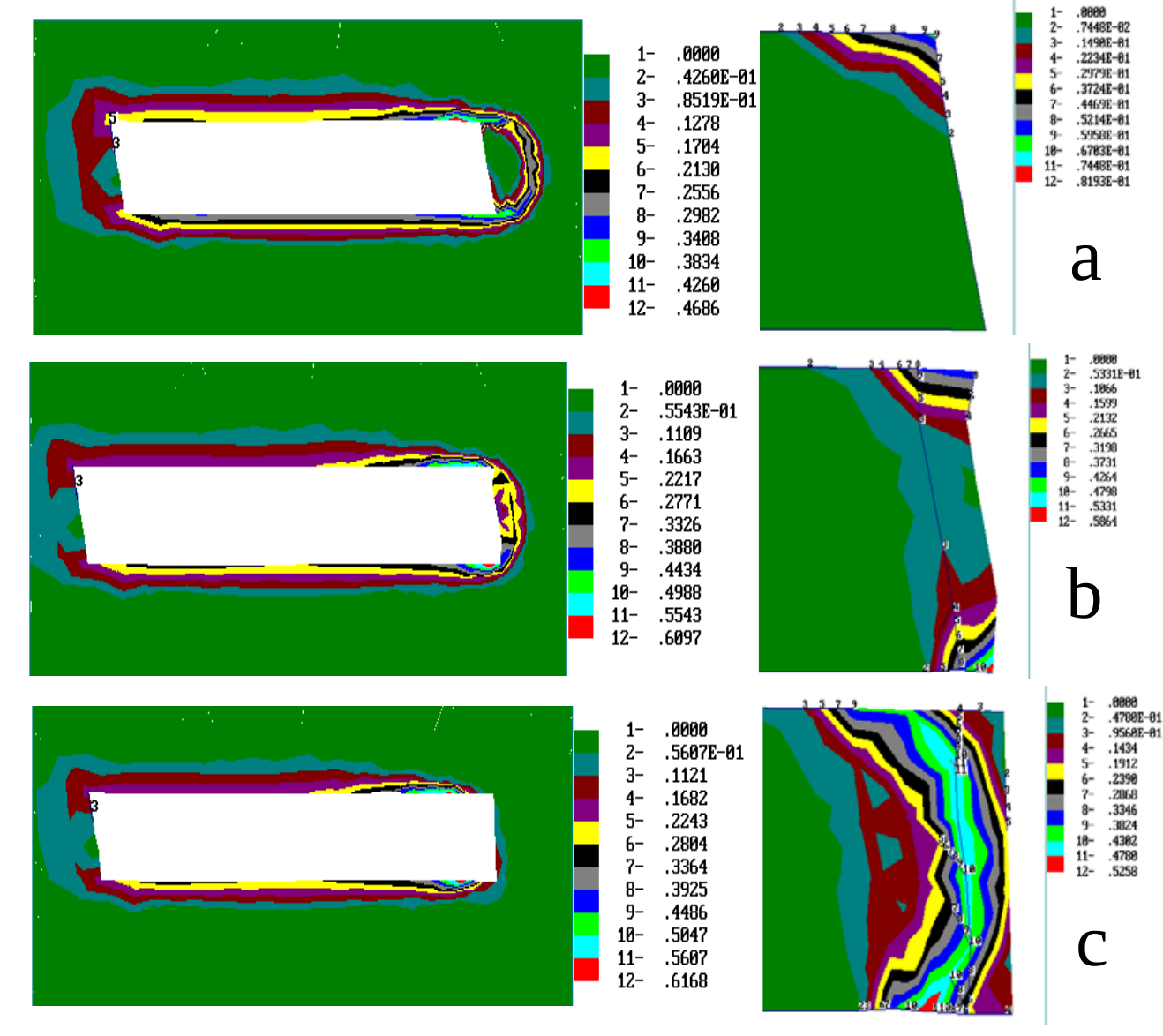}
\caption{ Distributions of the accumulated plastic strain $q$ in austenitic matrix near growing martensitic plate (white region in the left part of the figure) and in
the right part of the martensitic plate (right part of the figure) after nucleation (a) and  two different stages of growth (b) and (c)). Reproduced with permission from \cite{31},  Taylor \& Francis Ltd., www.tandfonline.com.
\label{Fig-martensite-plate}}
\end{figure}
Interface velocity during lengthening as a function of the temperature and the interface location has been
determined. The following conclusions were drawn in \cite{31}.
\par
1. One of the key parts of the solution was a complete inheritance of the plastic strain of the austenite during its transformation to martensite. Thus, plastic strain tensor in the austenite was taken as an initial condition while integrating the flow rule in the transforming material. This led to the very heterogeneous and unexpected   strain fields in the austenite
and martensite and their nonmonotonous change during the PT
process, see Fig. \ref{Fig-martensite-plate}. The transformation
deformation gradient was quite modest and the yield strength of the martensite was more than 3 times larger than that of the austenite (see  Eq. (\ref{75tr})). Still,
the plastic shear
strain at some points of the martensite  reaches 0.6 in the direction of the transformation
shear and, after an elastic stage, changes the  sign and increases  by 0.4 in the opposite direction. The edge
geometry  of the propagating interface evolves counterintuitively   (Fig. \ref{Fig-martensite-plate}):
the expected  transformation shear strain profile changes to a wave-like contour
with  reverse shears near the corners, and then to almost straight
vertical line. The plastic shear jumps by 0.9  across a phase interface.
\par
2. These results shed a light on  why the plastic deformation is localized in the much stronger
martensite rather than in the weaker austenite.
After nucleation, large plastic deformations are caused by the transformation strain near the interface mostly in the
austenite. This plastic strain field is  inherited by the growing martensite, causing, along with the transformation strain, significant  internal stresses, which are relaxed via  additional plastic flow (TRIP) in the martensite. After some growth increment, plastic
deformations in the austenite are much smaller (see Fig. \ref{Fig-martensite-plate} (c)) because of  small resultant
shape change of  the martensitic edge (see Fig. \ref{Fig-martensite-plate} (c)). Such a
complex strain variation in martensite changes significantly  the driving
force for the interface propagation  $X_\Sigma$ and athermal resistance $K^0$.
\par
3. After  martensite  nucleation and during
increase in the lath length by 10 \%, the transformation work
decreases from $\,- 50.35$ {\sl MPa} down to $\, -100.79 \,$ {\sl MPa} and the contribution to $K^0$ due to the dislocation forest hardening  $K_d$
 grows from zero to $\, 30.22 \,$ {\sl MPa} (compare to $K_\mu=2. 35\, MPa$ for the interfacial friction due to solute
hardening). As a result, the net thermodynamic driving force for the interface propagation is getting negative, interface stops, and lath martensite is formed rather than the plate.
It is also found that quite high
 internal shear stresses in the surrounding austenite may cause autocatalytic nucleation of
new martensitic laths  with the same or opposite transformation shear, which may coalesce, forming fine multi-lath structure.

The reduction of the yield strength of the austenite increases
the plastic deformation and the propensity to formation of the lath.
In particular, this happens for steels with
high martensitic temperature, which leads to the reduction in the yield strength. Thus, the
microstructure and mechanical properties of steels can be tailored by controlling the yield strength of austenite  by  alloying or preliminary plastic deformation.
\par
4.  A free surface does not essentially change the thermodynamic driving force
for  the PT until distance between the phase interface and surface is getting smaller than $\, 0.75h$. After this, the transformation work
grows and  the Odqvist parameter  $\, q \,$ and corresponding $K_d$ reduce  abruptly. If  not
stopped before this position, the phase interface accelerates and reaches the  surface. The
edge interface contour passes via a wave-like shape, but ends
with a shape produced by a transformation shear  slightly modified by plastic accommodation,
producing an asymmetric 'tent-shaped' profile.
\par
5. Obtained results allow one to understand the relationship
between thermally activated kinetics for a single interface and athermal
kinetics for a sample.
For athermal  nucleation postulated for martensite (see \cite{olson+cohen-72,olson+cohen-76,olson+cohen-1986,olson+roytburd-1995}), this transition  is related to the interface arrest due to reduction in $X_\Sigma$ and increase in $K_d$   independent of the interface kinetics for nonzero velocity.

Recent developments of modeling of dislocated lath martensite in steel can be found in \cite{Petersmann-Antretter-IJP-19}.

\section{Phase field approach to the interaction between plasticity and phase transformations}\label{Sec-PT-plast}

Thanks to the recent progress in nanotechnology and nanoscience, PT and plastic deformations are investigated in various nanoobjects:  wires, fibers, films, multilayered systems, and particles. This encourages research on  the coupled PT and discrete dislocation plasticity at the nanoscale. Thus, phase nucleation takes place at different dislocation configurations. Generally, nucleation   always occurs in  nanoscale volumes, even for bulk specimens.
 Loss of coherency of phase interface occurs via the dislocation nucleation and motion.
 Dislocations also produce athermal resistance to the interface motion.
 PFA is  an ideal  continuum method to address all of these problems.

 PFA utilizes  the concept of the order parameters, which  describe the crystal lattice instabilities:
  $\eta_i$ for PT between the {\sf A} and martensitic variants {\sf M}$_i$ and ${\xi}_{\alpha}$,
which describes variation of the magnitude of the Burgers vector in the ${\alpha}$ slip system from zero for perfect crystal to an integer number of dislocations.

  The free energy has the number of the  minima in the space of the order parameters $\eta_i$  and ${\xi}_{\alpha}$
   (separated by energy barriers) equal to the number of phases and variants, as well as the number of complete dislocations in the system.  Besides, the free energy depends on the gradients of the order parameters that are localized in the finite-width  interface and dislocation core regions. The gradient energy, together with local energy barriers between phases and dislocations, penalizes the interface and dislocation core energies.
   The solution of the thermodynamically consistent Ginzburg-Landau evolution equations for the order parameters  describes in a continuous way the evolution of multiphase and dislocation microstructures.
 There is no need to satisfy jump conditions at the interfaces and to track motion of interfaces and dislocations.
 Characteristic solutions for the Ginzburg-Landau equations are evolving phase interfaces that have a finite width and divide multiple phases,  and dislocation cores with a sharp but continuous variation of the Burgers vector, which separate sheared and non-sheared parts of a  crystals.

  In the previous PFAs on  PTs and dislocation evolution \cite{khachaturyan-01,Lookman-prl-08,Khachaturyan-actamat-97,Khachaturyan-actamat-01,wang01acta,wang01ph,wang01acta,huact01,kund11ph,kosl02}, these were the only constraints for choosing  the free energy.
Several other important conditions have been formulated in \cite{levitas+preston-I,levitas+preston-II} and then in \cite{Levitas-IJP-18} for PTs and in
\cite{levitas+preston-III,Levitas-Javanbakht-PRB-12,Levitas-Javanbakht-JMPS-14}  for dislocations.

\subsection{PFA to martensitic phase transformations}
\label{PFA-martensite}


{
PFA is widely used for modeling  martensitic PTs and evolution of corresponding microstructure, see \cite{khachaturyan-01,Lookman-prl-08,levitasetal-prl-07,Khachaturyan-actamat-97,Khachaturyan-actamat-01} and the reviews
\cite{chen-02,mohsen,Khachaturyan}.
We will concentrate  on the approach with the order parameters $\eta_i$ representing internal variables that  are related to the transformation strain or any other material properties \cite{khachaturyan-01,levitasetal-prl-07,Khachaturyan-actamat-97,Khachaturyan-actamat-01,Choetal-12,Chen-Material_sc,Chen-metals2003,chen-02} because for the total strain-related order parameters \cite{Lookman-prl-08,01,falk-83,reid+olson+moran-98,falk+konopko,Barsch-84,Jacob-84,abeyaratne-05}
 we are unable to meet all the desired conditions.
It is convenient to accept that $\eta_i=0$ in $\sf A$ and $\eta_i=1$ in ${\sf M}_i$.
The main requirements were introduced in \cite{levitas+preston-I,levitas+preston-II} for the description of the typical features  of stress-strain curves, which are
conceptually consistent with  experimental results for shape memory alloys, steels, and some ceramics,  and for the possibility to incorporate all thermomechanical properties of all involved phases. They were further developed in \cite{levitas-roy-ActaMat-16} for the temperature-induced PTs.
 These are the requirements:

(a) All material properties $M$  (transformation strain
tensor, elastic moduli of any rank, thermal expansion tensor, etc.) that follow from the {\it thermodynamic equilibrium conditions for homogeneous phases} should be 
equal to the corresponding values for {\sf A} and {\sf M}$_i$. 
Stress hysteresis should be controllable---in particular,  constant  or weakly temperature-dependent.

Any material property $M$ can be interpolated between phases {\sf P}$_0$ and {\sf P}$_1$ in the form
\bey
M (\eta ) = M_{0} + (M_1 -M_0 ) \varphi_m (\eta),
\label{FM5}
 \eey
 where $M_0$ and $M_1$ are the property $M$ of {\sf P}$_0$ and {\sf P}$_1$, respectively, and $\varphi_m (\eta)$ is the  interpolation function to be found. Requirement (a) leads to the following  conditions for  $\varphi_m (\eta)$:
\bey
\varphi_m (0)=0, \qquad \varphi_m (1)=1;  \qquad  \frac{d\varphi_m (0)}{d {\eta}}=\frac{d\varphi_m (1)}{d {\eta}}=0.
\label{FM5-1}
 \eey
If  Eq.(\ref{FM5-1}) is not satisfied, then
the thermodynamic equilibrium values of the order paramter $\eta_{eq}$ will depend on the temperature and stress tensor. Substituting $\eta_{eq} (\theta, \fsg)$
in Eq.(\ref{FM5}) will lead to the  artificial temperature- and stress-dependence of the property $M$, and one could not restore the properties $M_0$ and $M_1$ for bulk phases {\sf P}$_0$ and {\sf P}$_1$.

(b) The PT criteria that are derived from the formulated crystal lattice instability conditions for the homogeneous and defect-free phases should have a desired form in terms of the stress tensor. 
This requirement was significantly elaborated  in \cite{Levitas-IJP-18,Babaei-Levitas-IJP-18}
for satisfaction of the lattice instability conditions under general stress tensor obtained from the atomistic simulations
\cite{Levitasetal-Instab-17,Zarkevichetal-18}.

(c) All  properties of {\sf A} and {\sf M}$_i$ could be  included in the thermodynamic potential and transformation strain.

While these requirements  look quite natural, they were not satisfied in
any theory for martensitic PTs prior to works by  \cite{levitas+preston-I,levitas+preston-II,levitas+preston-III},
where the PFA that meets these requirements was developed.

The  large strain theory that meets these conditions was developed  in   \cite{levitas+preston-PL-05,levitasetal-prl-09,Levitas-IJP-13,Levitas-14b,Levitas-IJP-18,Babaei-Levitas-IJP-18}.
}
The theories that satisfy the above conditions were applied for FEM solutions  of various static \cite{levitasetal-prl-07,levitasetal-IJP-10} and
 dynamic  \cite{idesmanetal-APL-08,Cho-etal-IJSS-12} problems at small strains, as well as at large strains
 in \cite{levitasetal-prl-09,Levin-Levitas-IJSS-13,Babaei-Levitas-IJP-18,Babaei-Basak-Levitas-CM-19}.

\subsection{Multivariant martensitic phase transformations and transformations in multiphase materials}

\subsubsection{Twinning and transformations between martensitic variants}

Multivariant martensitic PTs are treated in most of the works within elasticity theory, see
\cite{khachaturyan-01,Lookman-prl-08,levitasetal-prl-07,Khachaturyan-actamat-97,Khachaturyan-actamat-01,Chen-Material_sc,Chen-metals2003,chen-02,mohsen,Khachaturyan,levitasetal-IJP-10,levitasetal-prl-09,Levin-Levitas-IJSS-13,Levitas-IJP-18,Babaei-Levitas-IJP-18}.
In particular, these approaches are applicable to the SMA. Since for most martensite lattices, some martensitic variants are in twin relationships to each other, transformation between them can be considered as twinning. However, twinning is also a mechanism of plastic deformation in general and represents a lattice-invariant shear producing invariant plane strain in martensite, in particular. That is why we shortly review the description of twinning and variant-variant transformations.

PFA for twinning that satisfies the above conditions (a)-(c) was developed in \cite{levitas+preston-I,levitas+preston-II,levitas+preston-III} for small strains and in \cite{levitas+preston-PL-05,Levitas-IJP-13,clayton-knap-11,Clayton-knap-11b} for large strains.
However, the main problem is in the description of variant-variant transformations.
Since this problem is a particular case of a more general problem of multiphase PFA
\cite{Steinbach-96,Steinbach-99,Garcke1999,Nestler2005,Bollada-2012,Moelans2008,Moelans2009,Kim2006PRE,Toth_et_al-2011a,Toth_et_al-2011b,Toth_et_al-2015},
which in most cases does not include mechanics and is devoted to grain structure evolution, we will include multiphase PFA in the discussion as well.
A critical analysis of the  multiphase PFA to PTs was presented, with further developments, in \cite{Toth_et_al-2015,Levitasetal-PRB-15,levitas-roy-ActaMat-16,Basak-Levitas-JMPS-18}.

\subsubsection{Phase field models with Cartesian order parameters}

\begin{figure}[h!]
\centering
\hspace{-2mm}
\vspace{-4mm}
\subfigure[]{
    \includegraphics[width=2.0in, height=1.5in] {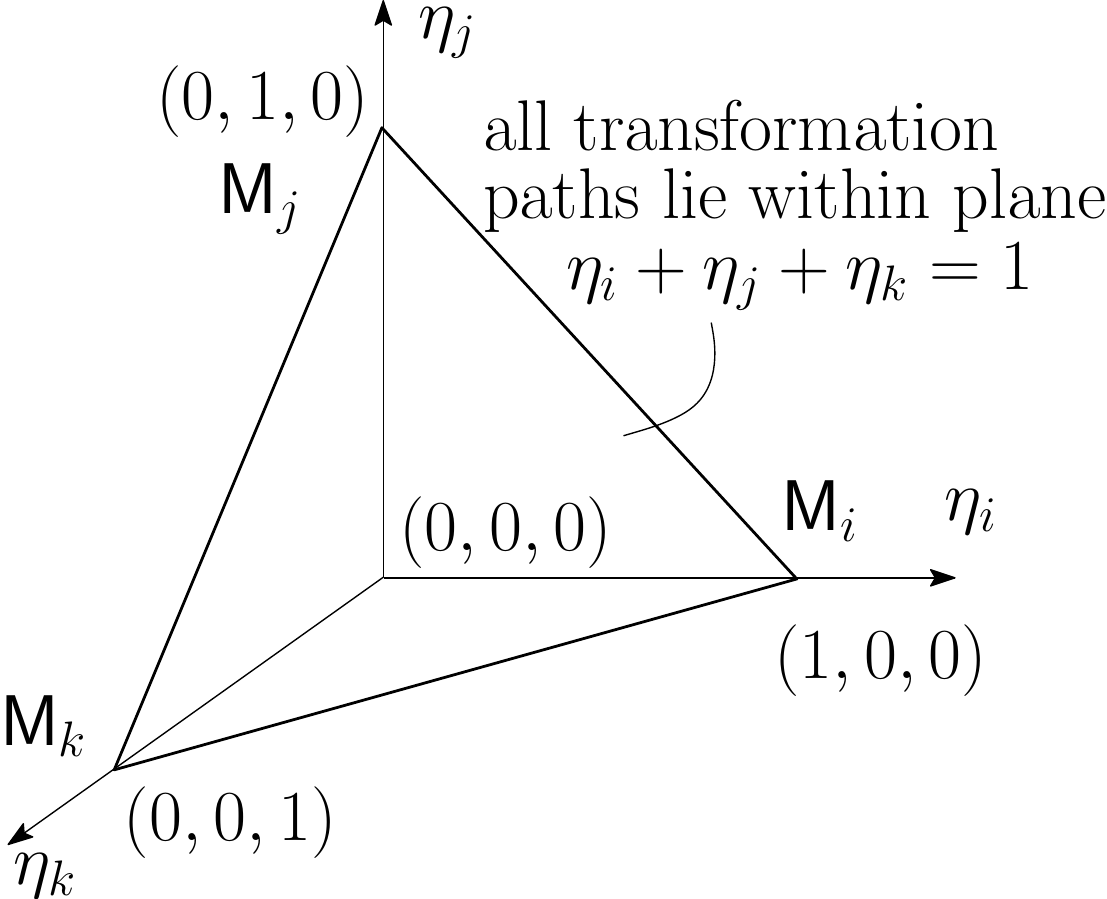}
}\hspace{1mm}
\subfigure[]{
  \includegraphics[width=1.3in, height=1.4in] {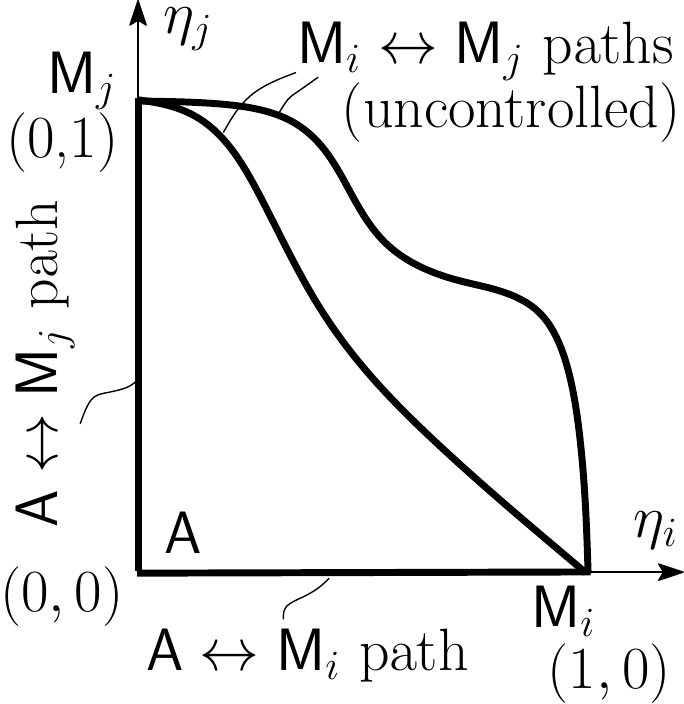}
	}\hspace{4mm}
	\subfigure[]{
    \includegraphics[width=1.3in, height=1.3in] {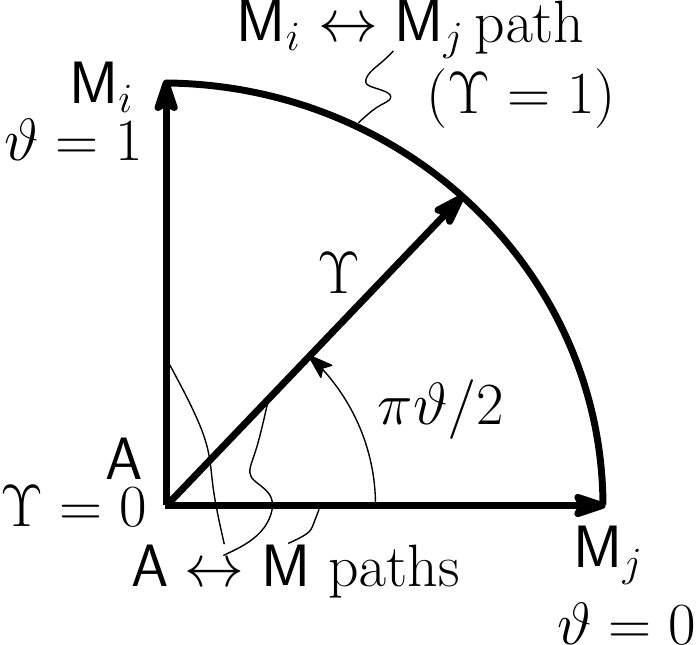}
}\hspace{-3mm}
\vspace{-1mm}
    \subfigure[]{
    \includegraphics[width=2.0in, height=1.5in] {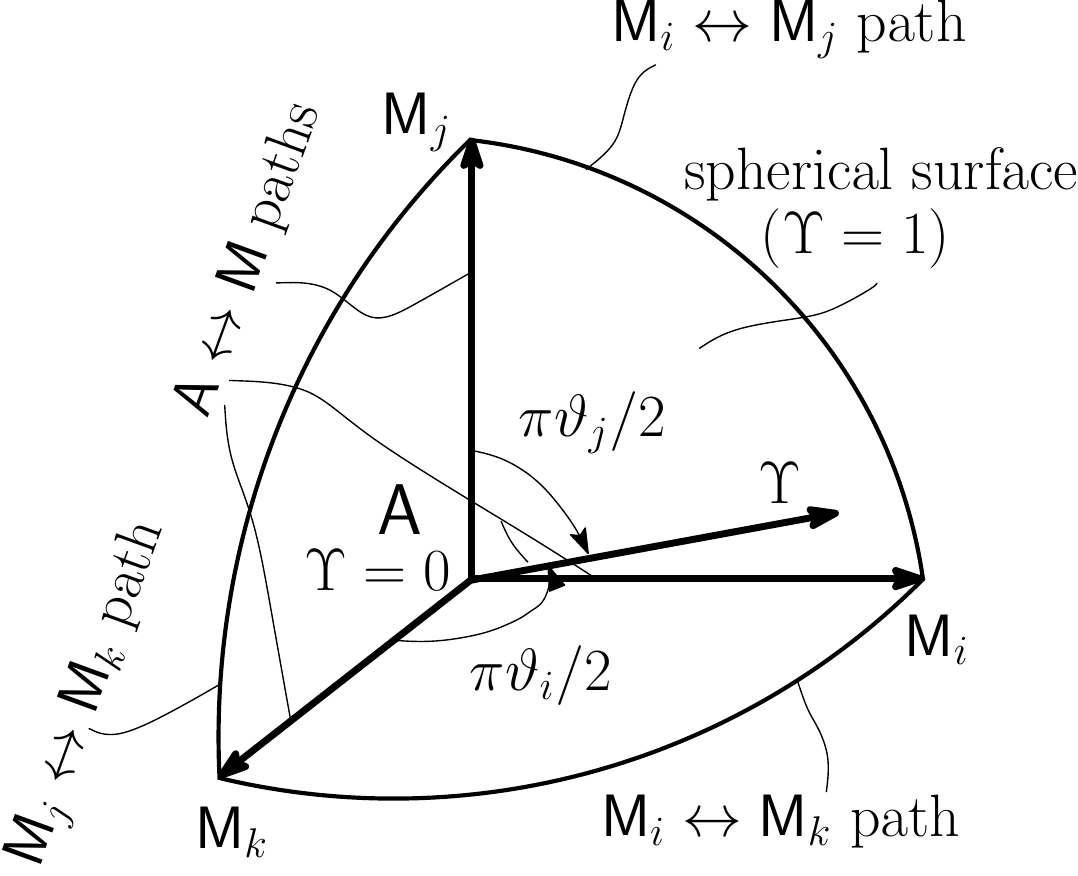}
}\hspace{1mm}
\subfigure[]{
    \includegraphics[width=2.5in, height=1.7in] {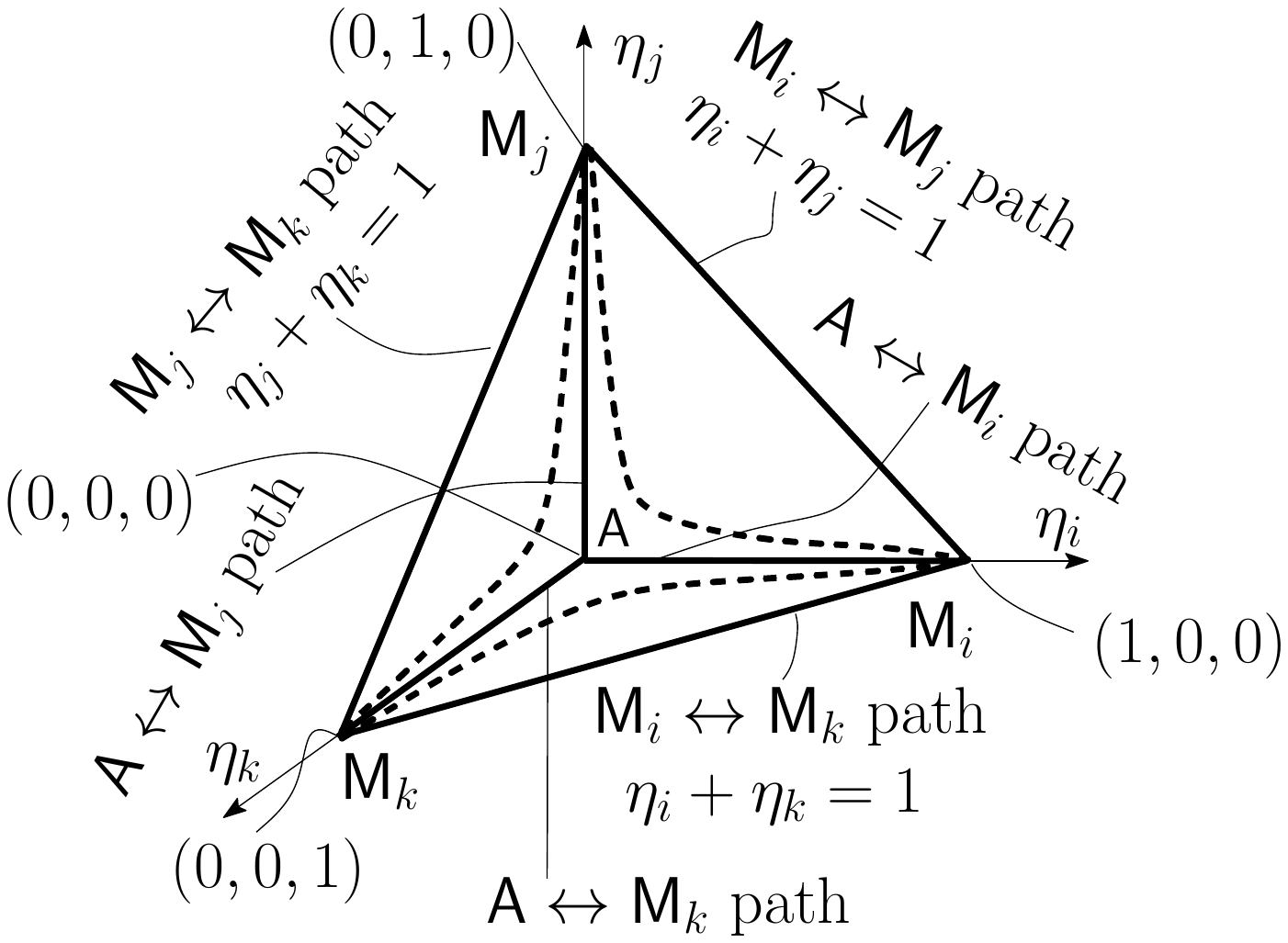}
}\hspace{1mm}
\subfigure[]{
    \includegraphics[width=1.8in, height=1.5in] {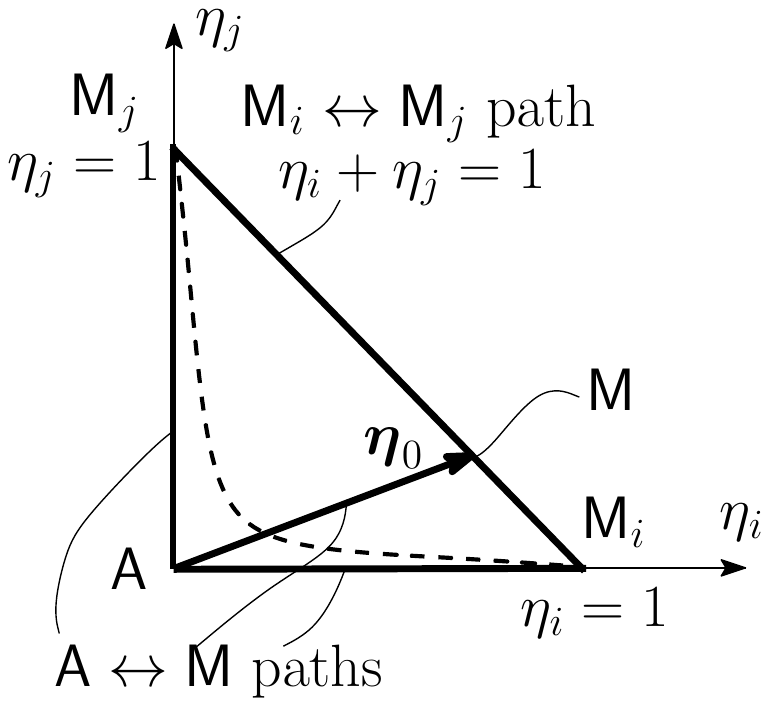}
}\hspace{1mm}
\subfigure[]{
    \includegraphics[width=2.2in, height=1.7in] {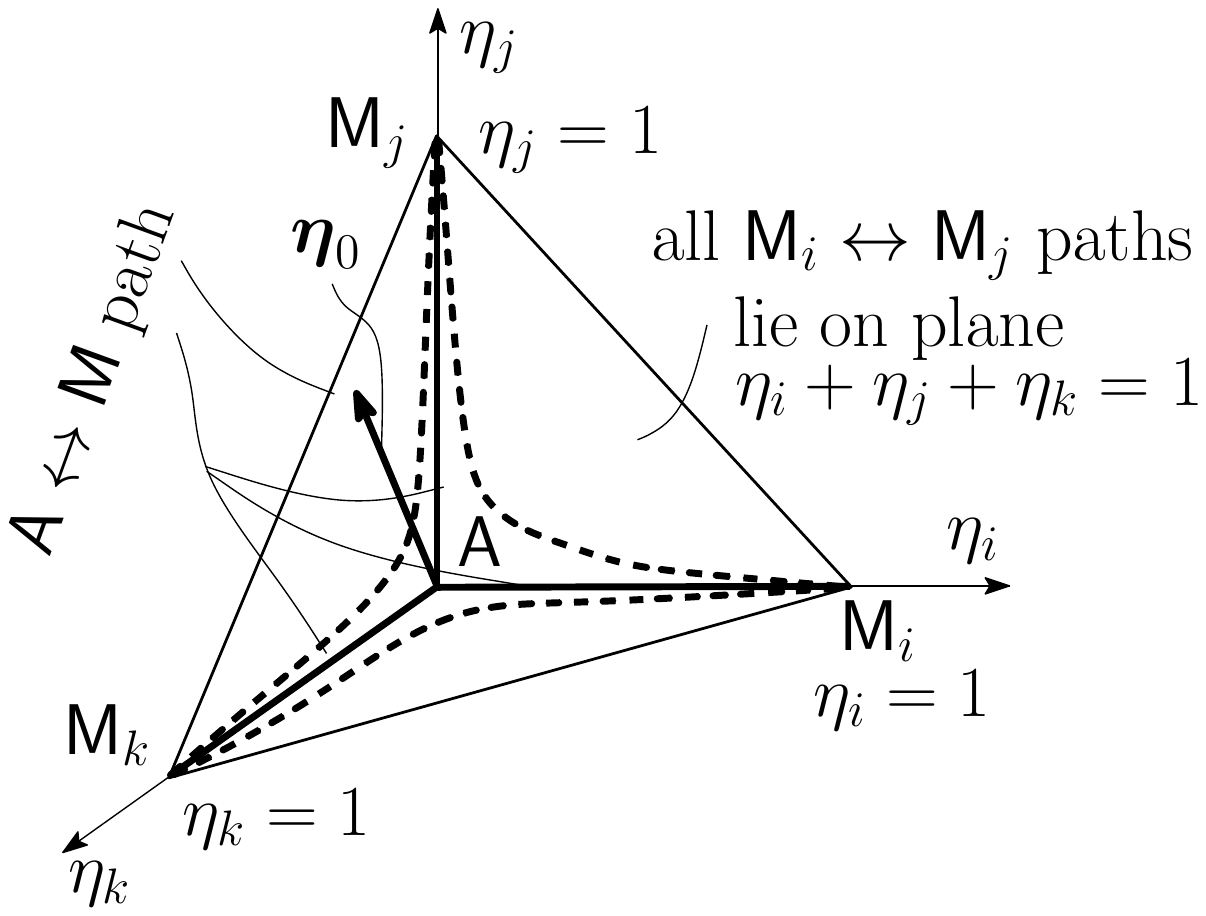}
}
\vspace{-4mm}
\caption{
Sketches  of the order parameter space and transformation paths for different PFAs.
(a) PFAs utilized in \cite{Steinbach-96,Steinbach-99,Garcke1999,Nestler2005,Bollada-2012,Moelans2008,Moelans2009,Kim2006PRE,Toth_et_al-2011a,Toth_et_al-2011b,Toth_et_al-2015}
 where   the transformation paths belong to the $\eta_i+\eta_j+\eta_k=1$ plane.
 (b) PFAs  in  \cite{khachaturyan-01,levitasetal-prl-07,Khachaturyan-actamat-97,Khachaturyan-actamat-01,Chen-Material_sc,Chen-metals2003,chen-02,levitas+preston-I,levitas+preston-II,Levitas-IJP-13},
for which variant$\leftrightarrow$variant transformation paths  are described by two order parameters in an uncontrolled way.
 (c) \& (d) PFAs with the polar (c) and the hyperspherical (d) order parameters \cite{levitas+preston-III,Levitasetal-PRB-twin-13}.
  (e) Model developed in  \cite{Levitasetal-PRB-15,levitas-roy-ActaMat-16}, for which transformation paths between different  martensitic variants are governed by additional penalizing energy terms.
   (f) \& (g)  PFA developed  in \cite{Basak-Levitas-JMPS-18} with two   and three variants, respectively.
Reproduced with permission from   \cite{Basak-Levitas-JMPS-18}.}
\label{etai_etaj_plane}
\end{figure}

In the first and most popular PFA  presented in
\cite{khachaturyan-01,Khachaturyan-actamat-97,Khachaturyan-actamat-01,Chen-Material_sc,Chen-metals2003,chen-02},
$N$ order parameters $\eta_i$ for $N$ martensitic variants were introduced,  and each ${\sf A}\leftrightarrow{\sf M}_i$ PT was described by  a single order parameter $\eta_i$ (Fig. \ref{etai_etaj_plane}(b)).
Then the
analytical solution for an ${\sf A}$-${\sf M}_i$ interface  allows one to perform a calibration of the
width, energy,  and mobility  of an ${\sf A}$-${\sf M}_i$ interface.
However, ${\sf M}_i\leftrightarrow{\sf M}_j$ transformations occur along some curvilinear line in the $\eta_i$-$\eta_j$ plane described by two order parameters (Fig. \ref{etai_etaj_plane}(b)) that  depends in some uncontrolled way on the temperature and stress tensor.
An analytical solution for an ${\sf M}_i$-${\sf M}_j$ interface cannot be found.
The numerically determined  energy, width, and mobility of ${\sf M}_i$-${\sf M}_j$ interfaces depend on the temperature and stress tensor in some
uncontrollable  way and  cannot be calibrated by experiment. Since these theories also do not meet the requirements (a)-(c),
even for ${\sf M}_i$ the equilibrium order parameter is not $\eta_i=1$ but depends on the temperature and stress tensor. Because of this, it is  difficult to obtain the thermodynamic ${\sf M}_i$-${\sf M}_j$ transformation criteria from the lattice instability conditions, and it is impossible to make them look like the criteria for twinning.

Theories that satisfy conditions (a)-(c) do not have  problems with $\eta_i=1$ for ${\sf M}_i$ and the thermodynamic ${\sf M}_i$-${\sf M}_j$ transformation criteria, but still ${\sf M}_i\leftrightarrow{\sf M}_j$ transformations occur along some curvilinear line in the $\eta_i$-$\eta_j$ plane, an analytical solution for ${\sf M}_i$-${\sf M}_j$ interface cannot be found, and the energy, width, and mobility of ${\sf M}_i$-${\sf M}_j$ interfaces cannot be calibrated by experiment.

\subsubsection{Phase field theory with  hyperspherical and polar order parameters}{\label{PFA-sphere}}

This problem was already recognized in \cite{levitas+preston-III}, where
an alternative PFA based on hyperspherical order parameters was developed (see Figs. \ref{etai_etaj_plane}(d)).  In this and more advanced theory in \cite{Levitasetal-PRB-twin-13}
 the radial order parameter
$\Upsilon$ describes ${\sf A}\leftrightarrow{\sf M}$ PTs and $N$
angles $\pi\vartheta_i/2$ characterizing the direction  of the radius-vector $\bf {\Upsilon}$
 describe ${\sf M}_i\leftrightarrow{\sf M}_j$ PTs along the hypersphere $\Upsilon=1$.
The angles $\vartheta_i$ meet a nonlinear constraint  $\sum_{i=1}^N\cos^2(\pi\vartheta_i/2)=1$.
Each
${\sf M}_i\leftrightarrow{\sf M}_j$ transformation represents a quarter of a circle and is
described by a single angular order parameter. Then the analytical solutions for all ${\sf M}_i\leftrightarrow{\sf M}_j$ exist, and  all interface widths, energies, and mobilities can be calibrated. While the desired crystal lattice instability
conditions in terms of the hyperspherical order parameters were proved in \cite{levitas+preston-III},
a flaw in this proof was found in \cite{Levitasetal-PRB-twin-13}, where also the nonlinear constraint was substituted by a
linear constraint. This, however, also does not resolve the problem, see \cite{Levitasetal-PRB-15,levitas-roy-ActaMat-16}.  No issues have remained only for two variants or three phases, the polar coordinates with a radial and single angular
order parameters  (Figs. \ref{etai_etaj_plane}(c)), without a  constraint.
In  \cite{Levitas-Momeni-ActaMat-14,Momeni-Levitas-15,Momeni-Levitas-PRB-14}, such a model  was developed  and applied to PTs between two solid phases and melt, in particular, for solid-solid PT via
intermediate (virtual) melt (see also Section \ref{virtual-melt}).

\subsubsection{Multiphase phase field approach utilizing linear constraint}\label{PFA-constr}

 In this approach the order parameters $\eta_i$ may be interpreted as the
 volume fraction of phases because they satisfy a  constraint $\sum_{i=1}^N\eta_i=1$, see  Fig. \ref{etai_etaj_plane}(a) and
\cite{Steinbach-96,Steinbach-99,Garcke1999,Nestler2005,Bollada-2012,Moelans2008,Moelans2009,Kim2006PRE,Toth_et_al-2011a,Toth_et_al-2011b,Toth_et_al-2015}.
These approaches were  mostly applied to  solidification/melting and grain growth without
coupling to mechanics; some applications involving stresses are presented in \cite{Steinbach-06,Schneider_etal_2015}.
The imposed  constraint does not guarantee  that each of the PTs occurs along straight lines connecting
${\sf M}_i$ and ${\sf M}_j$ variants (phases), which is necessary for avoiding  a third
(if spurious) phase between two others and description of a PT with  a single order parameter.
For  three phases, this  was accomplished in \cite{Folch-Plapp-2003,Folch-Plapp-2005} with some
restriction on the kinetic coefficients. It was not evident how to extend  the model for  more phases.
Also, since the appearance of  a third phase is observed in some experiments
(see \cite{Levitas-Henson-PRL-2004,Levitas-PRL-2005,Levitas-Henson-JPCB-2006,Levitas-Ren-PhysRevB-2012}
and Section \ref{virtual-melt}),
 an advanced model should be able to
control this process. Subsequent analysis and improvement of the multiphase models was performed  in
Ref. \cite{Bollada-2012,Toth_et_al-2015,levitas-roy-ActaMat-16,Basak-Levitas-JMPS-18}. However, none of the PFAs with the above
 constraint properly described the lattice instability criteria.

\subsubsection{Multiphase phase field approaches with penalizing functions}{\label{PFA-pen}}

The models  in  \cite{Levitasetal-PRB-15,levitas-roy-ActaMat-16} deal  with the  Cartesian  order
parameters $\eta_i$, without constraint but with the energy penalizing term that controls deviation
from the straight lines connecting   each ${\sf M}_i$ and ${\sf M}_j$ variant or two phases (see  Fig. \ref{etai_etaj_plane}(e)).  With the large (theoretically infinite)
 penalizing term,  the path between two phases is  the straight line that
 can be described by a single order parameter,
an analytical solution for the propagating interface exists, and  the interface energy, width, and mobility can be calibrated. No third phase is involved for the given case.
 At the same time, with the relatively small penalizing term, the third phase may exist within an interface, and the
transformation
path is  curved  (Fig. \ref{etai_etaj_plane}(e)).
 While this model  is more advanced, flexible, and consistent than the previous, some problems have been
 found and were overcome in \cite{Basak-Levitas-JMPS-18}.
 The main difference in  \cite{Basak-Levitas-JMPS-18} is that the  ${\sf A} \leftrightarrow{\sf M}$  PTs
are described by a separate order
parameter and other $N$ order parameters $\eta_i$ describe the martensitic variants within the plane
$\sum_i^N\eta_i=1$ (Fig. \ref{etai_etaj_plane}(f) and (g)).
The transformation path within this plane is controlled with the penalizing term similar to that in  \cite{Levitasetal-PRB-15,levitas-roy-ActaMat-16}. In addition, all   multiphase junctions are penalized.
Using separate order parameters for   the  ${\sf A} \leftrightarrow{\sf M}$ and variant-variant  PTs makes this theory much more flexible and removes contradictions found in \cite{Levitasetal-PRB-15,levitas-roy-ActaMat-16}. It was also developed for large strains and included interfacial stresses.
Numerical algorithms and FEM solutions for this model were presented in \cite{Basak-Levitas-CMAME-19}.

A reaction pathway approach for reconstructive PTs was developed in
\cite{Denoual-etal-JMPS-10,Vattre-Denoual-JMPS-16,Vattre-Denoual-JMPS-16-2}. Components of the transformation deformation gradient $\fg U_t$ are used as order parameters instead of $\eta_i$. Reverse PT may occur not only back to the initial {\sf A} lattice,
but to crystallographically equivalent variants  {\sf A'} of the parent phase, obtained by rotation matrices in the point group of the {\sf A} lattice. Each variant {\sf A'}  can be transformed to multiple crystallographically equivalent variants of {\sf M'}, and so on. Thus, a much larger number of variants (more than a hundred) is considered than in traditional theories for martensitic PTs described above. Gradient energy is dropped, making the theory and interface energy dependent on the mesh. This is justifiable for microscale theories (see Section \ref{microscale}),  but questionable for nanoscale theories.

\subsubsection{Microscale phase field models }\label{microscale}

The volume fraction of different phases as order parameters were utilized for martensitic
PTs, e.g. in \cite{Levitas-Id-Pres-PRL-04,Idesman2005,Lei2010,Tuma-Stupkiewicz-Petryk-16,Steinbach-06,Tuma-Stupkiewicz-16,Mosler_etal_JMPS-2014,Schneider_etal_2015,Esfahani-Gham-Levitas-IJSS-18,Levitas-etal-PRL-18,Esfahani-Levitas-AM-20}.
The simple mixture rule was used, all the interpolation functions
 are linear in volume fractions of phases, $c_k$, and an energy barrier   for PTs between phases $i$ and $j$ is $c_i c_j$. Such models are applicable for microscale simulations, when the interface width is either artificially increased from
 its actual size of  a few nanometers by  several orders of magnitude or processes in the interfaces are not important.
  In contrast to the  nanoscale, the requirements for the first derivative of the interpolation functions in Eq. (\ref{FM5-1})
  are not mandatory and not satisfied.

This model is combined with a contact problem formulation to include dislocation pileups and shear bands, see \cite{Levitas-etal-PRL-18,Esfahani-Levitas-AM-20} and Fig. \ref{fig:Microscale} for a scale-free modeling of nucleation and evolution of the HPP and discrete dislocations.

\subsection{PFA to dislocations}
 PFA to dislocations has many points borrowed from the PFA to martensitic PT in \cite{khachaturyan-01,Khachaturyan-actamat-97}; it is widely utilized for simulation of  plastic deformations in materials \cite{wang01ph,wang01acta,wang10acta,huact01,hucom02,jin01let,kund11ph,kosl02,hu04,hunter2,hunter1,lei11,Biscari-etall-JE-15}. Plastic strain is used instead of transformation strain, which has a similar expression to crystal plasticity: summation  of simple shears along all slip systems, with each shear represented  by the dyadic product of the Burgers vector and normal to the slip planes. The order parameters $\xi_\alpha$ for each slip system $\alpha$ describe the magnitude of the Burgers vector. Small strain approximation along with Hooke's law are utilized.
Spectral methods have been applied for the numerical solution of Ginzburg-Landau  equations for the order parameters and for   the elasticity theory (the Khachaturyan  microelasticity theory, see \cite{wang01ph,wang01acta,wang-apl-01,huact01,hucom02,rodney2}).

At the same time, these approaches inherited  the drawbacks of the  PFA for martensitic PTs, as well as  some additional drawbacks.
Some of them were resolved in
\cite{levitas+preston-III,Levitas-Javanbakht-PRB-12,Levitas-Javanbakht-JMPS-14}.
In particular,
 thermodynamic equilibrium and crystal lattice instability conditions for homogeneous stress-strain states have been formulated and
 the thermodynamic potential and the interpolation function for the Burgers vector were designed to meet these constrains.
  This allowed us to ensure that the magnitude of the thermodynamically equilibrium Burgers vector is stress-independent and the artificial dissipation does not take place in the course of elastic straining; the desired crystal lattice instability conditions and a resolved shear stress - order parameter dependence are reproduced, all in contrast to previous theories.
  Besides, PFA in \cite{Levitas-Javanbakht-JMPS-14} is developed for
  large strain using  the multiplicative decomposition of the deformation gradient into elastic and plastic contributions. Instead of interpolating the  plastic strain tensor versus order parameters in the small-strain approaches,
the relation for the plastic contribution to the velocity gradient versus the rate of the order parameters is postulated in the spirit of  crystal plasticity.
The height of a dislocation in \cite{Levitas-Javanbakht-PRB-12,Levitas-Javanbakht-JMPS-14} has been defined by constitutive equations rather than by computational mesh in the published theories.
 Some simplified equations are presented in Box 11. A comparison with previous approaches is presented in \cite{Levitas-Javanbakht-JMPS-14}.

One of the main ingredients of any phase field approach to dislocations is periodic crystalline energy $\psi_{xi}^c$, which is also called the generalized stacking fault energy or $gamma$-surface. It represents an excess energy per unit area for
any relative displacement of one part of a crystal with respect to another along the slip plane and direction. This energy has multiple zero-value minima corresponding to displacing a periodic lattice into a geometrically equivalent state, which determines the Burgers vectors of complete dislocations. We chose the simplest expression for   $\psi_{xi}^c$ that satisfies this condition.
 Partial dislocations and dislocation reactions are included in
\cite{shen02,wang10acta,Bulatov book,hunter1,Mianroodi-Svendsen-15,Beyerleinetal-IJP-20}. Their consideration is based on the approximation of the more complex 2D energy landscape $\psi_{xi}^c$ with additional intermediate minima, which is obtained using molecular static or first principle  simulations.
An alternative approach to dislocations based on the energy landscape in the total strain space is presented in \cite{Geslinetat-AM-14,Baggio-etall-PRL-19,Arbib-etall-IJP-20}.

Conditions on the interpolation functions imposed in \cite{Levitas-Javanbakht-JMPS-14} were not applied in the above papers, which should lead to stress-dependent equilibrium Burgers vectors and some undesirable features in the stress-strain curves.

\subsection{PFA to the interaction of phase transformation and dislocations}

Nucleation of {\sf M} on dislocations using PFA to PT was studied analytically by \cite{boulbitch+toledano-1998,Korzh-PRL03} and  numerically by \cite{reid+olson+moran-98,Khachaturyan,Malygin-Uspehi-01,Malygin-03,Zhang-Khachaturyan-AM-07,Zhang-Khachaturyan-PhilMag-07,Kashchenko-Uspehi-11,Babaei-Levitas-AM-19}. Stationary dislocations were introduced using their analytical stress field or transformation shear. In \cite{Xu-Khachaturyan-etal-NPJ-CM-18,Xu-Khachaturyan-etal-AM-19,Wang-etal-AM-19},
dislocations are introduced by preliminary plastic deformation, which  did not evolve during PTs.
Precipitates as nucleation cites were also included, and results showed low-hysteresis behavior of some SMAs, which
was observed in experiments in  \cite{Houetal-Science-19}.

In \cite{kund11ph} dislocations
are located at and  move together with a phase interface and  do not need separate phase-field equations.
In reality,
 dislocations may move away from the phase interface into any of the phases, and they become inherited by growing phases, see, e.g., \cite{levitas-javanbakht-Nanoscale-14,Javanbakht-Levitas-PRB-16,Javanbakht-Levitas-JMS-18}.

A continuum dislocation theory is coupled with the PFA for martensitic PT  in \cite{Groger-etal-PRB-08,Groger-etal-PRB-16,34}.
In \cite{34}
the plastic sliding was allowed in {\sf A} only and dislocations inherited by martensite retained the same eigen strain.
Finite-strain crystal plasticity is combined with  the PFA for martensitic PTs in \cite{Andersonetall-IJP-16}, allowing slip in
{\sf A} only. It is applied to SMA.
In \cite{Liu-Raabe-IJP-18}, a similar approach is applied to twinning rather than to PTs.

In \cite{Gou-etal-APL-05,Yeddu1,Yeddu3,Cottura,Yaman,Yamanaka-10,Cuietall-MMT-18,Cisse-Zaeem-CMS-20,Cisse-Zaeem-AM-20} the PFA to martensitic PTs is combined with the classical isotropic plasticity and applied to the growth of martensite.
All the above approaches to the interaction between PTs and plasticity were for small strains.
A reaction pathway approach for reconstructive PTs  developed in
\cite{Denoual-etal-JMPS-10,Vattre-Denoual-JMPS-16,Vattre-Denoual-JMPS-16-2} for large strain was supplemented by the elasoplastic or viscoplastic  models
in \cite{Vattre-Denoual-JMPS-16,Vattre-Denoual-JMPS-19}
and applied for studying PTs and plastic flow in iron under static and shock loading.

Description of martensitic PTs in \cite{Svendsen-JMPS-11} is based on the combination of  quasi-convexification of the nonconvex energy and classical von Mises-type
plasticity.
{
Micromechanical multiscale formulation and FEM solutions of some  problems on interaction between PT and plasticity are described in \cite{Turteltaub-Suiker-05,Kouznetsova-Geers-08}.
The nucleation and expansion  of an elliptic martensitic  inclusions with the prescribed aspect ratio combined with discrete dislocation evolution were analyzed in  \cite{Shi-Turteltaub-VanderGiessen-10}.}

It should be noted that a conceptual problem exists in combining PFA to PTs with the theory of continuously distributed dislocations and especially with phenomenological plasticity. Indeed, the spatial scale of the problem on PT is determined by the interface width, which is on the order of 1 nm. The width of a martensitic variant is a few to 10 nm. The averaged distance between dislocations $l_d = \rho_d^{-0.5}$ with $\rho_d$ as the dislocation density. For annealed materials $\rho_d= 10^{10} m^{-2}$, i.e., $l_d= 10^{-5} m$. For severe plastically deformed materials  $\rho_d= 10^{15} m^{-2}$, i.e.,  $l_d= 32 nm$. Continuum formulation requires  at least several dislocations in each direction in the representative volume, i.e., it is applicable at the scale from two to five orders of magnitude greater than the phase interface width. That means that a nanoscale PFA to PT is consistent  with discrete dislocation theories only. It is clear that any  way to include plasticity and corresponding stress relaxation is more realistic  than  elastic formulation; but it  should not be treated as a consistent  theory.

At the same time, continuum plasticity can be coupled to the microscale PFA, where interface width is much broader or is not considered as a physical parameter, see e.g., \cite{Levitas-Id-Pres-PRL-04,Idesman-Levitas-JMPS-05,Esfahanietal-18,Levitas-etal-PRL-18,Esfahani-Levitas-AM-20}. An important point in these works is that martensitic variants are not spatially resolved and are described in terms of their volume fractions as  internal variables. Another important feature in Ref. \cite{Levitas-etal-PRL-18,Esfahani-Levitas-AM-20} is that it includes a continuum description of
dislocations along discrete slip planes, which allows  reproduction of strong stress concentrators at the tip of the dislocation pileups that leads to the martensite nucleation, similar to the nanoscale approaches with discrete dislocations
in \cite{levitas-javanbakht-Nanoscale-14,Javanbakht-Levitas-PRB-16,Javanbakht-Levitas-JMS-18}

It is important to note that in \cite{Groger-etal-PRB-08,Groger-etal-PRB-16}
significant dislocation density was found  within interfaces between martensitic variants. This is contradictory, because the variant-variant interface is an invariant-plane interface, which should not generate elastic stresses. It is found in
\cite{Basak-Levitas-AM-17,Basak-Levitas-JMPS-18} that despite how the sharp variant-variant interface does not generate stresses, its  finite-width counterpart within PFA does possess significant  elastic stresses, which may lead to dislocations
within an interface.

In \cite{Levitas-Javanbakht-PRB-12},  a simplified version of PFA   for the interactions between martensitic PT and  discrete dislocations (without presenting detailed equations) was applied to the solution of  problems  on nucleation and propagation of misfit dislocations along the interface
and their effect on the athermal interface friction (see Section \ref{PFA-interface-dislocations}). For the stationary solution, both martensite and dislocations disappeared, illustrating reversible plasticity.

A simplified system of equations for the coupled evolution of dislocation and a single martensitic variant was suggested in \cite{levitas-javanbakht-APL-13,levitas-javanbakht-Nanoscale-14}, however,  without strict derivations.
It was applied to various FEM solutions including revealing
scale-dependent athermal semicoherent interface friction for direct and reverse PTs (see Section \ref{PFA-interface-dislocations} and Fig. \ref{Fig-semicoh-PFA}), reduction in PT pressure by up to an order of magnitude due to the dislocation pileup  generated by shear strain (see Fig. \ref{Fig-PFA-pressure-shear}),
 inheriting dislocations by a propagating phase interface, and the nucleation of dislocations by a growing martensitic plate, which leads to the plate arrest.

\subsection{Complete system of the phase-field equations for the interaction between phase transformation and discrete dislocations}\label{PT-disl}

\bec
{\large \bf Box 11. Phase-field equations for coupled phase transformation and discrete dislocations \cite{Levitas-Javanbakht-15-1,Javanbakht-Levitas-JMPS-15-2}}
\eec

{\bf 1. Kinematics}

I.  Finite strains

1.1. Multiplicative decomposition of the deformation gradient ${\fg F}$
\bey
{\fg F}  ={\fg F}_e {\fg \cdot}{\fg U}_t{\fg \cdot}{\fg F}_p.
\label{n1ag-2}
\eey

1.2. Jacobian determinants
\bey
&& J:=\frac{dV}{dV_0}=\frac{\rho_0}{\rho}=det {\fg F};  \quad J_e:=\frac{d V}{dV_t} =\frac{\rho_t}{\rho_0}=det {\fg F}_e;
\nonumber\\
&& J_{tp}:=\frac{d V_t}{dV_p} =\frac{\rho_p}{\rho_t}=det {\fg U}_t \, det{\fg F}_p = det {\fg U}_t = J_t; \quad J_p:=  det {\fg F}_p=1;
\qquad J=J_e  J_t,
\label{jac}
\eey
where  $dV_0$ ($\rho_0$), $dV_t$ ($\rho_t$),  $dV_p$ ($\rho_p$), and $dV$ ($\rho$) are the elemental volumes (mass densities) in the reference $\Omega_0$, transformed $\Omega_t$, plastically deformed $\Omega_p$,   and the actual ($\Omega$) configurations, respectively.

1.3. Symmetric transformation deformation gradient
\bey
  && {\fg U}_t =  {\fg I} +  \bar{\fvep}_{t} \varphi  (a, \eta) ; \quad \varphi  (a, \eta )  = a \eta_k^2 (1-\eta )^2 + (4 \eta^3 - 3
\eta^4) ;
\quad
0  <  a  <  6  ,
\label{ndau31d4a2}
\eey
where $\bar{\fvep}_{t}$ is the transformation strain of a complete {\sf M} variant  and $\eta$ is the order parameter that describes PT from {\sf A} ($\eta=0$) to {\sf M} ($\eta=1$).

1.4. Plastic part of the velocity gradient in the reference configuration
\bey
&&\fg l_{p}  :=
\sum_{\alpha=1}^{p }
 \frac{1}{H^{\alpha}}  {\fg  b}^{\alpha  } \otimes
{\fg  n}^{\alpha} \dot\Phi({\xi}_{\alpha })
  =
\sum_{\alpha=1}^{p }
 \gamma_{\alpha }  {\fg  m}^{\alpha  } \otimes
{\fg  n}^{\alpha} \dot\Phi({\xi}_{\alpha }),
\label{F6iir4-2}\\
& & \Phi (\xi_{\alpha}) = \phi(\bar{\xi}_{\alpha})+Int(\xi_{\alpha}); \quad \phi(\bar{\xi}_{\alpha})= \bar{\xi}^2_{\alpha} (3-2\bar{\xi}_{\alpha}),
\nonumber
\eey
where  ${\fg  b}^{\alpha}$ is the Burgers vector of a dislocation in the $\alpha^{th}$ slip system, ${\fg  n}^{\alpha}$ is the unit normal to the   slip plane; $\gamma_{\alpha}={|{\fg  b}^{\alpha}|}/{H^{\alpha}}$ is the plastic shear produced by a single dislocation in a dislocation band with the height $H^{\alpha}$, ${\fg  m}^{\alpha}= {\fg  b}^{\alpha}/|{\fg  b}^{\alpha}|$; ${\xi}_{\alpha}$ is the order parameter for a dislocation  in the $\alpha^{th}$ slip  system; and  $Int(\xi_{\alpha})$ and
$\bar{\xi}_{\alpha}:= {\xi}_{\alpha}-Int(\xi_{\alpha}) $
is the integer part (defining number of dislocations in a dislocation band) and fractional part of $\xi_{\alpha}$.

II.  Small strains
\bey
&& \fvep=({\fg  \nabla}  {\fg  u})_{\it s} = \fvep_e +\fvep_t +\fvep_p  ; \qquad  \fvep_t= \bar{\fvep}_{t} \varphi  (a, \eta_k);
\nonumber\\
&&\fg \omega= \fg \omega_e + \fg \omega_t + \fg \omega_p ; \qquad {\fvep}_p + {\fg \omega}_p = \sum_{\alpha=1}^{p }
 \frac{1}{H^{\alpha}}  {\fg  b}^{\alpha} \otimes
{\fg  n}^{\alpha} \Phi({\xi}_{\alpha}),
\label{F6iir4d-2s}
\eey
where ${\fg  u}$ is the displacement; $\fg  \nabla$ designates the gradient  in the $\Omega_0$; subscript $s$ designates  symmetrization; $\fvep$ and $\fg \omega$  are the small strain and rotations, respectively; and subscripts  $e$, $t$, and $p$ are for elastic, transformational, and plastic parts, respectively.

\noindent
{\bf 2. Helmholtz free energy  per unit mass}
\bey
&&
\psi =  J_t \psi^e+ \psi^\theta_{\eta}+\psi_{\xi}^c+\psi_{\xi}^{int}+ \psi^\nabla_{\eta}+ \psi^\nabla_{\xi}.
\label{helm-2}
\eey

2.1. {\it Elastic energy}
\bey
&&
\rho_0 \psi^e=\frac{1}{2} {\fg E_e} {\fg :}{\fg C}{\fg :} {\fg E_e},
\label{Eeneg-2}
\eey
where $\fg E_e= 0.5({\fg F}_e^T \cdot {\fg F}_e - {\fg I})$ is the elastic Lagrangian strain  tensor and  $\fg C$ is the fourth-rank tensor of elastic moduli, which are assumed here as the same for both  phases.
{Note that the definition of $\psi^e$ in Eqs. (\ref{helm-2}) and (\ref{Eeneg-2}) differs from that in Box 9.}

2.2. {\it Thermal energy}
\bey
&& \psi^\theta_{\eta}=  A \eta^2 (1-\eta)^2 + \Delta G^\theta (4 \eta^3
- 3 \eta^4);
\nonumber\\
&&
\Delta G^\theta \, = - \Delta s
\, \left(\theta - \theta_e \right) \, , \qquad A = A_0 \left(\theta -
\theta_c \right)\;
 , \qquad \, A_0 > 0,
\label{l7det2-2}
\eey
where $\Delta \, G^{\theta} $ and $\Delta s$ are the differences between the thermal part of the free energy and entropy for {\sf M} and {\sf A}, respectively; $A$ is the magnitude of the double-well barrier between {\sf A} and {\sf M} at phase equilibirum; $\, \theta_e \,$ is the thermodynamic equilibrium temperature for
stress-free {\sf A} and  {\sf M}; $A_0$ is a parameter; and $\, \theta_c \,$ is the
critical temperature at which stress-free {\sf A} loses its
thermodynamic
stability.

2.3. {\it Periodic in space crystalline energy}
\bey
&& \psi_{\xi}^c= \sum_{\alpha=1}^{p}  A_{\alpha} (\eta,\bar{y}^{\alpha}) (\bar{\xi}_{\alpha})^2 (1-\bar{\xi}_{\alpha})^2; \qquad
\label{F11s14-2}
\bar{A}_{\alpha}(\eta,\bar{y}^{\alpha})  = A_{\alpha}^A +(A_{\alpha}^{M}-A_{\alpha}^{A}) \eta^2 (3 - 2 \eta);
\nonumber\\
&& A_{\alpha}^{A,M}(y^{\alpha})=\begin{cases}
   \bar{A}^{A,M}_{\alpha} \qquad \bar{y}^{\alpha}\leq H^{\alpha} ;\\
  k \bar{A}^{A,M}_{\alpha} \qquad \bar{y}^{\alpha} > H^{\alpha}  .
\end{cases}
\quad
\bar{y}^{\alpha}=y^{\alpha}-Int(\frac{y^{\alpha}}{H^{\alpha}+w_{\alpha}})(H^{\alpha}+w_{\alpha}),
\eey
where $A_{\alpha}^A$ and $A_{\alpha}^M$ are the magnitudes of the multi-well crystalline energy in {\sf A} and {\sf M}, respectively, that define the critical shear stress for barrierless nucleation of a dislocation (i.e., the theoretical shear strength); $y^{\alpha}$ is the coordinate normal to the $\alpha_{th}$ slip plane; and $w_{\alpha}$ is the width of the thin layer between dislocation bands.

2.4. {\it Energy of  interaction of dislocation cores belonging to different slip systems}
\bey
&&
\psi_{\xi}^{int}=
\sum_{\alpha, k=1}^{p}
 {A}_{\alpha k} (\eta) (\bar{\xi}_{\alpha})^2 (1-\bar{\xi}_{\alpha})^2 (\bar{\xi}_{k})^2 (1- \bar{\xi}_{k})^2; \qquad {A}_{\alpha \alpha}=0;
 \nonumber\\
&& A_{\alpha k}(\eta) = A_{\alpha k}^A+ (A_{\alpha k}^M- A_{\alpha k}^A) \eta^2 (3 - 2 \eta),
\label{F11s15-2}
\eey
where ${A}_{\alpha k}^A$ and ${A}_{\alpha k}^M$ are the corresponding magnitudes for {\sf A} and {\sf M}, respectively.

2.5. {\it Gradient energies for PTs and dislocations}
\bey
&& \psi^\nabla_{\eta} = \frac{\beta^{\eta}}{2} |{\fg  \nabla} \eta|^2;
\quad
\psi^\nabla_{\xi}= 0.5 {\beta}_\xi(\eta) \sum_{\alpha=1}^{p}  \left(
(\nabla^m  \xi_{\alpha})^2  + Z(1-\bar{\xi}_{\alpha})^2
(\nabla^n \xi_{\alpha })^2 \right);
\label{F11s1-2}
\\
&& \beta_{\xi}(\eta)  = \beta_{\xi}^{A} +(\beta_{\xi}^{M}-\beta_{\xi}^{A})\eta^2 (3 - 2 \eta);
\quad
{  \nabla}^m  \xi_{\alpha} := {\fg  \nabla}  \xi_{\alpha} \cdot {\fg  m}^{\alpha};
\quad {  \nabla}^n  \xi_{\alpha} := {\fg  \nabla}  \xi_{\alpha} \cdot {\fg  n}^{\alpha},
\eey
where $\beta^{\eta}$ is the coefficient of the gradient energy  for PT; $\beta_{\xi}^{A}$ and $\beta_{\xi}^{M}$ are the coefficient of the gradient energy  for dislocations in  {\sf A} and {\sf M}, respectively; $Z$ is the ratio of the coefficients for the gradient energy normal to and along the slip plane; and superscripts $m$ and $n$ stand for the  directions along and the normal to the slip plane, respectively.

\noindent
{\bf 3. First Piola-Kirchhoff ${\fg P}$ and Cauchy $\fsg$ stress tensor}

I.  Finite strains
\bey
&&{\fg P}= \rho_0 J_t {\fg F_e}\fg\cdot \frac{\partial \psi^e}{\partial {\fg E_e}} \fg\cdot {\fg U^{-1}_t} \fg\cdot {\fg F^{T-1}_p} =  J_t
{\fg F_e}\fg\cdot {\fg C}{\fg :} {\fg E_e}  \fg\cdot {\fg U^{-1}_t} \fg\cdot {\fg F^{T-1}_p};
\label{Elrul1-2}
\eey
\bey
&& \fg \sigma = \rho J_t {\fg F_e}\fg\cdot \frac{\partial \psi^e}{\partial {\fg E_e}} \fg\cdot {\fg F^T_e} = \frac{1}{J_e} {\fg F_e}\fg\cdot {\fg C}{\fg :} {\fg E_e} \fg\cdot {\fg F^T_e}
\label{Elrul2-2}
\eey

II.  Small strains and linear elasticity
\bey
\fsg = \rho \frac{\partial \psi}{\partial \fvep_e} = {\fg C} {\fg :} {\fvep_e}.
\label{sigsls}
\eey

\noindent
{\bf 4. Ginzburg--Landau equations}

4.1.  Compact form in the $\Omega_0$ at finite strains
\bey
&& \dot{\eta}  =L^{\eta} X^{\eta} = L^{\eta} \left(\frac{1}{\rho_0} {\fg P}^{\ssst T} {\fg \cdot} \, {\fg F}_e {\fg :} \, \frac{\partial {\fg U}_t}{\partial \eta}
 {\fg \cdot} \, {\fg F}_p +   {\fg  \nabla}
            \cdot \left(  \frac{\partial \psi}{\partial   {\fg  \nabla} \eta}\right)
            -  \frac{\partial \psi}{\partial  \eta} \right);
\nonumber\\
&&  \dot{\xi}_{\alpha } =  L_{\alpha}({\eta}) X_{\alpha}^{\xi} = L_{\alpha} (\eta) \left(\frac{1}{\rho_0}\tau_{\alpha}\gamma_{\alpha} \frac{\partial \Phi }{\partial \xi_{\alpha }} +   {\fg  \nabla}
            \cdot \left(  \frac{\partial \psi}{\partial   {\fg  \nabla} \xi_{\alpha}}\right)
            -  \frac{\partial \psi}{\partial  \xi_{\alpha}} \right);
\label{-10-e0e1-2}
\nonumber\\
&& L_{\alpha} (\eta)= L_{\alpha}^A+ (L_{\alpha}^M-L_{\alpha}^A) \eta^2 (3 - 2 \eta);
\quad
\tau_{\alpha  }:={\fg  n}^{\alpha} {\fg \cdot} \, {\fg F}_p {\fg \cdot} \, {\fg P}^{\ssst T} {\fg \cdot} \, {\fg F}_e {\fg \cdot} \, {\fg U}_t {\fg \cdot} \, {\fg  m}^{\alpha  },
\eey
where $L^{\eta}$,  $L_{\alpha}^{A}$ and $L_{\alpha}^{M}$ are  the kinetic coefficients for PT and  dislocations in {\sf A} and {\sf M}, respectively;  $X^{\eta}$ and $X^{\xi}$ are the thermodynamic dissipative forces work-conjugate to $\dot{\eta}$ and $\dot{\xi}_{\alpha}$, respectively;  and $\tau_{\alpha}$ is the resolved shear stress for a dislocation.

4.2. Detailed form  at finite strains
\bey
&&
\dot{\eta}  =L^{\eta} \lbrace(\frac{1}{\rho_0} {\fg P}^{\ssst T} {\fg \cdot} \, {\fg F}_e {\fg :} \, \frac{\partial {\fg U}_t}{\partial \eta}
 {\fg \cdot} \, {\fg F}_p - J_{t}  {\fg U}_{t}^{-1} \fg : \frac{\partial {\fg U}_t }{\partial  \eta}\psi^e ( {\fg E_e},\eta) - J_t \frac{\partial \psi^e( {\fg E_e},\eta)}{\partial \eta} +
 \nonumber\\
 && - [ 2 A \eta (1-\eta) (1-2 \eta) + 12 \Delta G^\theta \eta^2 (1-\eta)] - \sum_{\alpha=1}^{p}  \frac{\partial A_{\alpha} (\eta,\bar{y}^{\alpha})}{\partial \eta} (\bar{\xi}_{\alpha})^2 (1-\bar{\xi}_{\alpha})^2 -
 \nonumber\\
 && \sum_{\alpha, k=1}^{p}  \frac{\partial {A}_{\alpha k} (\eta)}{\partial \eta}   (\bar{\xi}_{\alpha})^2 (1-\bar{\xi}_{\alpha})^2 (\bar{\xi}_{k})^2 (1- \bar{\xi}_{k})^2
\nonumber\\
 &&   - {0.5} \frac{\partial {\beta}_\xi (\eta) }{\partial \eta}  \sum_{\alpha=1}^{p}  \left(
(\nabla^m  \xi_{\alpha})^2  + Z (1-\bar{\xi}_{\alpha})^2
(\nabla^n \xi_{\alpha})^2 \right) +   {  \beta}^{\eta}    {\nabla}^2  \eta \rbrace.
\label{bi-8-10-022-2}
\eey

\bey
&& \dot{\xi}_{\alpha}  = L_{\alpha}  ({\eta}) \lbrace \frac{6}{\rho_0}\tau_{\alpha}  \gamma_{\alpha}   \bar{\xi}_{\alpha}  (1 - \bar{\xi}_{\alpha} )
+ \frac{1}{2} {\fg \nabla} \beta_{\xi} (\eta){\fg \cdot} {\fg \nabla} \bar{\xi}_{\alpha}  + \frac{1}{2}[Z(1-\bar{\xi}_{\alpha} )^2-1] ({\fg \nabla} \bar{\xi}_{\alpha} {\fg \cdot} {\fg n}^{\alpha}) ({\fg \nabla} \beta_{\xi} (\eta){\fg \cdot} {\fg n}^{\alpha})
 \nonumber\\
 && + \frac{1}{2} \beta_{\xi} (\eta) [\nabla^2 \bar{\xi}_{\alpha}  + (Z(1-\bar{\xi}_{\alpha} )^2-1) ({\fg \nabla} {\fg \cdot}{\fg n}^{\alpha}) ({\fg \nabla} \bar{\xi}_{\alpha}  {\fg \cdot}{\fg n}^{\alpha})]
 \nonumber\\
 &&  - 2 Z (1 - \bar{\xi}_{\alpha} ) ({\fg \nabla} \bar{\xi}_{\alpha}  {\fg \cdot}{\fg n}^{\alpha})^2 + [Z (1 - \bar{\xi}_{\alpha} )^2 - 1] {\fg \nabla} ({\fg \nabla} \bar{\xi}_{\alpha}  {\fg \cdot}{\fg n}^{\alpha}) {\cdot}{\fg n}^{\alpha}
 - 2 A_{\alpha}  (\eta,\bar{y}^{\alpha})  \bar{\xi}_{\alpha}  (1-\bar{\xi}_{\alpha} ) (1 - 2 \bar{\xi}_{\alpha} )
\nonumber\\
 &&  - 2 {A}_{\alpha k}  (\eta) \bar{\xi}_{\alpha}  (1-\bar{\xi}_{\alpha} ) (1 - 2 \bar{\xi}_{\alpha} ) (\bar{\xi}_{k})^2 (1- \bar{\xi}_{k} )^{2}
 + {\beta}_\xi  (\eta)  Z (1-\bar{\xi}_{\alpha} )(\nabla^n \xi_{\alpha} )^2 \rbrace.
\label{dislA-GL-2}
\eey

4.3.  Small strains, linear elasticity
\bey
&& \dot{\eta}  = L_{\eta} \lbrace(\frac{1}{\rho_0} \fsg {\fg :} \frac{\partial {\fg \fvep}_t}{\partial \eta} - \frac{J_t}{2\rho_0} (\fg I\fg :  \frac{\partial {\fg \fvep}_t}{\partial \eta}) \,  {\fg \fvep_e} {\fg :}{\fg C} {\fg :} {\fg \fvep_e}
\nonumber\\
 && - [ 2 A \eta (1-\eta) (1-2 \eta) + 12 \Delta G^\theta \eta^2 (1-\eta)] - \sum_{\alpha=1}^{p}  \frac{\partial A_{\alpha} (\eta,\bar{y}^{\alpha})}{\partial \eta} (\bar{\xi}_{\alpha})^2 (1-\bar{\xi}_{\alpha})^2 -
 \nonumber\\
 && \sum_{\alpha, k=1}^{p}  \frac{\partial {A}_{\alpha k} (\eta)}{\partial \eta}   (\bar{\xi}_{\alpha})^2 (1-\bar{\xi}_{\alpha})^2 (\bar{\xi}_{k})^2 (1- \bar{\xi}_{k})^2
\nonumber\\
 &&   -  {0.5}  \frac{\partial {\beta}_\xi (\eta) }{\partial \eta}  \sum_{\alpha=1}^{p}  \left(
(\nabla^m  \xi_{\alpha})^2  + Z (1-\bar{\xi}_{\alpha})^2
(\nabla^n \xi_{\alpha})^2 \right) +   {  \beta} ^{\eta}   {\nabla}^2  \eta \rbrace.
\label{bi-8-10-022-2g}
\eey
The Ginzburg-Landau equation for dislocations for small distortions looks like Eq.(\ref{dislA-GL-2}) with
the simplified expression for $\tau_{\alpha  }:={\fg  n}^{\alpha} {\fg \cdot} \,\fsg  {\fg \cdot} \, {\fg  m}^{\alpha  }$.

 In \cite{Levitas-Javanbakht-15-1} a large-strain and thermodynamically consistent PFA for combined discrete dislocations and  multivariant martensitic PTs is suggested. It synergistically  combined and extended the most mechanically advanced  PFA to  martensitic PT in terms of the order parameters $\eta_i$ \cite{Levitas-IJP-13} and dislocations  in terms of the order parameters $\xi_i$ \cite{Levitas-Javanbakht-PRB-12,Levitas-Javanbakht-JMPS-14}.
 Details for PFA to PT and dislocations separately can be found in these papers.
For compactness and simplicity, we   present in Box 9  the particular case of a complete system of equations from
\cite{Javanbakht-Levitas-JMPS-15-2}
with the following simplifications in comparison with the general theory in \cite{Levitas-Javanbakht-15-1}.

(a)  Slip systems of {\sf A} and {\sf M}, transformed back to {\sf A}, coincide, i.e., all slip systems are inherited during the direct and reverse PTs. {This is the case, e.g.,   for PTs between b.c.c.  and body centric tetragonal (b.c.t.) crystal lattices as well as for PTs between f.c.c and f.c.t. lattices. 

(b) We consider a single {\sf M} variant.

(c) Surface energy is independent of phase and dislocations, which results in the simplest zero-flux boundary conditions for  $\eta_i$ and $\xi_i$.

Box 11 also contains simplified equations for infinitesimal strains.
  The key problem was to justify the best kinematic decomposition.
  Several quite natural  and  logical options of multiplicative decomposition of the deformation gradient
 ${\fg F}$, e.g., ${\fg F}  
={\fg F}_e{\fg \cdot}{\fg F}_p^M {\fg \cdot}{\fg U}_t{\fg \cdot}
{\fg F}_p^A$, with ${\fg F}_p^A$ and ${\fg F}_p^M $ for the plastic deformation gradient in {\sf A} and {\sf M},
   were  rejected due to some undesired features.
 The thought experiments  considered  cyclic {\sf A}-{\sf M} PTs and plastic deformation of {\sf A} and {\sf M} after PTs, with focus on the inheritance and  evolution of dislocations during and after PTs, along the inherited slip systems that do not belong to the
 traditional ones for the product crystal lattice.

  The multiplicative decomposition (\ref{n1ag-2}) of the deformation gradient into elastic, transformational, and plastic contributions
  (exactly in this order) is justified
(Fig. \ref{kinematics}). Eq. (\ref{jac}) defines the corresponding Jacobian determinant describing ratios of elemental volumes in different configurations.  A symmetric transformation deformation gradient is interpolated in terms of the order parameter $\eta$ describing PT by Eq. (\ref{ndau31d4a2}), while more advanced expressions justified in \cite{Levitas-IJP-18,Basak-Levitas-JMPS-18} are currently available.

  Generally,  the plastic part $\fg l_p$ of the velocity gradient $\fg l= \dot{\fg F} \cdot {\fg F}^{-1}$ includes four different mechanisms: (a)  dislocation evolution in {\sf M} along the natural slip systems of {\sf M} and (b) slip systems of {\sf A} inherited during PT; (c) dislocation evolution in {\sf A} along the natural slip systems of {\sf A} and (d) slip systems of {\sf M} inherited during reverse PT.  Equations for transformation   of the parameters of the slip systems inherited by the crystal lattices during PT are presented in Fig. \ref{Figure1s} and caption.

\begin{figure}[htbp]
  \centering
\includegraphics[scale=.5]{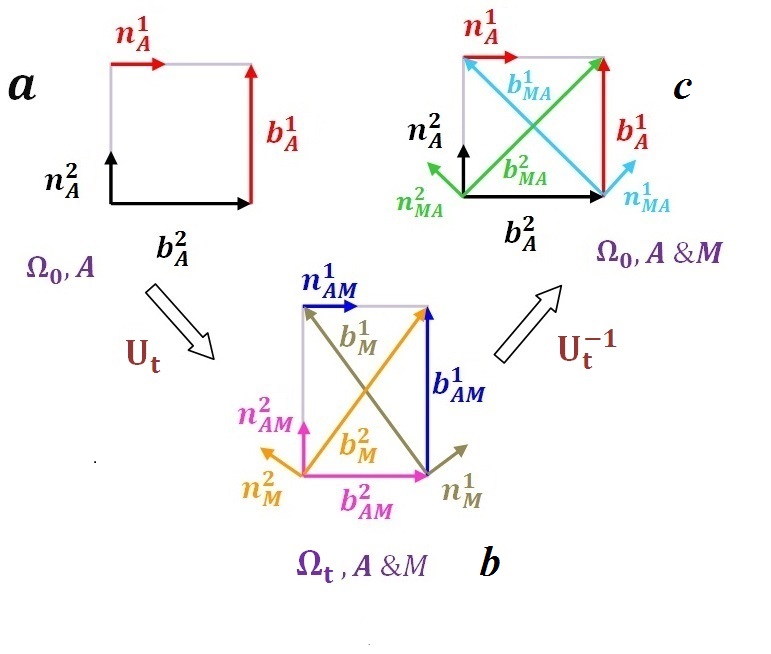}
\caption{  Sketches for Burgers vectors ${\fg  b}^{  \omega}$ and normals ${\fg  n}^{  \omega}$ to the slip planes in {\sf A} and {\sf M} in different configurations.
 (a) Two-dimensional  f.c.c. crystal lattice of {\sf A} with two slip systems (designated by ${\fg  b}^{\alpha }_A$ and ${\fg  n}^{\alpha }_A$)
 along the faces in the undeformed reference configuration $\Omega_0$. (b) Two-dimensional b.c.c. crystal lattice of {\sf M} with two slip systems (designated  by ${\fg  b}^{\omega }_{M}$, ${\fg  n}^{\omega }_M$) along the diagonals  in the transformed configuration $\Omega_t$.
 The slip systems of {\sf A} inherited by {\sf M}  after the PT and defined as ${\fg  b}^{\alpha }_{AM}= \fg U_t \cdot
{\fg  b}^{\alpha }_{A}$ and  ${\fg  n}^{\alpha }_{AM}= {\fg  n}^{\alpha }_A \cdot \fg U_t^{-1}/|{\fg  n}^{\alpha }_A \cdot \fg U_t^{-1}|$ are shown as well.
(c) The slip systems of {\sf M} inherited by {\sf A} after the reverse PT  and defined as ${\fg  b}^{\omega}_{MA}= \fg U^{-1}_t \cdot
{\fg  b}^{\omega }_{M}$ and  ${\fg  n}^{\omega }_{MA}= {\fg  n}^{\omega }_M \cdot \fg U_t/|{\fg  n}^{\omega }_M \cdot \fg U_t|$  are shown  in the  $\Omega_0$ along with the slip systems of  {\sf A}.
}
\label{Figure1s}
\end{figure}

  It is proved that the definition of $\fg l_p$ (a) in
{\sf M} expressed as a combination of plastic shear rates along the slip system of {\sf M} and (b)
in {\sf M} transformed back to {\sf A} expressed as a combination of plastic shear rates along the slip system of {\sf M}
back-transformed to {\sf A} are equivalent. This result leads to the description of $\fg l_p$ of {\sf M} in the crystal lattice of {\sf A} with the same
expression  as in the crystal lattice of {\sf M}, however,  with transformed-back crystallographic parameters of the slip systems of {\sf M} to the {\sf A}. This also led  to the additive combination
of $\fg l_p$ for all four mechanisms using crystal lattice of the {\sf A}.
  The $\fg l_p$ for all four mechanisms is expressed in the crystal lattice of the {\sf A} using  slip systems of {\sf A} and transformed-back slip systems of {\sf M}, and just two corresponding types of the order parameters.
  This is  a noncontradictory and economic decomposition, in contrast to, e.g., a multiplicative decomposition
  of ${\fg F}_p$ into ${\fg F}_p$ in the {\sf M} and {\sf A}. When the slip systems of {\sf A} and transformed-back slip systems of {\sf M} coincide, Eq. (\ref{F6iir4-2}) for $\fg l_p$ utilizes just one type of the order parameters, like in the
  PFA for dislocation without PT by \cite{Levitas-Javanbakht-JMPS-14}.

 The Helmholtz free energy  per unit mass consists of elastic, thermal, and periodic in space crystalline energies, the energy of  interaction of dislocation cores belonging to different slip systems, and gradient energies for PT and dislocations, see Eqs. (\ref{helm-2})-(\ref{F11s1-2}).
     Utilization of the  thermodynamic laws results in the elasticity rule (Eqs. (\ref{Elrul1-2})-(\ref{sigsls})) and expressions for the driving forces for PT $X^{\eta}$ and dislocation evolution $X_{\alpha}^{\xi}$
     (\ref{-10-e0e1-2}). The Ginzburg-Landau equations are obtained as the linear relationships between the thermodynamic driving forces and work-conjugate rates and are expressed in a compact and detailed form in  Eqs. (\ref{-10-e0e1-2})-(\ref{bi-8-10-022-2g}).

     Various types of coupling between PT and dislocations are included in the theory.
     The nonlinear kinematic decomposition (\ref{n1ag-2})  already contains such a coupling resulting
    in the presence of ${\fg F}_p$  in the transformation work and  ${\fg U}_t$  in the definition of the resolved shear stresses $\tau_{\alpha  }$. For infinitesimal  strains  this type of coupling disappears. All material parameters for dislocations depend on $\eta_i$ since they have different values in {\sf A} and {\sf M}.
     Due to such a dependence, additional  terms in the Ginzburg-Landau equations for PTs
     appear due to the change in dislocation structure and properties. Also, dislocations are
 inherited  during PT and they may further evolve along the nontraditional slip systems if corresponding critical shear strength is reached.
 In addition, one of the strongest interactions between PT and dislocations occurs through their eigen-stress fields, which is determined by a solution of the continuum mechanical boundary-value problem.
Application of the developed theory to the solution of some  important material problems is presented in  \cite{Javanbakht-Levitas-JMPS-15-2,Javanbakht-Levitas-PRB-16,Javanbakht-Levitas-JMS-18}.

\section{Phase transformations and chemical reactions induced by plastic shear under high pressure}\label{RDAC}

The main experimental phenomena related to the  PTs and CRs under pressure and shear are reviewed in
 \cite{bridgman-1937,bridgman-1947,zharov-1984,zharov-1989,Blank-Estrin-2014,Edalati-Horita-16,Levitas-chapter-04,Levitas-MT-19}.
Theoretical treatment was initiated in \cite{Levitas-PRB-04,Levitas-chapter-04}.  The current four-scale theory (from atomistic to macroscale behavior of a sample) is recently reviewed in \cite{Levitas-MT-19}.  To minimize
 repetitions, this Section will be short.

 \subsection{Main phenomena }\label{phenomena}

One of the most impressive effects of plastic deformation on the PTs and CRs is observed in experiments on high-pressure torsion under constant applied force. Initially, this work was performed in metallic or ceramic rotational Bridgman  anvils
(see \cite{bridgman-1937,bridgman-1947} for PTs and  \cite{zharov-1984,zharov-1989,chistotina+zharov+kissin+enikolopyan-1970,36,38} for CR), which are currently used for grain refinement and producing nanograined materials \cite{valiev+islamgaliev+alexandrov-2000,valievetal-MRL-2016,valievetal-JOM-2016,Edalati-Horita-16,Zhilyaev-2008}.
Currently, much more precise in-situ experiments are performed with the rotational diamond anvil cell (RDAC)
  \cite{Blank-Estrin-2014,Ji-Levitasetal-12,Gaoetal-17}, see inset in Fig. \ref{fig:pd-plasticstrain}.
  Among numerous phenomena that are observed in these experiments (see  \cite{bridgman-1937,bridgman-1947,Blank-Estrin-2014,Levitas-PRB-04,Levitas-chapter-04,Levitas-MT-19}), we enumerate just three
  the most important ones. Thus, plastic shear under high pressure:

{  (a)  leads to the formation
of new phases and reaction products that were {   not be}
produced without shear }\cite{bridgman-1937,bridgman-1947,Blank-Estrin-2014,Edalati-Horita-16,08,zharov-1984,zharov-1994,levitasetal-SiC-12,Levitas-chapter-04,Levitas-MT-19};

  (b)  reduces the transformation  pressure
 by a factor of 3 to 10  for some PTs \cite{bridgman-1937,bridgman-1947,Blank-Estrin-2014,Edalati-Horita-16,levitas+shvedov-2002,Ji-Levitasetal-12} and chemical reactions \cite{zharov-1984,zharov-1994}, and even by a factor of 100 for the PT from graphite to diamond
 \cite{Gaoetal-17}, and

 (c)  substitutes a reversible PTs
        with  irreversible PTs \cite{Blank-Estrin-2014,Levitasetal-JCP-BN-06,levitas+shvedov-2002}.

        We will discuss the most advanced results on characterization and general properties of strain-induced PTs in RDAC.
 The in situ quantitative synchrotron X-ray diffraction  investigation  of plastic strain-induced  $\alpha-\omega$ PT in Zr was  performed in \cite{Pandey-Levitas-ActaMat-20}. The most consistent results were obtained for strongly plastically predeformed Zr, when  strain hardening saturates and material hardness and yield strength and microstructure  do not evolve with further plastic deformation  \cite{levitas-book96,Edalati-Horita-16,valievetal-JOM-2016,Levitas-etal-NPJ-CM-19}.
 Polycrystalline materials for this case behave like perfectly plastic and isotropic with plastic strain history independent
 surface of perfect plasticity  \cite{levitas-book96}.
We assumed that since the plasticity theory  for such large strains is   significantly  simplified,  the kinetics of  strain-induced   PTs will be more tractable as well, and this is the best and repeatable  initial state to start with.
This was the case in experiments.
Working part of sample looks like a coin, with diameter of $500\, \mu m$ (diamond cullet) and thickness reducing from
$200\, \mu m$ down to $10\, \mu m$ and less, depending on the applied force and the rotation angle of an anvil.
    Distributions of pressure in each phase and in the mixture, and volume fraction of $\omega$-Zr along the radius, all averaged over the sample thickness, as well as thickness profile were measured using synchrotron X-ray diffraction and X-ray absorption.
    The simplified version of the  strain-controlled kinetic equation derived in \cite{Levitas-PRB-04,Levitas-chapter-04}
    (which is a particular case of  Eq. (\ref{str-ind-kineq-ss-a})) is
\bey
\frac{d c}{d q} =
a \; {\left(1-c \right)}\,\frac{\sg_{y2}^w}{\sg_a}\,
\frac{p - p_\vep^d}{p_h^d - p_\vep^d}H(p - p_\vep^d)
 -
b \; c \, \frac{\sg_{y1}^w}{\sg_a}\,
\frac{p_\vep^r - p}{p_\vep^r - p_h^r} H(p_\vep^r - p); \quad \sg_a = c   \sg_{y1}^w   +
\left(1   -   c \right) \sg_{y2}^w.
\label{l-g-4}
\eey
Here,
$ \sg_{yi} $ is the yield strength of $ i$-th phase;
$p_{\vep}^d$ and $p_{\vep}^r$  are the minimum pressure at which the direct strain-induced PT may occur and maximum pressure at which the reverse strain-induced PT proceeds, respectively, $H$ is  the Heaviside step function used to impose criteria for the direct ($p  >  p_\vep^d$) and reverse ($p<p_\vep^r$) strain-induced PTs;
$p_h^d$ and $p_h^r$ are the pressures for the direct and reverse PTs under hydrostatic loading;  all remaining symbols are material parameters.   Eq.(\ref{l-g-4}) includes
 the possibility of direct and reverse PTs and
 the different plastic strain in each phase due to different $ \sg_{yi} $.
For $\omega$-Zr, the reversed strain-induced PT was not observed in \cite{Pandey-Levitas-ActaMat-20}, i.e., the second term in Eq.(\ref{l-g-4}) disappears. Eq.(\ref{l-g-4}) was confirmed experimentally and all material parameters were identified.
   In particular,  the minimum pressure for the strain-induced $\alpha-\omega$ PT, $p^d_{\vep}$=1.2 GPa, is 4.5 times lower than under hydrostatic conditions  and 3 times lower than the phase equilibrium pressure. The $p^d_{\vep}$  is found to be independent of the compression-shear straining path. This means that the strain-induced PTs under compression in DAC and torsion in RDAC do not fundamentally differ for Zr.

Note that recent experimental advancements include possibility of measurements of the fields of all components of the stress tensor in the diamond in DAC at the contact surface with the sample \cite{{Hsiehetal-Science-19}}  and particle displacements at the contact surface of the sample in DAC and RDAC \cite{Pandey-Levitas-JAP-21}.

\subsection{Atomistic studies}\label{Atomistic studies}

Molecular dynamics \cite{Levitasetal-Instab-17,Levitasetal-PRL-17} and first principle simulations \cite{Zarkevichetal-18} were performed for an ideal lattice.  Under guidance of the analytical treatment of the martensitic PTs within PFA
\cite{Levitas-IJP-13}, they led to
  {\it an explicit expression for the PT (i.e., lattice instability) condition} for  cubic - tetragonal Si I$\leftrightarrow$Si II transformations under action of all six components of the stress tensor.
{  The strong effect of nonhydrostatic stresses was exhibited, in particular, in the following result: the pressure for Si I$\leftrightarrow$Si II PT  under uniaxial compression was reduced by a factor of 20 in comparison with hydrostatic loading \cite{Zarkevichetal-18}.} Stress-strain curves for different loadings were determined as well. A new phenomenon was predicted  \cite{Levitasetal-PRL-17}: unique homogeneous and hysteresis-free first-order phase transformations for which each intermediate crystal lattice along the transformation path is in indifferent thermodynamic equilibrium and can be arrested.
Elastic energy for Si I was analytically presented in terms of the fifth-degree polynomial of the Lagrangian strain in \cite{Chen-etal-arxiv-20}, for finite strains including lattice instability points.

 Because of different effects of the stress tensor on the PT conditions for the direct and reverse PTs, these atomistic results led to essential generalization of the PFA  in \cite{Levitas-IJP-18,Babaei-Levitas-IJP-18}.  This advanced theory
was applied to the nucleation and growth of Si II at a single dislocation in \cite{Babaei-Levitas-AM-19}, and the importance of the generalized PT criterion is demonstrated.

Generally, large-scale MD simulations are broadly used to study interaction of PTs and plasticity, especially in a shock wave, see examples for $\alpha-\vep$ PT in iron in
\cite{Kadau-etall-Science-02,Kadau-etall-PRB-05,Gunkelmann-etall-PRB-12,Gunkelmann-etall-PRB-14,Gunkelmann-etall-NJP-14,Wang-etall-IJP-15}.


\subsection{Nucleation and evolution of high-pressure phase at dislocation pileups}\label{pileup}

\begin{figure} [bp!] \centering
\includegraphics[width=0.5\textwidth]{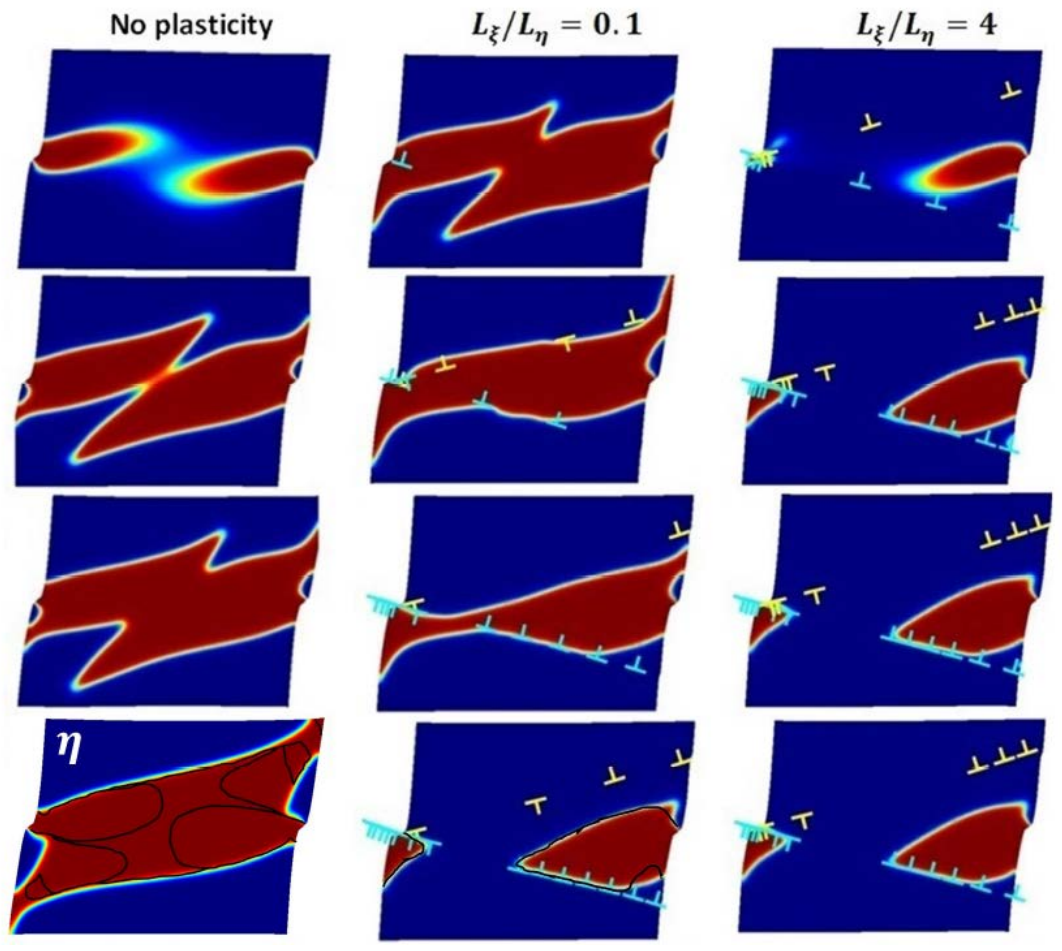}
\caption{Evolution of the $HPP$ (red) and dislocations obtained with  the PFA for coupled dislocations and PT \cite{levitas-javanbakht-APL-13,Levitas-Javanbakht-15-1,Javanbakht-Levitas-JMPS-15-2} within the right grain of a bicrystal under shear strain $\gamma$ at fixed vertical compressive stress.
The cases without dislocations in the right grain (left column) and with dislocations in the right grain (right column) are shown.
The last row,  corresponding to the  stationary structure for $\gamma=0.2$  includes  the  contour lines of the equilibrium transformation work $\fsg \fg{:}  \fvep_t (\eta)=\Delta \psi^\theta=1$ GPa. For the most of  the stationary  interfaces, this phase equilibrium condition is satisfied. Reproduced with permission from  \cite{Javanbakht-Levitas-JMS-18}.}
\label{Fig-PFA-pressure-shear}
\end{figure}

As the main nanoscale mechanism of drastic reduction in PT pressure in experiment due to plastic deformations, a
strong concentrator of the all components of the stress tensor at the tip of dislocation pileup was suggested in \cite{Levitas-PRB-04,Levitas-chapter-04} and treated analytically.
Much more precise results were obtained in
\cite{levitas-javanbakht-Nanoscale-14,Javanbakht-Levitas-JMPS-15-2,Javanbakht-Levitas-PRB-16,Javanbakht-Levitas-JMS-18}
using PFA approach to interaction between dislocations and PTs presented in Section \ref{PT-disl}
and \cite{Levitas-Javanbakht-15-1}. Example of the solution for a bicrystal  under compression and shear is presented in Fig. \ref{Fig-PFA-pressure-shear}. Dislocation  pileup located in the left grain (not shown)  creates a strong concentrator of all components of the stress tensor $\fsg$ in the right grain, which exceeds the crystal lattice instability limit and leads to a barrierless nucleation of a HPP. It is also important  that the deviatoric stresses in  the nanoscale defect-free regions are limited by the theoretical strength rather than the macroscopic yield strength and may be one to two orders of magnitude greater.
All this  increases  the local  driving force for PT, enabling drastic reduction of the applied pressure to initiate and run the PT. For the chosen parameters, the applied pressure is 3-20 times smaller than the PT pressure under hydrostatic conditions in the presence of a single dislocation and 2-12.5 times lower than the phase equilibrium pressure.  The unique highly deviatoric stresses that cannot be achieved for the macroscopic sample  may lead to new phases and phenomena.
 The  stationary geometry of the HPP is determined by the thermodynamic equilibrium
\bey
\fsg : \fvep_t=\Delta \psi^\theta
\label{PT-eq}
\eey
 either at point of the interfaces  (Fig. \ref{Fig-PFA-pressure-shear}) or in terms of the stress tensor  averaged over the  entire grain or polycrystal.
In the local approach, stresses due to dislocations are included in $\fsg$; if separated from the external stresses,
they will reproduce an athermal threshold for interface motion $K_\Sigma$ due to interaction with dislocations
(see also Section \ref{PFA-semicoherent}).

For larger scale, a scale-free PFA for the
coupled evolution of multivariant martensitic microstructure and discrete dislocation bands  was presented in \cite{Levitas-etal-PRL-18,Esfahani-Levitas-AM-20}. It includes a scale-free PFA for martensitic PTs
\cite{Levitas-Id-Pres-PRL-04,Idesman2005,Esfahani-Gham-Levitas-IJSS-18}. Dislocation pileups or shear bands
are introduced by the contact problem formulation, considering continuous sliding displacements (dislocations)
along the discrete slip systems. This allows one to produce stress concentrators required for  nucleation of the HPP, similar to the nanoscale approach described above. Scale-free model, while is much simpler than the nanoscale model, reproduces well all results obtained for nanoscale model for a bicrystal.
Also, this  model is applied in \cite{Levitas-etal-PRL-18,Esfahani-Levitas-AM-20} for FEM simulations of the  strain-induced PTs in a polycrystal sample under compression and shear (Fig. \ref{fig:Microscale}).
The phase equilibrium condition (\ref{PT-eq})
in  terms of  local transformation work at the interface
 and  transformation work averaged over the  entire polycrystal was confirmed.
 Also, similar to the nanoscale model, we obtained for the scale-free model that  stresses averaged over all martensitic and austenitic regions, as well as for the entire polycrystal are equal:
\bey
<\fsg>_M=<\fsg>_A=<\fsg>.
\label{stress-phases}
\eey
Note that result similar to  Eq. (\ref{stress-phases}) was obtained in \cite{leblond+devaux+devaux-1989} for traditional  continuum plasticity and sharp interface rather than for the PFA and  localized plasticity.
Also, evolution of the  volume fraction of HPP in each grain and in a sample, and averaged over the sample the volume fraction of HPP, each martensitic variant,  pressure, and shear stress versus  shear strain are determined in \cite{Levitas-etal-PRL-18,Esfahani-Levitas-AM-20}. All the obtained info is planned to be applied for   derivation of more precise strain-controlled kinetic equation than  Eq. (\ref{l-g-4}), which is currently used.

\begin{figure}[ht]
\vspace{-2mm}
\centering
\includegraphics[width=0.7\linewidth]{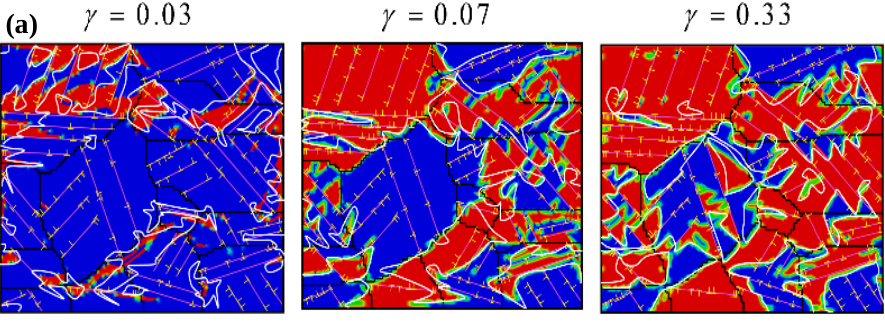}
\caption{ Evolution of high-pressure phase (red) and dislocation structure in a polycrystalline sample under shear at constant compressive stress obtained with  the scale-free PFA.
Reproduced with permission from  \cite{Levitas-etal-PRL-18}.}
\label{fig:Microscale}
\vspace{-2mm}
\end{figure}

MD simulations in \cite{Chen-etal-AM-19} on amorphization of Si I at the tip of the $60^o$ shuffle dislocation pileup against three different grain  boundaries under shear confirms main the analytical \cite{Levitas-PRB-04,Levitas-chapter-04} and PFA
\cite{levitas-javanbakht-Nanoscale-14,Javanbakht-Levitas-JMPS-15-2,Javanbakht-Levitas-PRB-16,Javanbakht-Levitas-JMS-18,Levitas-etal-PRL-18,Esfahani-Levitas-AM-20}
results. Thus resolved shear stress for initiation of amorphization reduces from $\sim 9$ GPa for perfect crystal down to $\sim 1.5$ GPa for 8 dislocations in a pileup. At the same time, screw shuffle dislocations in Si I
pass through all three grain boundaries ($\Sigma3$, $\Sigma9$, and $\Sigma19$) and cannot pileup, and consequently affect a PT \cite{Chen-etal-CMM-19}. Atomistic mechanism of strain-induced nucleation of amorphous phase, different for different grain boundaries is elucidated.  While $\Sigma3$ grain boundary
amorphous band propagates
 along the $(112)$ plane directly, for $\Sigma9$, and $\Sigma19$ boundaries the stacking faults and an intermediate phase precede the formation of an amorphous band along the $(110)$ and $(111)$ planes, respectively.

MD simulations of the interactions of plasticity and $\alpha-\epsilon$ PT
under shock in iron bicrystals in \cite{Zhang-etal-JAP-19} also confirms the reduction in PT pressure due to the effect of the grain boundary and dislocations induced by grain boundary and pre-existing dislocations. This work also gives important atomistic detail of the promotion of the PT and distinguishes between strain- and stress-induced PTs.

At the same time, in the MD studies of shock loading of a polycrystalline iron sample
\cite{Kadau-etall-Science-02,Gunkelmann-PRB-12,Gunkelmann-PRB-14,Wang-etal-IJP-15} dislocation activity is suppressed by small grain size ($<10$ nm), or PT precedes plastic flow, and nucleation of the high-pressure phase is promoted by grain boundary and triple junctions.

The simultaneous occurrence of dislocation bursts \cite{Meyers-etal-Si-SciRep-16} and
amorphization in SI has been found in shock experiments
\cite{Meyers-etal-Si-EurPhysJ-16,Meyers-etal-Si-ActaMat-16,Meyers-etal-ExteMechL-15}. In these works, both MD simulations and  transmission electron microscopy
offer that the stacking faults along
$\{111\}$ planes and their intersections serve as nucleation sites for
amorphization. For larger shock intensity, the
amorphous band broadens and deviates from the $\{111\}$ plane toward
the  maximum-shear-stress plane. Shock-induced amorphization in SiC
was studied in a similar way in
\cite{Meyers-etal-SiC-ActaMat-18}.  Amorphization in $SmCo_5$ without dislocations under uniaxial loading was studied with MD and experimentally  in \cite{Szlufarska-etal-VM-amorph-band-20}.

{
Inheritance of dislocations by martensite in shape memory alloys was considered in experiments and simple models in \cite{likhachev,lovey99,lovey04}.
  There is also huge literature on  PT under indentation, and compression of nanospehere and nanopilar, which we will not consider here.  }

 \subsection{ Macroscale theory and FEM modeling of strain-induced transformations}\label{macroscale}

\begin{figure}[ht]
\vspace*{-5mm}
\centering
\includegraphics[width=0.5\linewidth]{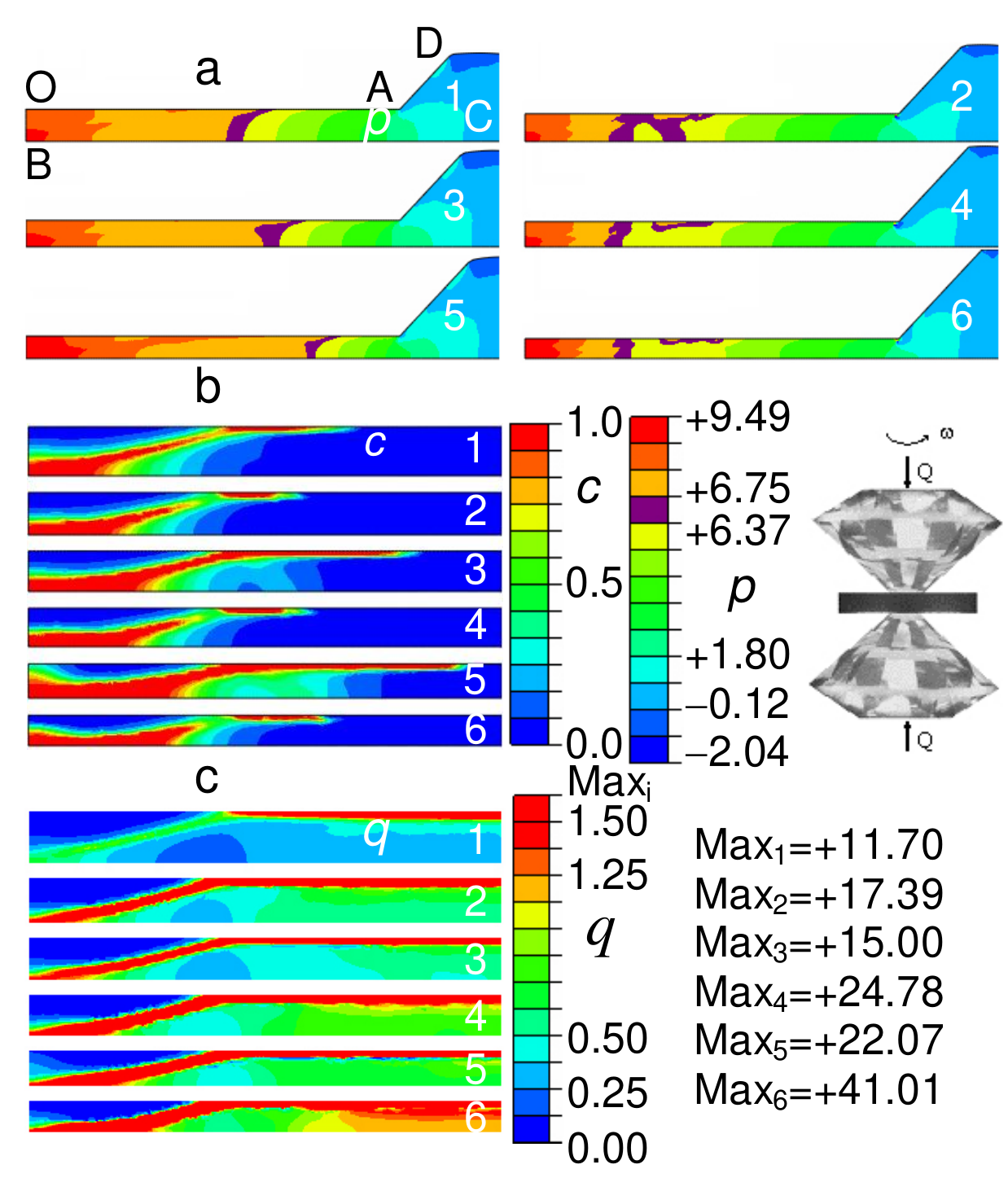}
\caption{Comparison of distributions of pressure $p$ (a),
volume fraction of the HPP $c$ (b), and accumulated plastic strain $q$ (c) for compression (1, 3, 5) and torsion (2, 4, 6) for the case when the yield strength of the HPP is five times smaller than that for the low-pressure phase \cite{pieces9,pieces10}. Due to transformation softening, localized deformation-transformation bands are observed. }
\label{fig:pd-plasticstrain}
\vspace{-3mm}
\end{figure}

The main goal of a macroscopic modeling and simulations is to
determine  the evolution of the distributions of the components of the stress tensor,  accumulated plastic strain, 
and the volume fraction of the HPP \cite{pieces9,pieces10,Feng-Levitas-IJP-BN-DAC-17,Feng-Levitas-IJP-BN-RDAC-19}
 in a sample compressed in DAC and twisted in RDAC, see example in Fig.  \ref{fig:pd-plasticstrain}.
The first results were obtained  analytically in \cite{Levitas-PRB-04,Levitas-chapter-04} with many simplifications, but still some important conclusions were  made. The first FEM results in \cite{pieces9,pieces10} were much more comprehensive, but still the material and contact friction models were simplified.  The most  advanced model for coupled large elastoplasticity and strain-induced PT under megabar pressure and
corresponding  FEM algorithms were developed and applied to study PT in BN under compression
 in DAC \cite{Feng-Levitas-IJP-BN-DAC-17} and torsion in
 RDAC \cite{Feng-Levitas-IJP-BN-RDAC-19} within rhenium gasket.
 Finite-strain kinematics for a polycrystal included multiplicative decomposition of the deformation gradient into elastic and inelastic parts, and then inelastic deformation rate was additively decomposed into plastic and transformational parts, different from the multiplicative decomposition   (\ref{n1ag-2}) for a single crystal. Murnaghan  elasticity rule and  pressure-dependent Prager-Drucker yield condition for each phase along with the simplest mixture rule were utilized.
Kinetic Eq. (\ref{l-g-4}) for strain-induced PT was implemented as well. Third- and fourth-degree anisotropic elastic energy were utilized for diamond \cite{Feng-Levitas-Hemley-IJP-16,Levitas-etal-NPJ-CM-19}. Combined Coulomb and plastic friction between sample, diamond, and gasket were considered. Without PTs, elastoplastic models were calibrated and verified by independent experiments for rhenium \cite{Feng-Levitas-Hemley-IJP-16,Feng-Levitas-IJP-17,Feng-Levitas-PRA-18} and  tungsten \cite{Levitas-etal-NPJ-CM-19} up to 300 GPa and 400 GPa, respectively.
 Various experimental phenomena have been reproduced and explained, some have been predicted, methods of controlling PTs were suggested, and some possible misinterpretations of experimental results have been demonstrated, see \cite{Levitas-chapter-04,Levitas-MT-19}.

\section{Scale transitions and phenomenological theories for interaction between phase transformation and plasticity}\label{phenomenology}

\par
An averaged description of phase transformations in terms of the volume fraction of phases is
presented in \cite{levitas-1990,levitas-kniga-1992,raniecki+bruhns-1991,bhattacharyya+weng-1994,cherkaoui+berveiller+sabar-1998,cherkaoui+berveiller-aam2000} at small
strain and in  \cite{levitas-kniga-1992,petryk-98} at large strain.
An averaging procedure for PT with semicoherent interfaces at large strains is developed in
\cite{levitas-iutam96}. A computational studies of martensite formation and averaging, including multicale approaches,
can be found in \cite{leblond+devaux+devaux-1989,ganghoffer+etal-2001,marketz+fischer-1994,marketz+fischer-cms1994,marketz+fischer-1995,simonsson-1995,levitas+idesman+stein-ijss1998,Turteltaub-Suiker-05,Turteltaub-Suiker-06,Kouznetsova-Geers-08,Iwamoto-04}.
Strain-induced PTs are described in \cite{olson+cohen-75,63a,diani+parks-1998}.
Significant  progress in the study of TRIP is presented in \cite{fischer+sun+tanaka-1996,fischer+reisner+werner+etal-2000,cherkaoui+berveiller+sabar-1998,cherkaoui+berveiller-sms2000,olson-1996}.

More detailed studies, e.g.,  \cite{Hecker-etal-82,Iwamoto-etal-98,Lebedev-Kosarchuk-00,Beese-Mohr-11,Kim-etal-15,Mansourinejad-Ketabchi-17}
demonstrate dependence of PT kinetics on strain/stress mode and path, texture, temperature, and strain rate; some of them include constitutive modeling.
Simultaneous occurrence of stress-induced and plastic strain-induced PTs was modeled in \cite{Ma-Hartmaier-15,Knezevicetal-IJP-20}. In the last paper, these two types of PTs are distinguished based on the transformation mechanism: for stress-induced PT $\gamma $-austenite transforms directly to $ \alpha '$-martensite, but
for  strain-induced PT these transformation occurs through appearance of bands of $\vep$-martensite, and $\alpha'$-martensite nucleates at intersection of these bands (alternatively, martensite may nucleate at intersection of twins \cite{Das-etal-16}).  An advanced version of elastoplastic self-consistent approach to the behavior of polycrystalline aggregate, initially developed in \cite{Turner-Tome-94} was implemented.

Simulations of the interaction between PTs and plasticity, utilizing macroscscale constitutive equations  can be found in \cite{pieces9,pieces10,Sitko-Skoczen-12,Mahkenetal-09,Mahnkenetal-12,Budnitzki-Kuna-JMPS-16,Sallami-etal-19}.
Models and simulations
 for single crystals and dynamic loading can be found in \cite{Barton-etal-05,Fengetal-JMPS-18,Fengetal-IJP-20}.
Some theories and FEM approaches to strain-induced PTs coupled to large elastoplasticity under high pressure were analyzed in Section \ref{macroscale}.

\section{Fracture and interaction between fracture and phase transformation in inelastic materials}\label{Sec-fracture}

\par
A general theory summarized in Boxes 1,  2, and 6-8 is applicable to fracture,
including  crack and void nucleation and growth   \cite{Levitas-fracture-MRC-98,levitas-ijp2000-2,idesman-levitas+stein-ijp-2000}.
{\it Fracture is defined as a thermomechanical process of change in some region
of tensile and shear
elastic moduli and
yield strength from their initial values to zero.} This process cannot be
arrested at a material point in the intermediate state. After complete fracture,
the tensile stresses in the fracture region are zero. Formally, fracture here is considered as a particular case of a PT without transformation strain and a specific properties of the product phase (vacuum), which was coined in \cite{Levitas-fracture-MRC-98,levitas-ijp2000-2} a generalized second-order PT.
With such a definition, local driving force   Eq. (\ref{ijss1-180}) for an isothermal fracture,   neglected internal variables, and independent of $\fvep_p$ free energy, reduces to
\bey
\rho X & = &  \,
 - \, 0.5
\int\limits^{\scriptscriptstyle {\fg 0}}_{\scriptscriptstyle {\fg E}_1}
\fvep_e \, {\fg :} \, d \, {\fg E} \, {\fg :} \, \fvep_e  \, - \, \rho \; \Delta \psi^{\theta}
=
\int\limits^{\fvep_{e2}}_{\fvep_{e1}} \fsg \, {\fg :} \, d \,
\fvep_e \, + \, 0.5 \, \fvep_{e \, 1} \, {\fg :} \, {\fg E}_1
\, {\fg :} \, \fvep_{e \, 1} \, - \, \rho \; \Delta \psi^{\theta} \; .
\label{ijss1-20b}
\eey
It is not clear where to get $\Delta \psi^{\theta}$. In applications, it can be included in athermal friction or just neglected.
But we will keep it for generality.
\par
Similar to PTs,  the developed approach is valid for an arbitrary
inelastic material, because the  constitutive equations were not
used in the process of derivation. To illustrate the method,  we  consider analytical solutions
to two simple problems.

\subsection{ Crack propagation in elastoplastic material}

\begin{figure}[htp]
\centering
\includegraphics[width=0.45\textwidth]{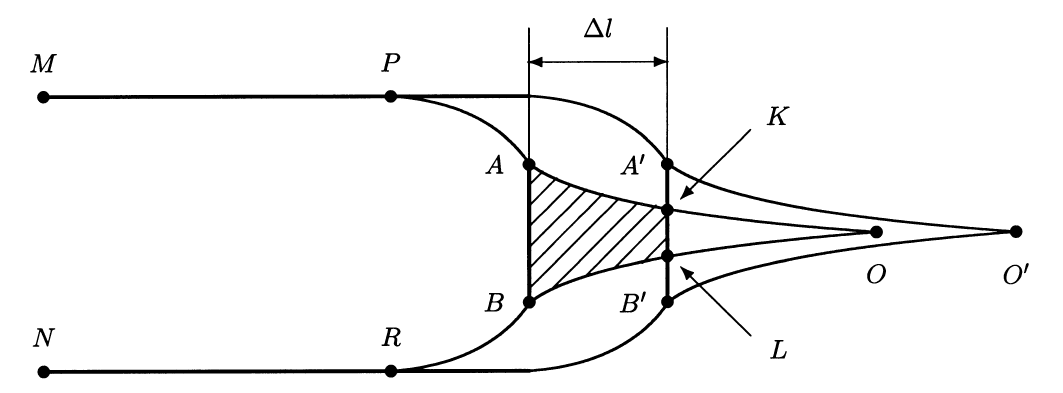}
\caption{ The positions of a crack $\, MABN \,$ before and
$\, MA'B'N \,$ after crack advance by distance $\, \Delta l \,$;
$\, AOB \,$ and $\, A'O'B' \,$ are the localized plastic regions
before and after crack advance by distance $\, \Delta l \,$;
$\, AKLB \,$ is the fracture zone at the beginning of fracture process.
Reproduced with permission from \cite{levitas-ijp2000-2}.
\label{crack}}
\end{figure}

Equations for crack propagation in an
elastic-perfectly-plastic material were derived in \cite{Levitas-fracture-MRC-98,levitas-ijp2000-2} in a framework similar to the
\cite{Dugdale-60} framework  for the plane stress.
The localized plastic region $\, OAB \,$ (Fig. \ref{crack}) ahead of the crack
is subjected to  cohesive  stresses
$\, \sigma \, = \, \sigma_y \,$ on the extended crack surfaces
$\, OA \,$ and $\, OB \,$; $\, AKLB \,$ is the fracture zone.
The transforming volume before fracture
$\, V_1 \, = \, b \; \Delta \, l \; \bar{\delta} \,$, where
$\, \bar{\delta} \,$ is some intermediate value of the crack opening
at the length $\, \Delta \, l \,$ and $\, b \,$ is the thickness of
the sample.

Neglecting all
the stresses and elastic strains $\, \vep_e \,$ but normal to the crack plane only and  assuming
that during the fracture process
\bey
E \left(\xi \right) \, = \, E \left(1 - \xi \right)
\quad {\rm and} \quad
\sigma_y \left(\xi \right) \, = \, \sigma_y \left(1 - \xi \right) \, ,
\quad {\rm i.e.} \quad
\vep_e \, = \, \frac{\sigma_y \left(\xi \right)}{E \left(\xi \right)}
\; = \; \frac{\sigma_y}{E_1} \; = \; {\sl const},
\label{esh-16+}
\eey
we obtain that  the work integral in Eq.(\ref{ijss1-20b})
is zero. Evaluating elastic energy, we obtain
\bey
\rho \, X \, = \, 0.5 \; \frac{\sigma_y^2}{E_1} \; - \, \rho \;
\Delta \, \psi^{\theta},
\label{esh2-16+89}
\eey
which  is uniform in the
fracture region and is independent of its geometry; thus,
we have to use equations from Boxes 6-8.
The thermodynamic fracture criterion  and the
principle of the minimum of transforming mass can be expressed as
\bey
\left(  0.5 \; \frac{\sigma_y^2}{E_1} \; - \; \rho \; \Delta \,
\psi^{\theta} \, - \, \rho \, K^0 \, \right) \, b \; \Delta \, l
\; \bar{\delta} \, - \, 2 \, \Gamma \; \Delta \, l \; b \, = \,
0 \; ; \qquad \Delta \, l \; \bar{\delta}
\;\; \rightarrow \;\; {\sl min} \; .
\label{esh-18+}
\eey
The minimum value of $\, \bar{\delta} \,$ is determined from the
 fracture criterion (\ref{esh-18+}).
Condition $\, \Delta \, l \; \rightarrow \; {\sl min} \,$ leads to
$\, \Delta \, l \, = \, a \,$, where $\, a \,$ is the interatomic
distance in the direction of crack propagation. Then $\, \bar{\delta}
\, = \, \delta \,$ is the crack opening
displacement between points $\, A \,$ and $\, B \,$ and the thermodynamic
fracture criterion (\ref{esh-18+}) can be transformed to the criterion of critical
crack opening
\bey
\delta \; = \; \delta_c \; := \; \frac{2 \, \Gamma}{0.5 \;
\displaystyle{ \frac{\sigma_y^2}{E_1}} \, - \, \rho \; \Delta \,
\psi^{\theta} \, - \, \rho \, K^0} \; .
\label{esh-19+}
\eey
In contrast to previous publications, the critical crack
opening is defined in terms of known material parameters.

For  the Dugdale model  $\, \displaystyle{
\delta \; = \; \frac{J}{\sigma_y} } \,$ \cite{Rice-Chapter-68},
where $\, J \,$ is the path independent $\, J \,$-integral for the pathes that do not cross  the plastic region. Then criterion
(\ref{esh-19+}) can be expressed in term of the critical value of
$\, J \,$-integral
\bey
J \; = \; J_c \; := \; \sigma_y \, \delta_c \, = \, \sigma_y \;
\frac{2 \, \Gamma}{0.5 \, \displaystyle{\frac{\sigma_y^2}{E_1}} \, - \,
\rho \; \Delta \, \psi^{\theta} \, - \, \rho \, K^0 } \; .
\label{esh-20+}
\eey
 Eq.  (\ref{meyers1-126a-c})  for transformation time can be expressed as a kinetic equation
for a crack growth:
\bey
\dot{l} \, = \; \frac{a}{t_s} \; = \;  \frac{ a }{t_0}
 \; {\sl exp} \; \left( \,
- \; \frac{\rho \, E_a}{R \, \theta} \; \frac{N}{n} \;
b \, a \, \delta_c \right) \; .
\label{mrc98-1-4c}
\eey
Eq.(\ref{mrc98-1-4c})  can be easily transformed to the traditional
Arrhenius equation
formulated in  \cite{Zhurkov-65} kinetic concept of strength based on
experimental regularities, see also  \cite{Cherepanov-97}. In
our solution the crack propagates atom by atom, i.e., almost continuously, in contrast to the discrete finite crack advance
in the model by \cite{Kfouri-Rice-77}. The characteristic size in our
model is  the thickness  $\, \delta \,$.

\begin{figure}[htp]
\centering
\includegraphics[width=\textwidth]{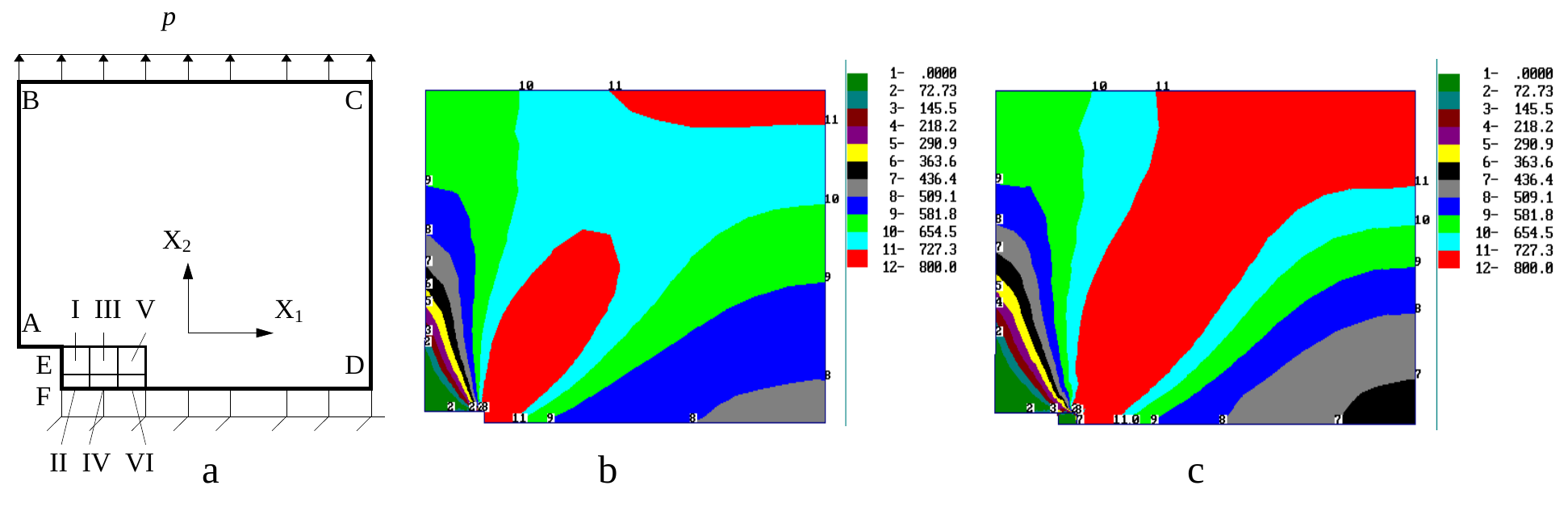}    
\caption{(a) The cross-section of an elastoplastic  sample with notch under tension with six possible fracture regions; (b) and (c) distribution of equivalent von Mises stress before fracture and after  fracture in regions I-VI, respectively.  Reproduced with permission from
 \cite{idesman-levitas+stein-ijp-2000}.
\label{fig-fracture}}
\end{figure}

\par
FEM  study of ductile fracture in a sample with edge
notch under tension (Fig. \ref{fig-fracture} a) based on the same theory is given in \cite{idesman-levitas+stein-ijp-2000}.
Six regions in a small rectangular near the edge
notch was considered in different combinations as a possible fracture zone. The distribution of von Mises stress before fracture and after  fracture in regions I-VI is shown in   Figs. \ref{fig-fracture} b and c, respectively.
After solving multiple elastoplastic problems with fracture in different regions (modeled as incremental reduction of elastic moduli down to zero), the mechanical part of the driving force $\varphi= \rho (X-\Delta \, \psi^{\theta}) $
was evaluated, and various scenarios were compared (Table \ref{TABLE1}).

For example, for fracture in  narrow  regions II, IV and VI, $\varphi$ is larger and its growth is faster during
 the fracture propagation than for thick regions I, III and V.
 However, the resistance due to  the surface energy for a thin layer is larger than for a thick layer.
 When fracture occurs simultaneously in  regions II + IV or II + IV + VI rather
 than first in a small region II and then spreads through regions IV and VI, the
 $\varphi$ is larger,
 i.e. it is thermodynamically more favorable scenario.
 However, the damaging volume  $V_n$ grows by factors 2 and 3,
 respectively. Thus, the subsequent growth is more favorable kinetically,
 if  the thermodynamic criterion is met.

 At a relatively small surface energy, a void nucleation in  region
VI
 occurs instead of  crack propagation. The mechanical driving force is
significantly larger  than for any damage scenarios.
After  pore nucleation fracture propagates  from
pore to notch.
This corresponds to a  well-accepted experimental result that the void nucleation and bridging
ahead of the crack tip is an actual physical mechanism of ductile fracture.

 An analytical approximation of the computational results for $\varphi$ allowed us to analyze the application of macroscopic
 thermally activated kinetics from Boxes 5, 6, and 8.
 Below are the found typical cases in
the determining  two characteristic sizes of the fracture region:
(a) from the principle of minimum of fracture time  without any
constraints; (b) from the thermodynamic fracture criterion; (c) as an interatomic
distance, and (d) as the sample size. In some cases crack advance is
finite; however, different from  \cite{Kfouri-Rice-77},
the advance is  determined from the
extremum principle or the thermodynamic fracture criterion rather than being  a chosen material parameter.
\begin{table}[htp]
\centering
\caption{Mechanical driving force $\varphi= \rho (X-\Delta \, \psi^{\theta}) $  for different fracture scenarios; $V_n$ is the volume of fracture zone.  Red color designates the regions where fracture has already occurred; dashed regions show the fracture zone for the current simulation step. Reproduced with permission from
 \cite{idesman-levitas+stein-ijp-2000}.}
 \includegraphics[width=0.6\textwidth]{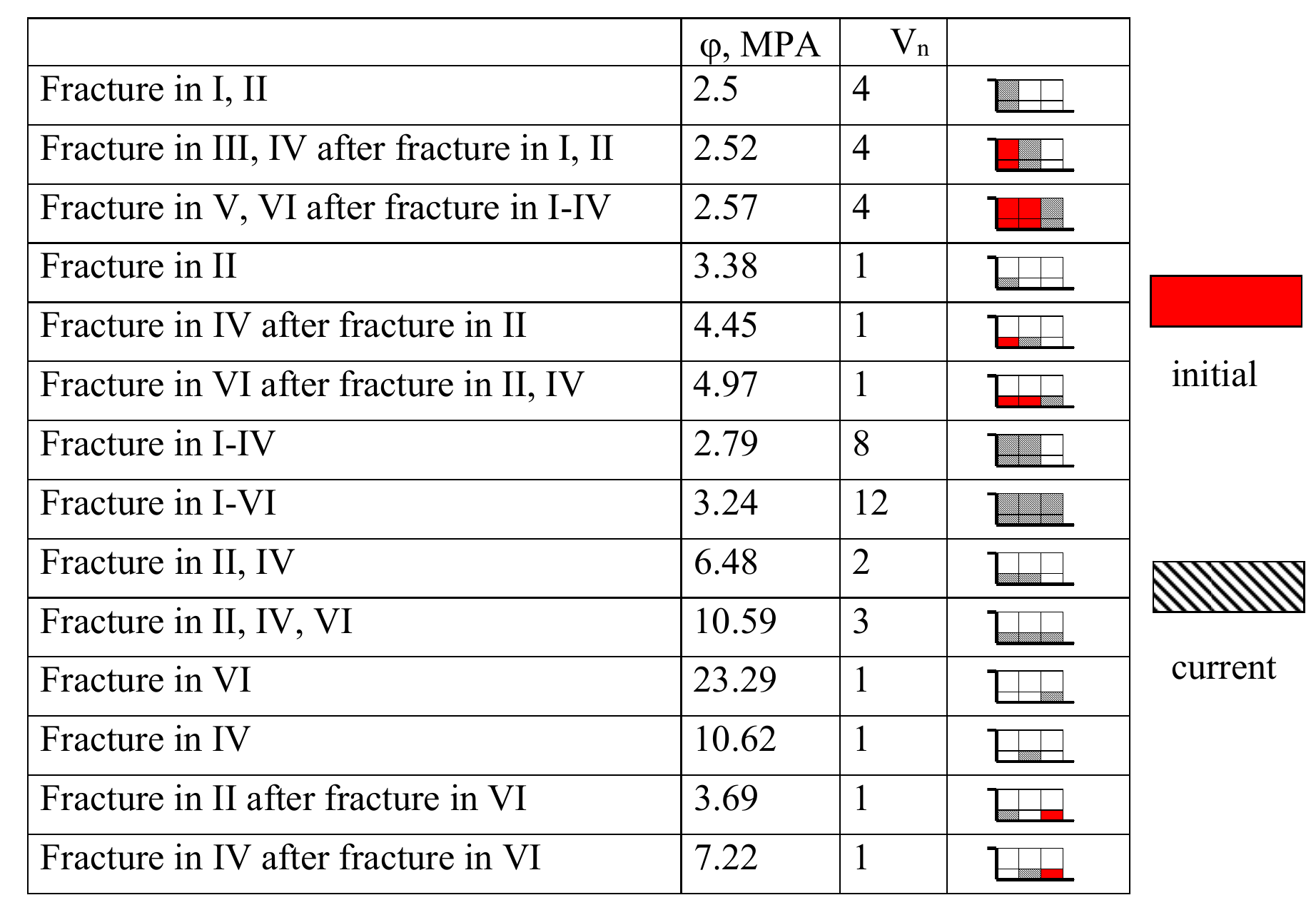}    
\label{TABLE1}
\end{table}

\subsection{Interaction between fracture and phase transitions}\label{PT-fracture}

Relatively large transformation strain induces high stresses, that may relax by  crack and void nucleation and growth instead of plasticity,
see examples in \cite{Hornbogen-91}. At the same time, a large stress concentrator
at the crack tip may induce PT \cite{Stump-Budiansky-89,Bulbich-92}, which may increase material toughness; this is the so-called
transformation-toughening phenomenon.

To illustrate some types of interaction between fracture and PT in elastoplastic material,
several model problems were  solved in \cite{idesman-levitas+stein-ijp-2000} for sample and loading shown in Fig. \ref{fig-fracture}a.
Obtained results are presented  in Table \ref{TABLE2}. For fracture in region II,   $\,\varphi \, = \, 3.38 MPa \,$ before PT and $\, \varphi= 1.35 MPa \,$   after PT  in  region I + II, due to stress relaxation caused by dilatational transformation strain. Fracture in region
II + IV decreases the mechanical driving force for PT in region I down to $\, - 8.06$
{\sl MPa} due to stress release and increases $\, \varphi$
for PT in regions V and   VI by moving stress concentrator closer.

Next, consider competition between PT and fracture
for athermal (time-independent) kinetics, see Box 4. We compare four  processes: PT
region I + II or in I--IV, and fracture in  region II or in region II + IV.
Let each of these processes be thermodynamically admissible, i.e.
the thermodynamic SC criterion   (\ref{ijss1-19a}) is met for the selected values of the surface
energy and $K^0$. One has to select which thermomechanical process will
take place in reality, i.e., to choose the single solution among all possible ones.
This is situation described in Section \ref{global},  i.e., the best unique solution is the stable one.

According to the extremum principle (\ref{eq:yyyyy-u}), which is the global SC criterion   for the determination of
the stable deformation process, for the prescribed  normal stress the
larger the normal displacement averaged over line BC is the more stable is the
SC process. Based on Table \ref{TABLE3}, fracture in  region II + IV should occur as the
most stable SC process. If for fracture in  region II + IV the thermodynamic
fracture criterion  (\ref{ijss1-19a}) is not met, then PT in region  I--IV should take place.
For the case when PT in  region I + II and fracture in region II are the only thermodynamically
possible, the fracture will occur. Consequently, similar to PTs (see Section \ref{Shear-band-int}),  in addition to the
thermodynamic fracture and PT criteria (\ref{ijss1-19a}), the global fracture and PT criteria (\ref{eq:yyyyy}) and (\ref{eq:yyyyy-u})
based on stability analysis should be applied for some cases.

 For time-dependent kinetics we do not need additional global fracture
 and PT criteria for this case. Indeed,
 the principle of minimum of transformation time allows us to choose
 which process -- PT or fracture -- will occur in the shortest time.
 However, in the case of time-dependent kinetics of PT or fracture and
 time independent plasticity we may need the global criterion of SC again, because plastic flow without PT and fracture
 may be the most stable process.

\begin{table}[htp]
\centering
\caption{Mechanical driving force $\varphi= \rho (X-\Delta \, \psi^{\theta}) $  for different scenarios of fracture and phase transformation.  Red color designates the PT regions; dashed regions show the fracture zones. Reproduced with permission from
 \cite{idesman-levitas+stein-ijp-2000}.}
 \includegraphics[width=0.6\textwidth]{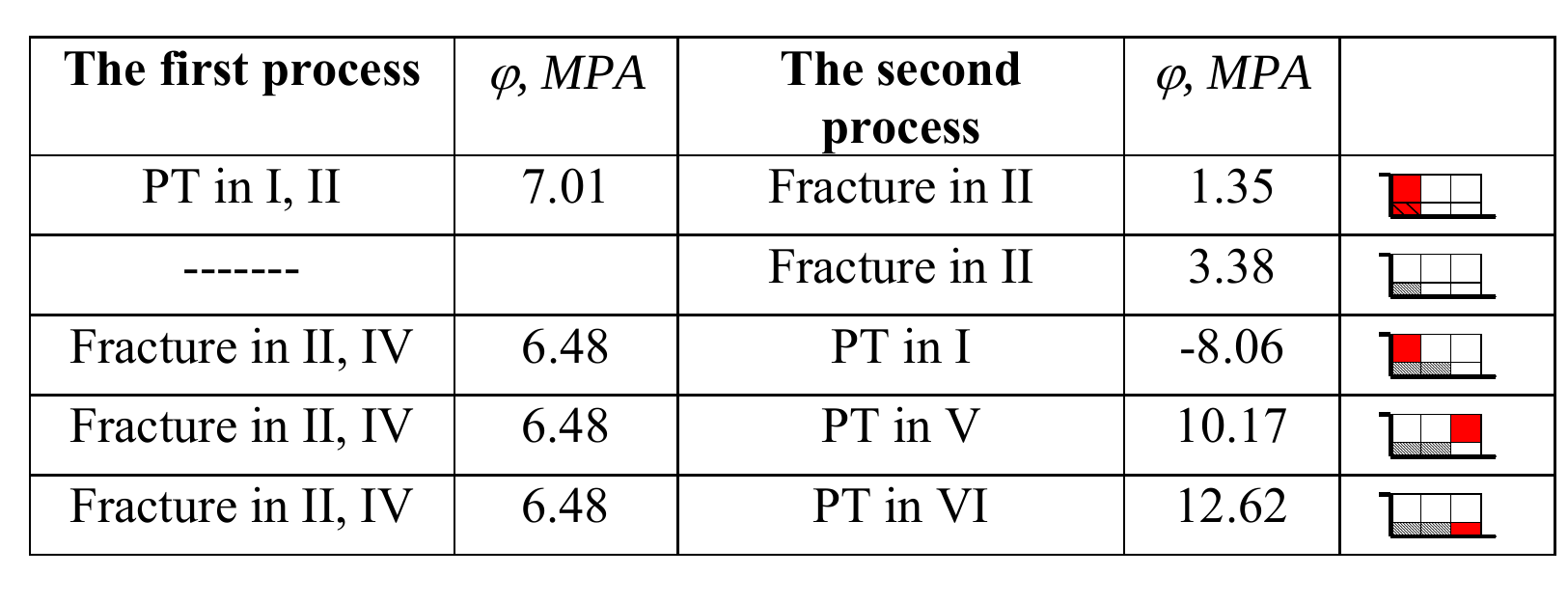}    
\label{TABLE2}
\end{table}

\begin{table}[htp]
\centering
\caption{Normal displacement, averaged over line BC, and mechanical driving force $\varphi $ for various processes.  Red color is for the PT regions; dashed regions designate the fracture zones. Reproduced with permission from
 \cite{idesman-levitas+stein-ijp-2000}.}
 \includegraphics[width=0.6\textwidth]{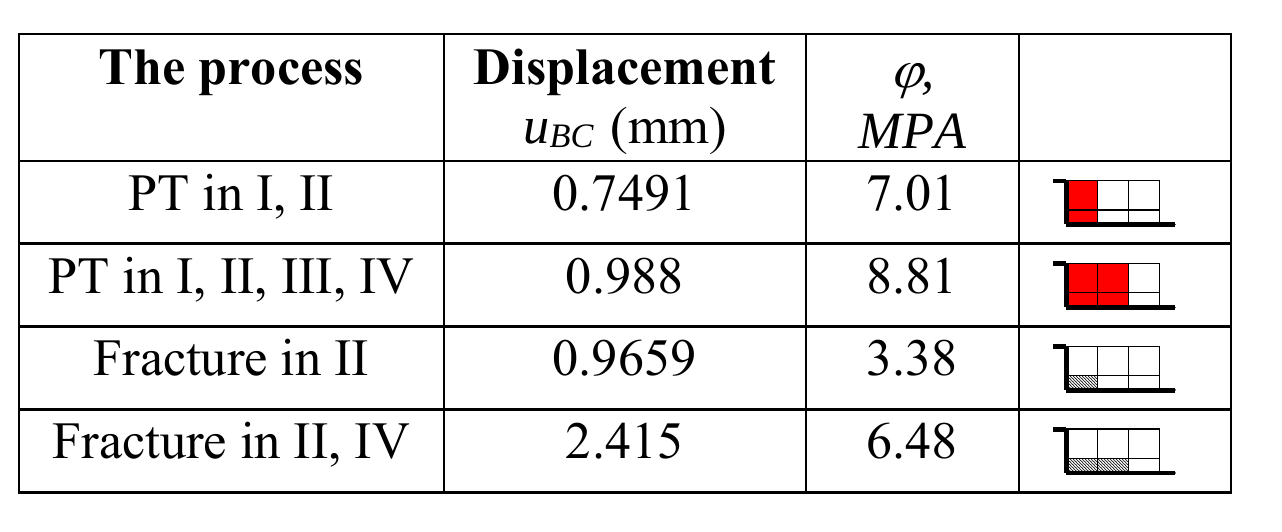}    
\label{TABLE3}
\end{table}

\par
One of the remaining problems is the mass balance for fractured material.
In the energy methods of fracture mechanics the fractured region
is  removed. From physical point of view, fracture means that atoms are separated
by the distance when their interaction is negligible. For such a formulation,
the volume of the fracture region increases during the fracture
process from zero to some  value determined by a solution of a mechanical problem. For modes II and III, crack
surfaces can be in contact.
\par
Other problem is related to definition of fracture: which components
of the elasticity tensor $\, {\fg E} \,$ tend to zero and in which
sequences? The stress state and thermodynamic driving force may depend on the chosen process.

\subsection{Void nucleation in infinite elastoplastic sphere}\label{void-sph}

Spherical
void nucleation in
an infinite elastic-perfectly-plastic sphere
under a homogeneously distributed
tensile stress $\, p \,$ was considered in \cite{levitas-ijp2000-2} in small strain approximation.
Fracture  process was modelled by decreasing the
 bulk modulus $K$ in a small sphere with a radius $\, r \,$ from an
initial value $K_1$ to zero accompanied by jump $\, \Delta \psi^{\theta} \,$
under isothermal conditions. The stress state for
this problem was taken from  \cite{roitburd+temkin-1986}.
Surface stresses were neglected.  For an elastic material, the mean tensile stress
 inside the transforming zone is
\bey
\tilde{p} \, = \, p \; \frac{S_1 + q}{S + q} \; ,
\label{esh1-1+}
\eey
where
$\, \displaystyle{ S_1 \, = \, \frac{1}{3 \, K_1}} \,$ and
$\, \displaystyle{ S \, = \, \frac{1}{3 \, K}} \,$ are the elastic
bulk compliance of undamaged material and during the fracture,
$\, \displaystyle{q \, = \, \frac{1}{4 \, \mu_1}} \,$, $\, \mu_1 \,$
is the shear modulus of undamaged material. For a void after fracture
$\, K_2 \, = \, 0 \,$, $\, S_2 \, \rightarrow \, \infty \,$.
After integration one obtains
\bey
\rho \, X_e \, = \, 0.5 \, p^2 \left(
\frac{1}{3 \, K_1} \; + \; \frac{1}{4 \, \mu_1} \right)
\, - \, \rho \, \Delta \psi^{\theta} \, = \,
p^2 \; \frac{3}{4} \; \frac{\left(1 - \nu_o \right)}{E} \, -
\, \rho \, \Delta \psi^{\theta} \; ,
\label{esh1-3+}
\eey
where $\, E \,$ and $\, \nu_o \,$  are Young's modulus and  the
Poisson ratio  of the undamaged material.

For  void nucleation in an elastoplastic material, during
the increase in compliance $\, S \,$,  the
material initially deforms elastically. Plastic deformation
near the damaging region starts at $S=S_p$ determined from equation
\bey
\tilde{p}_p \; = \; p \; \frac{S_1 + q}{S_p + q} \; = \;
p \, - \; \frac{2}{3} \; \sigma_y \; ,
\label{esh1-5+}
\eey
According to  \cite{roitburd+temkin-1986}, in the elastoplastic regime variation of stress $\tilde{p}$ is determined by equation
\bey
\tilde{p} \, \left(S \, - \, S_1 \right) \; = \; \frac{2}{3} \;
\sigma_y \left(S_1 \, + \, q \right) \, {\sl exp} \; \left( \,
\frac{p \, - \, \tilde{p}}{2/3 \, \sigma_y} \; - 1 \right) \; .
\label{esh1-4+}
\eey
After manipulations, one obtains  for elastoplastic regime
\bey
\rho \, X_p \, =\; \frac{3 \left( 1 - \nu_o \right)}{4 \, E} \;
\left( \frac{2}{3} \; \sigma_y \right)^2 \left[ 2 \, {\sl exp}
\left(\frac{3 \, p}{2 \, \sigma_y} \; - 1 \right) - 1 \right] -
\rho \, \Delta \psi^{\theta} \; .
\label{mrc98-1-1}
\eey
Since $\, X \,$  is independent of
radius $\, r \,$, we can use the same equations and conclusions like for  PT in a
spherical nucleus in Section \ref{sphere}. For macroscale thermally-activated kinetics,
 the   principle of minimum mass results  in condition $\, r \, \rightarrow \,
{\sl min} \,$ and equation from Box 7 are applicable for the radius of the thermodynamically admissible void and the fracture
time. One can also apply equations in Box 4 to described thermally activated void nucleation, like for PT in Section \ref{sphere}.

For an elastic sphere, the thermodynamic criterion for void nucleation
was presented in \cite{Cherepanov-97}. It was obtained that the
stress required  for void nucleation   reduces with the
increasing void radius, i.e., nucleation of an infinite void needs the smallest
$\tilde{p}$. The same should be true for elastoplastic sphere. In contrast, our kinetic approach allows determination of the
explicit  void radius and nucleation time versus applied tensile stress.

Numerous papers are devoted to the void nucleation and growth due to diffusion of vacancies, Kirkendall effect,
and chemical reaction, see e.g., \cite{Fischer-Antretter-IJP-09,Levitas-Attariani-void-12} and references herein.

Thermodynamic driving force for a void growing in elastoplastic material

\subsection{Alternative approach to void nucleation}\label{void-altern}

Completely different definition of the void nucleation due to fracture is accepted and used in \cite{Levitas-Alt-void-ActaMat-11}: it is  a thermomechanical process of the growth of a cavity
  inside the solid from zero to some critical size due to atomic or molecular bond breaking.
Applying our thermodynamic approach to such a definition,  taking into that dissipation due to fracture only is independent of the volume under consideration,  and considering sample infinitesimally larger than  a void, it was obtained for the thermodynamic driving force
\bey
X_v =
\int\limits^{{\fg u}_2}_0 \int\limits_{\Sigma} {\fg p}_\Sigma {\fg \cdot}
d{\fg u}_\Sigma   d \Sigma
  - \int\limits^{u_{n2}}_0 \int\limits_{\Sigma} \frac{2\gamma}{ R}
d{u}_n   d \Sigma,
\label{11-void}
\end{eqnarray}
 where ${\fg p}_\Sigma$ is the traction vector
in the solid at the variable void surface $\Sigma$, $\fg u$ and $u_n$ is the displacement vector and its normal component at the void surface, and $1/R$ is the mean curvature of the void surface.
Thus, the thermodynamic driving force for the void nucleation (i.e., the dissipation due to fracture only) is localized at the void surface, is independent of plastic strain in solid, and represents the difference
 between work produced by external
traction acting on the void surface ${\fg p}_\Sigma$ and work produced by the Laplacian pressure $\frac{2\gamma}{ R}$.  For a spherical void of a radius $r$ under axisymmetric normal tensile stress $\sg_n$ in solid
 at the void surface,
 Eq. (\ref{11-void}) simplifies to
\bey
X_{v}  =
\int\limits^{r_c}_0 \int\limits_{\Sigma} \left(\sg_n-\frac{2\gamma}{ r}\right)
dr   d \Sigma =  \sg_n \frac{4}{3} \pi r^3_c - \gamma 4 \pi r^2_c.
\label{11frkj}
\end{eqnarray}
Here $r_c$ is the radius of the critical nucleus, which can be found from
minimization of $X_v$ with respect to $r_c$, or equivalently, from the  mechanical
equilibrium  equation $\sg_n= \frac{2\gamma}{r_c}$. Since for a subcritical nucleus $\sg_n< \frac{2\gamma}{r_c}$,
the thermodynamic driving force $X_v$  is negative, and  determines
 the energy necessary for void nucleation, which is supplied by the thermal fluctuations.
The large-strain solution to the perfectly plastic problem on expansion of a spherical cavity
 from zero size \cite{hill-1950,Levitas-Alt-PRL-08} is
\begin{eqnarray}
\sg_n = \sg - \sg_c; \quad
\sg_c:=
\frac{2}{3}\sg_y \left(1 + \ln\left(
\frac{2\mu \alpha}{3  \sg_y}  \right)\right).
\label{s7}
\end{eqnarray}
Here, $\alpha=\frac{1+\nu}{1-\nu}$ and $\sg_c$ is the cavitation pressure, i.e., the possible maximum value of tensile pressure that
solid can sustain for neglected surface energy.
 Utilizing for such a driving force a kinetic theory from Box 4, we obtain
 the  explicit relationship between tensile stress  for nanovoid nucleation
$\sg$ vs.  temperature:
\begin{eqnarray}
\sg_n=\sg - \sg_c   = \left(\frac{16 \pi \gamma^3}{3 \beta k \theta}\right)^{1/2}.
\label{s3oi}
\end{eqnarray}
In addition to fracture,
    void nucleation may also occur due to  PT-related mechanisms \cite{Levitas-Alt-PRL-08,Levitas-Alt-VM-PRB-09,Levitas-sublim-IJP-12,Levitas-Alt-sublim-IJP-12}, namely
  due to direct sublimation  (i.e., transformation of a critical volume of solid to gas), sublimation via virtual melting, and kinetic   melting and evaporation of a stable liquid. Fig. \ref{fig.8} from \cite{Levitas-Alt-void-ActaMat-11} shows a complete temperature-tensile stress kinetic diagram for void nucleation due to different processes.

\begin{figure}
\includegraphics[width=160mm]{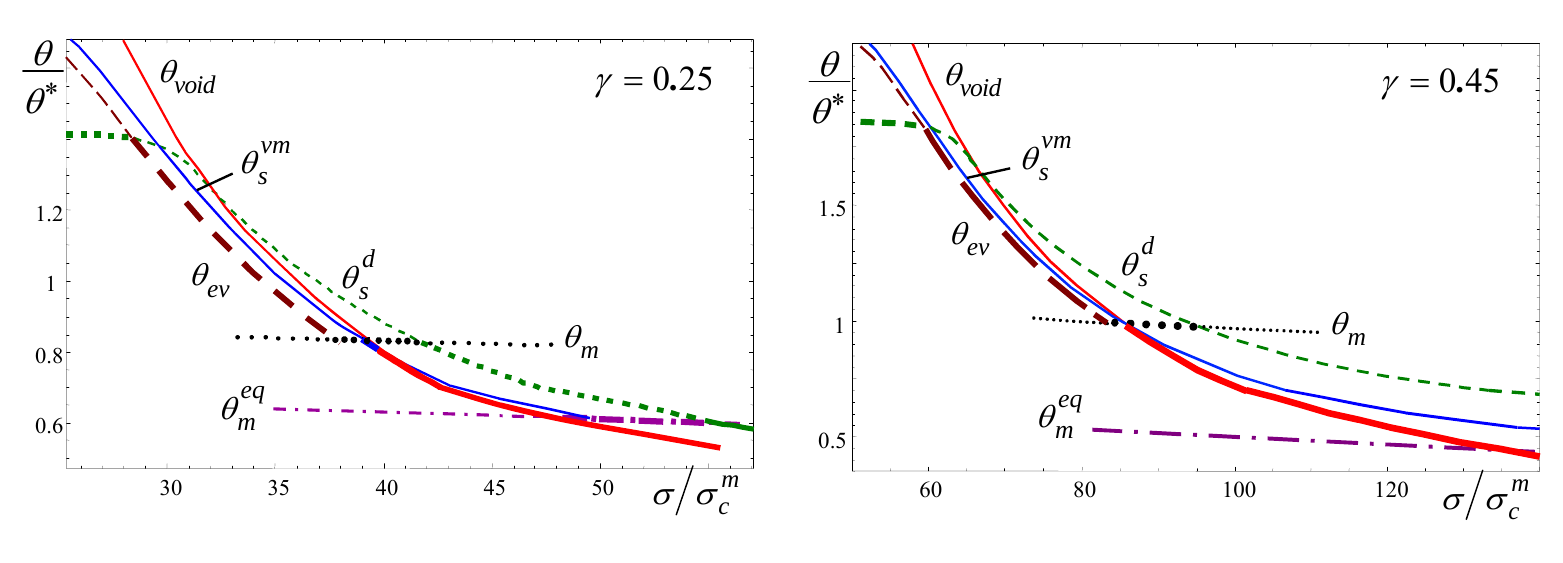}
\caption{Kinetic dimensionless temperature-tensile stress criteria for the nucleation of a spherical void
due to fracture, sublimation, sublimation via virtual melting, and evaporation of a liquid drop
within an elastoplastic solid for two values of a specific surface energy $\gamma$ (in $J/m^2$).
Temperature is normalized by the sublimation temperature $\theta^*$ for  $\sg=0$ and neglected surface energy,
and stress is normalized by a cavitation stress at melting temperature $\sg_c^m$.
The following designations of the curves are utilised: for fracture $\theta_{void}$ (red solid line), for direct
sublimation $\theta_s^d$ (green dashed line), for sublimation via virtual melting $\theta_s^{vm}$
 (blue solid line),
for kinetic melting $\theta_m$ (black dotted line), for  evaporation of melt $\theta_{ev}$ (brown dash-style line), and for
thermodynamic equilibrium melting $\theta_m^{eq}$
 (magenta dash-dot-style line). Bold parts of curves
correspond to the lowest temperature for a given tensile stress at which critical void can
appear. With increasing tensile stress, they correspond first to direct sublimation, then to
evaporation of liquid, sublimation via virtual melting (for
$\gamma = 0.25J/m^2$ only), and fracture due to bond breaking. Reproduced with permission from \cite{Levitas-Alt-void-ActaMat-11}.}
\label{fig.8}
\end{figure}

\subsection{Phase field approach to fracture}\label{PFA-fracture}

It is clear from solutions in Section \ref{Sec-fracture} that search for crack trajectory in elastoplastic materials based on the presented sharp interface theory and extremum principle is very computationally expensive. Similar to PT, PFA with finite-width interface is much more effective for this purpose.  The PFA has been widely used for the simulation of fracture \cite{Aransonetal-00,Karmaetal-01,Henry-Levine-04,Hakim-Karma-05,Hakim-Karma-09,Spatscheketal-06,Kuhn-Muller-10,Mieheetal-10,Bourdinetal-00,Bourdinetal-11,Bordenetal-14,Ambatietal-15,Dudaetal-15,Wu-17} and  damage  \cite{Mozaffari-Voyiadjis-15,Mozaffari-Voyiadjis-16,Wu-17,Voyiadjis-Song-19}.

The order parameter $\phi$ in the majority of PFAs to fracture characterizes atomic bond breaking in a solid and separates a sample into the   undamaged solid ($\phi=0$), fully broken atomic bonds ($\phi=1$), and the finite-width crack surface, in which the material is partially broken ($0<\phi<1$). The evolution of the damage order parameter occurs mainly in the crack tip zone and is described as a solution of  the corresponding Ginzburg-Landau equation. The Ginzburg-Landau equation coupled with mechanics, i.e. kinematics, constitutive rules, and equilibrium/motion equations, are used to obtain the evolution of the order parameter and stress-strain fields.
All PFAs to fracture can be divided into two groups:  with a double-well potential, like for PTs,   and  with a single-well potential.

{\it PFAs with double-well potential}. The first PFA for crack propagation utilizing  the Ginzburg-Landau equation for PT was presented in \cite{Aransonetal-00} with  the  concentration of point defects as the order parameter. The coupled Ginzburg-Landau and elastodynamic equations were utilized for finding  the evolution of the order parameter  and  displacement fields  for mode I crack propagation. The KKL (Karma-Kessler-Levine) PFA for mode III of fracture \cite{Karmaetal-01} was similar to the conventional PFA to dendritic solidification, i.e. to PT. The double-well  energy barrier between the gaseous and the solid states mimiced the fracture energy. This model overcame some limitations in \cite{Aransonetal-00}, which could not completely release  the nonphysical bulk stresses. The KKL model \cite{Karmaetal-01} has been generalized to modes I and II of crack propagation in \cite{Henry-Levine-04}  and for description of  quasi-static crack growth paths in elastically anisotropic materials in \cite{Hakim-Karma-05,Hakim-Karma-09}. In \cite{Hakim-Karma-09}, the double-well term was substituted with the critical elastic energy and 1D analytical solution was presented. Some drawbacks of the KKL model was discussed in \cite{Levitasetal-fracture-IJP-18} by analyzing the stress-strain curves for the homogenous states.
Some other  PFA approaches with double-well term were developed in \cite{Spatscheketal-06,Spatscheketal-07,Jafarzadehetal-CMT-20} and for modeling of damage in \cite{Mozaffari-Voyiadjis-15,Mozaffari-Voyiadjis-16}.

The PFAs with a double-well barrier treat fracture as a solid-gas transformation while it is an atomic bond breaking;
they lead to crack widening and lateral expansion during its propagation, see \cite{Levitasetal-cavitation-11,Bourdinetal-11}.

{\it PFAs with single-well potential} are not related to PTs, see \cite{Francfort-Marigo-98,Bourdinetal-00,Amoretal-09,Kuhn-Muller-10,Mieheetal-10,Bourdinetal-11,Bordenetal-14,Ambatietal-15,Levitasetal-fracture-IJP-18}. However, they do not possess sufficient degrees of freedom  to reproduce the complex stress-strain curves obtained, e.g., in atomic simulations. Requirements to interpolation functions based on the desired stress-strain curve were formulation and applied in \cite{Levitasetal-fracture-IJP-18}.

Besides the  works describing damage and the nonlinear stress-strain curves via weakening elastic moduli, in \cite{Jinetal-01,Wangetal-02,Levitasetal-cavitation-11,Jafarzadehetal-TAFM-20}  they are included using eigen strain. Some contradictions in these models were discussed in \cite{Levitasetal-fracture-IJP-18}. Surface stresses were introduced in PFA for fracture in \cite{Levitasetal-fracture-IJP-18}. This model also includes scale-dependency and is applicable from the atomistic to the macroscopic scales.

There is significant literature on combining PFA to fracture with plasticity for description of  phenomenological or dislocation plasticity, e.g., \cite{Ambatietal-15,Ambatietal-16,Dudaetal-15,Mieheetal-16a,Mieheetal-16b,Mieheetal-16c,Mieheetal-17,Ruffini-Finel-15,Spatscheketal-07,Dittmannetal-18,Voyiadjis-Song-19,Mozaffari-Voyiadjis-16}   or twinning \cite{Clayton-Knap-13,Clayton-Knap-16}, which requires separate detailed consideration. Detailed review on strain gradient enhanced  plasticity and damage theories in presented in \cite{Voyiadjis-Song-19}.

\subsection{Phase field approach to interaction between phase transformation and fracture}\label{PFA-PT-fracture}

\begin{figure}[htp]
\centering
\includegraphics[width=0.4\textwidth]{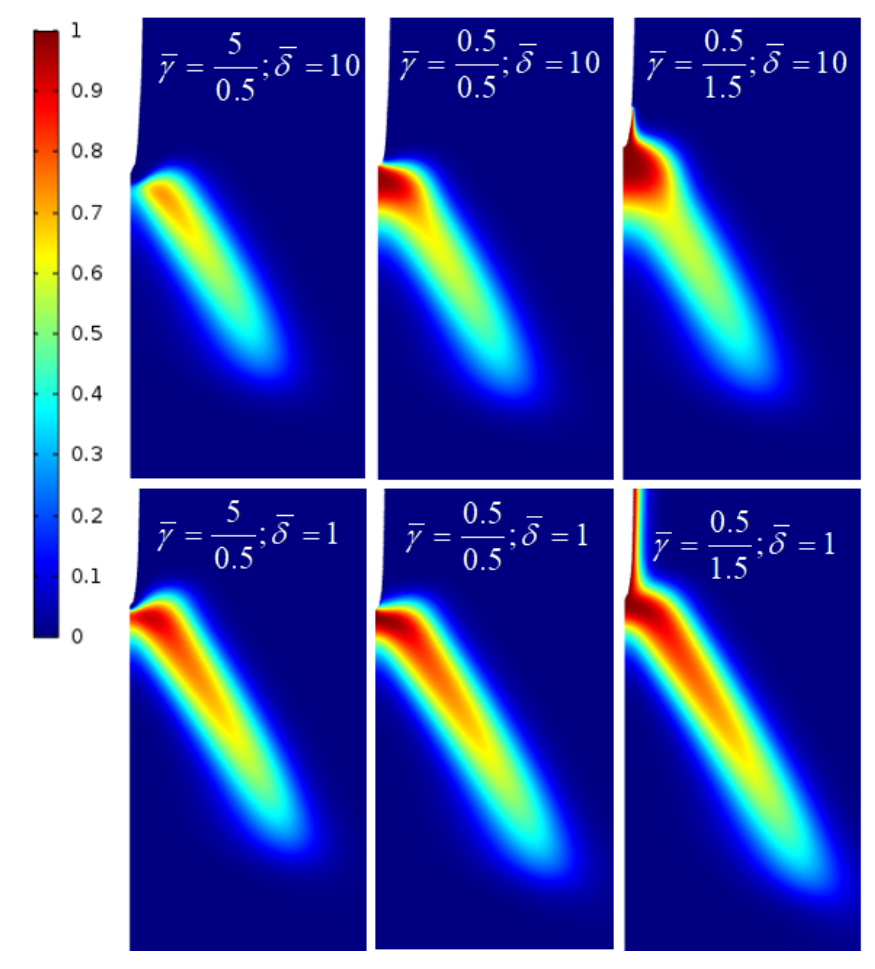}
\caption{ The distribution of the order parameter describing PT ahead of the moving crack tip  for different $\bar\gamma=\gamma_M/ \gamma_A$  and $\bar\delta=\delta_c/ \delta_p$  (shown in figures) for the pseudoelastic regime. Reproduced with permission from \cite{Jafarzadeh-Levitas-etal-Nanosclae-19}.
 \label{PFA-PT-crack}}
\end{figure}
Interaction between fracture and martensitic PTs is a very  important problem in the mechanics and physics of strength, transformational, and deformational materials properties. For example, the strong stress concentrator at the crack tip may cause nucleation of the product phase and initiate PTs \cite{Stump-Budiansky-89,Bulbich-92,Hangen-Sauthoff-99,Creuzigeretal-08,Gollerthanetal-09,Mamivandetal-14}. PT dissipates  energy and  produces transformation strain  as a mechanism of inelastic deformation and elastic stress relaxation. This increases the resistance to the crack propagation and material ductility, which is coined the transformation toughening. Besides, stresses produced in the course of a PT may cause fracture, in particular during cyclic loading of SMAs \cite{Hornbogen-91}.   PFA has been widely used for modeling  fracture (see Section \ref{PFA-fracture}) and  PTs (see Sections \ref{PFA-semicoherent}, \ref{PFA-interface-dislocations} and \ref{Sec-PT-plast}), and their interaction \cite{Mamivandetal-14,Bulbich-92,Boulbitch-Korzhenevskii-16,Boulbitchetal-17,Schmittetal-15,Zhaoetal-16,Clayton-19}. However, only several papers \cite{Schmittetal-15,Zhaoetal-16,Clayton-19,Jafarzadeh-Levitas-etal-Nanosclae-19,Karmaetal-NPJCM-20} study both fracture and PT within the PFA. The most advanced PFA \cite{Jafarzadeh-Levitas-etal-Nanosclae-19} is based on combination of PFAs to PT  from \cite{levitas+preston-II},  to fracture from \cite{Levitasetal-fracture-IJP-18}, but most important, to surface-induced pre-transformations and transformations \cite{Levitas-PRL-11,Levitas-Samani-14}, which leads to   new nanoscale effects. In the PFA  to surface-induced pre-transformations and transformations, numerous effects were found
after transition from sharp external surface \cite{Levitas-PRL-10,Levitas-Samani-11} to a finite-width external surface described  with a separate order parameter in \cite{Levitas-PRL-11,Levitas-Samani-14}. In PFA to fracture, free surfaces of finite width appear naturally,
which makes it natural to integrate PFA to PT and fracture with PFA to surface-induced PT with a finite-width external surface  in \cite{Levitas-PRL-11,Levitas-Samani-14}.
  Such a  theory possesses two characteristic nanoscale parameters: widths of the crack surface $\delta_c$ and the $A-M$ interface width $ \delta_p$. Then the dimensionless scale parameter $\bar\delta=\delta_c/ \delta_p$
  significantly affects PT and fracture, similar to other PFAs with two scale parameters,
  see Section \ref{PFA-interface-dislocations}  and \cite{Levitas-Javanbakht-PRB-12,levitas-javanbakht-APL-13},   Section \ref{virtual-melt}  and  \cite{Momeni-Levitas-PRB-14,Levitas-Momeni-ActaMat-14,Momeni-Levitas-15,Momeni-Levitas-PCCP-16,levitas-roy-ActaMat-16}, as well as review \cite{Levitas-ScriptaMat-18}.
  It was found  that the lower surface energy of $M$ than that of $A$ (i.e.,   $\bar\gamma=\gamma_M/ \gamma_A<1$) promotes nucleation of $M$ at the crack tip, its stabilization at the crack surface as a nanolayer ("wetting" by martensite), as well as nucleation of the pre-martensite or $M$ at the crack surfaces, even in the pseudoelastic regime, when stress release near the crack surface has to lead to the reverse PT (Fig. \ref{PFA-PT-crack}). In the opposite case, growth in the surface energy during PT inhibits the PT near the crack tip and at the crack surfaces, and
 displaces $M $  away from the crack tip  in the pseudoelastic regime, and leads to reverse PT   at the crack tip in the pseudoplastic regime.

   From the other side, different surface energies of $A$ and $M$  influence the crack evolution through the change in cohesion and gradient energies, which affects crack nucleation location and propagation trajectory
   (branching) and the process of interfacial damage evolution (Fig. \ref{PFA-crack-interface}), as well as transformation toughening. All these variations are significantly affected by the dimensionless width $\bar\delta$ and  the surface energy  $\bar\gamma$.
 Consequently,   these are two new parameters controlling coupled fracture and PTs.

\begin{figure}[htp]
\centering
\includegraphics[width=0.4\textwidth]{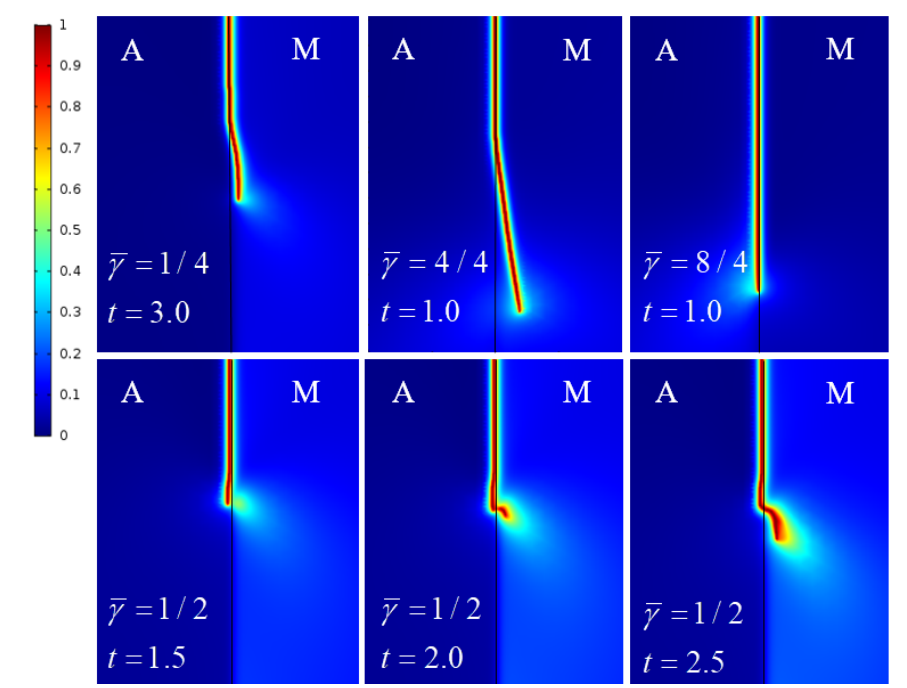}
\caption{ The field  of the order parameter describing damage distribution  within and outside the $A-M$ phase interface for $\bar\delta=1$ and different $\bar\gamma$ and time $t$ shown in figures. Reproduced with permission from \cite{Jafarzadeh-Levitas-etal-Nanosclae-19}.
\label{PFA-crack-interface}}
\end{figure}

PT at the preexisting void at finite strain was studied in \cite{levitasetal-prl-09,Levin-Levitas-IJSS-13,Javanbakhtetal-void-PT-MMS-20}.

\section{Concluding remarks}\label{Concl-rem}

The presented review focuses on the various fundamental problems of PTs, CRs, and fracture in inelastic materials, which the author works last 30  years at.
Within sharp interface approach, explicit expression for the thermodynamic driving force for SCs within a finite region in an inelastic material as well as extremum principle for finding all unknown parameters (position, shape, orientation of the transformed region, i.e., position of phase interfaces) are derived and discussed in Section \ref{general-theory}.
When a finite region represents volume covered by a moving interface during infinitesimal time increment, theory describes interface propagation in inelastic materials. This theory represents a nontrivial conceptual generalization of the theory for SCs in elastic materials. For elastic materials (see Section \ref{PT-elastic}),   the thermodynamic driving force for SCs is the gain in the Gibbs energy and all unknown parameters are determined by minimization of the Gibbs energy. While there were attempts to apply slightly modified Gibbs-energy-based theory to inelastic material, this is contradictory.
It was demonstrated in  Section \ref{general-theory} that  the thermodynamic driving force for SCs  in an inelastic material
is the dissipation increment for the complete SC in the region under study due to SC only, i.e., the difference between
the total dissipation increment and the dissipation increment due to all other dissipative processes but
SC (e.g., plastic flow,  evolution of internal variables, etc.).  Instead  of a surface-independent
Eshelby integral in the theory of defects and path-independent
J-integral in fracture mechanics, we introduced a region-independent integral for arbitrary inelastic materials.
For an interface propagation, the driving force is not the Eshelby driving force but difference between the Eshelby driving force and dissipation due to other processes than SCs, in particularly, plasticity.

Special attention is devoted to kinetics of SCs, and three type of kinetics is considered: athermal or rate-independent kinetics, thermally-activated kinetics of appearance of the critical nucleus, and "macroscale" thermally-activated kinetics, for which time for the appearance of a macroscale
region is postulated. For each type of kinetics, the kinetic equation and extremum principle for determination of all unknown parameters are formulated. The athermal kinetics is the most close to the traditional approach to SCs in elastic materials:
when all types of dissipation are neglected, the theory reduces to the  Gibbs-energy-based theory for elastic materials.
However, for inelastic materials, two solutions are always possible: with and without SCs. It was suggested to choose an actual solution with the help of the extremum principle for choosing the stable solution, which represents the global SC criterion. This additional principle does not have counterparts in elastic material, because for elastic materials solution corresponding to the minimum of the Gibbs energy is the actual one.

For other two cases, the Arrhenius-type kinetics for the time of SCs is utilized, and all unknown parameters are
determined by the principle of the minimum of transformation time or its particular cases (e.g., the principle of minimum of activation energy or transformation mass (volume)). Since the  Gibbs energy is not a driving force for inelastic materials, new definition of the activation energy was suggested. Again, kinetic approaches, when all types of dissipation are absent, reduce to kinetics of SCs in elastic materials. Since this kinetics was not broadly used in continuum approaches, it was presented in Section \ref{PT-elastic} as well. All extremum principles in this theory were derived using the postulate of realizability, presented in Section \ref{post-realiz} for athermal kinetics and Section \ref{macro-act-kin} for a macroscale thermally activated
 kinetics. In particular, the principle of the minimum of transformation time, which follows from the postulate of realizability, was intuitively used in material science and physical literature.

 To illustrate the theory, various problems were solved analytically (even for large strains) or numerically in Sections   \ref{sphere} - \ref{interface-numerics}, \ref{Shear-band-int}, and \ref{sec-lath} for PTs and CRs, and in Section \ref{Sec-fracture}
for fracture and interaction between fracture and PT.
Nontrivial points were related to various conditions at the phase interface (coherence, with sliding and/or decohesion),
new approach to the incoherent phase interface,
inheritance of plastic deformation during PTs and its effect on PT, introducing RIP, deriving analytical expression for TRIP and RIP at the propagating interface and in shear band,  and martensite nucleation at the intersection of the shear bands.
Virtual or intermediate melting, much (from 100 to 5000 K) below the thermodynamic melting temperature as a new mechanism of plastic deformation and stress relaxation during various PTs and high strain rate loading was recently revealed and is discussed in Sections  \ref{virtual-melt} and \ref{VM-shock}.

Phase field approaches to PTs, twinning, dislocations and their interaction is presented in Section \ref{Sec-PT-plast};
PFA to fracture and interaction between fracture and PT is described in Section \ref{Sec-fracture}.
 PFAs  described SCs  in a continuous way by solving Ginzburg-Landau evolution
equations for the order parameters. Special attention was devoted to strict continuum thermomechanical treatment of the PFA, which include formulation of new conditions for interpolation functions for all material properties, and satisfaction of these conditions. In contrast to sharp-interface approach, interfaces and defects have a finite width; there are no discontinuities and
no needs to satisfy jump conditions across interfaces and develop special numerical procedures to track interface and defect  motion. That is why very complex microstructure evolution (including splitting tips of martensitic variants, branching of cracks, and dendrite formation) can be reproduced by direct simulations without any a priory information. The
PFA also includes additional information about stability and instability of phases and different states.
All these items constitutes advantages of the PFA. Multiple  solutions to various PFA problems are presented and new effects are revealed.
At the same time,
the sharp-interface approach gives specific expressions for the thermodynamic driving forces for nucleation and evolution of defects. It is convenient for solution of problems with relatively simple geometry of interfaces and defects, and allows analytical solutions for some problems.

PTs and CRs induced by large plastic shear under high pressure are reviewed in Section \ref{RDAC}.
It includes new material phenomena and four-scale approaches for their analyses, from atomistic studies
to nano- and scale-free PFAs to microscale and macroscale modeling. This method of plastic treatment has strong potential for technological applications since plastic shear drastically (up to one to two order of magnitude) reduces
PT and CR pressure and lead to new phases and reaction products that are not accessible under hydrostatic conditions.

Due to multifaceted physics and mechanics  of the interaction between PTs, CRs, and fracture with plasticity, many aspects were
not covered properly or covered at all. In particular, such an interaction at the crystal plasticity and macroscopic levels, while shortly discussed in Section \ref{phenomenology}, requires much more attention. Sharp-interface approach to fracture in elastoplastic materials is covered as an illustration of our general thermodynamic and kinetic approaches only,
while many other approaches are developed in literature.
Experimental, especially modern in situ methods, at different scales, should be discussed separately.
There is also huge literature on lithium-ion batteries, in particular, silicon based, which includes large elastoplasticity
 of Si during lithiation/delithiation and corresponding chemical reactions.

 Most of the problems discussed in the review are far from being fully resolved; many of them are in their infancy.
Complexity and multidisciplinary and multifaceted character of the problems on interaction between various SCs and plasticity as well as their applied significance will definitely attract a lot of attention from researches from various disciplines working at multiple scales.
\\
\\
\noindent {\large\bf Acknowledgements}
\\
\par
The author worked for more than three decades on various aspects of the topic of the current review.
This work was initiated in 1988 at the Institute for Superhard Materials of the Ukrainian Academy of Sciences, Kiev, Ukraine with support from the  Ukrainian Academy of Sciences. Special thanks is to Prof. Nikolay V. Novikov, director of the Institute,
supervisor of the author's PhD work (1978-1981), who played very important role in formation of the author's interests, and with whom authors continued working till 1993 and closely communicating after.
This work was continued during 1992-1999 at the University of Hannover (Germany) with support from the Alexander von Humboldt Foundation, Volkswagen Foundation, and German Research Society.
In the USA, at Texas Tech University (1999-2008) and currently at Iowa State University, this work was supported by
National Science Foundation, Army Research Office, Office of Naval Research, Defense Advanced Research Projects Agency, Defense Threat Reduction Agency, Los Alamos National Laboratory, National Institute of Standards and Technology, Geophysical Laboratory of  Carnegie
Institution of Washington,  and Air Force Office of Scientific Research, as well as Schafer 2050 Challenge Professorship (2008-2017) and Vance Coffman Faculty Chair Professor in Aerospace Engineering (2017- ). Current support used for writing this review was obtained from
NSF (CMMI-1943710 and DMR-1904830), ONR
(N00014-19-1-2082), ARO (W911NF-17-1-0225), and Iowa State University (Vance Coffman
Faculty Chair Professorship).
The author greatly appreciates and enjoyed collaboration with all his coauthors of papers cited here, including graduate students, post docs, and faculty and researchers from different organizations.

\renewcommand{\baselinestretch}{1.1}

\newpage


\end{document}